\documentclass{article}

\usepackage[final,nonatbib]{neurips_2022}

\usepackage[utf8]{inputenc} % allow utf-8 input
\usepackage[T1]{fontenc}    % use 8-bit T1 fonts
\usepackage{url}            % simple URL typesetting
\usepackage{booktabs}       % professional-quality tables
\usepackage{amsfonts}       % blackboard math symbols
\usepackage{nicefrac}       % compact symbols for 1/2, etc.
\usepackage{microtype}      % microtypography
\usepackage{tabu}
\usepackage{graphicx, amsmath, amssymb, caption, subcaption, multirow, overpic, textpos}
\usepackage{wrapfig}
\usepackage{tabulary}
\usepackage[table]{xcolor}
\usepackage{tablefootnote}
\usepackage{afterpage}
\usepackage{xspace}
\makeatletter
\DeclareRobustCommand\onedot{\futurelet\@let@token\@onedot}
\def\@onedot{\ifx\@let@token.\else.\null\fi\xspace}
\def\eg{\emph{e.g}\onedot} 
\def\ie{\emph{i.e}\onedot} 
 
\def\etc{\emph{etc}\onedot}

\makeatother
\newcommand{\app}{\raise.17ex\hbox{$\scriptstyle\sim$}}

\newlength\savewidth
\newcommand{\tablestyle}[2]{\setlength{\tabcolsep}{#1}\renewcommand{\arraystretch}{#2}\centering\footnotesize}
\renewcommand{\paragraph}[1]{\vspace{1.25mm}\noindent\textbf{#1}}

\newcolumntype{x}[1]{>{\centering\arraybackslash}p{#1pt}}
\newcolumntype{y}[1]{>{\raggedright\arraybackslash}p{#1pt}}
\newcolumntype{z}[1]{>{\raggedleft\arraybackslash}p{#1pt}}

\definecolor{defaultcolor}{gray}{0.9}
\definecolor{citecolor}{HTML}{0071BC}
\definecolor{linkcolor}{HTML}{ED1C24}
\definecolor{hrefcolor}{rgb}{0.93, 0.0, 0.55}
\usepackage[pagebackref=false, breaklinks=true, letterpaper=true, colorlinks, citecolor=citecolor, linkcolor=linkcolor, bookmarks=false]{hyperref}

\usepackage{tikz}
\def\checkmark{\tikz\fill[scale=0.4](0,.35) -- (.25,0) -- (1,.7) -- (.25,.15) -- cycle;}

\title{Masked Autoencoders that Listen}

\author{
Po-Yao Huang\textsuperscript{$1$} \quad Hu Xu\textsuperscript{$1$} \quad Juncheng Li\textsuperscript{$2$} \quad Alexei Baevski\textsuperscript{$1$}   \\ \textbf{Michael Auli\textsuperscript{$1$}} \quad \textbf{Wojciech Galuba\textsuperscript{$1$}} \quad \textbf{Florian Metze\textsuperscript{$1$}} \quad \textbf{Christoph Feichtenhofer\textsuperscript{$1$}}
\\
\\
\textsuperscript{1}Meta AI\qquad \textsuperscript{2}Carnegie Mellon University \vspace{-1em}
}

\begin{document}

\maketitle

\begin{abstract}

\vspace{-0.8em}
This paper studies a simple extension of image-based Masked Autoencoders (MAE)~\cite{mae} to self-supervised representation learning from audio spectrograms.
Following the Transformer encoder-decoder design in MAE, our Audio-MAE first encodes audio spectrogram patches with a high masking ratio, feeding only the non-masked tokens through encoder layers.
The decoder then re-orders and decodes the encoded context padded with mask tokens, in order to reconstruct the input spectrogram.
We find it beneficial to incorporate local window attention in the decoder, as audio spectrograms are highly correlated in local time and frequency bands. 
We then fine-tune the encoder with a lower masking ratio on target datasets. Empirically, Audio-MAE sets new state-of-the-art performance on six audio and speech classification tasks, outperforming other recent models that use external supervised pre-training.
Our code and models is available at \url{https://github.com/facebookresearch/AudioMAE}.
\vspace{-0.8em}

\end{abstract}

%%%%%%%%%%%%%%%%%%%%%%
% \vspace{-0.75em}
\section{Introduction}
\vspace{-0.5em}
%% 1. The grand trend: Transformer and self supervised learning 
Transformers~\cite{Vaswani:2017:AYN:3295222.3295349} and self-supervised learning~\cite{bert,gpt,roberta,moco,mocov3,mae} are dominating computer vision (CV) and natural language processing (NLP) research. 
The revolution firstly started in NLP with the invention of the Transformer architecture and self-attention~\cite{paulus17intra}. 
Masked autoencoding with BERT~\cite{bert} set a new state-of-the-art on various NLP tasks by self-supervised pre-training on large-scale language corpus.
Similarly in the CV community, Vision Transformers (ViT)~\cite{dosovitskiy2020image} have become popular for CV tasks, and, for self-supervised image representation learning, Masked Autoencoders (MAE)~\cite{mae} have brought the CV community closer to the success of BERT in NLP.
In addition to the existing masked autoencoders that can read (BERT) or see (MAE), in this work we study those that can \textit{listen}.

%% 2. Vision to audio
Transformer-based models have recently refreshed leaderboards for audio understanding tasks.
For example, AST~\cite{gong2021ast} and MBT~\cite{Nagrani21c} improved the audio classification performance on the  AudioSet~\cite{gemmeke2017audio}, Event Sound Classification~\cite{piczak2015dataset}, etc.
The key technique behind this is initialization of audio model weights with ImageNet pre-trained supervised models (\eg, DeiT~\cite{touvron2021training})
by deflating patch embeddings and interpolating positional embeddings for encoding audio spectrograms.
%% 3. Cons of audio models with ImgNet-PT
However, exploiting ImageNet pre-trained models could be sub-optimal.
% 3.a Con 1: heterogeneity
Unlike initializing video models with weights from image models (\eg, the initial weights of I3D~\cite{i3d} or 3D-ResNets~\cite{feichtenhofer2016spatiotemporal} are inflated from ImageNet pre-trained image models), there are clear and notable discrepancies between spectrograms representing audio content and natural images.
It remains unclear why such heterogeneous image-to-audio transfer is useful beyond arguably similar low-level semantics such as shapes of spectrograms and shapes of visual objects. 
% 3.b Con 2: dependency to labeled data and thus non-scalable
Further, any label bias would inevitably be transferred to audio models.

%%% B. Audio-SSL and challenge
%% 1. Why SSL: large-scale training as no labeled data required
Addressing these concerns, self-supervised audio representation learning has recently attracted much research attention. 
Based on BEiT~\cite{beit} that learns to reconstruct image patches or learnt patch tokens, SS-AST~\cite{ssast} extends to the audio domain and exploits spectrograms (akin to 1-channel 2D images) and use both contrastive and reconstruction objective as self-supervision. 
%% 3. Issue of SSL: large-scale data and need efficient training with (efficient Transformer)
Without using any labels, the key enabler to effective self-supervised representation learning is large-scale pre-training data.
In this work we use AudioSet~\cite{gemmeke2017audio} for pre-training, a common dataset containing $\app$2 million audio recordings.
%%% C. Efficient self-supervised learning with Transformers
%% 1. N^2 complexity, challenging with large scale data
Performing large-scale training with Transformer architectures is challenging % with their massive computational and memory overhead.
as self-attention in Transformers has quadratic complexity w.r.t. the length of input sequence.
%% 2. possible solutions

This computational burden has been addressed in different ways. A popular approach is to reduce the sequence length in self-attention. 
Various ViT-based architectures have been developed to alleviate such issues for image and video understanding.
%% 2.a swin window for localized attention
For example, Swin-Transformer~\cite{swin} only performs local attention within windows that shift across layers.
%% 2.b MViT for pooling and hierarchy
MViT~\cite{maskedfeat} employs pooling attention to construct a hierarchy of Transformers where sequence lengths are downsampled.
%% 2.c MAE for image
For self-supervised learning, MAE~\cite{mae} efficiently encodes only a small portion (25\%) of visual patches while the majority of patches is discarded.
%% 3 Intuition why MAE-like models
The simplicity and scalability in MAE make it a promising framework for large-scale self-supervised learning.

\begin{figure*}[t]
    % \vspace{-1em}
    \centering
    \includegraphics[width=1.0\linewidth]{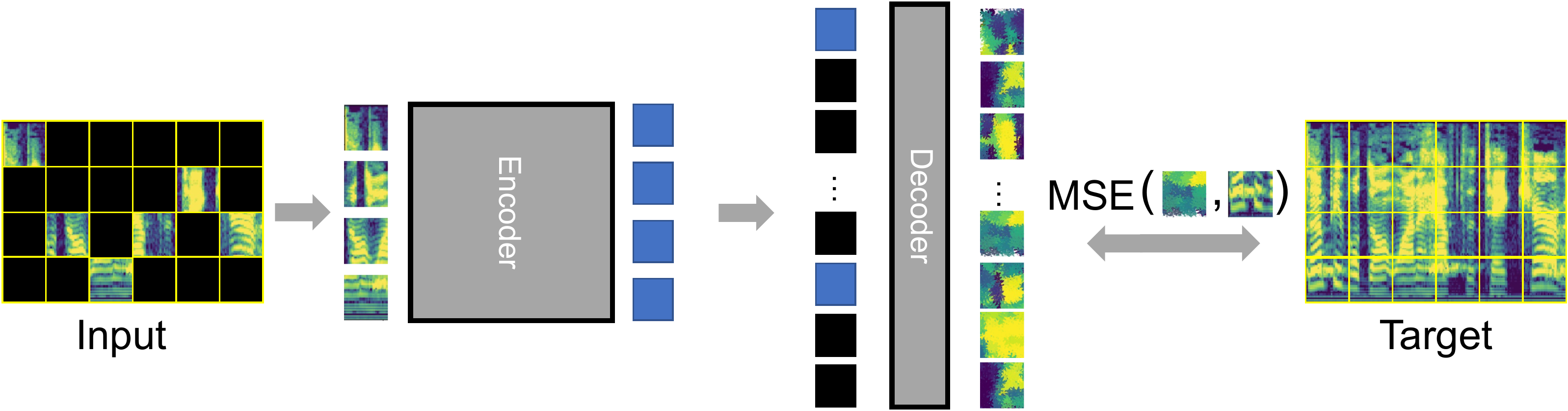}
    \caption{\textbf{Audio-MAE for audio self-supervised learning}. An audio recording is first transformed into a spectrogram and split into patches. 
    We embed patches and mask out a large subset (80\%).
    An encoder then operates on the visible (20\%) patch embeddings.
    Finally, a decoder processes the order-restored embeddings and mask tokens to reconstruct the input.
    Audio-MAE is minimizing the mean square error (MSE) on the masked portion of the reconstruction and the input spectrogram.
    }
    \vspace{-1em}
    \label{fig:amae}
\end{figure*}

%%% D. Our work: Audio-MAE for audio SSL
In this work, we study MAE for sound recognition and the unique challenges of the audio domain.
%% 1. High-level of PT-FT with Audio-MAE
We present Audio-MAE (Fig.~\ref{fig:amae}) as unified and scalable framework for learning self-supervised audio representations.
%% 2. Details of Audio-MAE
Similar to MAE, it is composed of a pair of a Transformer encoder and decoder.
%% 2.a input
Sound is first transformed and embedded into spectrogram patches.
%% 2.b encoder
Before feeding them into the Transformer encoder, we mask and discard the majority and only feed a small number of non-masked embeddings into the encoder for efficient encoding.
%% 2.c decoder
After padding encoded patches with learnable embeddings to represent masked patches, it then restores the order of these patches in frequency and time and propagates them through a Transformer decoder to reconstruct the audio spectrogram. 

Different from image patches, spectrogram patches are comparably local-correlated.
For example, formants, the vocal tract resonances, are typically grouped and continuous locally in the spectrogram.
The location in frequency and time embeds essential information that determines the semantics of a spectrogram patch and how it sounds like.
To this end, we further investigate using localized attention and a hybrid architecture in the Transformer decoder to properly decode for reconstruction.
This simple-yet-effective upgrade leads to improved performance for Audio-MAE.

%% 2.d objective
Similar to MAE for images, we minimize the patch-normalized mean square error.
At the fine-tuning stage, we discard the decoder and fine-tune the encoder with patch-masking.
%% 3. Summary of experiments
Empirically, Audio-MAE sets a new state-of-the-art performance on six audio and speech classification tasks. 
It is the first audio-only self-supervised model that achieves state-of-the-art mAP on AudioSet-2M, outperforming other recent models with external supervision.
We further provide the visualization and audible examples to qualitatively demonstrate the effectiveness of the Audio-MAE decoder.

%%%%%%%%%%%%%%%%%%%%%
%\vspace{-0.5em}
\section{Related Work}
%\vspace{-0.5em}

\paragraph{Visual masked pre-training.}
Masked/Denoising autoencoders~\cite{dae,sae,bert} are 
a general representation learning methodology by 
reconstructing source from masked or corrupted inputs.
In CV, visual masked pre-training has made recent progress~\cite{PathakKDDE16,chen2020,mae,maskedfeat}.
Based on ViT~\cite{dosovitskiy2020image} that applies Transformers to image patches, BEiT~\cite{beit} and MAE~\cite{mae} present masked image modeling frameworks.  BEiT~\cite{beit} learns to predict discrete visual tokens generated by VAE~\cite{dalle} in masked patches. 
MAE~\cite{mae} reduces  sequence length by masking a large portion of image patches randomly and encoding only non-masked ones for reconstruction of pixel color information.
MaskFeat~\cite{maskedfeat} studies features for masked pre-training and finds that Histograms of Oriented Gradients (HoG)~\cite{Dalal2005}, which are in turn related to spectrogram features, perform strongly for image and video classification models.
Our work extends the MAE framework for representation learning with audio spectrograms.

\paragraph{Out-of-domain pre-training for audio.}
Transferring ImageNet supervised pre-trained ViT~\cite{dosovitskiy2020image} or ResNet~\cite{He2015DeepRL} has become a popular practice for audio models ~\cite{gong2021ast,paast,Nagrani21c,chen2022hts,gong2021psla,cmkd}. 
After pre-training, these models operate over audio spectrograms by deflating from 3-channels (RGB) into 1-channel (spectrogram) in the pre-trained patch embedding in ViT and employing the rest of the transformer blocks on top. 
For example, 
HTS-AT~\cite{chen2022hts} encodes spectrograms with hierarchical Transformer initialized from the Swin Transformer~\cite{swin}.
MBT~\cite{Nagrani21c} uses ImageNet-21K pre-trained ViT; AST~\cite{gong2021ast} and PaSST~\cite{paast} employ DeiT~\cite{touvron2021training} as the Transformer backbone.
Without using out-of-domain (non-audio) data, the proposed Audio-MAE focuses on audio-only self-supervised pre-training from scratch.

%\vspace{-0.5em}
\paragraph{In-domain pre-training for audio.}
%% 2. Types of Audio-SSL: inputs: Contrastive vs Reconstruction; objectives: Time signal vs Spectrogram
Existing in-domain (\ie, audio-only) self-supervised methods can be broadly categorized by the input signal type (\eg, raw waveform~\cite{wav2vec,wav2vec2,d2v}, frame-level features~\cite{Hsu2021HuBERTSS,avhubert,srivastava2021conformer}, or spectrogram patches~\cite{ssast,baade2022}); and the objective used for self-supervision (\eg, contrastive~\cite{cpc,wav2vec2,object_sound,stica,Hsu2021HuBERTSS} or prediction/reconstruction~\cite{ssast,d2v,srivastava2021conformer,avhubert}).
% 2.a example
For example, wav2vec 2.0~\cite{wav2vec2} takes raw waveform as inputs and exploits contrastive learning to discriminate contextualized representations in different time segments.
Mockingjay~\cite{mockingjay} proposed a masked acoustic model pretext task to reconstruct frame-level Mel-features of masked time frames.
% 2.b example
SS-AST~\cite{ssast} is the closest work to Audio-MAE and is our main benchmark. 
Inspired by the success of BERT~\cite{bert}, SS-AST proposed a self-supervised learning method which operates over spectrogram patches and employs joint contrastive and reconstructive objectives on masked patches.
These previous methods generate audio representations by encoding full-view of both masked and non-masked time or spectrogram segments for self-supervised pre-training.
In contrast, Audio-MAE encodes only the non-masked spectrogram patches.

Our work is done independently and concurrently with~\cite{baade2022,chong2022masked,niizumi2022masked} related methods.
We also compare our model to these concurrent works in the experiments and showcase the superiority of Audio-MAE.

%%%%%%%%%%%%%%%%%%%%%
%\vspace{-0.2em}
\section{Audio Masked Autoencoders (Audio-MAE)}
%\vspace{-0.2em}

Audio-MAE is a conceptually simple extension of MAE to learn self-supervised representations from audio spectrograms. Fig.~\ref{fig:amae} depicts an overview. The details of each component are as follows.

\noindent \textbf{Spectrogram Patch Embeddings}. 
Following~\cite{gong2021ast,ssast}, we transform audio recordings into Mel-spectrograms and divide them into non-overlapped regular grid patches. These patches are then flattened and embedded by a linear projection.
Similar to MAE~\cite{mae}, 
we add fixed sinusoidal positional embeddings to the embedded patches.

\begin{figure}[t!h!]
    %\vspace{-0.5em}
    \centering
    \begin{subfigure}[b]{0.19\linewidth}
        \includegraphics[width=0.98\textwidth]{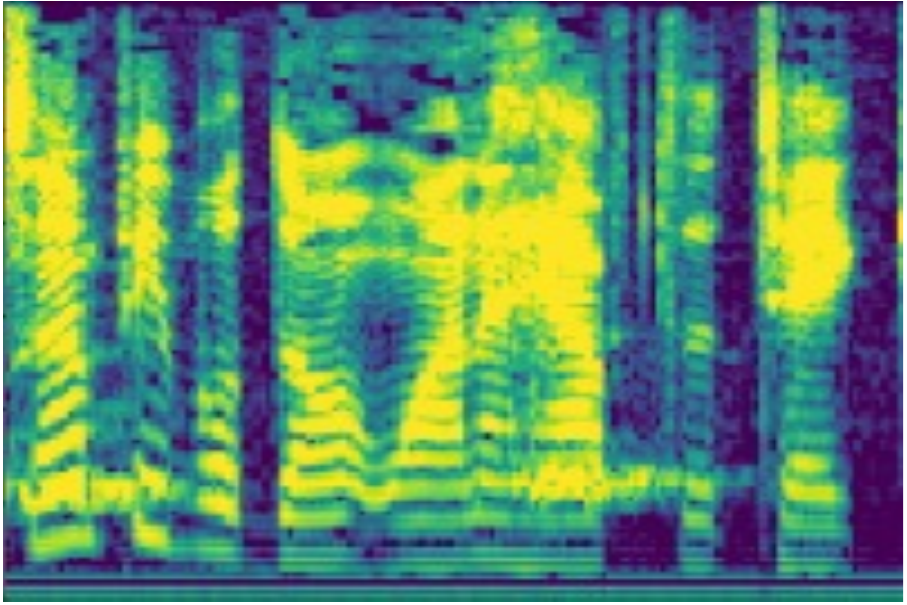}
        \caption{Original}
        \label{fig:masking:org}
    \end{subfigure}
    \begin{subfigure}[b]{0.19\linewidth}
        \includegraphics[width=0.98\textwidth]{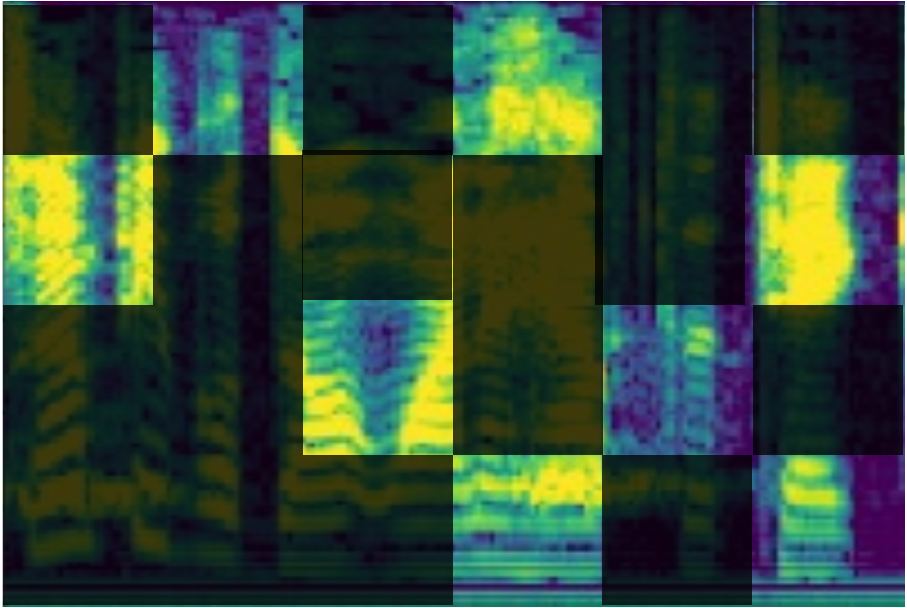}
        \caption{Unstructured}
        \label{fig:masking:random}
    \end{subfigure}
    \begin{subfigure}[b]{0.19\linewidth}
        \includegraphics[width=0.98\textwidth]{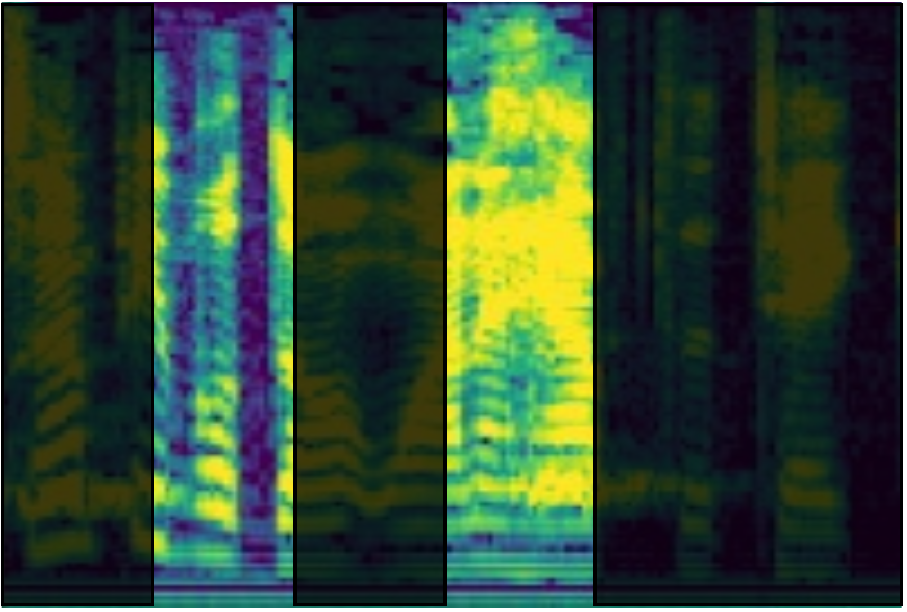}
        \caption{Time}
        \label{fig:masking:time}
    \end{subfigure}
    \begin{subfigure}[b]{0.19\linewidth}
        \includegraphics[width=0.98\textwidth]{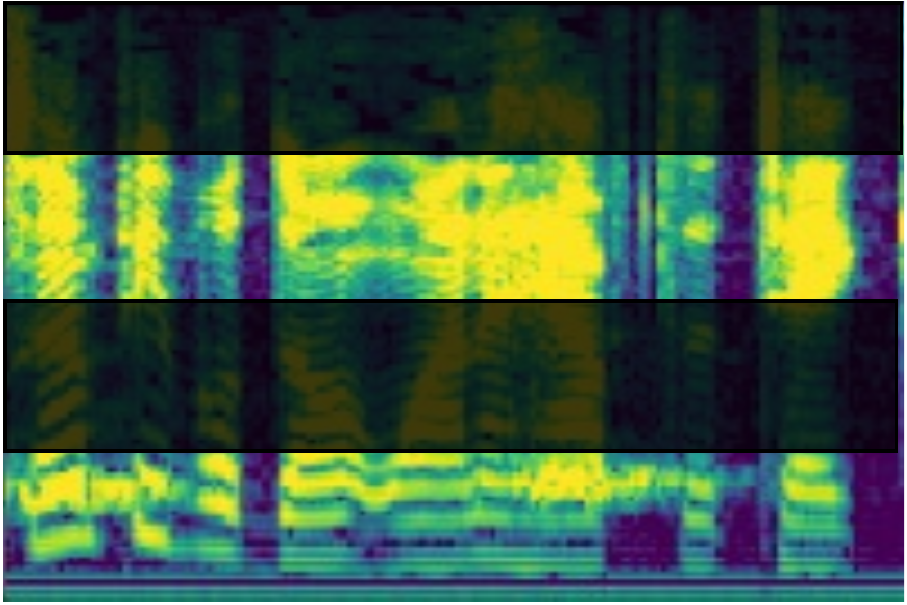}
        \caption{Frequency}
        \label{fig:masking:freq}
    \end{subfigure}
    \begin{subfigure}[b]{0.19\linewidth}
        \includegraphics[width=0.98\textwidth]{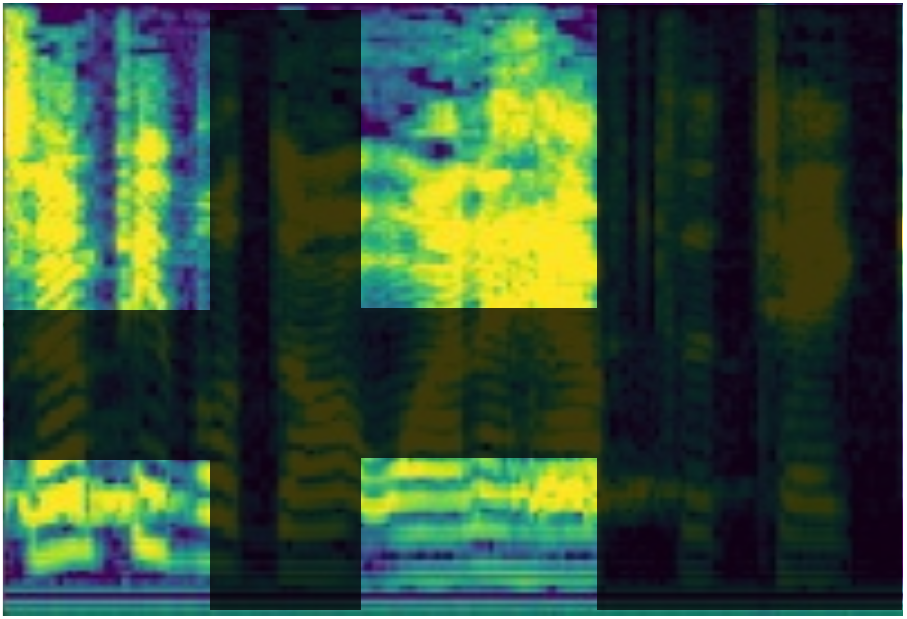}
        \caption{Time+frequency}
        \label{fig:masking:time_freq}
    \end{subfigure}
    %\vspace{-0.2em}
    \caption{Audio-MAE's masking strategies~\label{fig:masking} on Mel-spectrograms.
    }
    %\vspace{-0.5em}
\end{figure}

\noindent \textbf{Masking Strategies}.
Audio-MAE masks out a large subset of spectrogram patches.
As a spectrogram can be viewed as a 2D representation of time and frequency components of a sound, it is reasonable to explore treating time and frequency differently during masking. 
In this work, we investigate both the \textit{unstructured} (\ie, random masking without any prior) and \textit{structured} (\ie, randomly masking a portion of time, frequency, or time$+$frequency of a spectrogram) in the pre-training and fine-tuning phase.
Illustrative examples are shown in Fig.~\ref{fig:masking}. We show masked regions with dark overlay. 

The masking mechanism, as introduced in MAE~\cite{mae}, is the key ingredient for efficient self-supervised learning.
For a input patch sequence, this can be regarded as a Bernoulli process where each patch is masked/dropped with probability $p$ (masking ratio).
Masking reduces input patch sequence length and encourages learning global, contextualized representations from limited ``visible'' patches.
We observe that akin to images, a large masking rate (80\% in our experiments for spectrogram patches, which is similar to 75\% in MAE for images) is feasible for learning self-supervised audio representations. 
Unlike BERT~\cite{bert} that uses 15\% masking rate for self-supervised learning in NLP, most of the tokens/patches can be discarded for spectrograms as well as images due to high redundancy in these modalities.
Beyond self-supervised pre-training, we further explore the effectiveness of masking in the supervised fine-tuning stage. 
Empirically, we found unstructured (random) masking at a higher ratio for pre-training and structured (time$+$frequency masking) at a lower ratio for fine-tuning provide best accuracy (ablations are in~\S\ref{sec:ablations}).

\noindent \textbf{Encoder}.
Audio-MAE uses a stack of standard Transformers~\cite{Vaswani:2017:AYN:3295222.3295349} as its encoder. 
The encoder only processes (20\%) non-masked patches to reduce computation overhead which is quadratic to the input sequence length.
We use the 12-layer ViT-Base (ViT-B)~\cite{dosovitskiy2020image} Transformer as our default.

% remark: make sure terminology is consistent: window/local. 
\noindent \textbf{Decoder with Local Attention}.
The decoder is also composed of standard Transformer blocks.
The encoded patches from the encoder are padded with trainable masked tokens.
After restoring the original time-frequency order in the audio spectrogram, we add the decoder's (fixed sinusoidal) positional embeddings and feed the restored sequence into the decoder. 
At the top of the decoder stack, we add a linear head to predict and reconstruct the input spectrogram.

To address the unique characteristics of audio spectrograms, our work investigates an enhancement to the vanilla MAE decoder.
Image-based MAE uses \emph{global self-attention} in the Transformer decoder which is appropriate for visual context, because visual objects are typically invariant under translation or scaling, and their exact position may not affect the semantics of an image.
In contrast, the position, scale, and translation of spectrogram features however \textit{directly affects} the sound or semantics of an audio recording.
Consequently, global self-attention is sub-optimal for spectrograms if the time-frequency components is predominantly local.
For instance, 
we would have better success to use the harmonics (\eg, Fig.~\ref{fig:masking:org}) in lower bands of a vowel to predict the spectrogram patch vertically in a higher frequency band rather than horizontally in the time domain.
Similarly, a frictional sound of a consonant likely only correlates to other part of the consonant, and is without dependency to other silence segments in the audio recording.
Compared to images, the spectrogram patches are more similar to speech or text tokens where its order and position is more relevant.

\begin{wrapfigure}{tr}{5cm}
    \vspace{-0.5em}
    \centering
    \includegraphics[width=\linewidth]{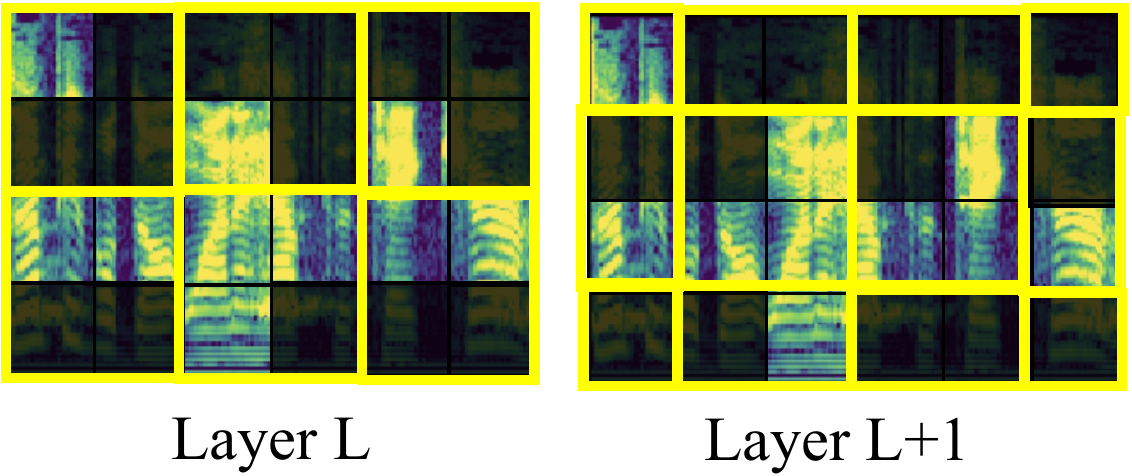}
    \caption{Decoder's local attention and shifted window (right). 
    }\label{fig:decoder_attn}
    \vspace{-0.5em}
\end{wrapfigure}

To address the nature of audio spectrograms, in addition to using Transformers with global self-attention as in vanilla MAE, we incorporate the \emph{local attention mechanism} which groups and separates the spectrogram patches in to local windows in self-attention for decoding.
We investigate two types of local attention:
(1) Shifted window location: Inspired by the shifted-window in Swin Transformers~\cite{swin}, we  shift window attention by 50\% between consecutive Transformer decoder layers. 
For padding the margin when shifting, we cyclically shift the spectrogram to the top-left direction. 
Fig.~\ref{fig:decoder_attn} illustrates the localized decoder attention by shifted windows.
(2) Hybrid window attention (global$+$local attention): Inspired by~\cite{li2021improved}, to add better cross-window connections, we design a simple hybrid (global+local) attention that computes local attention within a window in all but the last few top layers. In this way, the input feature maps for the final reconstruction layer also contain global information.
For simplicity, we use \textit{no} pooling or hierarchical structure. Decoders with different attention types are compared in~\S\ref{sec:ablations}.

\noindent \textbf{Objective}. 
The Audio-MAE decoder learns to reconstruct the input spectrogram by predicting the values in the spectrogram patches or their per-patch normalized ones. 
The objective is the mean squared error (MSE) between the prediction and the input spectrogram, averaged over unknown patches.
Empirically we found employing the reconstruction loss alone is sufficient while including additional contrastive objectives (\eg, InfoNCE loss~\cite{info_nce}) does not improve Audio-MAE.

\noindent \textbf{Fine-tuning for Downstream Tasks}. 
In the fine-tuning stage, we only keep and fine-tune the Audio-MAE encoder and discard the decoder.
Different from the original MAE, and inspired by~\cite{effective_ssl_speech,paast}, we also explore to employ masking in the fine-tuning stage to remove a portion of patches to further regularize learning from a limited view of spectrogram inputs, which, as a side effect, also reduces computation during fine-tuning. 
Compared to SpecAug~\cite{Park2019SpecAugmentAS} which takes full-length input with the masked portion set to zero as data augmentation, Audio-MAE sees only a subset of real-valued input patches without the nullified ones.
Audio-MAE then encodes these non-masked patches and applies an average pooling layer followed by a linear layer on top for fine-tuning in  classification tasks.

%\vspace{-0.75em}
\section{Experiments}
%\vspace{-0.75em}
We perform an extensive evaluation on six tasks, 
including audio classification on AudioSet (AS-2M, AS-20K) and Environmental Sound Classification (ESC-50), and speech classification on Speech Commands (SPC-1 and SPC-2) and VoxCeleb (SID).
We use AudioSet for ablation studies.

%\vspace{-0.75em}
\subsection{Datasets and Tasks\label{sec:exp:dataset}}
%\vspace{-0.75em}

\noindent \textbf{AudioSet}~\cite{gemmeke2017audio} (AS-2M, AS-20K) contains $\app$2 million 10-second YouTube clips for audio classification.
%, summing up to 5,800 hours of videos. 
527 types of audio events are weakly annotated~\cite{tagging_right,vggish,hershey2021benefit} for each clip. 
There could be multiple events in a clip.
The \emph{full} training set has 2 subsets: A class-wise \emph{balanced} (22,176 clips) and an \emph{unbalanced} (2,042,985 clips) set.
The \emph{eval} set has 20,383 clips. 
We downloaded and processed around 1.96M unbalanced training, 21K balanced training, and 19K evaluation clips. 

For the AS-2M experiments, we use the union of unbalanced and balanced training audio for pre-training and fine-tuning. 
For the AS-20K experiments, we use AS-2M for pre-training and the 20K balanced set for fine-tuning.
We report the testing mAP on the 19K \emph{eval} set used by AST~\cite{gong2021ast}.

\noindent \textbf{Environmental Sound Classification} 
(ESC-50)~\cite{piczak2015dataset} is an audio classification dataset consists of 2,000 5-second environmental sound recordings. There are 50 classes in ESC. We report accuracy under 5-fold cross-validation with the same split used by~\cite{gong2021ast}. 

\noindent \textbf{Speech Commands}
(SPC-2, SPC-1)~\cite{speechcommandsv2} are two keyword spotting tasks.
In SPC-2, there are 35 speech commands. 
The training/validation/testing set has 84,843/9,981/11,005 1-second recordings, respectively.
In SPC-1, there are 10 classes of keywords, 1 silence class, and 1 unknown class that includes all the other 20 common speech commands. We use the data and split provided in the SUPERB~\cite{yang21c_interspeech} benchmark to report the testing accuracy.
% The training set is 84,843 and testing and validation sets are with size 11,005, 9,981 respectively.

\noindent \textbf{VoxCeleb}
(SID)~\cite{Nagrani2020VoxcelebLS} contains 150K utterances from 1,251 speakers. The speaker identification task (SID) is to classify the utterances to identify its original speaker. We use the V1 standard train (138,361), validation (6,904), testing (8,251) sets and  report the testing accuracy.

%\vspace{-0.75em}
\subsection{Implementation Details\label{sec:exp:impl}}
%\vspace{-0.75em}

We use a vanilla 12-layer ViT-B by default as the Transformer encoder. 
For the decoder, we use a 16-layer Transformer with shifted local attention.
We investigate the vanilla (global attention) and hybrid (global$+$local attention) decoder variants (see Table.~\ref{tab:ablation:attn_type}).

Following~\cite{gong2021ast,Nagrani21c}, we transform raw waveform (pre-processed as mono channel under 16,000 sampling rate) into 128 Kaldi~\cite{povey2011kaldi}-compatible Mel-frequency bands with a 25ms Hanning window that shifts every 10 ms. 
For a 10-second recording in AudioSet, the resulting spectrogram is of 1$\times$1024$\times$128 dimension.

For patch embedding, we use convolutional kernels with $(16,16)$ size and stride in time and frequency (thus, patches are non-overlapping) to avoid short-cuts via overlap in self-supervision (though, at high masking ratios such short-cuts are less severe). 
By default, we use a masking ratio of $0.8$ with (unstructured) random masking for pre-training. 
During fine-tuning, we employ a lower masking ratio ($0.3$ in time and $0.3$ in frequency). Ablations on these design choices are given in~\S\ref{sec:ablations}.

%\vspace{-0.75em}
\subsection{Pre-training and Fine-tuning\label{sec:exp:setup}}
%\vspace{-0.75em}
We use AudioSet-2M for pre-training and randomly iterate over all audio recordings. 
We train for 32 epochs with a batch size of 512 and a 0.0002 learning rate. 
We distribute the training load over 64 V100 GPUs and the total training time is $\app$36 hours.
For each audio, we randomly sample the starting time, cyclically extract 10-second audio, and randomly jitter its magnitude by up to $\pm$ 6dB. 
We use only natural audio spectrograms and apply \emph{no} augmentations (\eg,~\cite{Park2019SpecAugmentAS,cutmix,mixup}) as we do not find these strong augmentations helpful in the pre-training phase.

In the fine-tuning phase, we remove the decoder and only fine-tune the encoder.
For the supervised fine-tuning on AudioSet-2M, since the size of training samples are uneven across classes (unbalanced), we follow the common practice of using a weighted sampling to balance the classes during training. 
In each epoch, we sample 200K instances ($\app$10\% of AudioSet-2M) without replacement. We  fine-tune for 100 epochs, which aggregate to $\app$10 full epochs of AudioSet-2M.
The probability of sampling an instance is inversely proportional to the dataset-wise occurrences of its classes. 
Fine-tuning on 64 GPUs takes $\app$12 hours.
For the smaller balanced AudioSet-20K, we fine-tune on 4 GPUs for 60 epochs without weighted sampling.
Please see Supplementary for the details on other datasets.

%\vspace{-0.75em}
\subsection{Ablations and Model Properties \label{sec:ablations}}
%\vspace{-0.75em}

\begin{figure}[t]
    %\vspace{-1.0em}
    \centering
    \begin{subfigure}[b]{0.48\linewidth}
        \includegraphics[width=0.98\textwidth]{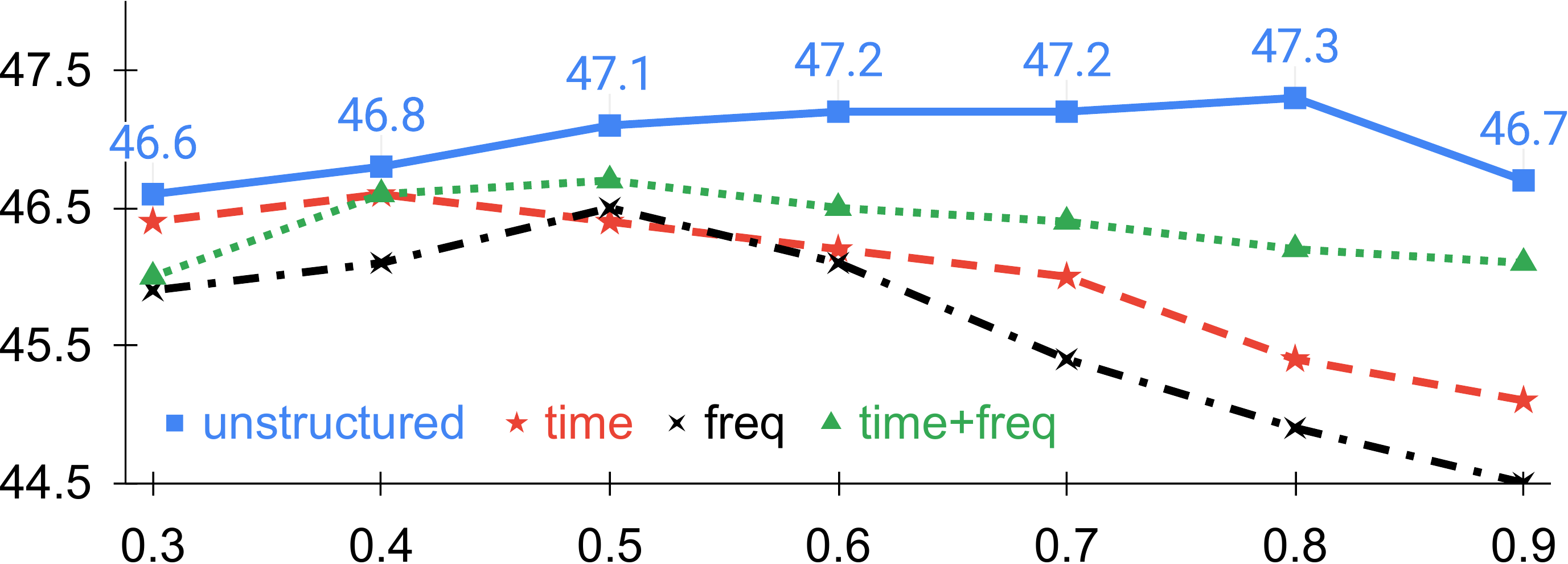}
        %\vspace{-0.2em}
        \caption{Pre-training masking}
        %\vspace{-0.2em}
        \label{fig:pt_masking_comparison}
    \end{subfigure}
    \begin{subfigure}[b]{0.48\linewidth}
        \includegraphics[width=0.98\textwidth]{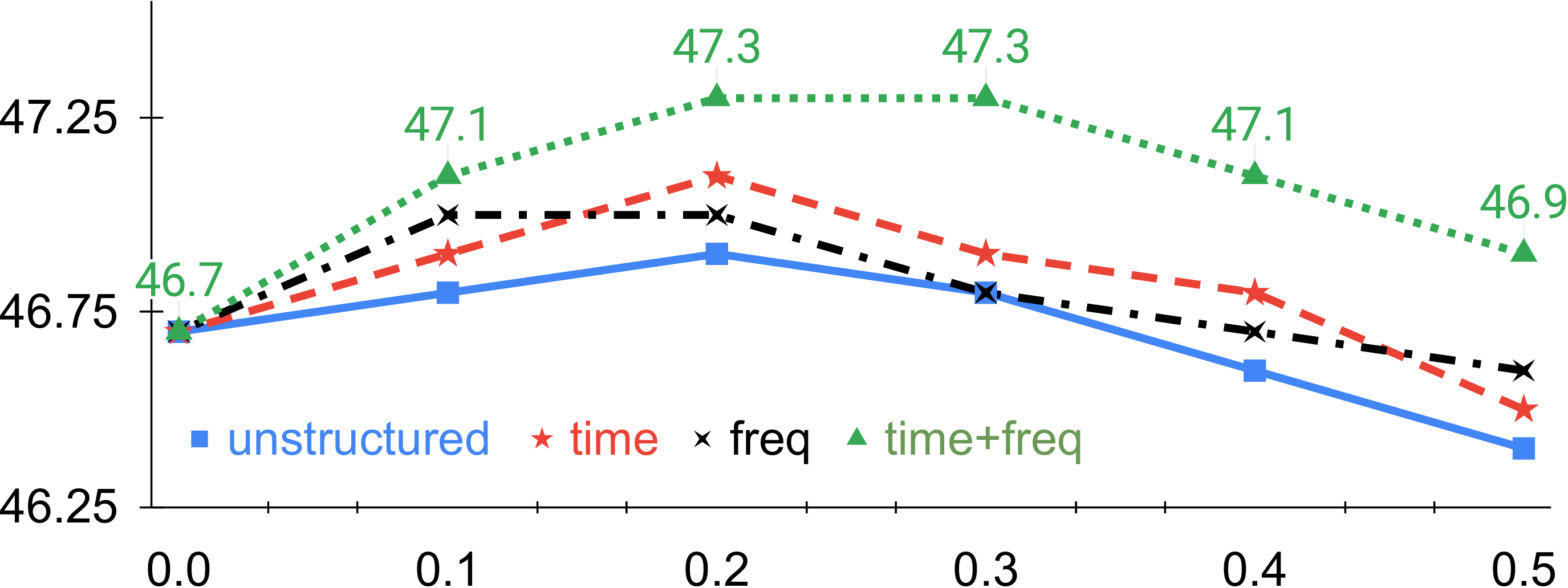}
        %\vspace{-0.2em}
        \caption{Fine-tuning masking}
        %\vspace{-0.2em}
        \label{fig:ft_masking_comparison}
    \end{subfigure}
    %\vspace{-0.2em}
    \caption{\textbf{Masking strategy}. 
    % Address the insight for masking
    For pre-training, a \textit{higher} ratio and \textit{unstructured} masking (random) is preferred.
    For fine-tuning, a \textit{lower} ratio and \textit{structured} masking (time$+$frequency) is better. 
    The y-axes are mAP on AS-2M and the x-axes are masking ratio. This ablation format follows \cite{mae}. 
    }
    \label{fig:masking_comparison}
    \vspace{-1.0em}
\end{figure}

\paragraph{Masking Strategies in Pre-training and Fine-tuning.}
In Fig.~\ref{fig:masking_comparison}, we compare different pre-training and fine-tuning masking strategies for Audio-MAE.
First, in Fig.~\ref{fig:pt_masking_comparison} we explore the \textit{pre-training masking ratio}. We observe, similar as in MAE for images~\cite{mae}, that a high pre-training masking ratio (80\% in our case) is optimal for audio spectrograms. 
This is due to the fact that both audio spectrograms and images are continuous signals with significant redundancy.
Further, we find the unstructured random masking works the best for self-supervised pre-training over more structured masking (\eg, time$+$frequency).

\begin{figure}[h]
% \vspace{-15pt}
    \centering
    \begin{subfigure}[b]{0.245\linewidth}
        \includegraphics[width=0.98\linewidth,height=0.4\linewidth]{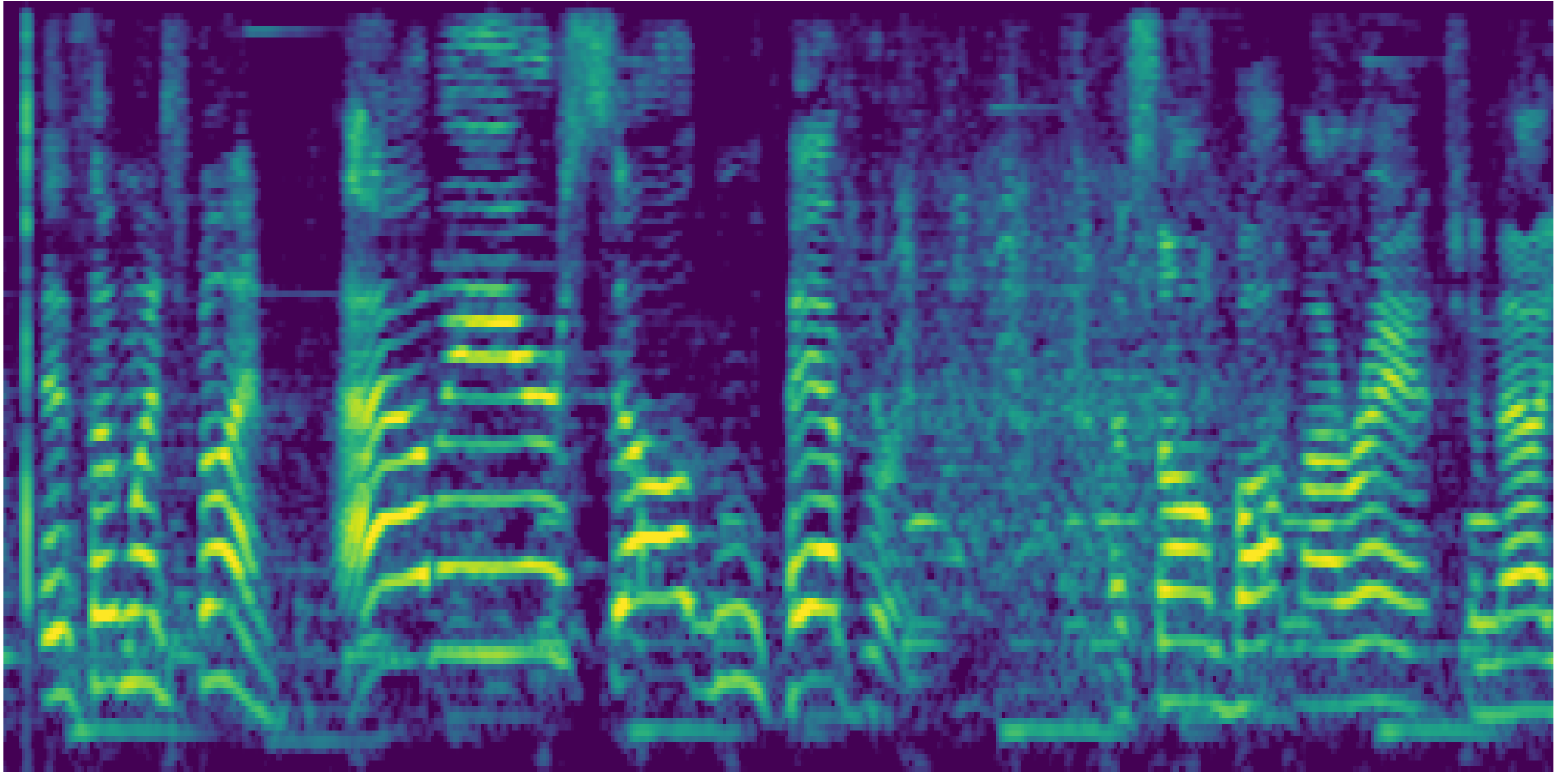}
    \end{subfigure}
    \begin{subfigure}[b]{0.245\linewidth}
        \includegraphics[width=0.98\linewidth,height=0.4\linewidth]{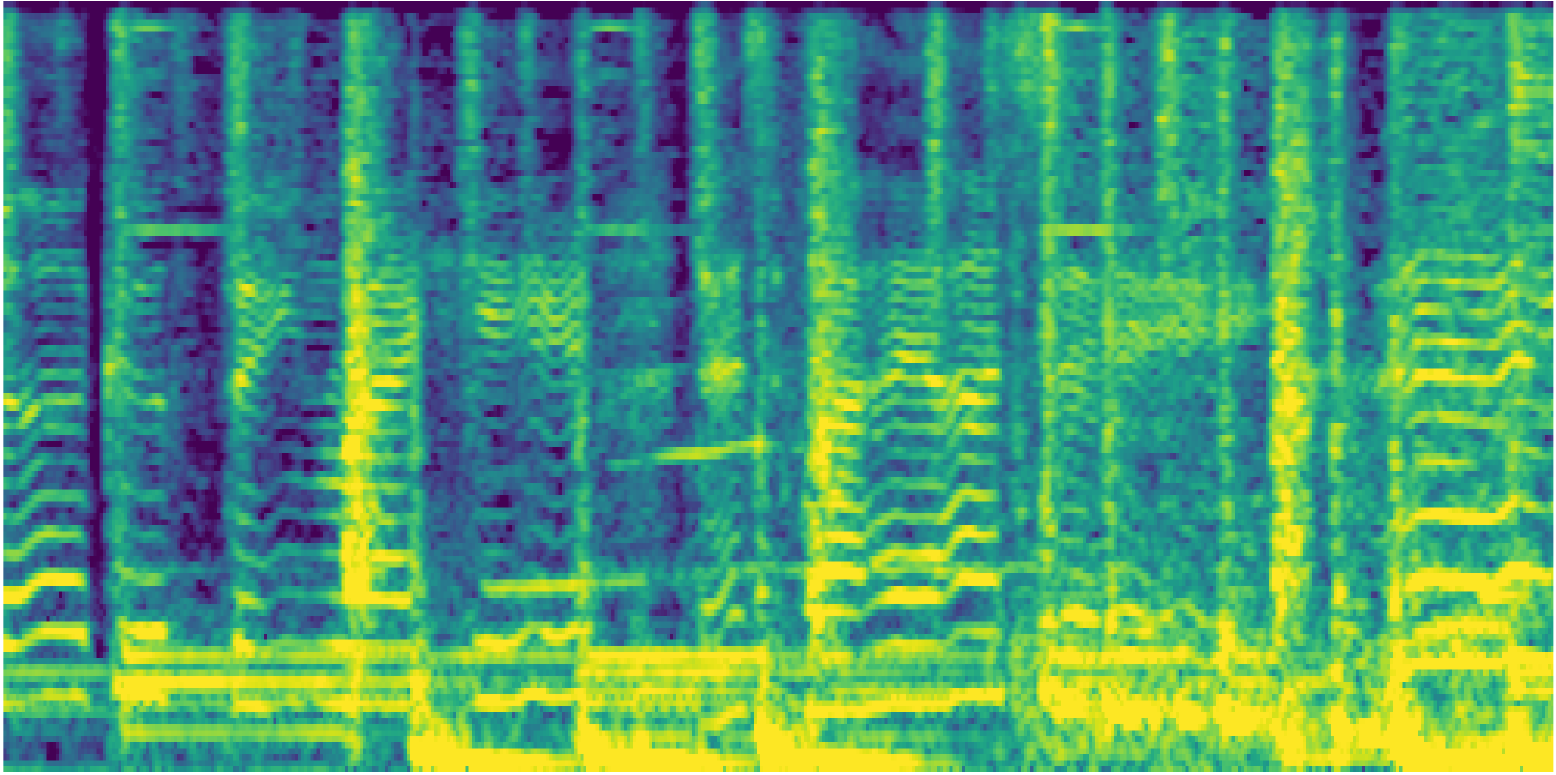}
    \end{subfigure}     
    \begin{subfigure}[b]{0.245\linewidth}
        \includegraphics[width=0.98\linewidth,height=0.4\linewidth]{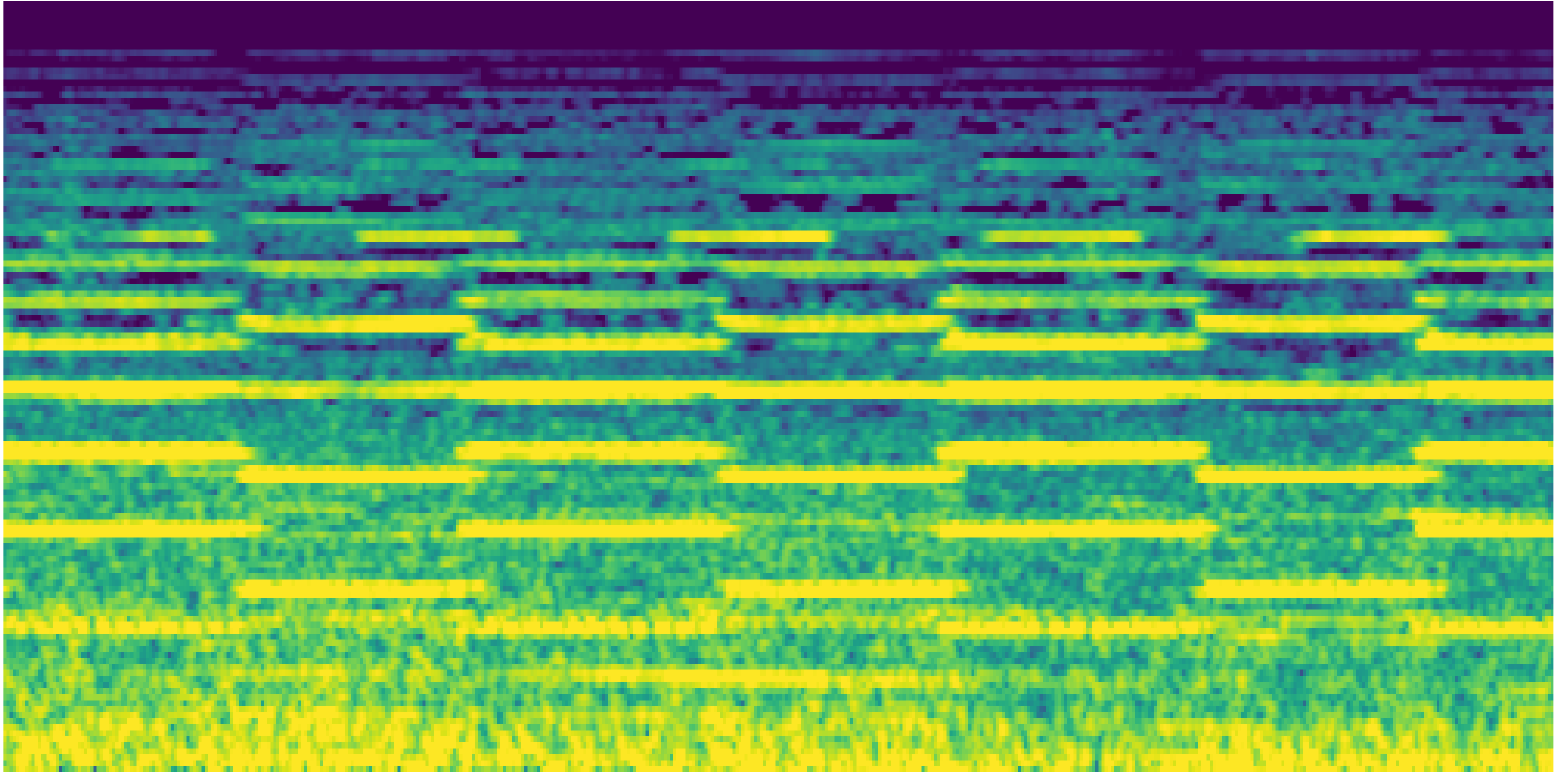}
    \end{subfigure}
    \begin{subfigure}[b]{0.245\linewidth}
        \includegraphics[width=0.98\linewidth,height=0.4\linewidth]{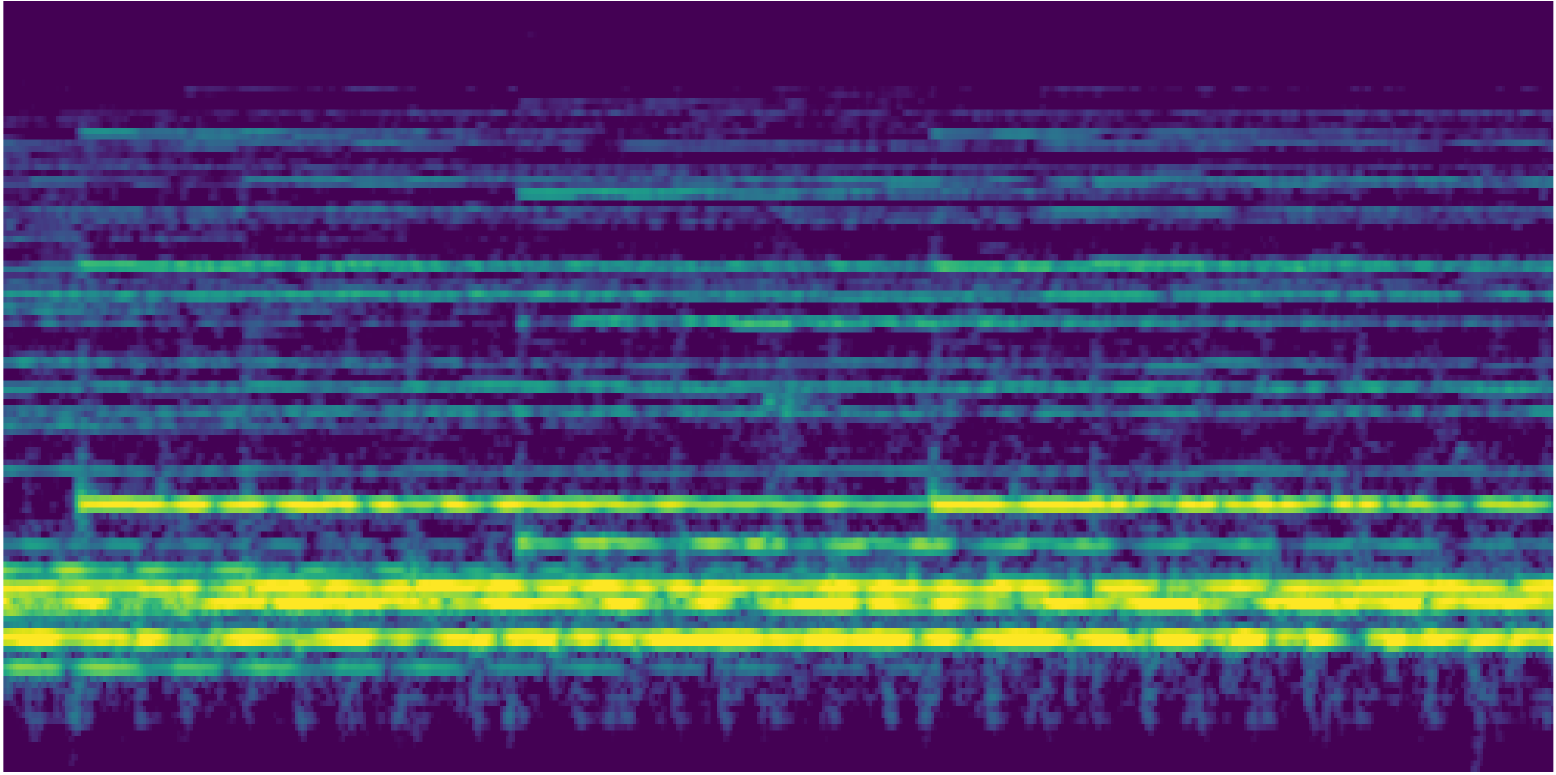}
    \end{subfigure}
    \\
    \begin{subfigure}[b]{0.245\linewidth}
        \includegraphics[width=0.98\linewidth,height=0.4\linewidth]{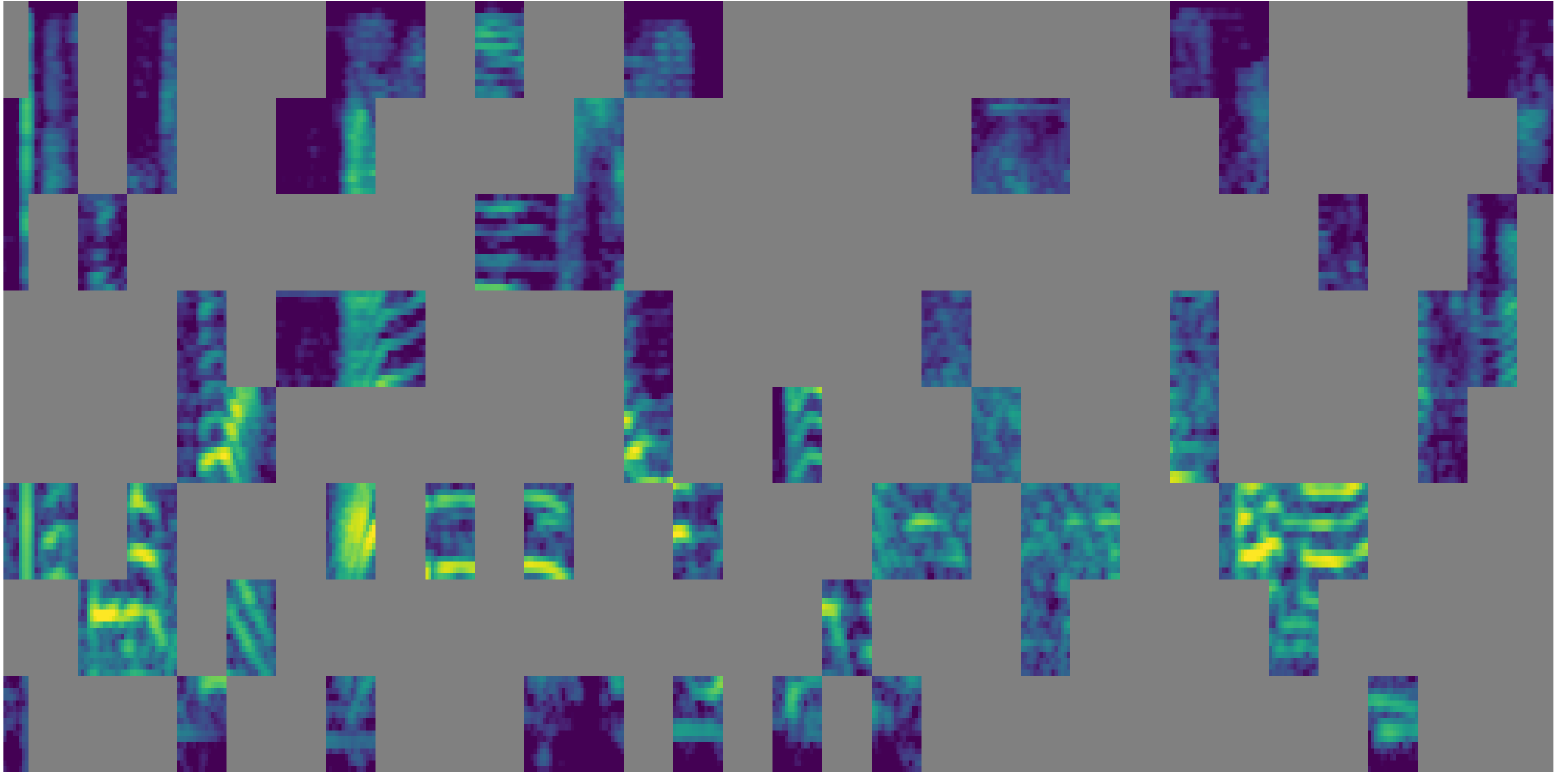}
    \end{subfigure}
    \begin{subfigure}[b]{0.245\linewidth}
        \includegraphics[width=0.98\linewidth,height=0.4\linewidth]{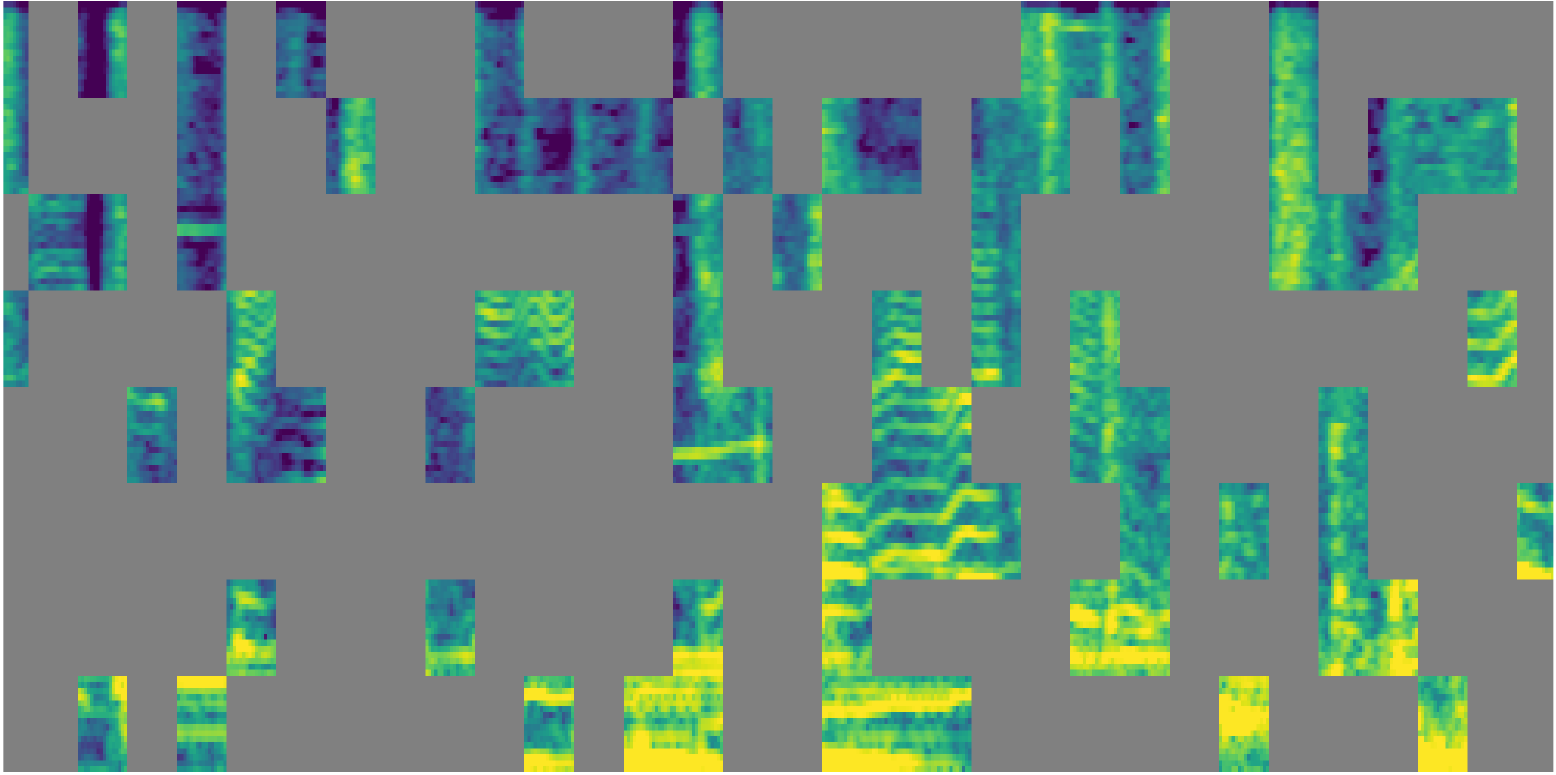}
    \end{subfigure}     
    \begin{subfigure}[b]{0.245\linewidth}
        \includegraphics[width=0.98\linewidth,height=0.4\linewidth]{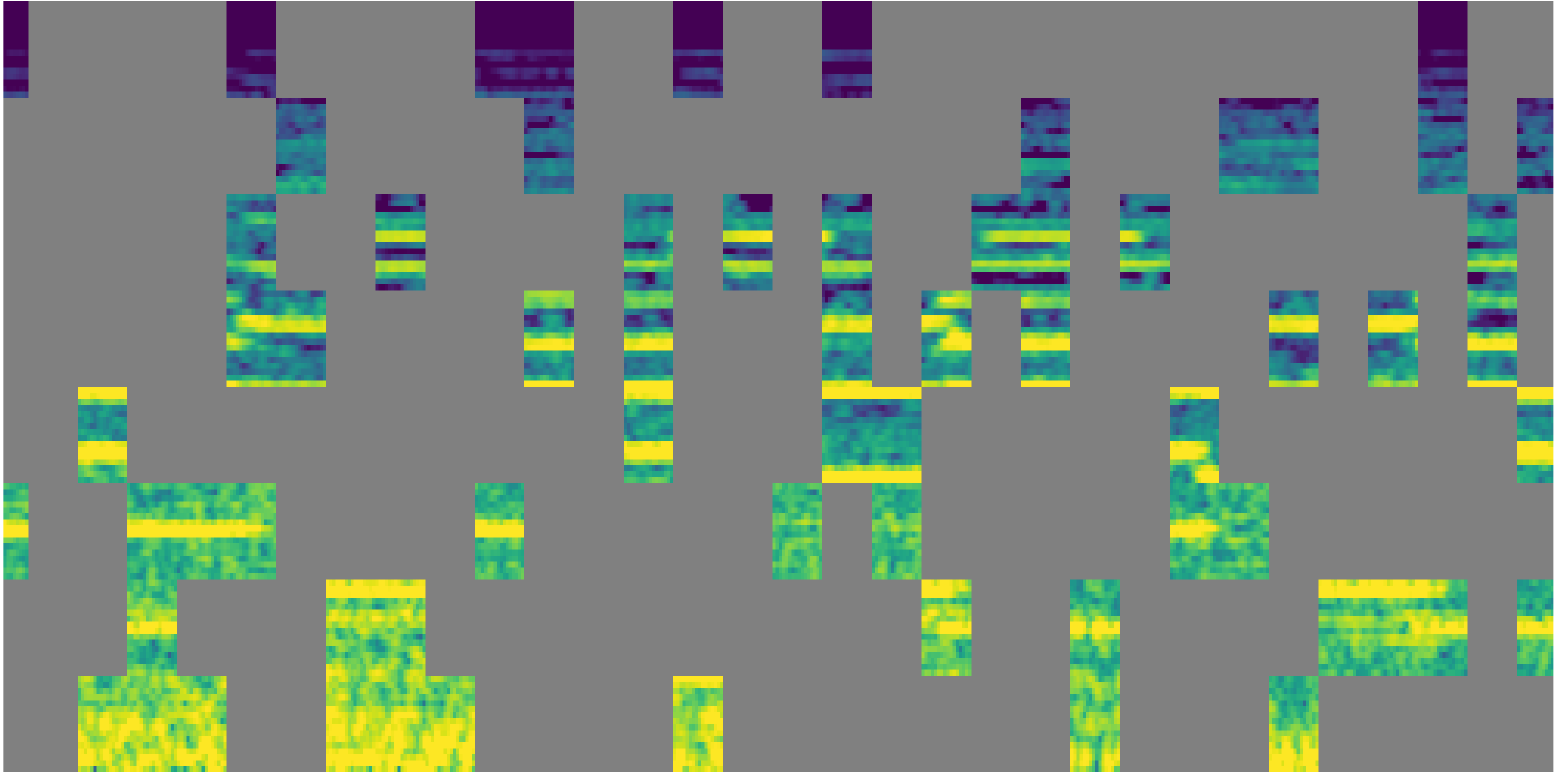}
    \end{subfigure}
    \begin{subfigure}[b]{0.245\linewidth}
        \includegraphics[width=0.98\linewidth,height=0.4\linewidth]{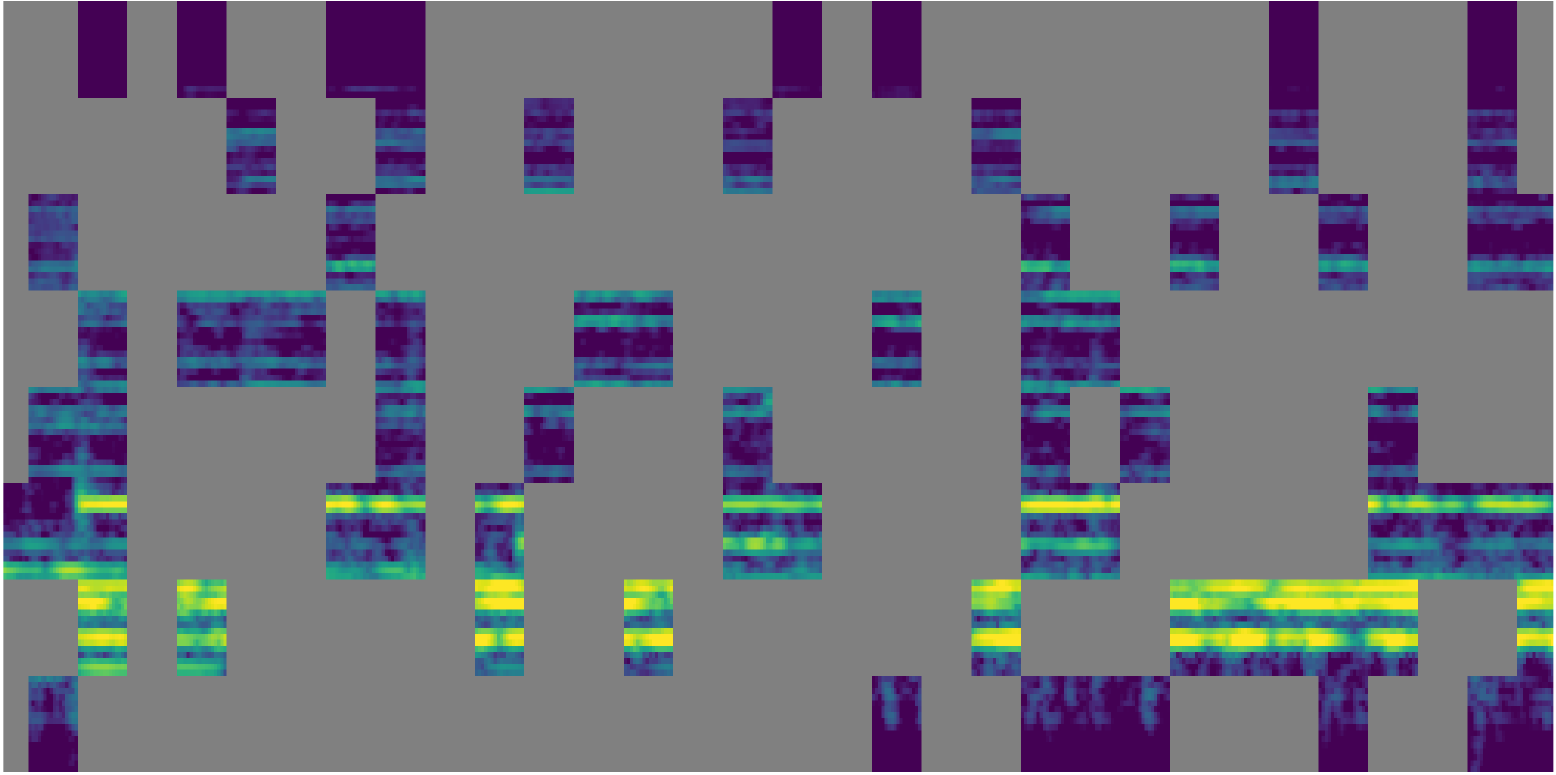}
    \end{subfigure}
    \\ 
    \begin{subfigure}[b]{0.245\linewidth}
        \includegraphics[width=0.98\linewidth,height=0.4\linewidth]{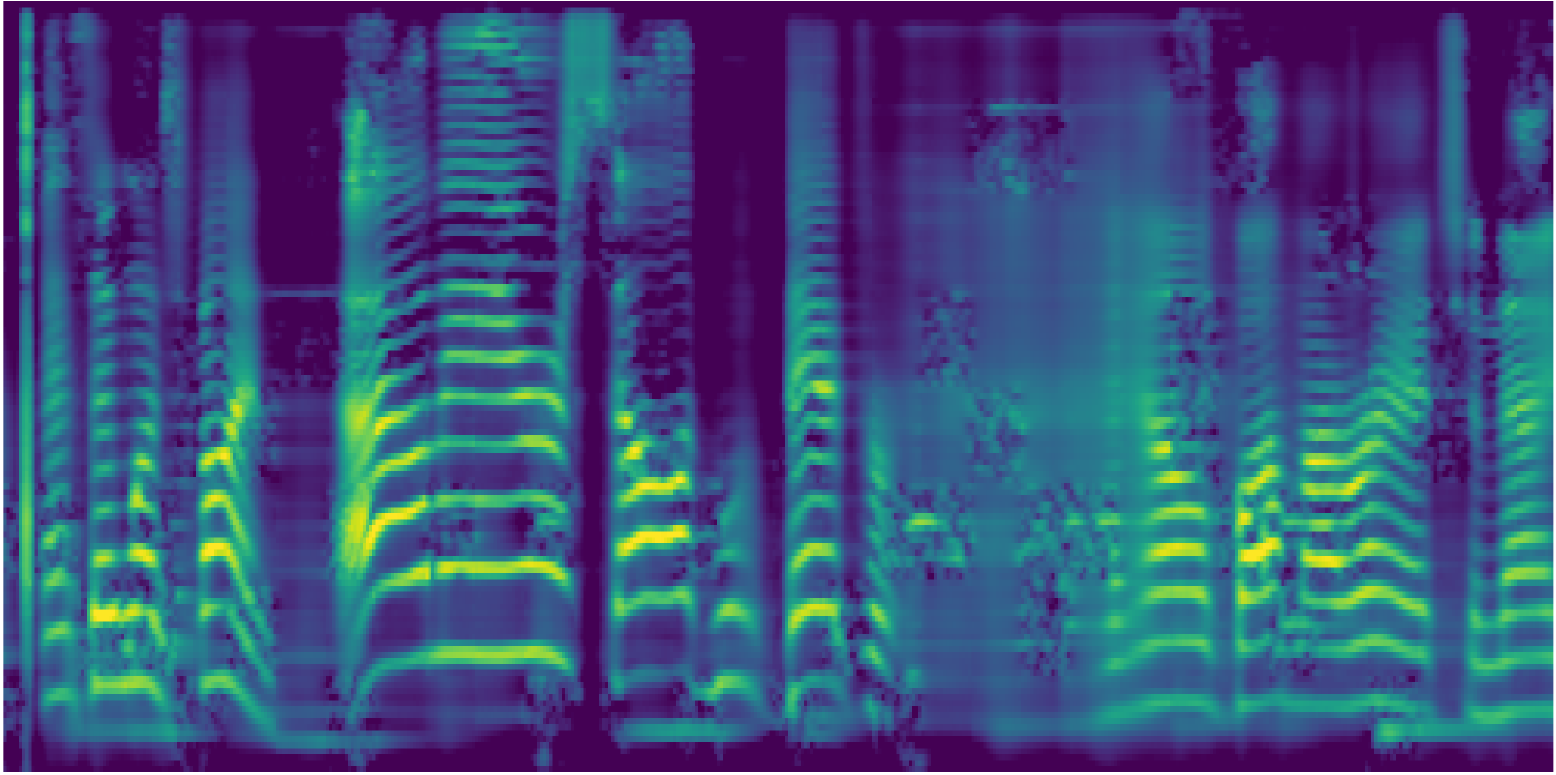}
        \vspace{-0.2em}
        \subcaption{
            Unstructured
            \href{https://www.dropbox.com/s/938ghmnlb2yddw1/-hJ1YTZ5AGI_0.7_org.mp4?dl=0}{1}
            \href{https://www.dropbox.com/s/8em45di4kyka50b/-hJ1YTZ5AGI_0.7_masked.mp4?dl=0}{2}
            \href{https://www.dropbox.com/s/qzamski5g0lgms5/-hJ1YTZ5AGI_0.7_restored.mp4?dl=0}{3}            
        }
        \label{fig:vis:a}
    \end{subfigure} 
    \begin{subfigure}[b]{0.245\linewidth}
        \includegraphics[width=0.98\linewidth,height=0.4\linewidth]{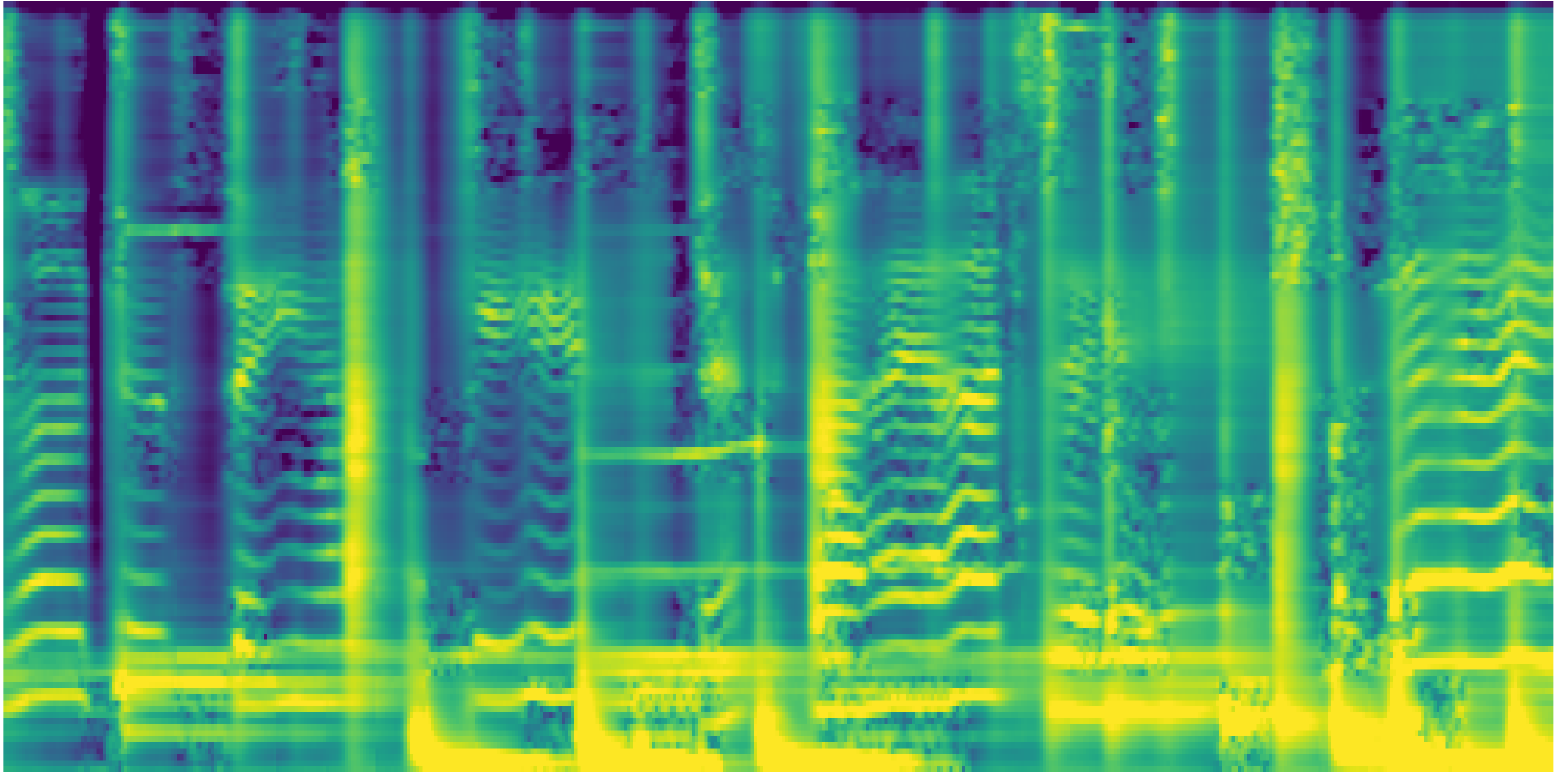}
        \vspace{-0.2em}
        \subcaption{
            Unstructured
            \href{https://www.dropbox.com/s/2721v67ud4qxbur/DB38NRSHw9A_0.7_org.mp4?dl=0}{1}
            \href{https://www.dropbox.com/s/g7xl0o8hczecqm2/DB38NRSHw9A_0.7_masked.mp4?dl=0}{2}
            \href{https://www.dropbox.com/s/onmdvpgykphlfx6/DB38NRSHw9A_0.7_restored.mp4?dl=0}{3}
        }
        \label{fig:vis:b}
    \end{subfigure}    
    \begin{subfigure}[b]{0.245\linewidth}
        \includegraphics[width=0.98\linewidth,height=0.4\linewidth]{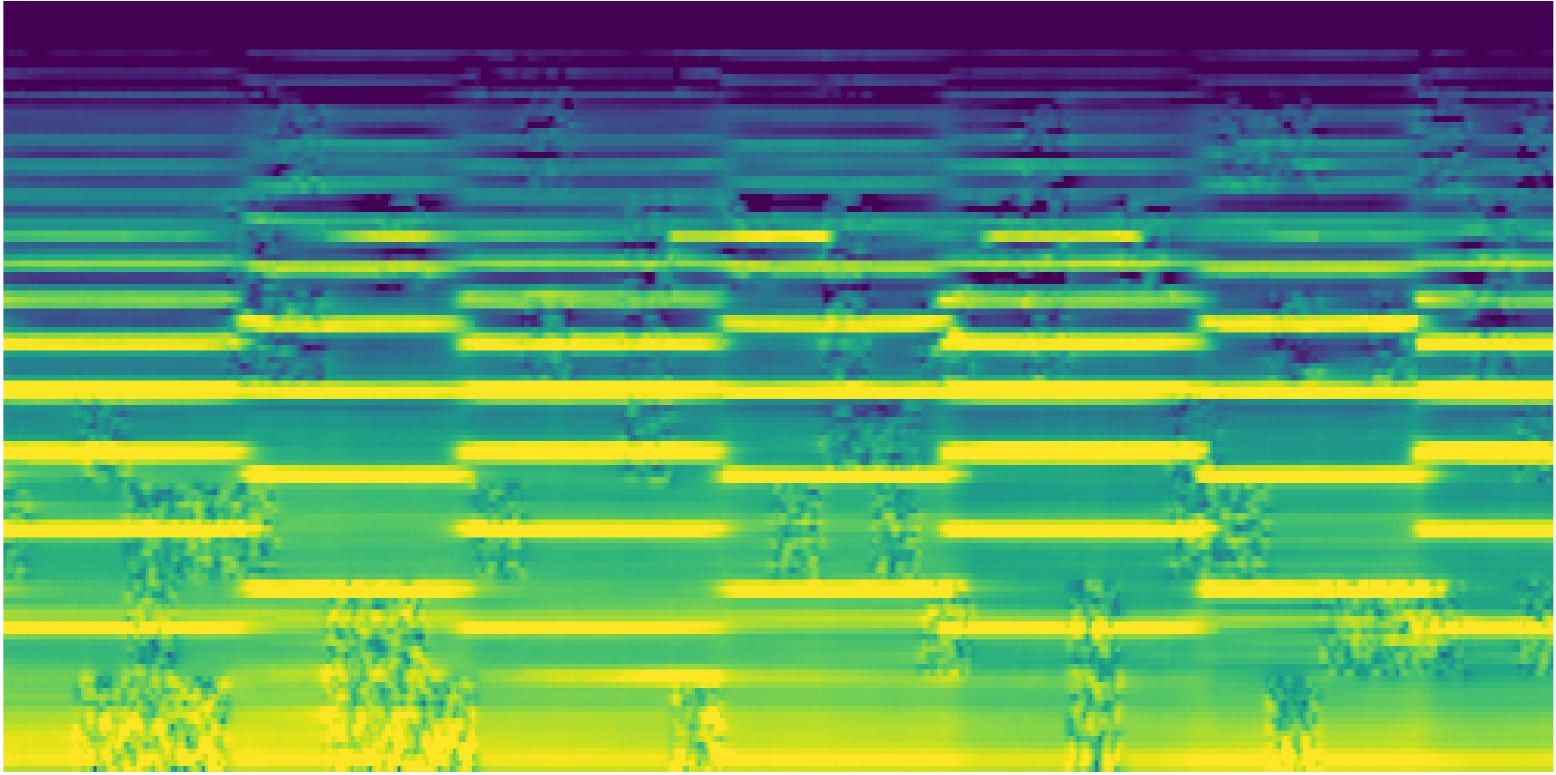}
        \vspace{-0.2em}
        \subcaption{
            Unstructured 
            \href{https://www.dropbox.com/s/7iybu7afhd1jbcf/02Ak1eIyj3M_0.7_org.mp4?dl=0}{1}
            \href{https://www.dropbox.com/s/ywa94w4dw9mdog0/02Ak1eIyj3M_0.7_masked.mp4?dl=0}{2}
            \href{https://www.dropbox.com/s/avpg02hjikw8o6e/02Ak1eIyj3M_0.7_restored.mp4?dl=0}{3}
        }
        \label{fig:vis:c}
    \end{subfigure}
    \begin{subfigure}[b]{0.245\linewidth}
        \includegraphics[width=0.98\linewidth,height=0.4\linewidth]{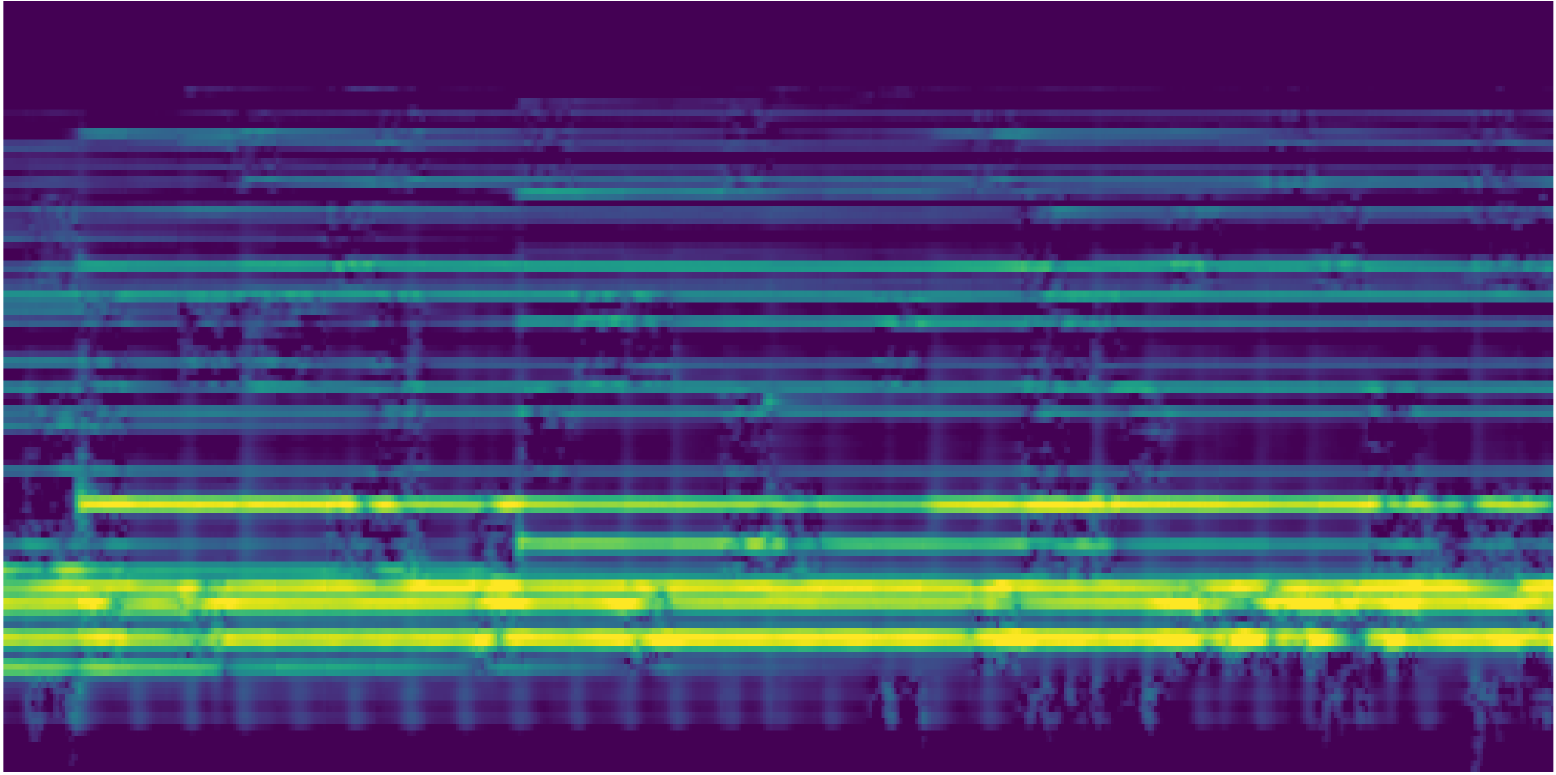}
        \vspace{-0.2em}
        \subcaption{
            Unstructured 
            \href{https://www.dropbox.com/s/h1b1j54prlwdb3h/hRbukCd6N68_0.7_org.mp4?dl=0}{1}
            \href{https://www.dropbox.com/s/z318mayqsgucw5u/hRbukCd6N68_0.7_masked.mp4?dl=0}{2}
            \href{https://www.dropbox.com/s/2ej44ovp7f86gcn/hRbukCd6N68_0.7_restored.mp4?dl=0}{3}            
        }
        \label{fig:vis:d}
    \end{subfigure}
    \\ \vspace{2pt}
    %%%%%%%%%%%%%%%%%%% second row %%%%%%%%%%%%%%%%%%%%%%%%
    \begin{subfigure}[b]{0.245\linewidth}
        \includegraphics[width=0.98\linewidth,height=0.4\linewidth]{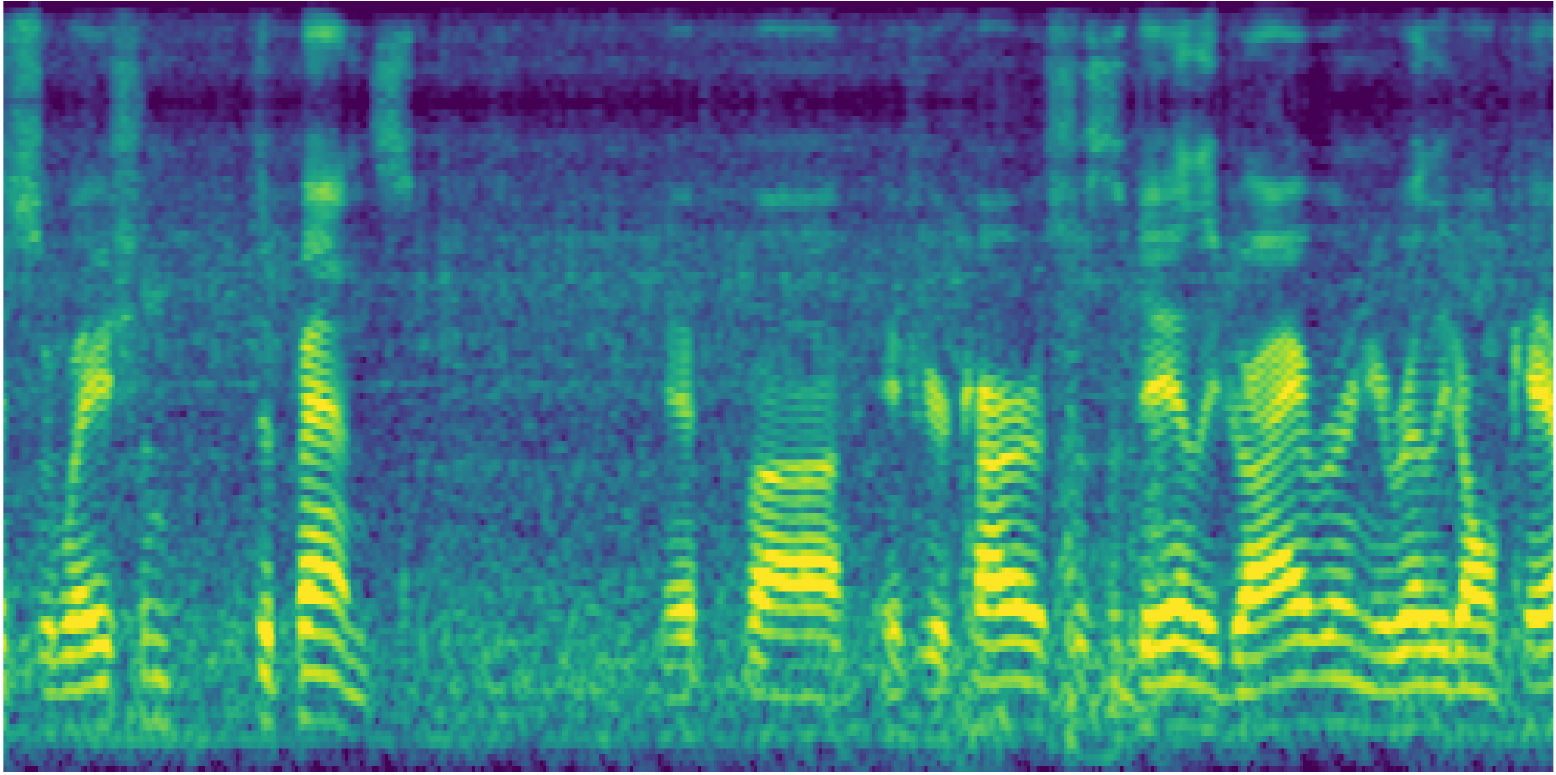}
    \end{subfigure}    
    \begin{subfigure}[b]{0.245\linewidth}
        \includegraphics[width=0.98\linewidth,height=0.4\linewidth]{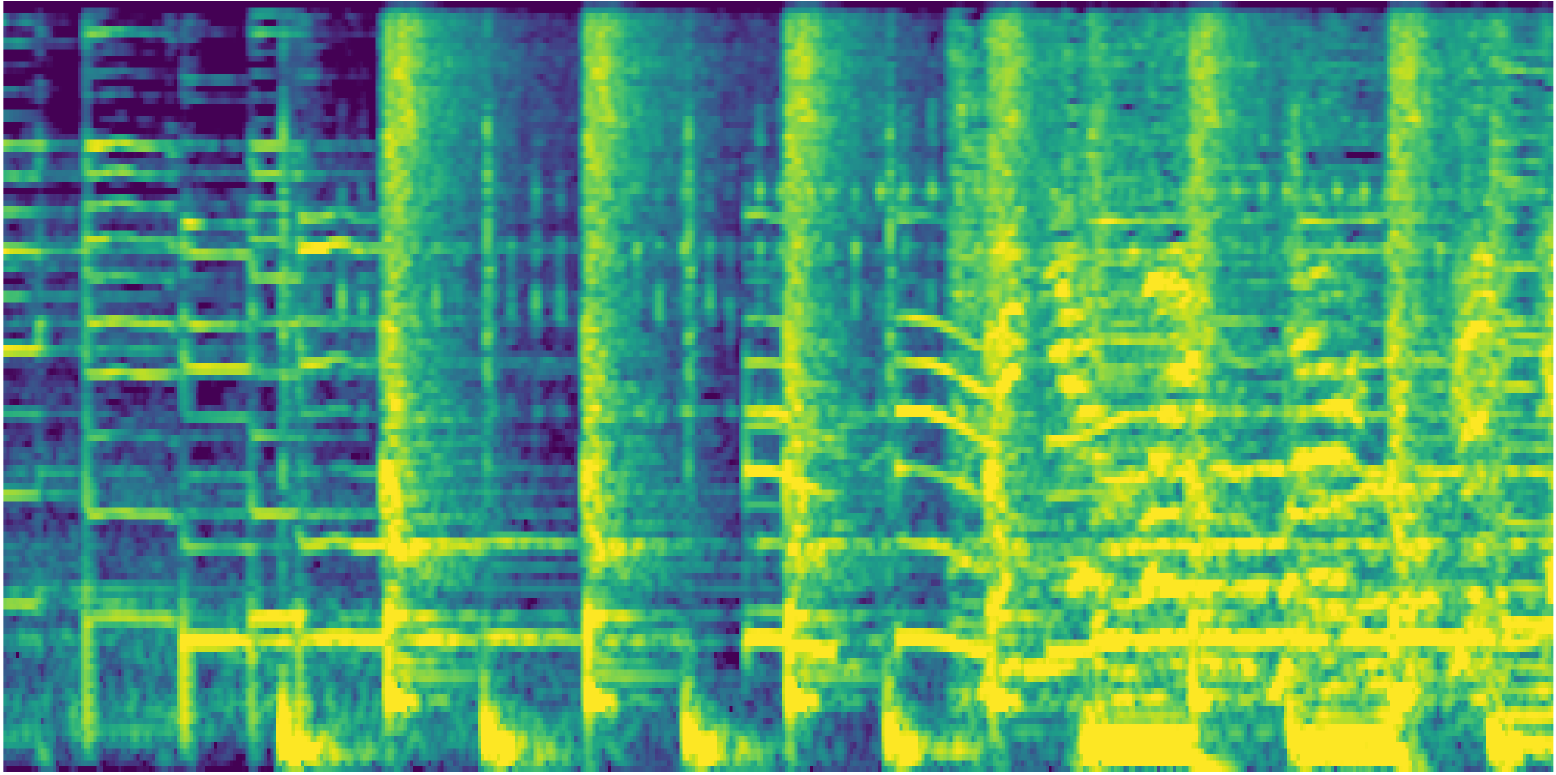}
    \end{subfigure}     
    \begin{subfigure}[b]{0.245\linewidth}
        \includegraphics[width=0.98\linewidth,height=0.4\linewidth]{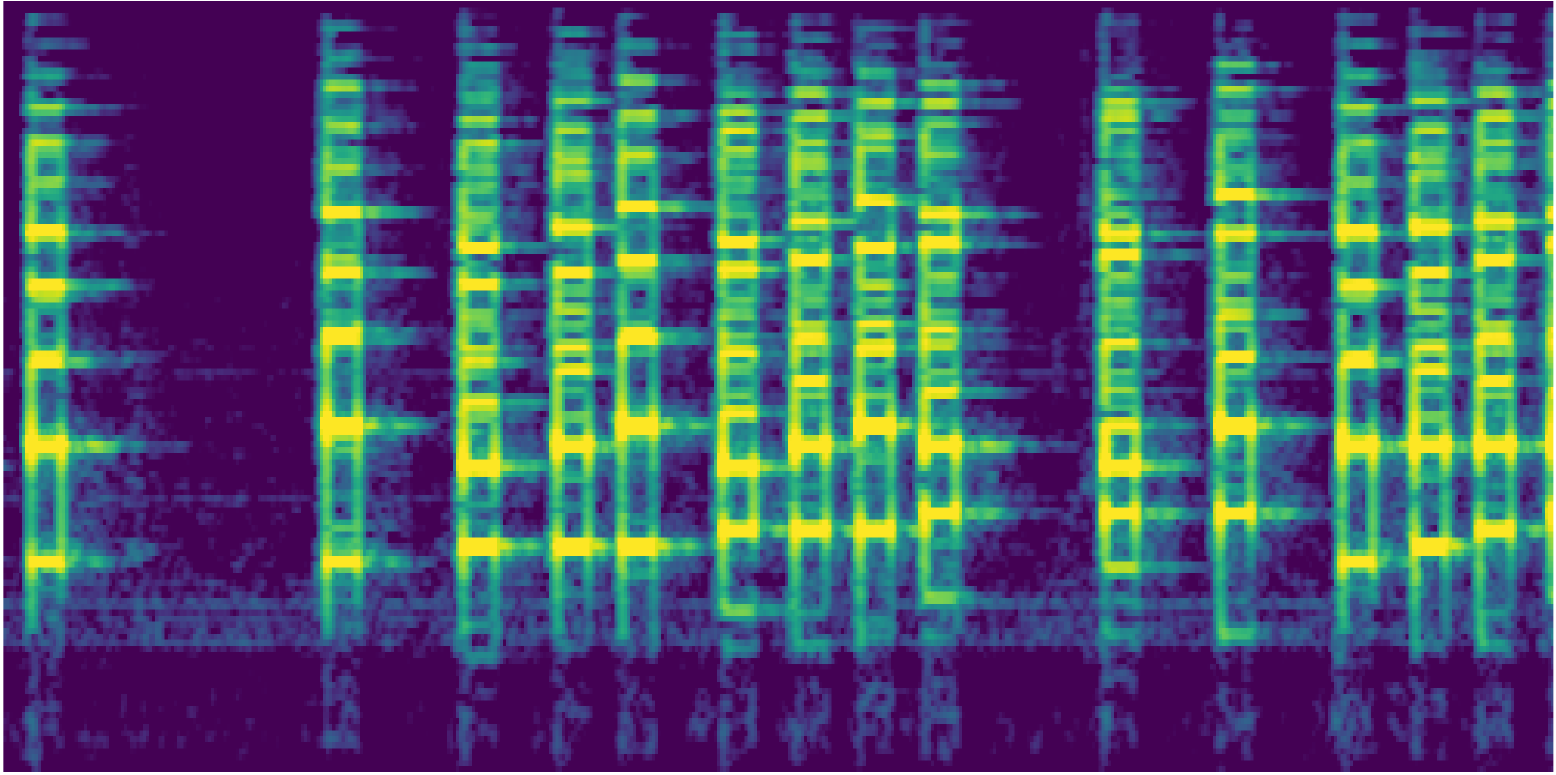}
    \end{subfigure}
    \begin{subfigure}[b]{0.245\linewidth}
        \includegraphics[width=0.98\linewidth,height=0.4\linewidth]{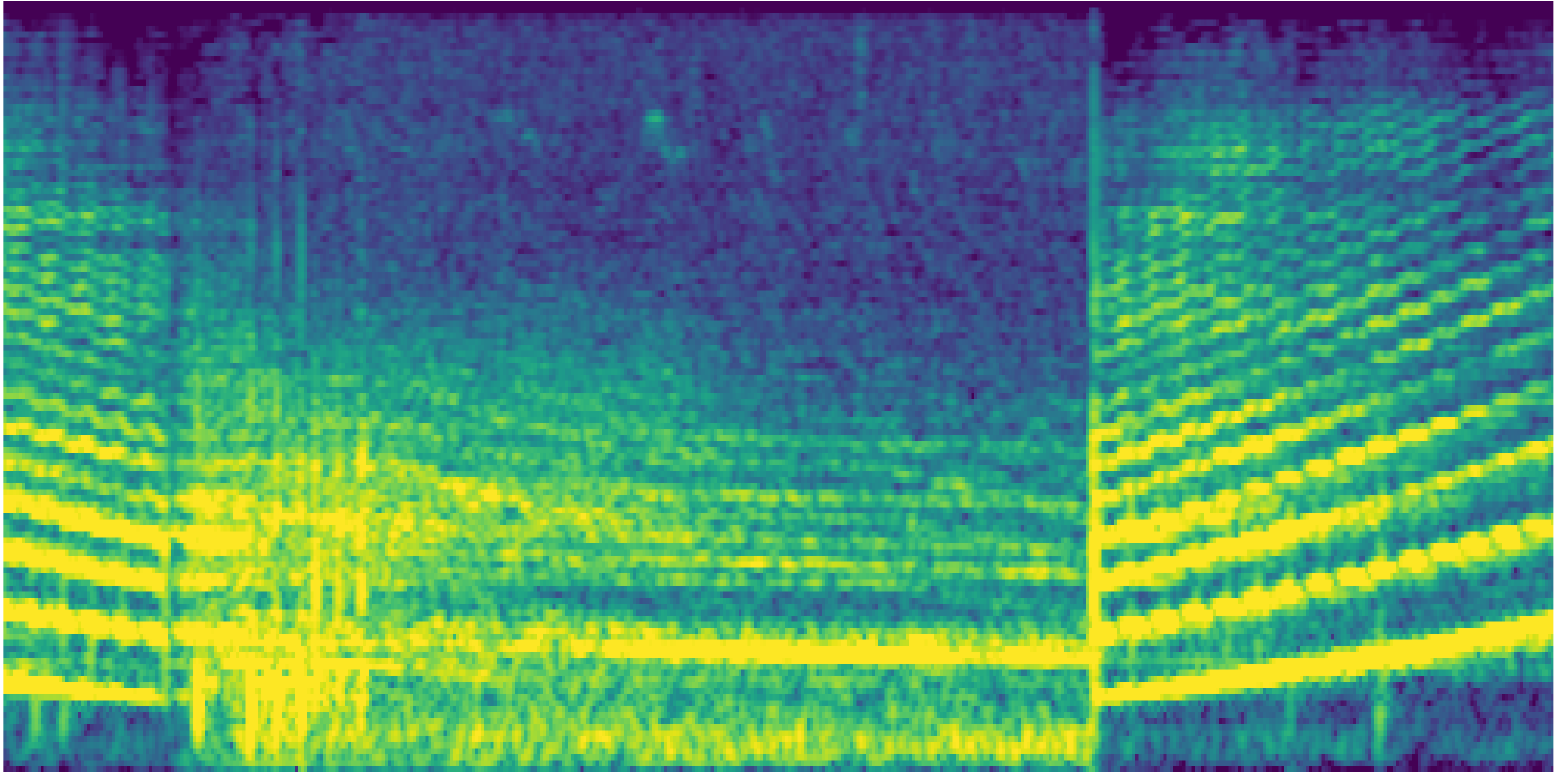}
    \end{subfigure}   
    \\
    \begin{subfigure}[b]{0.245\linewidth}
        \includegraphics[width=0.98\linewidth,height=0.4\linewidth]{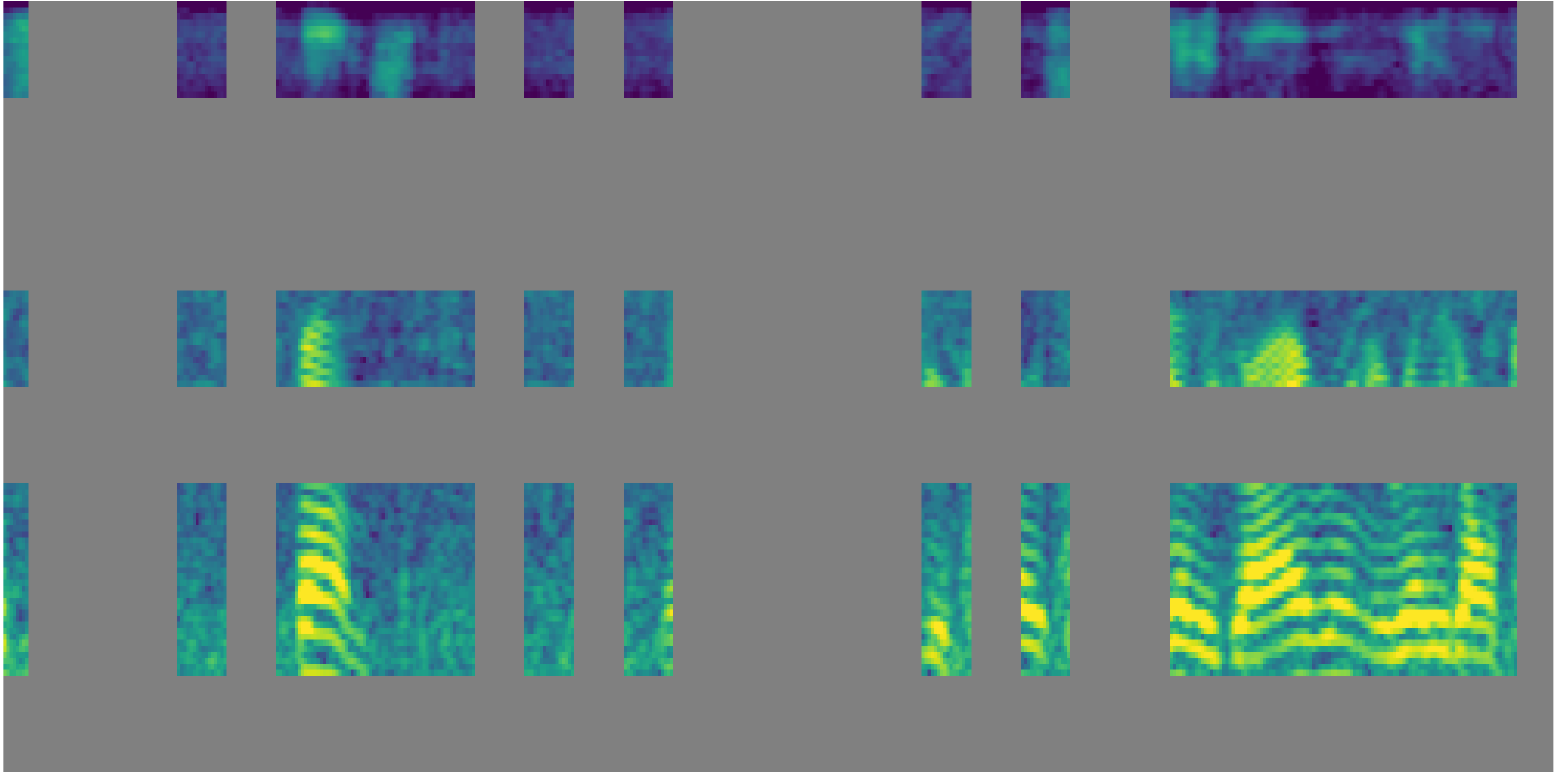}
    \end{subfigure}    
    \begin{subfigure}[b]{0.245\linewidth}
        \includegraphics[width=0.98\linewidth,height=0.4\linewidth]{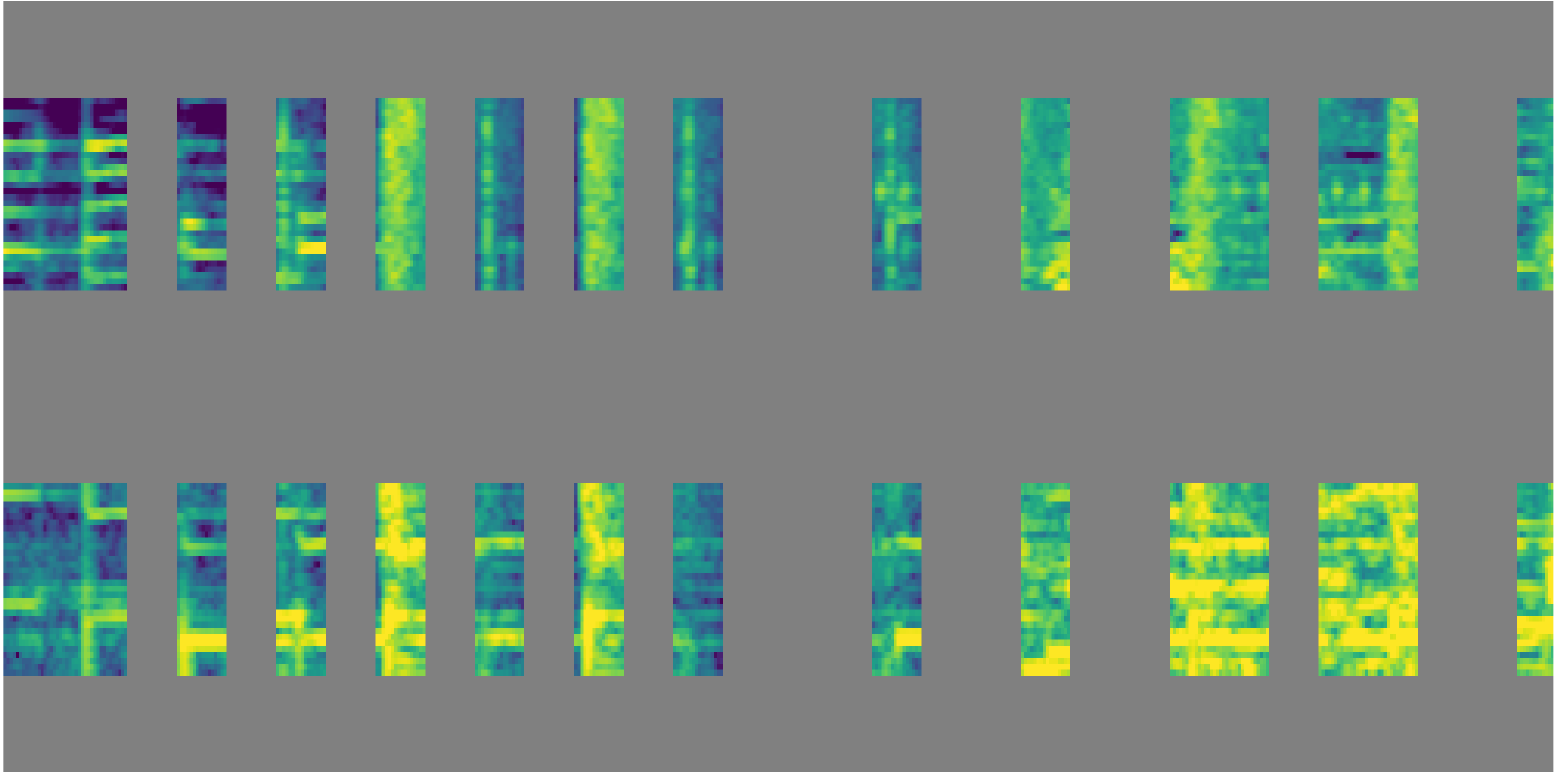}
    \end{subfigure}      
    \begin{subfigure}[b]{0.245\linewidth}
        \includegraphics[width=0.98\linewidth,height=0.4\linewidth]{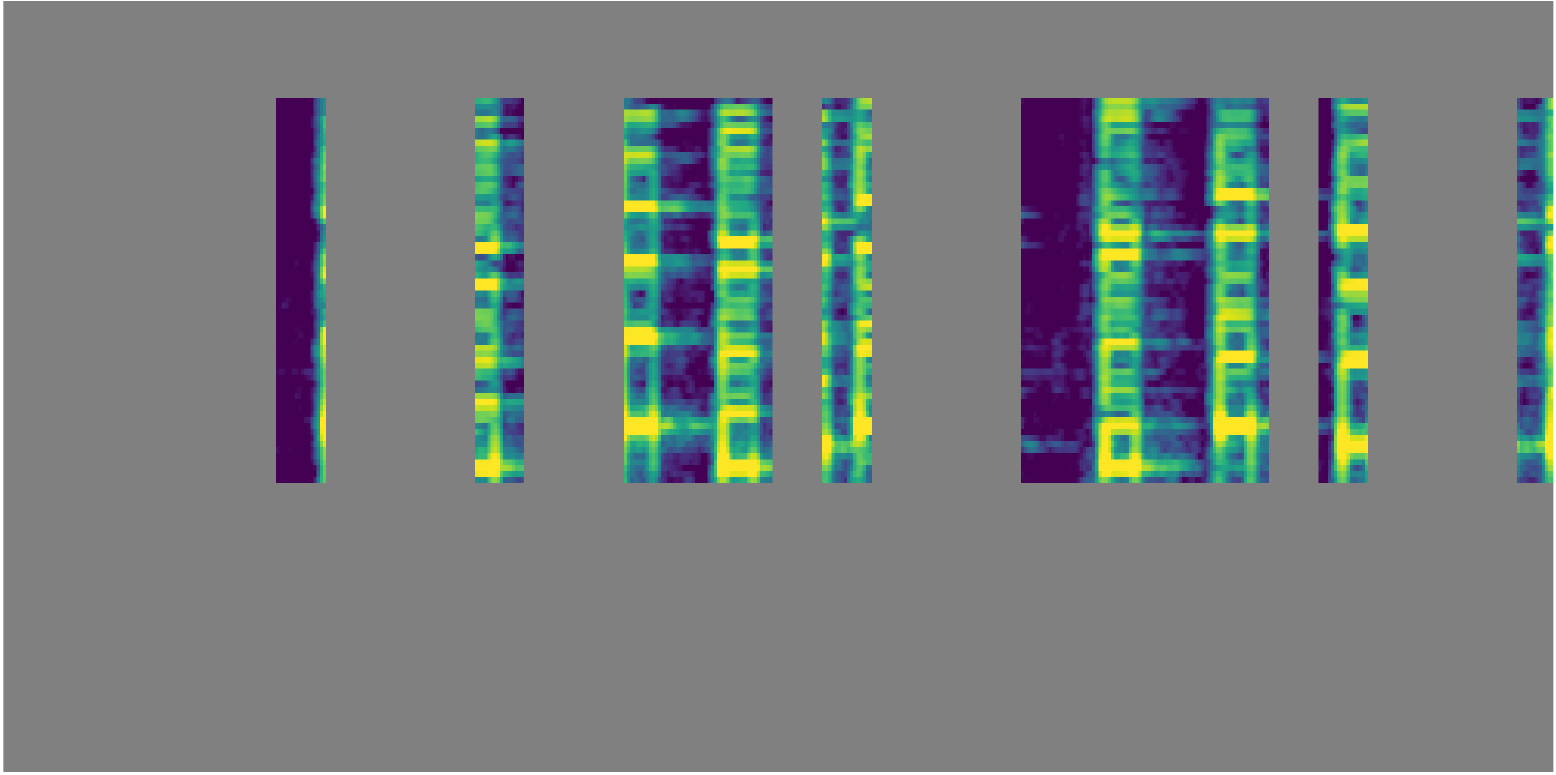}
    \end{subfigure}
    \begin{subfigure}[b]{0.245\linewidth}
        \includegraphics[width=0.98\linewidth,height=0.4\linewidth]{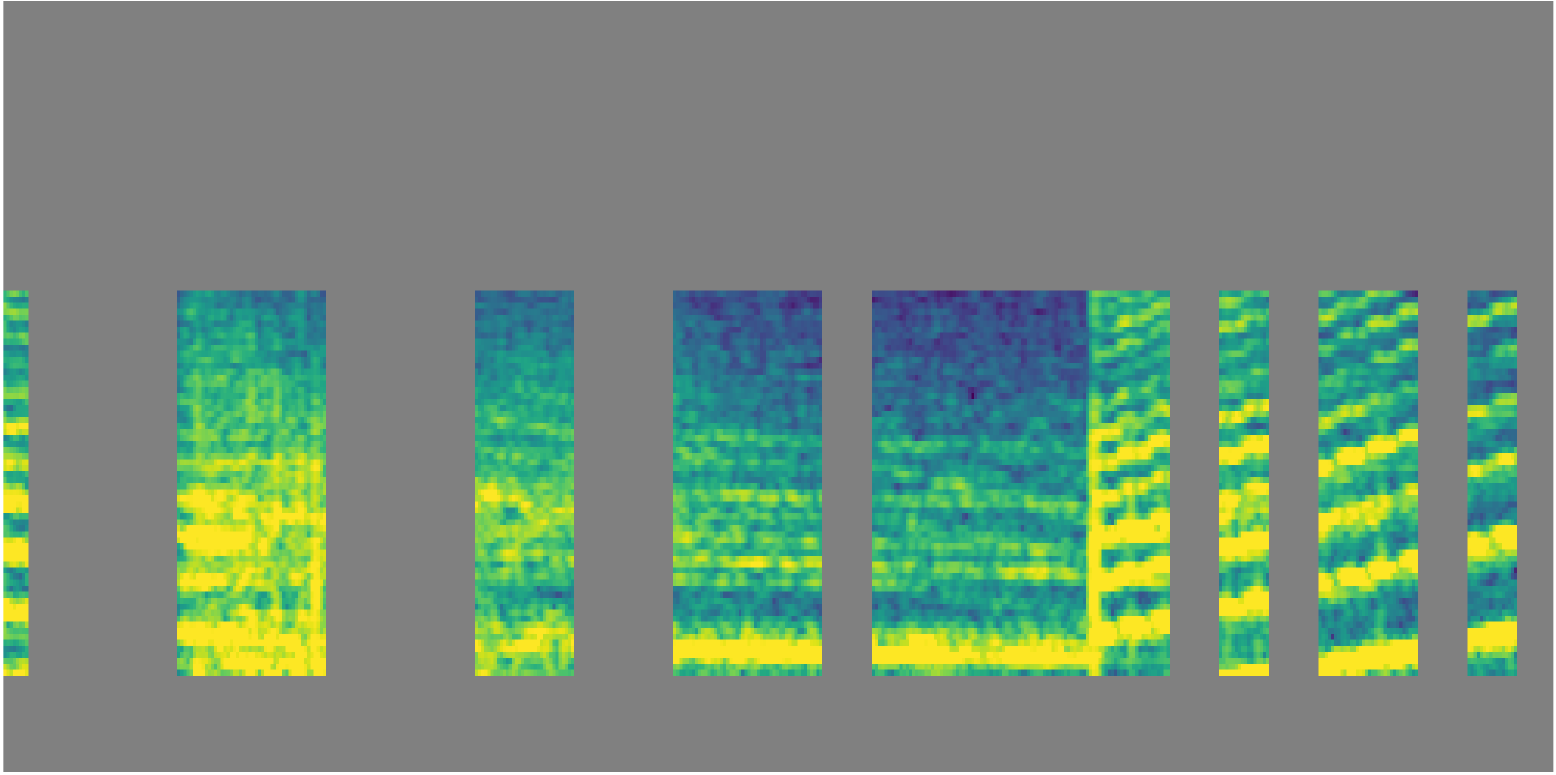}
    \end{subfigure} 
    \\
    \begin{subfigure}[b]{0.245\linewidth}
        \includegraphics[width=0.98\linewidth,height=0.4\linewidth]{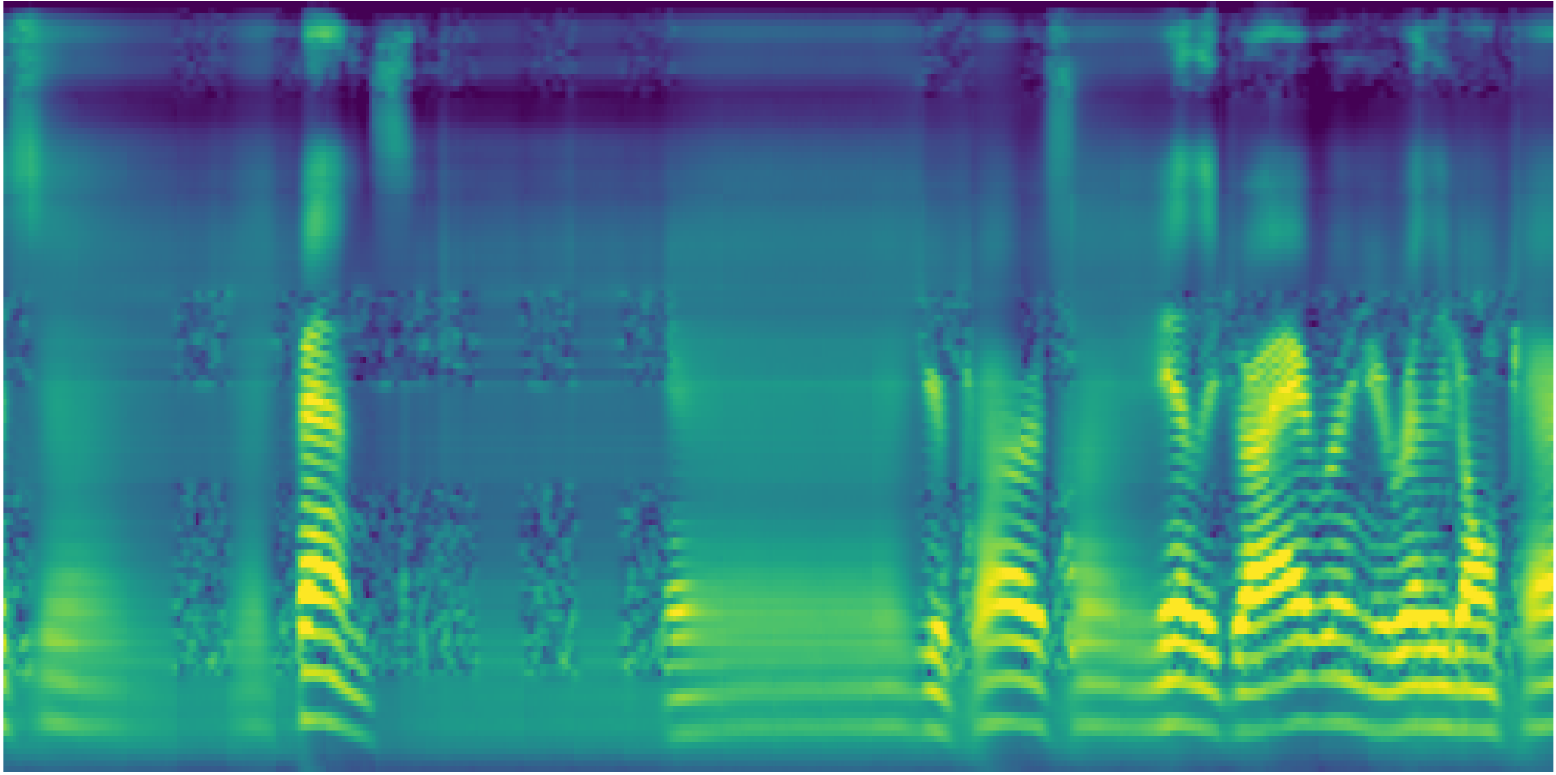}
        \vspace{-0.2em}
        \subcaption{
            Structured
            \href{https://www.dropbox.com/s/oep6dq2hdkmphn3/tzJH5FfR9j8_0.7_2d_org.mp4?dl=0}{1}
            \href{https://www.dropbox.com/s/4wcfhcbb3pibn7n/tzJH5FfR9j8_0.7_2d_masked.mp4?dl=0}{2}
            \href{https://www.dropbox.com/s/712o434428i4kdv/tzJH5FfR9j8_0.7_2d_restored.mp4?dl=0}{3}            
        }
        %\vspace{-0.5em}
        \label{fig:vis:e}
    \end{subfigure} 
    \begin{subfigure}[b]{0.245\linewidth}
        \includegraphics[width=0.98\linewidth,height=0.4\linewidth]{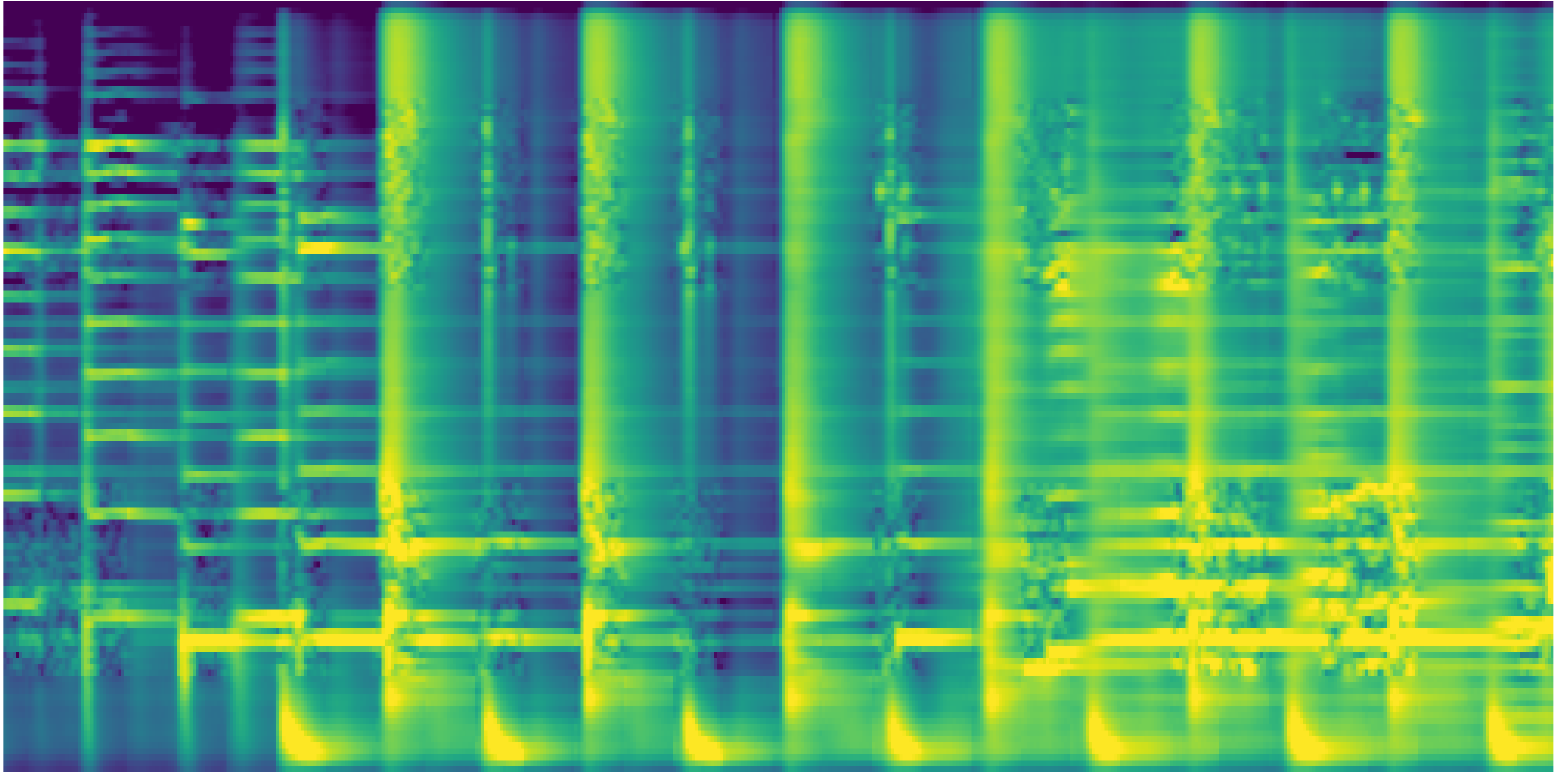}
        \vspace{-0.2em}
        \subcaption{
            Structured
            \href{https://www.dropbox.com/s/6m7amgbd58z9a9h/CWQvCCRuU6k_0.7_2d_org.mp4?dl=0}{1}
            \href{https://www.dropbox.com/s/467zm11n8a8n5jd/CWQvCCRuU6k_0.7_2d_masked.mp4?dl=0}{2}
            \href{https://www.dropbox.com/s/93yobqlo0brgol7/CWQvCCRuU6k_0.7_2d_restored.mp4?dl=0}{3}            
        }
        %\vspace{-0.5em}
        \label{fig:vis:f}
    \end{subfigure}   
    \begin{subfigure}[b]{0.245\linewidth}
        \includegraphics[width=0.98\linewidth,height=0.4\linewidth]{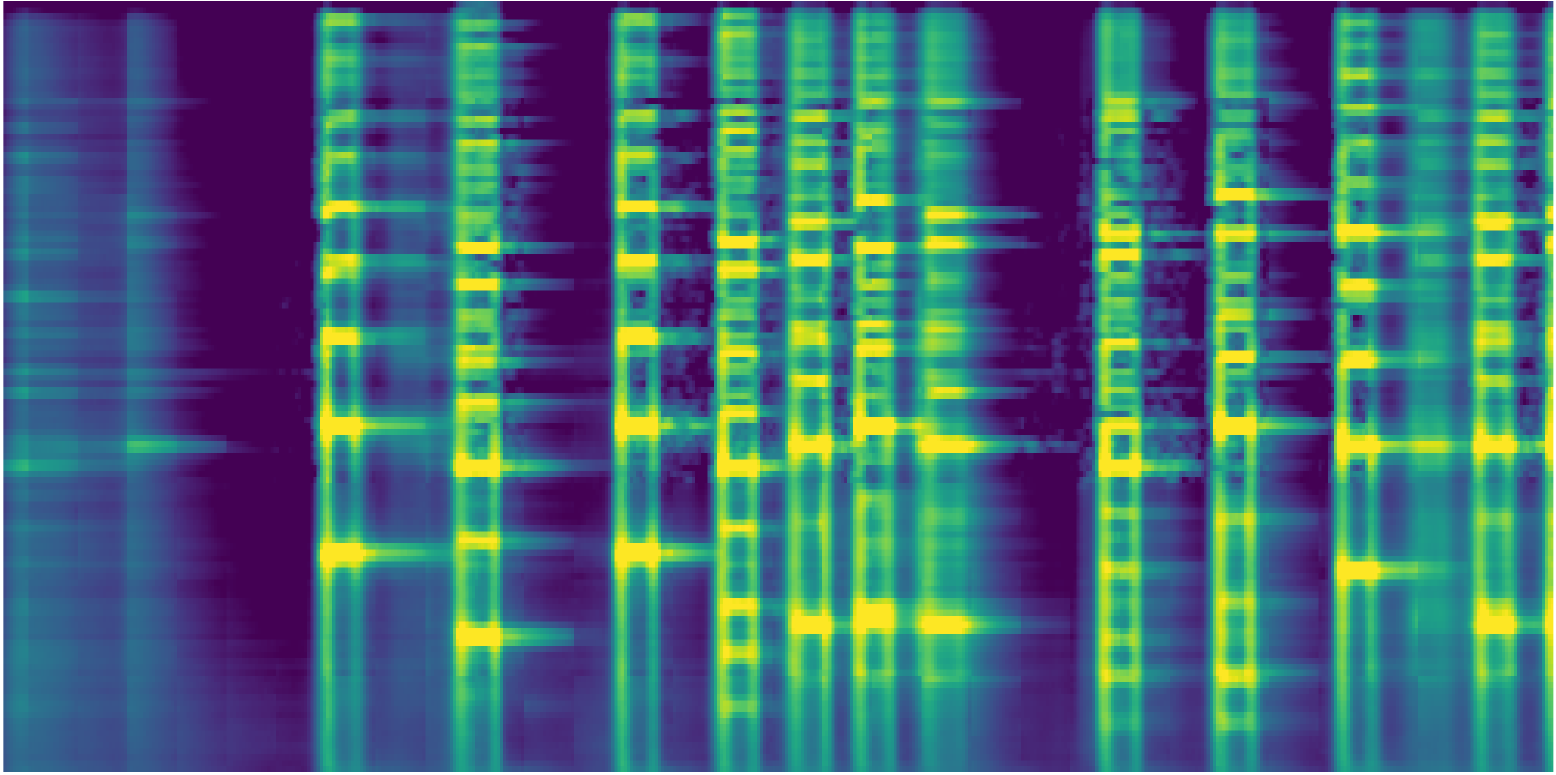}
        \vspace{-0.2em}
        \subcaption{
            Structured
            \href{https://www.dropbox.com/s/t7fowmyfs4jpcuq/MhUjXrKVVwQ_0.7_2d_org.mp4?dl=0}{1}
            \href{https://www.dropbox.com/s/463gxr882fci2rz/MhUjXrKVVwQ_0.7_2d_masked.mp4?dl=0}{2}
            \href{https://www.dropbox.com/s/dmzelh9h59td5tl/MhUjXrKVVwQ_0.7_2d_restored.mp4?dl=0}{3}            
        }
        %\vspace{-0.5em}
        \label{fig:vis:g}
    \end{subfigure}
    \begin{subfigure}[b]{0.245\linewidth}
        \includegraphics[width=0.98\linewidth,height=0.4\linewidth]{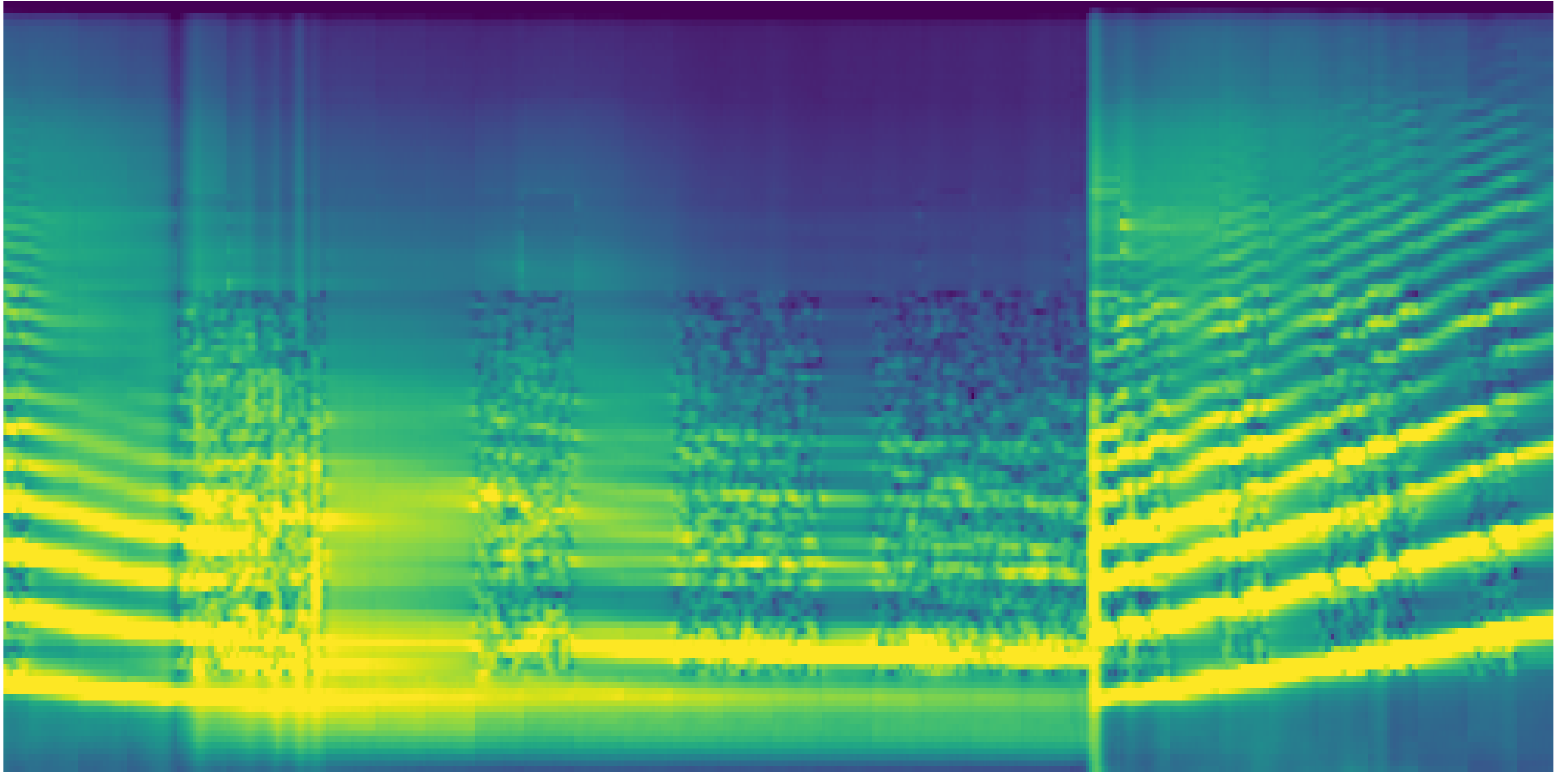}
        \vspace{-0.2em}
        \subcaption{
            Structured
            \href{https://www.dropbox.com/s/j0sqrgtky7i6ih5/3PVT0BcK-2I_0.7_2d_org.mp4?dl=0}{1}
            \href{https://www.dropbox.com/s/wsobte05m92a5x9/3PVT0BcK-2I_0.7_2d_masked.mp4?dl=0}{2}
            \href{https://www.dropbox.com/s/gkd5rmzq6k7eer9/3PVT0BcK-2I_0.7_2d_restored.mp4?dl=0}{3}            
        }
        %\vspace{-0.5em}
        \label{fig:vis:h}
    \end{subfigure} 
    \\
%%%%%%%%%%%%%%%%%%%%%%%%% Third Row %%%%%%%%%%%%%%%%%%%%%%%%%%%%%%%%%    
    % removed, add back if space permits
    %\vspace{2pt}
    \caption{
    \textbf{Spectrogram reconstruction visualizations on the AudioSet \textit{eval} set}. 
    Column-wise type: speech, music, event, others. 
    Masking type: (a-d) unstructured (random); (e-h) structured (time$+$frequency).
    Masking Ratio: 70\%.
    In each group, we show the original spectrogram (\href{https://www.dropbox.com/s/2721v67ud4qxbur/DB38NRSHw9A_0.7_org.mp4?dl=0}{1}, top), masked input (\href{https://www.dropbox.com/s/g7xl0o8hczecqm2/DB38NRSHw9A_0.7_masked.mp4?dl=0}{2}, middle), and MAE output (\href{https://www.dropbox.com/s/onmdvpgykphlfx6/DB38NRSHw9A_0.7_restored.mp4?dl=0}{3}, bottom). 
    The spectrogram size is 1024$\times$128; patch size is 16$\times$16. Each sample has 64$\times$8=512 patches with 154 (70\% masked) patches being visible to Audio-MAE. 
    Please click ({\color{hrefcolor} 1 2 3})
    for audible \emph{.wav}s. More audible examples are in Supplementary.
    }
    \label{fig:vis}
    %\vspace{-1.em}
\end{figure}

Unlike MAE for images, there are clear performance differences among masking strategies when pre-training with audio spectrograms.
Comparing Audio-MAE reconstructions between Fig.~\ref{fig:vis:a} to~\ref{fig:vis:e} and \ref{fig:vis:d} to~\ref{fig:vis:h}, under the same masking ratio, we observe the unstructured random masking is comparably easier than structured masking (\ie, time and/or frequency) as the model can guess the missing component by extrapolating nearby context (\eg, formants in vowels and frictional sounds in consonants around). 
We also observe that for higher masking ratios, the structured masking alternatives drop in performance, presumably because the task becomes too difficult while random masking improves steadily up to 80\%.
This result show that designing a pretext task with \emph{proper hardness} is important for effective self-supervised learning of audio representations. 
%We find the \textit{easier} random masking optimal and use as default for pre-training.
We therefore use random masking with ratio of 80\% as our default for pre-training.

Fig.~\ref{fig:ft_masking_comparison} studies the effect of masking during the \textit{fine-tuning} phase. We see that in this case, it is more beneficial to use structured masking: time$+$frequency performs better than time- or frequency-based masking, and these perform better than unstructured masking. 
Overall, we see that the optimal masking ratios are \textit{lower} than for pre-training and we use 0.3 as our default in the fine-tuning phase.

In general, we observe that for task-agnostic pre-training, unstructured masking with a higher ratio is preferred. While in task-specific fine-tuning, structured masking with lower ratios performs better.

%\vspace{-0.5em}
\paragraph{Impact of Patch Size and Stride.}
We compare the performance of Audio-MAE trained with different patch sizes and strides in Table~\ref{tab:ablation:patch}. 
A non-zero overlap (\ie, stride $<$ patch size) between patches will increase the number of patches and quadratically increase computation in floating point operations (FLOPs), as reported in the table. 
Most prior works follow AST~\cite{gong2021ast} to use overlapped patches (patch~$=16$ and stride~$=10$) to boost end task performance.
As shown in Table~\ref{tab:ablation:patch}, 
we do not observe a performance improvement using overlapped patches for Audio-MAE (both 47.3 mAP), presumably because due to overlap, the patch embedding can leak information into the masked patches.
The non-overlapped 16$\times$16 patches achieve a good balance between computation and performance. 
By default, we use this setup in our experiments. 

%\vspace{-0.5em}
\paragraph{Encoder.}
We investigate the design choices of encoder and decoder architectures in Audio-MAE.
Table~\ref{tab:ablation:model_size} shows the trade-off between encoder model size and performance. 
As expected, larger models achieve better performance, at a cost of computation and memory. 
The accuracy gain of ViT-L over ViT-B/S is more significant on the smaller and balanced AS-20K.
For ViT-S, the performance gap to ViT-B can be significantly closed (5.0 $\rightarrow$ 2.3 mAP) when fine-tuning with more in-domain data (AS-20K $\rightarrow$ AS-2M).
% add small model discussion
% Audio-MAE with ViT-S is already pretty powerful (45.0 mAP on AS-2M)

\begin{wrapfigure}{tr}{6cm}
    %\vspace{-0.5em}
    \centering
    \includegraphics[width=\linewidth]{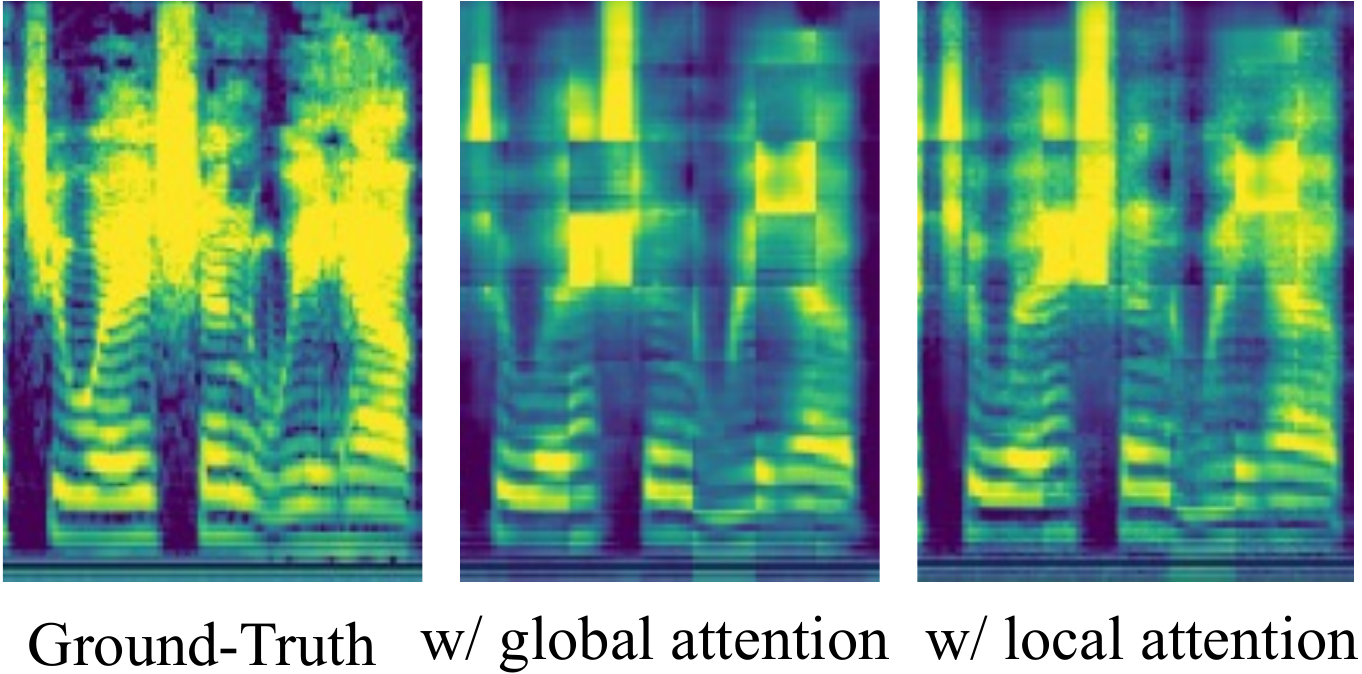}
    %\vspace{-1.2em}
    \caption{\small{Decoder reconstruction comparison.}
    \label{fig:reconstruction_comparison}
    }
    %\vspace{-1em}
\end{wrapfigure}

\paragraph{Decoder.}
Table~\ref{tab:ablation:attn_type} compares decoder attention types in Audio-MAE.
Note that decoders are discarded after pre-training and only the equal-sized ViT-B encoders are fine-tuned for the end task.
Our results show that \textit{local attention} with shifted window achieves the best performance. 
Combining local and global attention (\ie, hybrid attention, Hwin) also improves vanilla global self-attention.
Fig.~\ref{fig:reconstruction_comparison} shows the qualitative reconstruction comparison. 
In the spectrogram of vowels, the decoder with local attention reconstructs better harmonics and recovers more context in the spectrogram. Similar phenomena are observed in the frictional sound in the middle consonant.

Table~\ref{tab:ablation:decoder_depth} ablates the impact of decoder depth on mAP. 
A deeper 16-layer decoder achieves better performance against its shallower  variants. Note that our decoder uses local window attention by default where only a fraction of tokens (4$\times$4 local windows \textit{vs.} 64$\times$8 with global attention) are attended.  For global attention we find 8-layer decoders to perform better than 16-layer.
Table~\ref{tab:ablation:decoder_width} compares decoder width (embedding dimension). A 512-dimension decoder achieves a good trade-off between computation and performance as a wider one is not better.

\begin{table*}[t]\centering%\vspace{-0.5em}
    \small
   %\vspace{-5pt}
    \subfloat[\textbf{Patch size and stride}\label{tab:ablation:patch}]{  
        \vspace{-0.5em}
        \tablestyle{2pt}{1.05}
        \setlength\tabcolsep{2.0pt}
        \begin{tabular}{cccc}
            Patch size, stride & Seq shape & FLOPs & mAP \\
            \toprule
            \rowcolor{defaultcolor} (16,16), (16,16) & 64$\times$8 & 48.6 &\textbf{47.3} \\
            (16,16), (10,10) & 101$\times$12 & 130.5 & \textbf{47.3} \\
            (32,16), (16,16) & 63$\times$8 & 47.8 & 46.6 \\
            (16,32), (16,16) & 64$\times$7 & 42.1 & 46.8 \\
        \end{tabular}
    }
    \hspace{3mm}
    \subfloat[\textbf{Model size     (encoder)}\label{tab:ablation:model_size}]{
        \vspace{-0.5em}
        \makebox[0.5\linewidth][c]{
            \tablestyle{2pt}{1.05}
            \setlength\tabcolsep{1.0pt}
            \begin{tabular}{cccc}
                Backbone & \#Params & AS-20K & AS-2M  \\
                \toprule
                ViT-S &  22M & 32.1 & 45.0  \\
                \rowcolor{defaultcolor}ViT-B & 86M & 37.1 & 47.3  \\
                ViT-L & 304M & \textbf{37.6} & \textbf{47.4}  \\
                \multicolumn{4}{c}{} \\ % space holder
                %\multicolumn{4}{c}{} \\ % space holder
            \end{tabular}
        }
    }
    \\
    \vspace{10pt}
    \subfloat[\textbf{Decoder attention comparison}. Attn type\textsuperscript{(depth)}\label{tab:ablation:attn_type}]{
        \vspace{-0.5em}
        \setlength\tabcolsep{2.0pt}
        \tablestyle{2pt}{1.05}
        \begin{tabular}{ccccc}
            Attention type &  AS-20K & AS-2M & ESC-50 & SID \\
            \toprule
            Global\textsuperscript{(8)} (vanilla) & 36.6 & 46.8  & 93.6 & 94.1 \\
            \rowcolor{defaultcolor} Local\textsuperscript{(16)} (shifted) & \textbf{37.1} & \textbf{47.3}  & \textbf{94.1} & 94.8 \\
            Hwin (local\textsuperscript{(8)}+ global\textsuperscript{(4)}) & 36.8 & \textbf{47.3} & 93.8 & \textbf{95.0} \\
            % \multicolumn{4}{c}{} \\ % space holder
        \end{tabular}    
    }
    \subfloat[\textbf{Decoder depth}\label{tab:ablation:decoder_depth}]{
        \vspace{-0.5em}
        \makebox[0.2\linewidth][c]{
            \tablestyle{2pt}{1.05}
            \setlength\tabcolsep{2.0pt}
            \begin{tabular}{ccc}
                Depth & mAP \\
                \toprule
                2 & 46.8 \\
                %4 & 47.1 \\
                8 & 47.2 \\
                \rowcolor{defaultcolor} 16 & \textbf{47.3} \\
            \end{tabular}
        }
    }
    \subfloat[\textbf{Decoder width}\label{tab:ablation:decoder_width}]{
        \vspace{-0.5em}
        \makebox[0.2\linewidth][c]{
            \tablestyle{2pt}{1.05}
            \setlength\tabcolsep{1.0pt}
            \begin{tabular}{cc}
            Width  & mAP \\
            \toprule
            256 & 46.9 \\
            \rowcolor{defaultcolor} 512 & \textbf{47.3} \\ 
            768 & 47.3 \\
            %\multicolumn{2}{c}{} \\ % space holder
            \end{tabular}        
        }
    }
    \\ 
    \vspace{10pt}
    \subfloat[\textbf{Pre-training size}\label{tab:ablation:data_size}]{
        \vspace{-0.5em}
        \makebox[0.22\linewidth][c]{
            \small
            \setlength\tabcolsep{2.0pt}
            \tablestyle{2pt}{1.05}
            \begin{tabular}{lcc}
                \% of AS-2M & mAP\\
                \toprule
                1\% \tiny{(AS-20K)} & 39.4 \\
                1\% \tiny{(AS-2M)} & 39.6 \\
                10\% &  42.6 \\
                %20\% &  45.0 \\
                50\% &  46.4 \\
                \rowcolor{defaultcolor} 100\% &  \textbf{47.3}\\
            \end{tabular}    
        }
    }
    \subfloat[\textbf{Pre-training epoch}\label{tab:ablation:epoch}]{
        \vspace{-0.5em}        
        \makebox[0.25\linewidth][c]{
            \setlength\tabcolsep{2.0pt}
            \tablestyle{2pt}{1.05}
            \begin{tabular}{cc}
                epoch & mAP\\
                \toprule
                8 &  46.5 \\
                16 & 46.8 \\
                24 & 47.2 \\
                \rowcolor{defaultcolor} 32 & \textbf{47.3} \\
                40 & 47.3 \\
                %\multicolumn{2}{c}{} \\ % space holder
            \end{tabular}        
        }
    }
    \subfloat[\textbf{External ImageNet (IN) pre-training}. SSL: %initialization 
    w/ self-supervised MAE. SL: w/ supervised (fine-tuned) MAE.\label{tab:ablation:pt_dataset}]{
        \vspace{-0.5em}
        \tablestyle{2pt}{1.0}
        \makebox[0.5\linewidth][c]{
            \setlength\tabcolsep{2.0pt}
            \begin{tabular}{cccccc}
                scenario & IN-SSL & IN-SL & AS-SSL & AS-20K & AS-2M \\
                \toprule
                \rowcolor{defaultcolor} (1) & & & \checkmark & \textbf{37.1}~\tiny{(-0.0)} &\textbf{47.3} \tiny{(-0.0)} \\ %\hline
               (2) & \checkmark &  & & 32.1~\tiny{(\color{red}-5.0)}& \color{gray} 45.4 \tiny{(\color{red}-1.9)} \\
               (2) & \checkmark & \checkmark & & 32.5~\tiny{(\color{red}-4.6)} & \color{gray} 45.9 \tiny{(\color{red}-1.4)} \\
               (3) & \checkmark & & \checkmark & 36.9~\tiny{(\color{red}-0.2)}& \color{gray} 47.1 \tiny{(\color{red}-0.2)} \\
               (3) & \checkmark & \checkmark & \checkmark & 36.2~\tiny{(\color{red}-0.9)}& \color{gray} 46.9 \tiny{(\color{red}-0.4)} \\
                %\multicolumn{5}{c}{} \\ % space holder
                % \bottomrule
            \end{tabular}        
        }
    }
    \vspace{-.20em}
    \caption{\textbf{Ablation studies on  AS-2M}. The \colorbox{gray!30}{gray} entries are the default Audio-MAE setup (ViT-B encoder, decoder with shifted local attention, pre-trained for 32 epochs). Table format follows \cite{mae}. 
    \label{tab:ablation}}
    \vspace{-1.5em}
\end{table*}

%\vspace{-0.5em}
\paragraph{Pre-training Data and Setup.}
Table~\ref{tab:ablation:data_size} summarizes the impact of pre-training dataset size. 
Overall the model performance is monotonically increasing when using more data for pre-training.
Comparing the performance of using 1\% well-annotated AS-20K balanced data to using randomly sampled 20K unbalanced data for pre-training, the similar mAPs (39.4 vs 39.6) suggest that the \textit{distribution} of data classes (balanced vs.\ unbalanced) is \textit{less} important for pre-training.
Meanwhile, as shown in Table~\ref{tab:ablation:epoch},
training for longer is beneficial yet the performance saturates after the 24-\textit{th} epoch.

%\vspace{-0.5em}
\paragraph{Out-of-domain Pre-training on ImageNet.} 
Initializing audio models from ImageNet pre-trained weights has become popular for audio classification. 
However, as there are significant discrepancies between image and audio modalities, 
it is questionable if out-of-domain pre-training benefits audio representation learning.
In Table~\ref{tab:ablation:pt_dataset} we design 3 scenarios to investigate this for Audio-MAE:
(1) Audio-only pre-training (AS-SSL) from scratch. 
We consider this the ideal schema for learning audio representations as it is a simple and clean setup that prevents uncontrollable bias transfer from other modalities.
(2) Directly using self-supervised ImageNet MAE models (IN-SSL) and its fine-tuned variant (IN-SL).
(3) Audio-MAE self-supervised pre-training on top of these ImageNet weights.

The results show that (1) from-scratch \textit{audio-only} pre-training is the best.
For scenarios (2) and (3), we observe that ImageNet pre-training alone (2) is not sufficient (especially when the downstream data is smaller, AS-20K), and, in self-supervised pre-training on AudioSet, ImageNet initialization (3) does not help but degrades accuracy. 
Also in (3), supervised ImageNet pre-training (IN-SL) seems harmful.
Consequently, the result suggests that out-of-domain pre-training (\ie, ImageNet) is not helpful for Audio-MAE, possibly due to domain shift.

\begin{table}[t!]
\vspace{-1em}
\setlength\tabcolsep{3.0pt}
\small
\begin{center}
\begin{tabular}{llllllllll} 
% \toprule
%\multicolumn{2}{c}{}  &\multicolumn{7}{c}{Tasks}\\
\multicolumn{1}{l}{Model} & \multicolumn{1}{l}{Backbone} & PT-Data & AS-20K & AS-2M & ESC-50 & SPC-2 & SPC-1 & SID \\
\midrule
\multicolumn{3}{l}{\textbf{No pre-training}} & \multicolumn{5}{l}{}\\
ERANN~\cite{verbitskiy2021eranns} & CNN & -& - & 45.0 & 89.2 & - & -& - \\
PANN~\cite{kong2019panns} & CNN & - & 27.8 & 43.1 & 83.3 & 61.8 &- &-  \\
\hline & \\[-2ex]
\multicolumn{3}{l}{\textbf{In-domain self-supervised pre-training}}& \multicolumn{5}{l}{} \\
wav2vec 2.0~\cite{wav2vec2} & \footnotesize{Transformer} & LS & -  & - & - & - & 96.2\textsuperscript{*} & 75.2\textsuperscript{*}\\
HuBERT~\cite{Hsu2021HuBERTSS}  & \footnotesize{Transformer} & LS & - & - & - & - & 96.3\textsuperscript{*} & 81.4\textsuperscript{*} \\
%Data2vec~\cite{d2v}  & VIT-B& AS & 34.5 & - & - & - & - & -  \\
Conformer~\cite{srivastava2021conformer} & \footnotesize{Conformer} & AS & - & 41.1 & 88.0 &  -  & -  & -\\
SS-AST~\cite{ssast} & ViT-B & AS+LS & 31.0 & - & 88.8 & 98.0 & 96.0 & 64.3 \\
%\hline
\multicolumn{3}{l}{\textit{Concurrent MAE-based works}}& \multicolumn{5}{l}{}\\ 
MaskSpec~\cite{chong2022masked} &ViT-B & AS & 32.3 & 47.1 & 89.6 & 97.7 & - & - \\
MAE-AST~\cite{baade2022} & ViT-B & AS+LS & 30.6 & - & 90.0 & 97.9 & 95.8 & 63.3 \\
\rowcolor{gray!20}\textbf{Audio-MAE} (global)  & ViT-B & AS & 36.6\tiny{$\pm .11$} & 46.8\tiny{$\pm .06$} & 93.6\tiny{$\pm .11$} & \textbf{98.3}\tiny{$\pm .06$} & \textbf{97.6}\tiny$\pm .06$ & 94.1\tiny$\pm .06$ \\
\rowcolor{gray!30} \textbf{Audio-MAE} (local) & ViT-B & AS & \textbf{37.0}\tiny$\pm .11$ & \textbf{47.3}\tiny{$\pm.11$} & \textbf{94.1}\tiny{$\pm.10$}&  \textbf{98.3}\tiny{$\pm.06$} & 96.9\tiny$\pm.00$ & \textbf{94.8}\tiny$\pm.11$ \\
\hline & \\[-2ex]
\multicolumn{3}{l}{\textbf{Out-of-domain supervised pre-training}}& \multicolumn{5}{l}{} \\
\color{gray}PSLA~\cite{gong2021psla} & \color{gray}EffNet~\cite{efficient} & \color{gray}IN & \color{gray}31.9 & \color{gray}44.4 & - & \color{gray}96.3 & - & \color{gray}- \\
\color{gray}AST~\cite{gong2021ast} & \color{gray}DeiT-B & \color{gray}IN &   \color{gray}34.7 & \color{gray}45.9 & \color{gray}88.7 & \color{gray}98.1 & \color{gray}95.5 & \color{gray}41.1 \\ 
\color{gray}MBT~\cite{Nagrani21c} & \color{gray}ViT-B & \color{gray}IN-21K &  \color{gray}31.3 & \color{gray}44.3 & - & - & - & - \\
\color{gray}HTS-AT~\cite{chen2022hts} & \color{gray}Swin-B & \color{gray}IN & - & \color{gray}47.1 & \color{gray}97.0\textsuperscript{$\dagger$} & \color{gray}98.0 & - & - \\
\color{gray}PaSST~\cite{paast} & \color{gray}DeiT-B & \color{gray}IN & - & \color{gray}47.1 & \color{gray}96.8\textsuperscript{$\dagger$} & - & - & - \\
% \bottomrule
\end{tabular}
\vspace{5pt}
\caption{
\textbf{Comparison with other state-of-the-art models} on audio and speech classification tasks.
Metrics are mAP for AS and accuracy (\%) for ESC/SPC/SID.
For pre-training (PT) dataset, AS:AudioSet, LS:LibriSpeech, and IN:ImageNet. 
\textsuperscript{$\dagger$}: Fine-tuning results with additional supervised training on AS-2M. 
% check Table~\ref{tab:supp:ft_esc} in Supplementary for Audio-MAE\textsuperscript{$\dagger$}
We {\color{gray}gray-out} models pre-trained with external non-audio datasets (\eg, ImageNet).
Best single models in AS-2M are compared (no ensembles). \textsuperscript{*}: linear evaluation results from~\cite{yang21c_interspeech}.
}
\label{tab:sota}
\end{center}
%\vspace{-2em}
\end{table}

%\vspace{-0.75em}
\subsection{Comparison with the State-of-the-art }
%\vspace{-0.75em}

Table~\ref{tab:sota} compares Audio-MAE (with 3-run error bars) to prior state-of-the-art. We categorize the comparison into 3 groups. 
For fair comparison, our main benchmark is the models in the middle group with self-supervised pre-training on in-domain (audio) datasets (AudioSet and LibriSpeech). 
For reference we also list other models without pre-training (the top group) and other models with supervised pre-training on out-of-domain ImageNet (the bottom group), where the latter contains previous best systems on the datasets.

Pre-trained on AudioSet, Audio-MAE achieves the best performance across all tasks compared to other models with in-domain self-supervised pre-training. 
On AudioSet-20K, its 37.1 mAP significantly outperforms all other approaches including concurrent works and other models with out-of-domain pre-training.
On AudioSet-2M and ESC-50, our method also outperforms Conformer~\cite{srivastava2021conformer} and SS-AST~\cite{ssast}. 
Notably, unlike SS-AST and concurrent MAE-AST~\cite{baade2022}, which trained with additional 1,000 hours of speech in Librispeech, we use only AudioSet for pre-training.

%%% comparison to supervised models
In the bottom group of Table~\ref{tab:sota}, Audio-MAE also outperforms previous state-of-the-art models with ImageNet supervised pre-training. 
Note that the proposed Audio-MAE does not rely on any out-of-domain data and labels, nor using knowledge distillation (\eg, DeiT) from additional CNN-based models. 
Also, compared to HTS-AT~\cite{chen2022hts} and PaSST~\cite{paast}, Audio-MAE is trained with audio under 16K sampling rate. As experimented in~\cite{kong2019panns}, there could be up to 0.4 potential mAP improvement for Audio-MAE if audio with 32K sampling rate are available.

%%%%%% Speech tasks
For the speech tasks (SPC-1, SPC-2, and SID), Audio-MAE  outperforms other models without pre-training (ERANN~\cite{verbitskiy2021eranns}, PANN~\cite{kong2019panns}), supervised (AST) and self-supervised models (SS-AST, MAE-AST).
%Unlike the findings in SS-AST and MAE-AST, we do not observe benefits of pre-training with frame/time masking for speech tasks but find pre-training with unstructured random masking generalizes across end tasks (for comparison, see Supplementary).
We further list other works (marked with~\textsuperscript{*}) to include the latest results introduced in the SUPERB~\cite{yang21c_interspeech} benchmark. But note that these results are not strictly comparable since SUPERB employs linear evaluation where the underlying pre-trained models are not end-to-end fine-tuned.

In summary, with audio-only from-scratch pre-training on AudioSet, our Audio-MAE performs well for both the audio and speech classification tasks.

%\vspace{-0.5em}
\subsection{Visualization and Audible Examples by Audio-MAE Decoder}
%\vspace{-0.5em}

For better visualization, we follow MAE~\cite{mae} to use MSE over non-normalized spectrograms as the self-supervised objective. We use ViT-L as the Audio-MAE encoder for visualization.
Fig.~\ref{fig:vis} illustrates the reconstruction results sampled from the AudioSet-2M \textit{eval} set.
We further reconstruct~\emph{.wav}s using the Griffin-Lim~\cite{griffin_lim} algorithm, audible under the anonymous links (accessible in respective {\color{hrefcolor} 1 2 3}).

As can be seen and heard, 
for various masking strategies and different sounds, our Audio-MAE generates reasonable reconstruction.
It works well for noisy event sounds 
(\eg, the reconstructed siren in Fig.~\ref{fig:vis:c}-\href{https://www.dropbox.com/s/avpg02hjikw8o6e/02Ak1eIyj3M_0.7_restored.mp4?dl=0}{3}), 
as well as speech and music 
(\eg, the reconstructed singing in Fig.~\ref{fig:vis:b}-\href{https://www.dropbox.com/s/onmdvpgykphlfx6/DB38NRSHw9A_0.7_restored.mp4?dl=0}{3}).
Notably, unlike visual contents that are typically scale/translation/position invariant~\cite{swin}, absolute positions and arrangement of spectrogram components are critical for humans to understand  sound~\cite{suzuki2004equal}. 
For example, shifting a pitch will make an audio sounds completely different. Also, phoneme sequences in time are important cues for speech understanding. 
Consequently,
unstructured masking produces better aligned outputs that are closer to the ground-truth (top row in each subfigure) as the model can make better predictions based on nearby spectrogram patches; while structured masking is harder (less accurate or with words missing), especially when masking is performed over the time axis.
A failure example (missing words) is the reconstructed speech in Fig.~\ref{fig:vis:e}-\href{https://www.dropbox.com/s/712o434428i4kdv/tzJH5FfR9j8_0.7_2d_restored.mp4?dl=0}{3}.

% \vspace{-1em}
\section{Conclusion\label{sec:conclusion}}
 %\vspace{-1em}
We have explored a simple extension of MAE~\cite{mae} to audio data. 
Our Audio-MAE learns to reconstruct masked spectrogram patches from audio recordings and achieves state-of-the-art performance on six audio and speech classification tasks. 
We have drawn four interesting observations: First, a simple MAE approach works surprisingly well for audio spectrograms. 
Second, we find that it is possible to learn stronger representations with local self-attention in the decoder. % proper domain knowledge regarding audio spectrograms. 
Third, we show that masking can be applied to both pre-training and fine-tuning, improving accuracy and reducing training computation. The optimal strategy depends on the nature of the data (audio, image, \etc) and the learning type (self-/supervised).
Fourth, the best performance can be achieved by pre-training and fine-tuning under the same modality, without reliance on cross-modality transfer learning.
In future work, we aim to explore multimodal self-supervised learning with a joint audio-visual MAE approach as these domains share natural correspondences in video data. 

\paragraph{Acknowledgements.} We thank Kaiming He and Luke Zettlemoyer for their feedback and discussions.

\bibliographystyle{IEEEtran_bernie}
\bibliography{main}

%\newpage
\vspace{2em}

\section*{Checklist}

%%% BEGIN INSTRUCTIONS %%%
%The checklist follows the references.  Please
%read the checklist guidelines carefully for information on how to answer these
%questions.  For each question, change the default \answerTODO{} to \answerYes{},
%\answerNo{}, or \answerNA{}.  You are strongly encouraged to include a {\bf
%justification to your answer}, either by referencing the appropriate section of
%your paper or providing a brief inline description.  For example:
%\begin{itemize}
%  \item Did you include the license to the code and datasets? \answerYes{See Section~\ref{gen_inst}.}
%  \item Did you include the license to the code and datasets? \answerNo{The code and the data are proprietary.}
%  \item Did you include the license to the code and datasets? \answerNA{}
%\end{itemize}
%Please do not modify the questions and only use the provided macros for your
%answers.  Note that the Checklist section does not count towards the page
%limit.  In your paper, please delete this instructions block and only keep the
%Checklist section heading above along with the questions/answers below.
%%% END INSTRUCTIONS %%%

\begin{enumerate}

\item For all authors...
\begin{enumerate}
  \item Do the main claims made in the abstract and introduction accurately reflect the paper's contributions and scope?
    \answerYes{}
  \item Did you describe the limitations of your work?
    \answerYes{} Ans: Please refer to the limitation discussion in the supplemental material.
  \item Did you discuss any potential negative societal impacts of your work?
    \answerNA{}
  \item Have you read the ethics review guidelines and ensured that your paper conforms to them?
    \answerYes{}. Ans: Our paper conforms the ethics requirement.
\end{enumerate}

\item If you are including theoretical results...
\begin{enumerate}
  \item Did you state the full set of assumptions of all theoretical results?
    \answerNA{}
        \item Did you include complete proofs of all theoretical results?
    \answerNA{}
\end{enumerate}

\item If you ran experiments...
\begin{enumerate}
  \item Did you include the code, data, and instructions needed to reproduce the main experimental results (either in the supplemental material or as a URL)?
    Data: \answerYes{}- the datasets we used are publicly available~\S\ref{sec:exp:dataset}. 
    Code and model: \answerYes{}- the code and pre-trained model will be released at the url specified in the Abstract.
  \item Did you specify all the training details (e.g., data splits, hyperparameters, how they were chosen)?
    \answerYes{} Ans: part of them are in ~\S\ref{sec:exp:impl} and ~\S\ref{sec:exp:setup} and the rest are specified in the Appendix.
    \item Did you report error bars (e.g., with respect to the random seed after running experiments multiple times)? 
    \answerYes{} Ans: Please see Table~\ref{tab:sota}.
        \item Did you include the total amount of compute and the type of resources used (e.g., type of GPUs, internal cluster, or cloud provider)?
    \answerYes{} Ans: please check ~\S\ref{sec:exp:setup}
\end{enumerate}

\item If you are using existing assets (e.g., code, data, models) or curating/releasing new assets...
\begin{enumerate}
  \item If your work uses existing assets, did you cite the creators?
    \answerYes{}
  \item Did you mention the license of the assets?
    \answerYes{}
  \item Did you include any new assets either in the supplemental material or as a URL?
    \answerYes{}
  \item Did you discuss whether and how consent was obtained from people whose data you're using/curating?
    \answerNA{}
  \item Did you discuss whether the data you are using/curating contains personally identifiable information or offensive content?
    \answerNA{}
\end{enumerate}

\item If you used crowdsourcing or conducted research with human subjects...
\begin{enumerate}
  \item Did you include the full text of instructions given to participants and screenshots, if applicable?
    \answerNA{}
  \item Did you describe any potential participant risks, with links to Institutional Review Board (IRB) approvals, if applicable?
    \answerNA{}
  \item Did you include the estimated hourly wage paid to participants and the total amount spent on participant compensation?
    \answerNA{}
\end{enumerate}

\end{enumerate}

\vspace{2em}
\appendix

\section*{Appendix}
The appendix is organized as follows: 
In \S\ref{sec:app:vis}, we first demonstrate additional audible visualizations with anonymous URL links. 
In \S\ref{sec:app:hyper}, we provide the complete experimental details and hyperparameter configurations for pre-training and fine-tuning on each dataset.
Then in~\S\ref{sec:app:exp}, we conduct extra experiments on ESC-50 (\S\ref{sec:app:exp:esc}) with additional supervised pre-training on AudioSet to complete the comparison with the models marked with \textsuperscript{$\dagger$} in Table 2 of the main paper.
We then study a case how Audio-MAE could be applied to a practical speech generation task (\S\ref{sec:app:exp:plc}); and share some negative results and insights on directions we tried that did not work well (\S\ref{sec:app:exp:negative}). 
Finally, we discuss the limitations (\S\ref{sec:app:limitations}) 
%and potential negative social impacts (\S\ref{sec:app:impact}) 
of Audio-MAE.

%Please also find the attached \textit{.mp4} files for corresponding visualizations of Fig. 6 in the main paper.

\afterpage{%
\begin{figure}[h!]
% \vspace{-15pt}
    \centering
    \begin{subfigure}[b]{0.245\linewidth}
        \includegraphics[width=0.98\linewidth,height=0.4\linewidth]{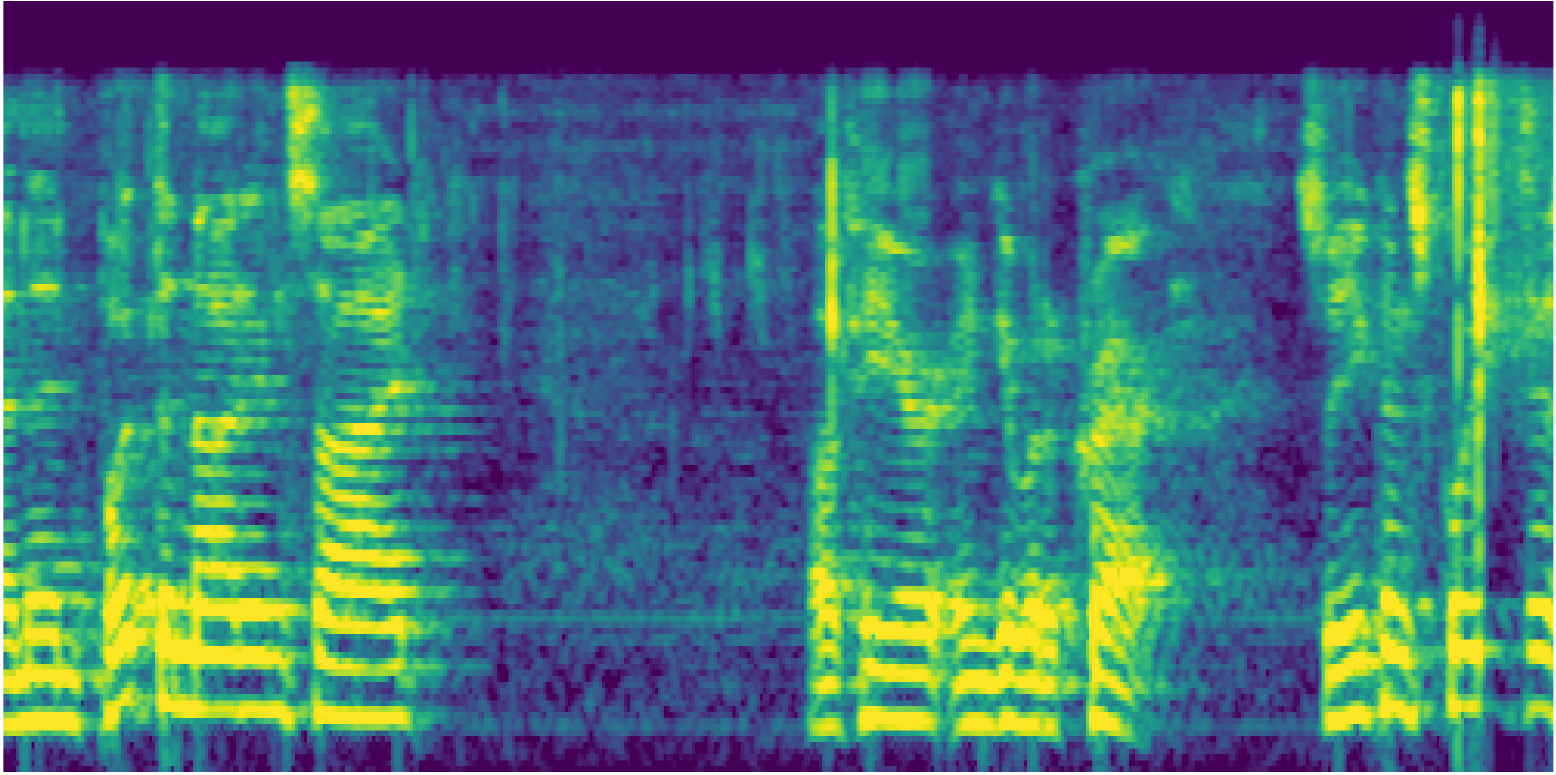}
    \end{subfigure}
    \begin{subfigure}[b]{0.245\linewidth}
        \includegraphics[width=0.98\linewidth,height=0.4\linewidth]{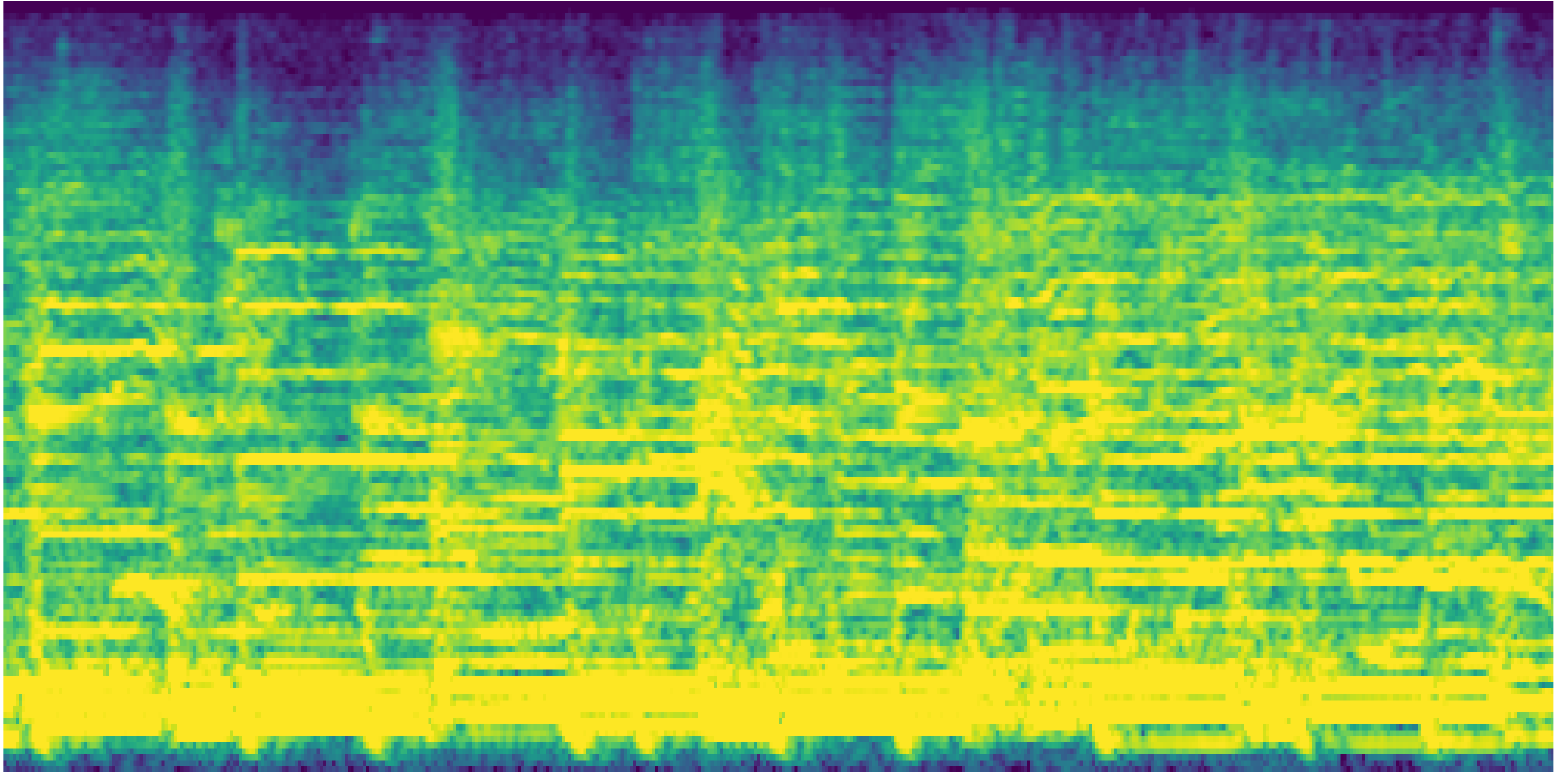}
    \end{subfigure}     
    \begin{subfigure}[b]{0.245\linewidth}
        \includegraphics[width=0.98\linewidth,height=0.4\linewidth]{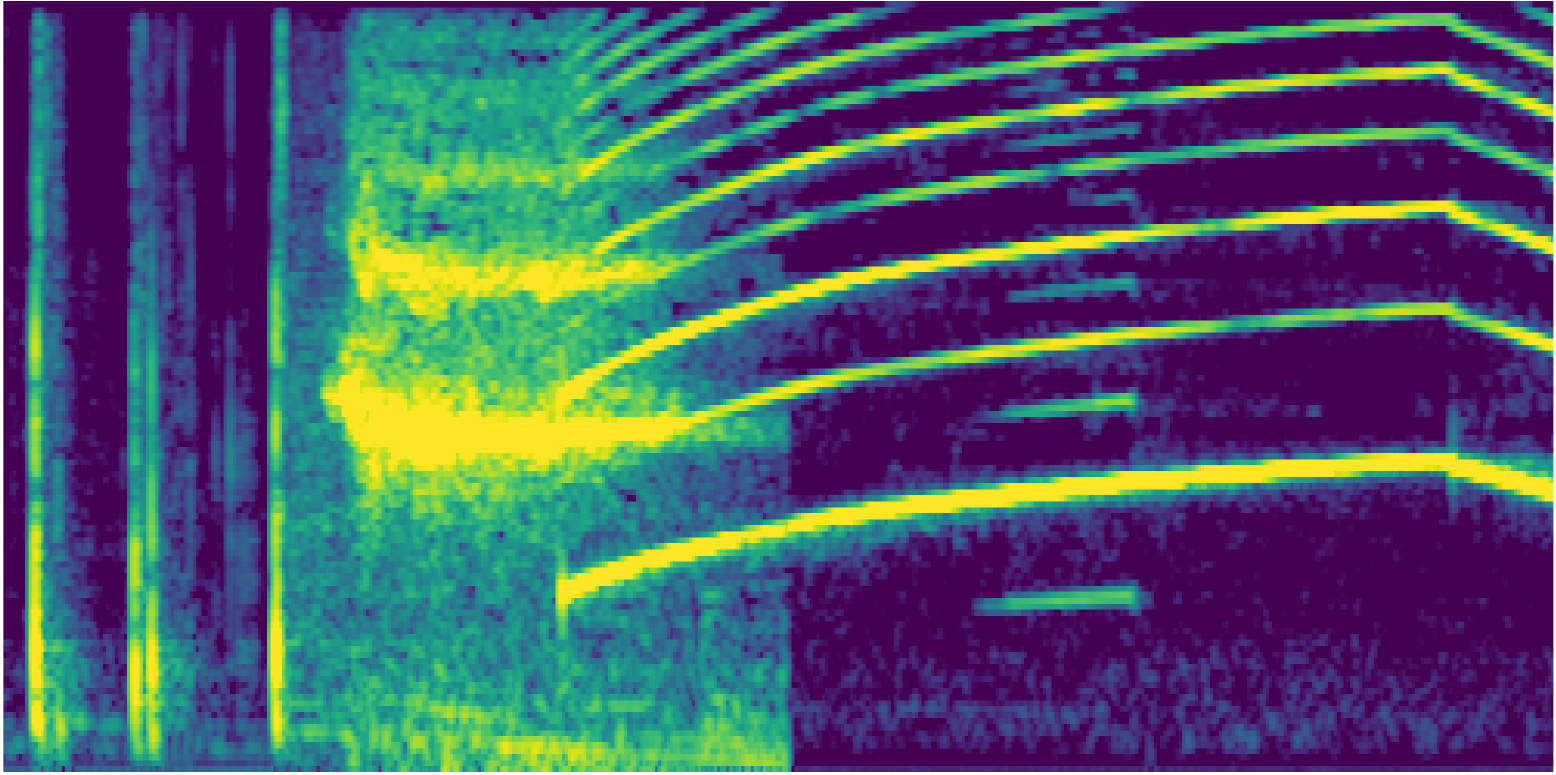}
    \end{subfigure}
    \begin{subfigure}[b]{0.245\linewidth}
        \includegraphics[width=0.98\linewidth,height=0.4\linewidth]{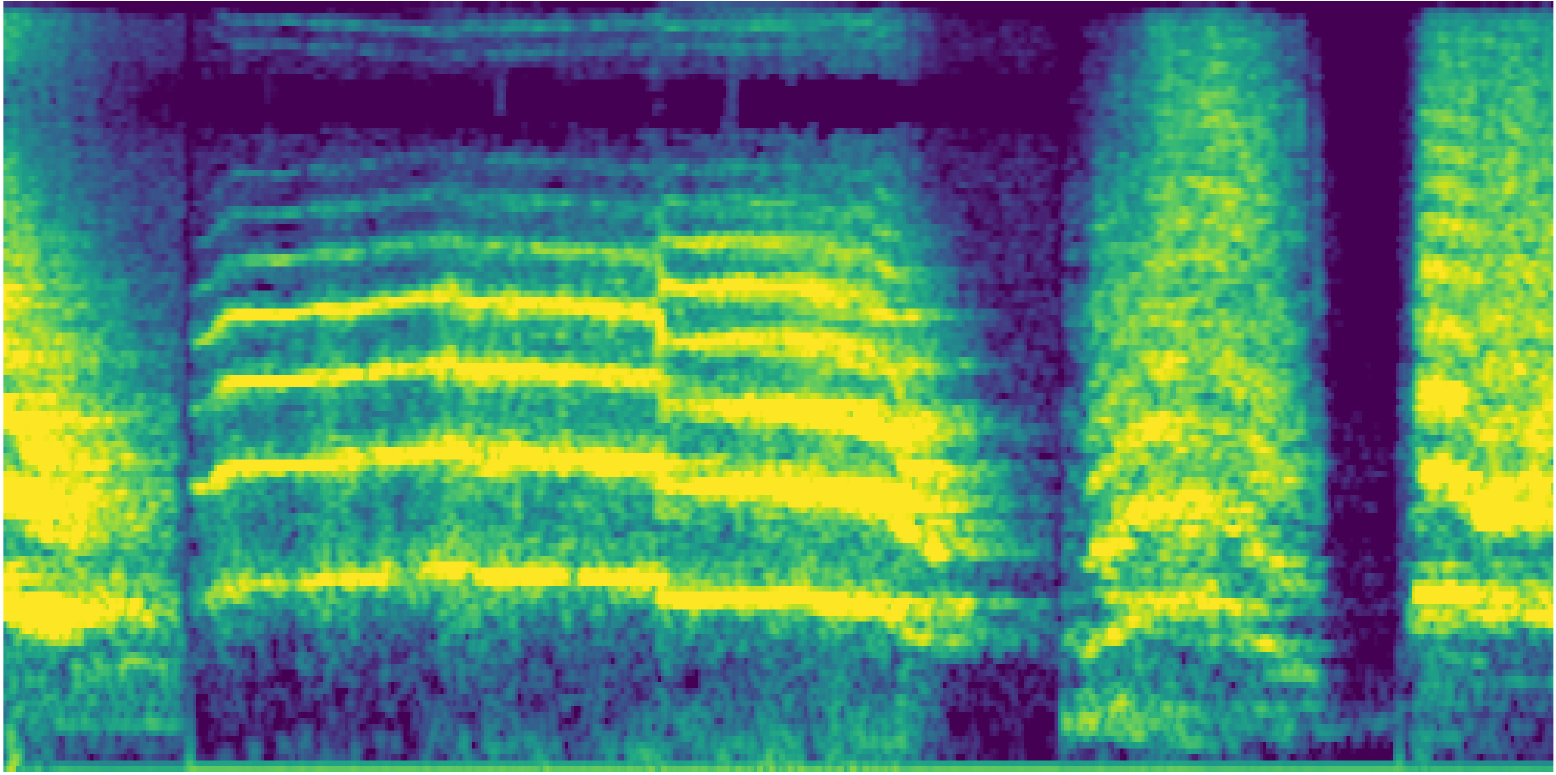}
    \end{subfigure}
    \\
    \begin{subfigure}[b]{0.245\linewidth}
        \includegraphics[width=0.98\linewidth,height=0.4\linewidth]{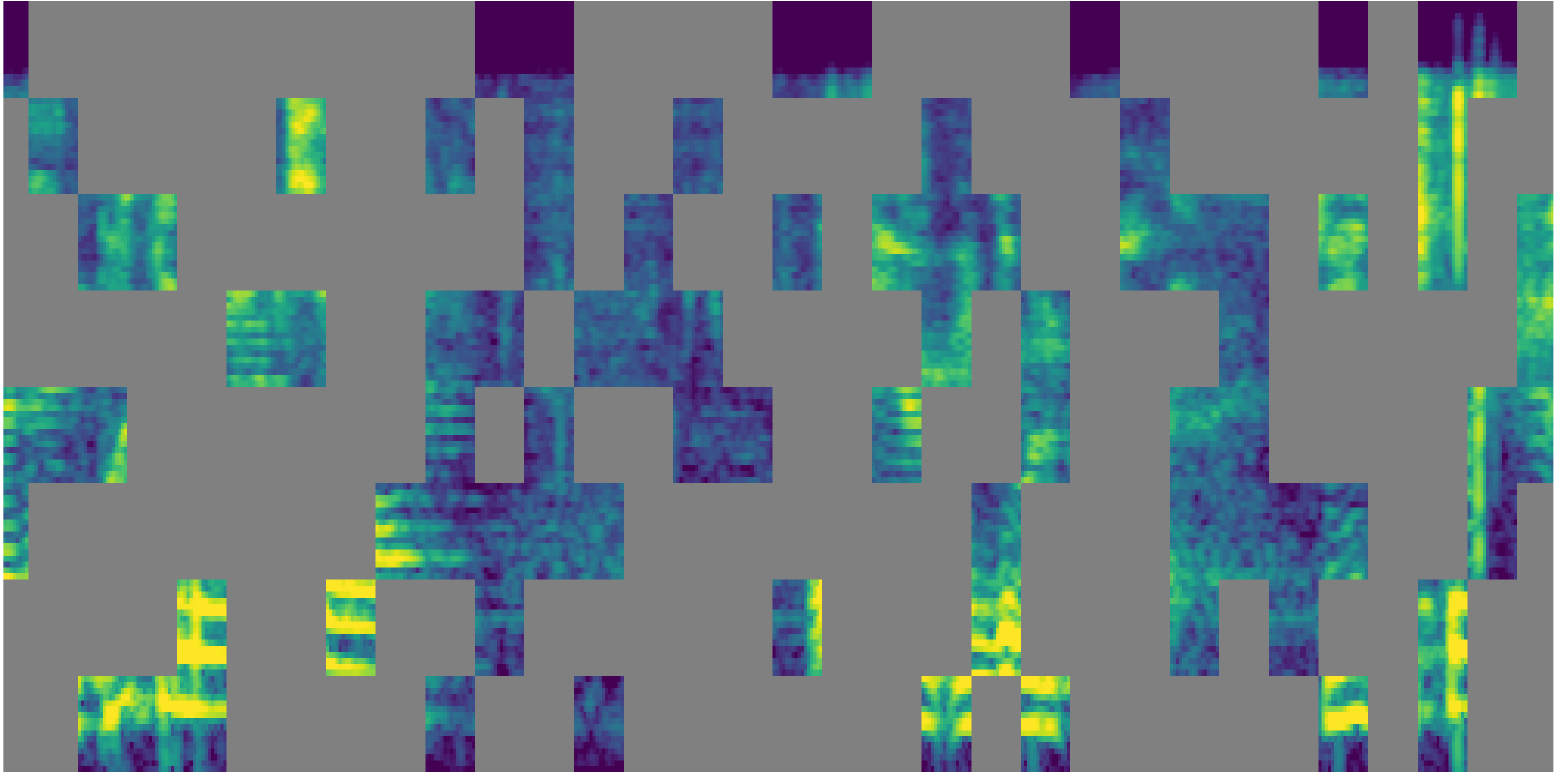}
    \end{subfigure}
    \begin{subfigure}[b]{0.245\linewidth}
        \includegraphics[width=0.98\linewidth,height=0.4\linewidth]{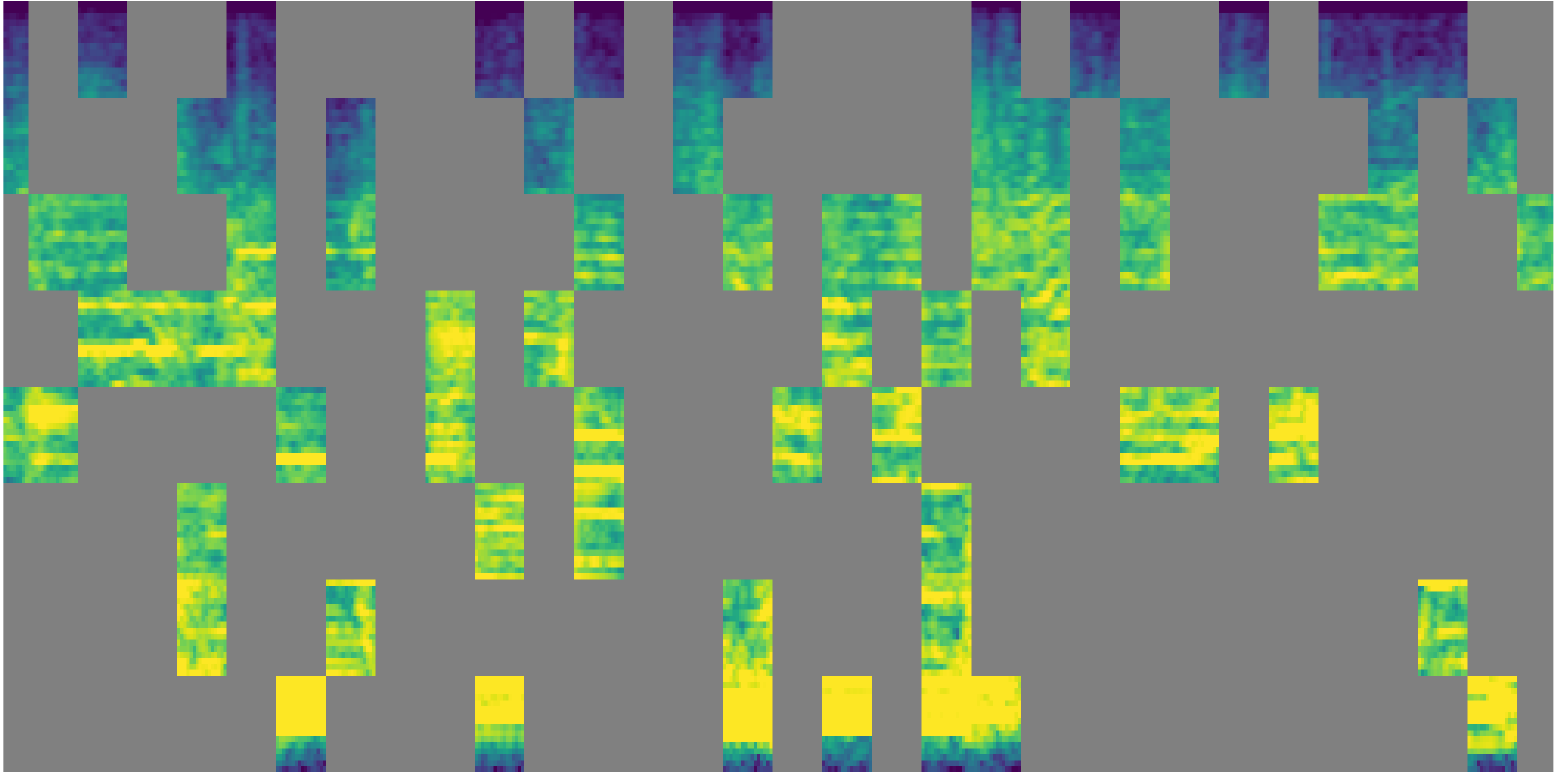}
    \end{subfigure}     
    \begin{subfigure}[b]{0.245\linewidth}
        \includegraphics[width=0.98\linewidth,height=0.4\linewidth]{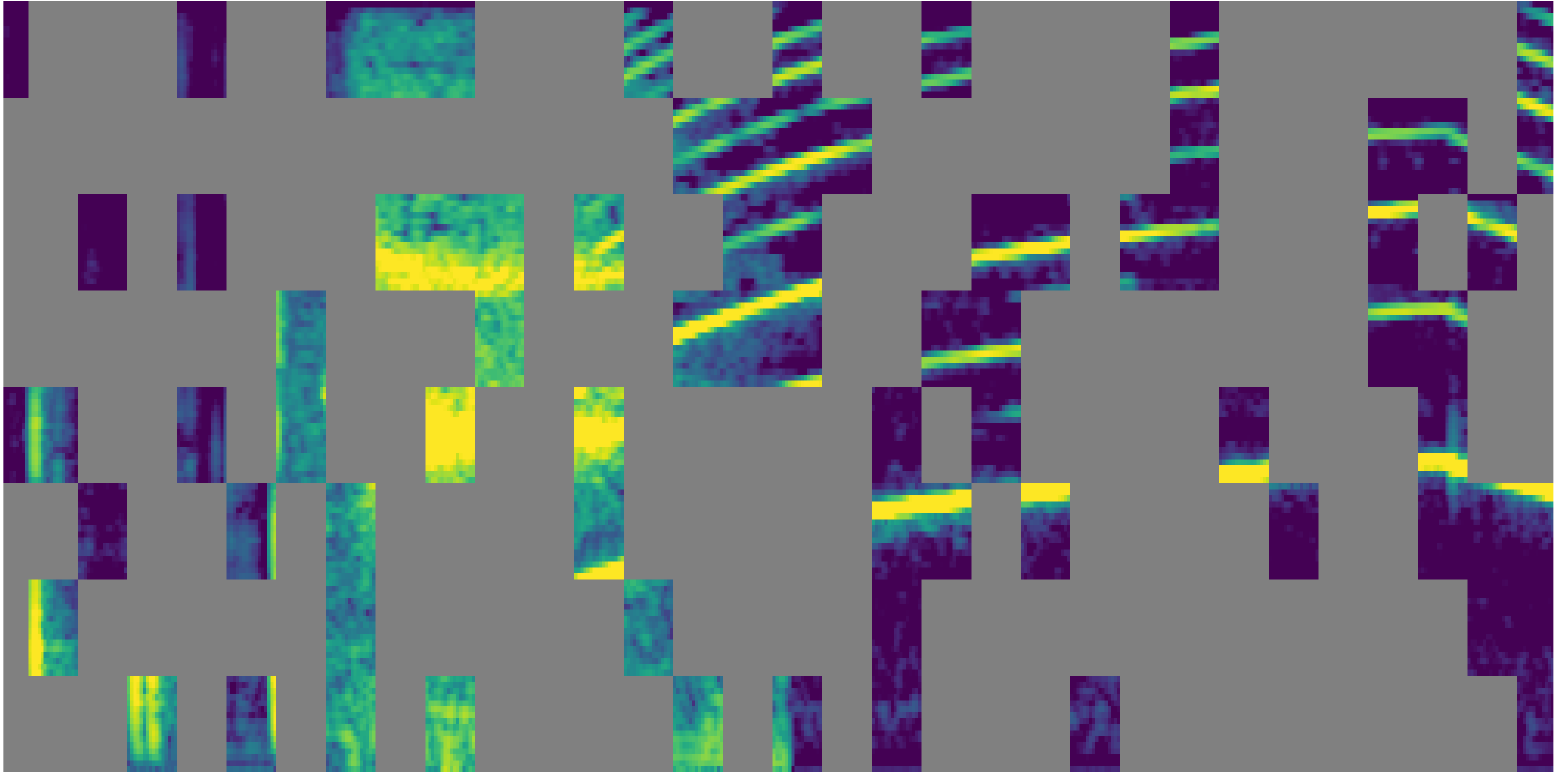}
    \end{subfigure}
    \begin{subfigure}[b]{0.245\linewidth}
        \includegraphics[width=0.98\linewidth,height=0.4\linewidth]{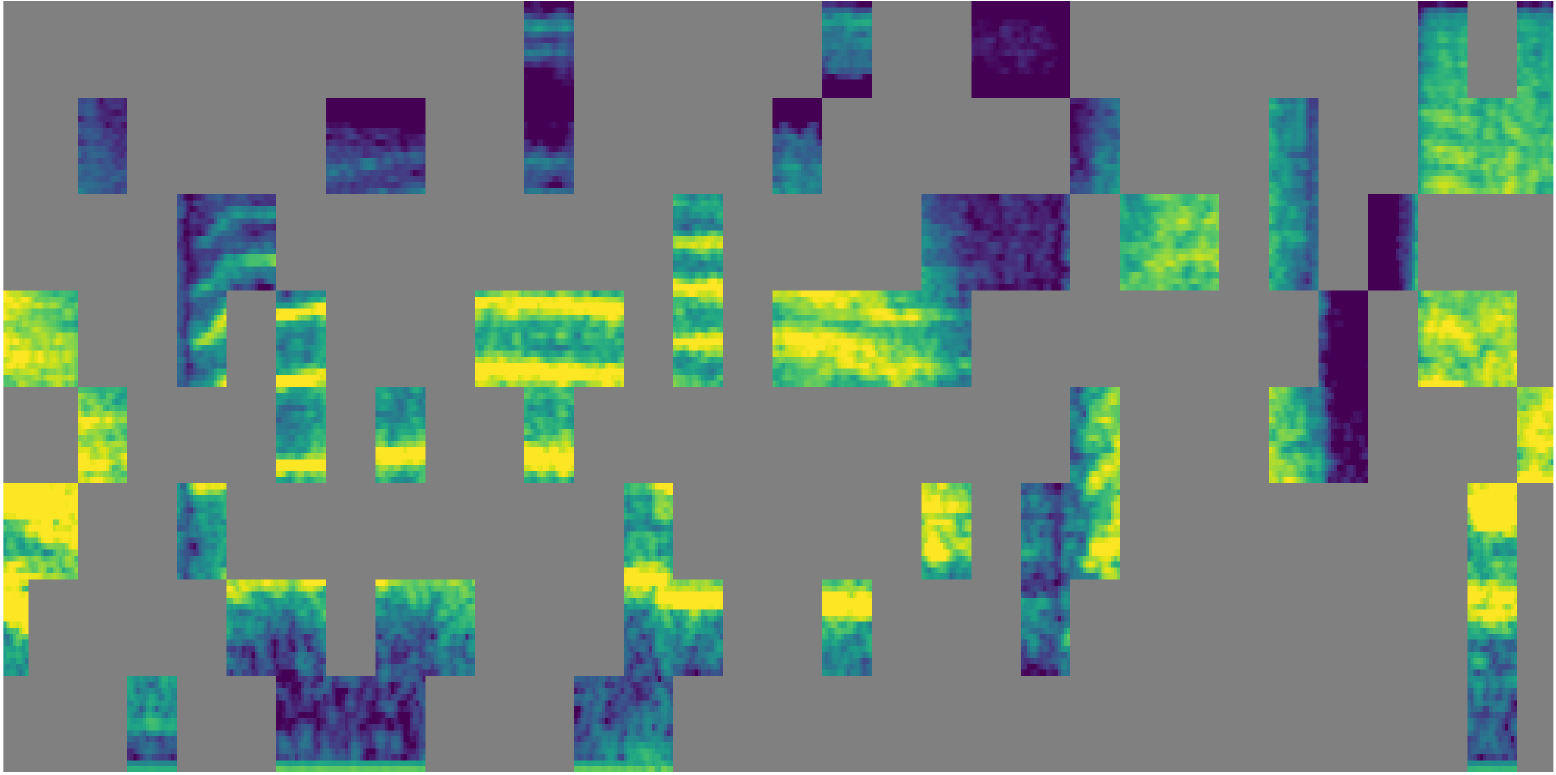}
    \end{subfigure}
    \\ 
    \begin{subfigure}[b]{0.245\linewidth}
        \includegraphics[width=0.98\linewidth,height=0.4\linewidth]{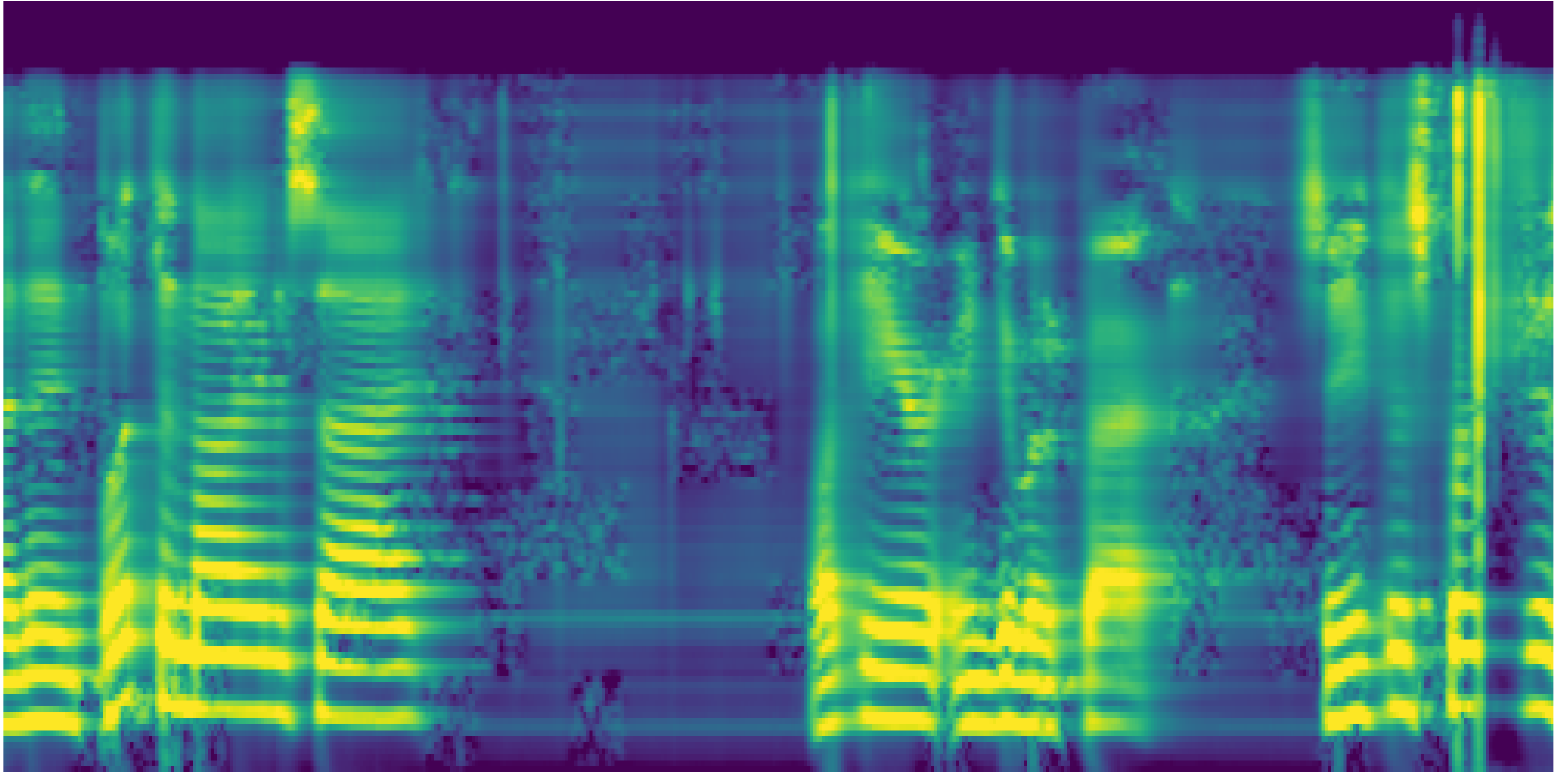}
        %\vspace{-0.2em}
        \subcaption{
            70\% Unstructured
            %\href{https://www.dropbox.com/s/kbmo6wfl1rhnx5b/wSw8k1qysoE_0.8_org.mp4?dl=0}{1}
            %\href{https://www.dropbox.com/s/rnpwvjlkbofsb8t/wSw8k1qysoE_0.8_masked.mp4?dl=0}{2}
            %\href{https://www.dropbox.com/s/7afd9ey9ois6c6n/wSw8k1qysoE_0.8_restored.mp4?dl=0}{3}
            \href{https://www.dropbox.com/s/st9ho31unybmd8k/wSw8k1qysoE_0.7_org.mp4?dl=0}{1}
            \href{https://www.dropbox.com/s/ysehu67folxwywf/wSw8k1qysoE_0.7_masked.mp4?dl=0}{2}
            \href{https://www.dropbox.com/s/owrfj5qm6azaa9s/wSw8k1qysoE_0.7_restored.mp4?dl=0}{3}            
        }
        \label{fig:app:vis:a}
    \end{subfigure} 
    \begin{subfigure}[b]{0.245\linewidth}
        \includegraphics[width=0.98\linewidth,height=0.4\linewidth]{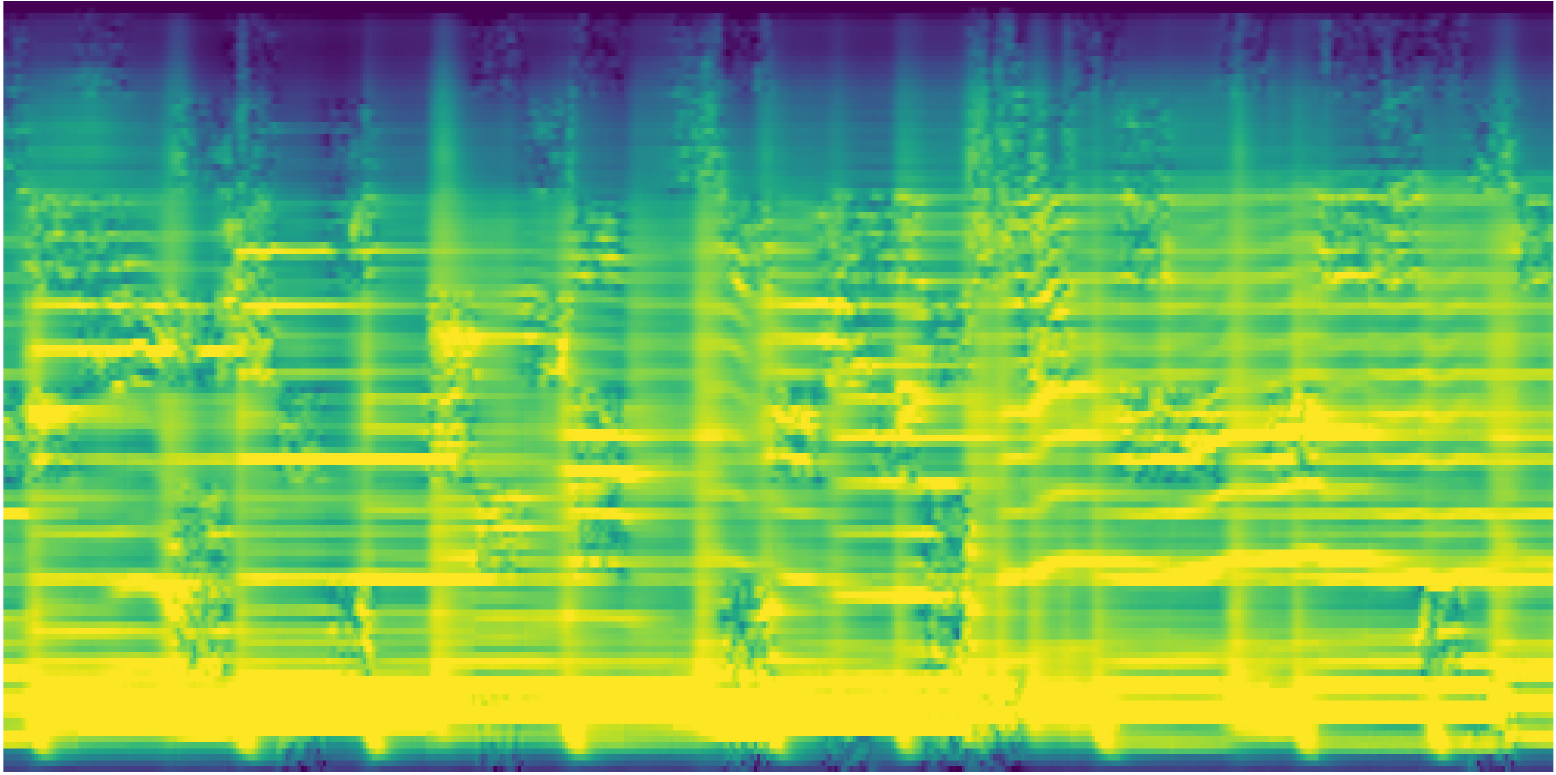}
        %\vspace{-0.2em}
        \subcaption{
            70\% Unstructured
            \href{https://www.dropbox.com/s/q44uoqokfqv243d/zWJC_qr2610_0.7_org.mp4?dl=0}{1}
            \href{https://www.dropbox.com/s/ovxu7dmp3joav4k/zWJC_qr2610_0.7_masked.mp4?dl=0}{2}
            \href{https://www.dropbox.com/s/vblj6tqdoj8t8mk/zWJC_qr2610_0.7_restored.mp4?dl=0}{3}            
        }
        \label{fig:app:vis:b}
    \end{subfigure}    
    \begin{subfigure}[b]{0.245\linewidth}
        \includegraphics[width=0.98\linewidth,height=0.4\linewidth]{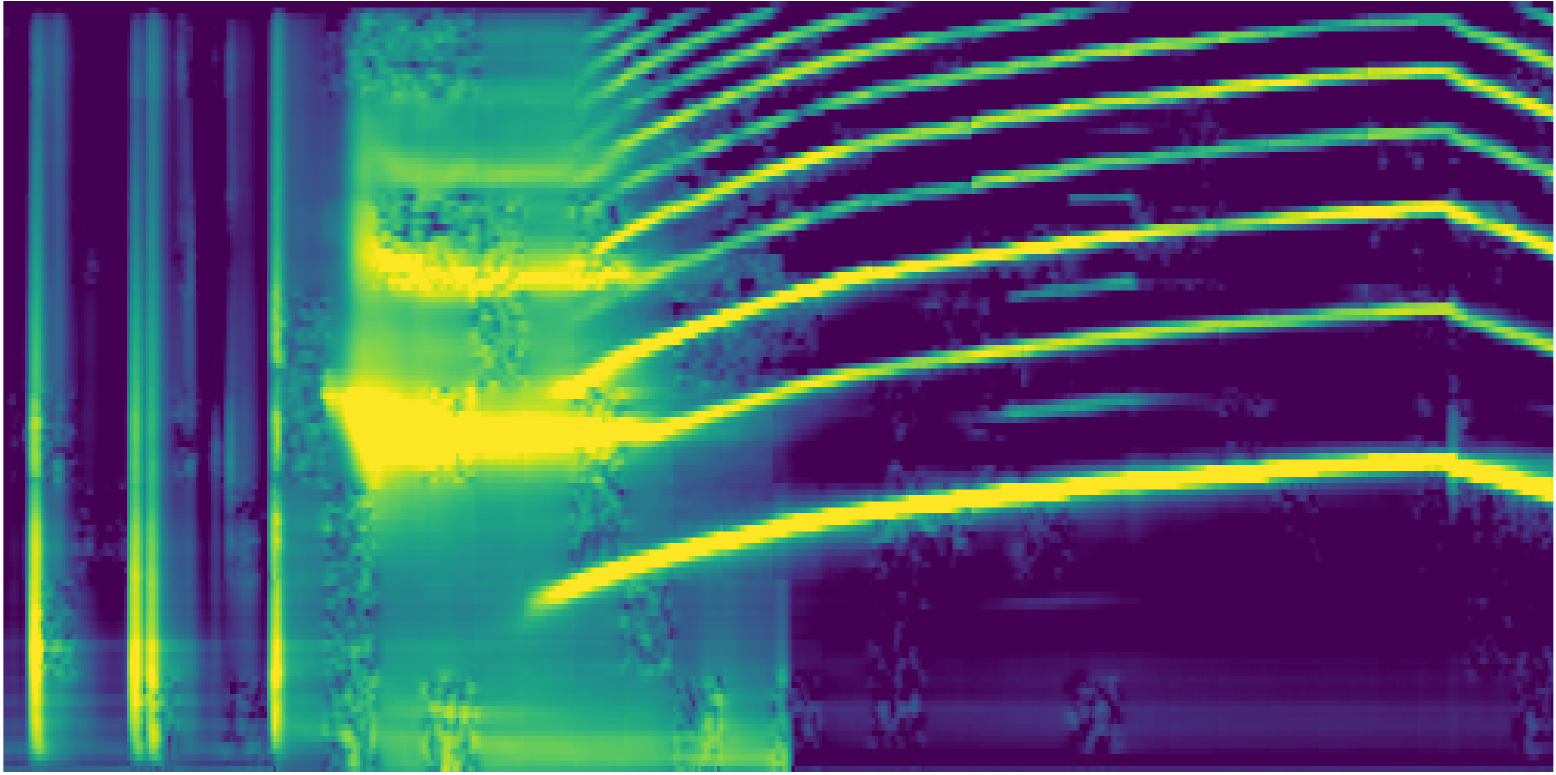}
        %\vspace{-0.2em}
        \subcaption{
            70\% Unstructured 
            \href{https://www.dropbox.com/s/saa4p61rfqnpp9v/oZoJ26C6LrU_0.7_org.mp4?dl=0}{1}
            \href{https://www.dropbox.com/s/6f2az0aean4kqr2/oZoJ26C6LrU_0.7_masked.mp4?dl=0}{2}
            \href{https://www.dropbox.com/s/uklfvgfpb9ulyqs/oZoJ26C6LrU_0.7_restored.mp4?dl=0}{3}            
        }
        \label{fig:app:vis:c}
    \end{subfigure}
    \begin{subfigure}[b]{0.245\linewidth}
        \includegraphics[width=0.98\linewidth,height=0.4\linewidth]{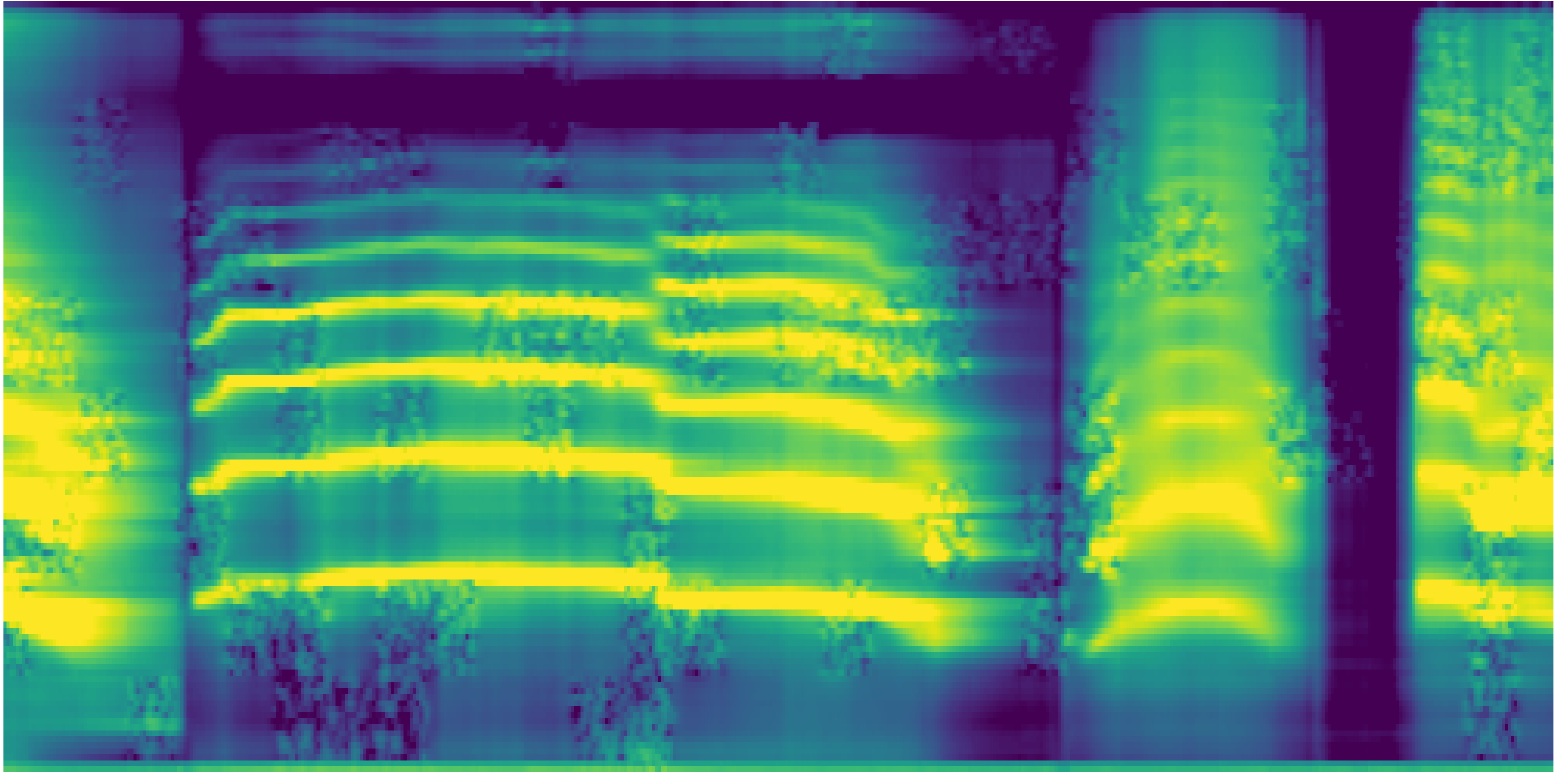}
        %\vspace{-0.2em}
        \subcaption{
            70\% Unstructured
            \href{https://www.dropbox.com/s/tb35xirgp4op0an/153xbn1k2H8_0.7_org.mp4?dl=0}{1}
            \href{https://www.dropbox.com/s/w95cuk3k86yj9uj/153xbn1k2H8_0.7_masked.mp4?dl=0}{2}
            \href{https://www.dropbox.com/s/rs8rzaocsme5gtm/153xbn1k2H8_0.7_restored.mp4?dl=0}{3}            
        }
        \label{fig:app:vis:d}
    \end{subfigure}
    \\ \vspace{5pt}
    
    %%%%%%%%%%%%%%%%%%%%%%%%% 2nd Row %%%%%%%%%%%%%%%%%%%%%%%%%%%%%%%%%  

    \begin{subfigure}[b]{0.245\linewidth}
        \includegraphics[width=0.98\linewidth,height=0.4\linewidth]{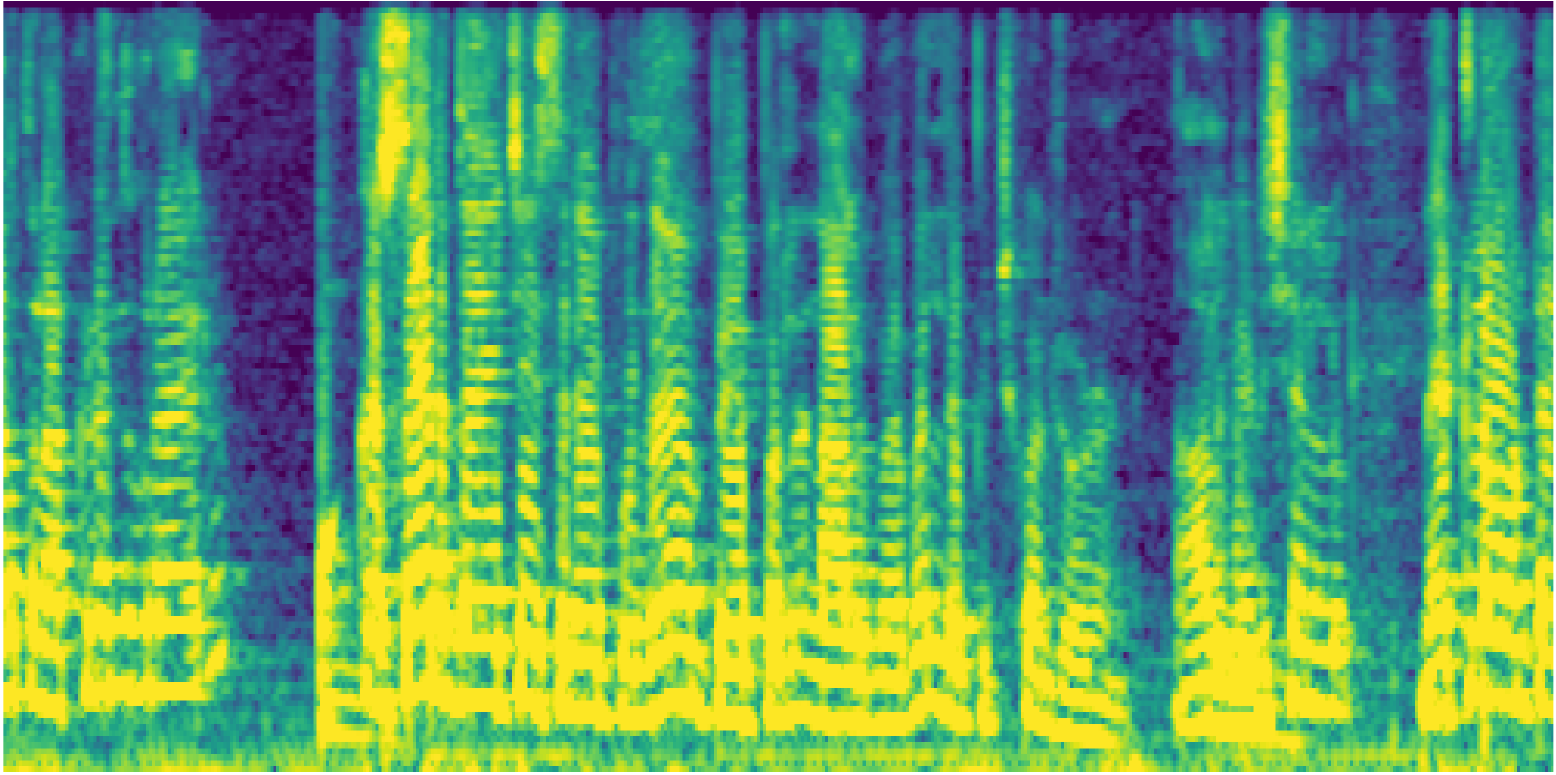}
    \end{subfigure}     
    \begin{subfigure}[b]{0.245\linewidth}
        \includegraphics[width=0.98\linewidth,height=0.4\linewidth]{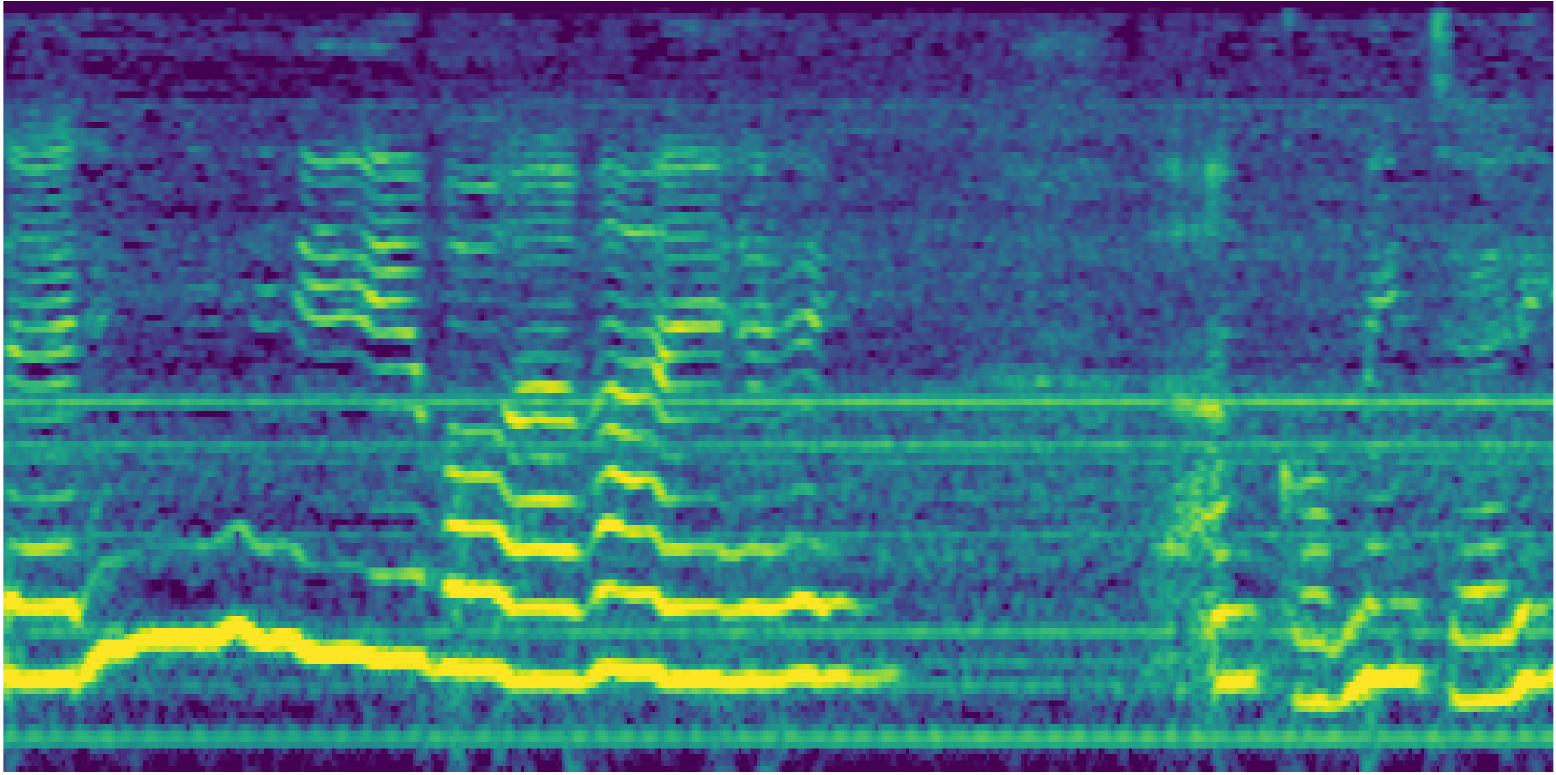}
    \end{subfigure}    
    \begin{subfigure}[b]{0.245\linewidth}
        \includegraphics[width=0.98\linewidth,height=0.4\linewidth]{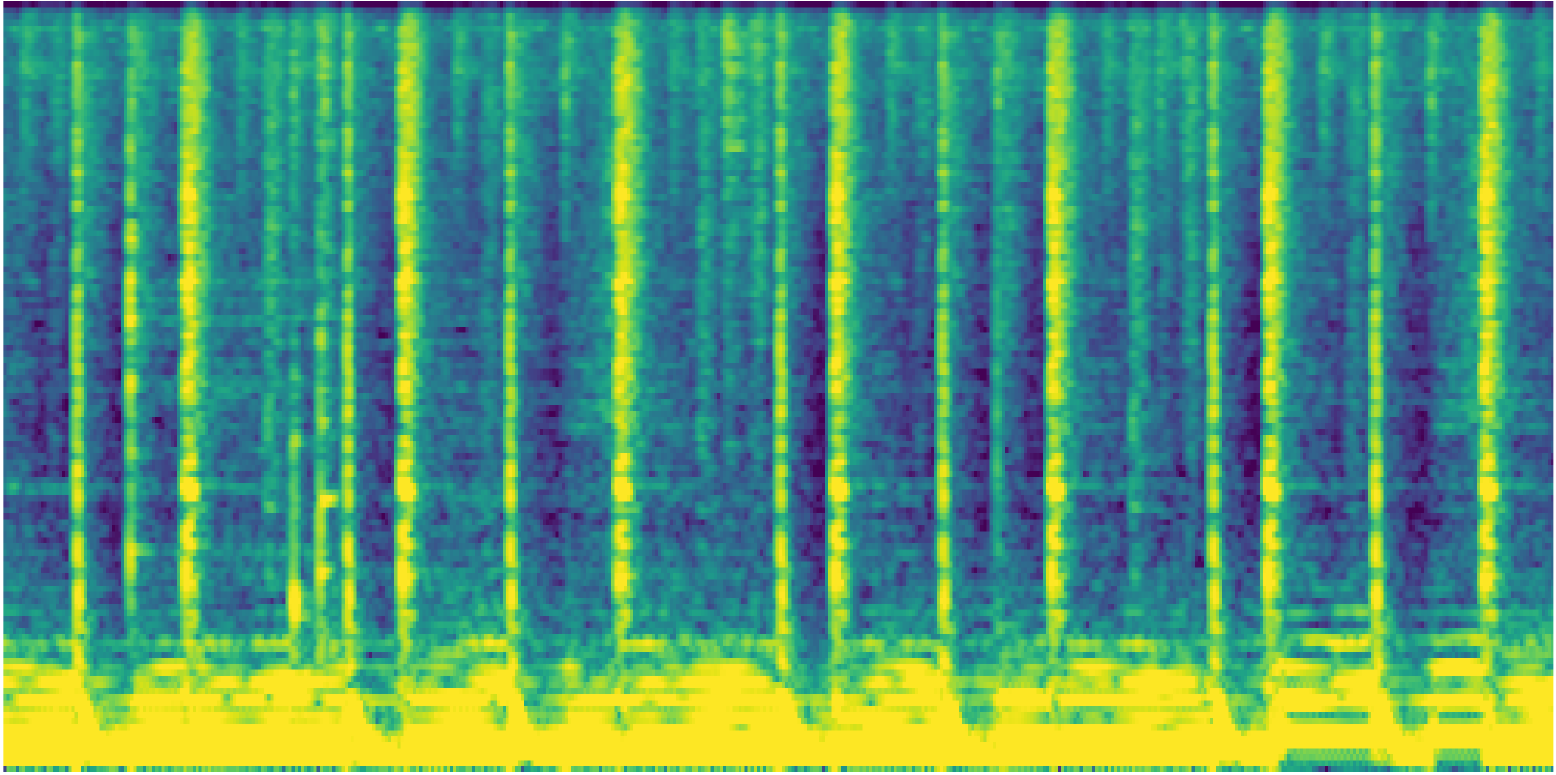}
    \end{subfigure}
    \begin{subfigure}[b]{0.245\linewidth}
        \includegraphics[width=0.98\linewidth,height=0.4\linewidth]{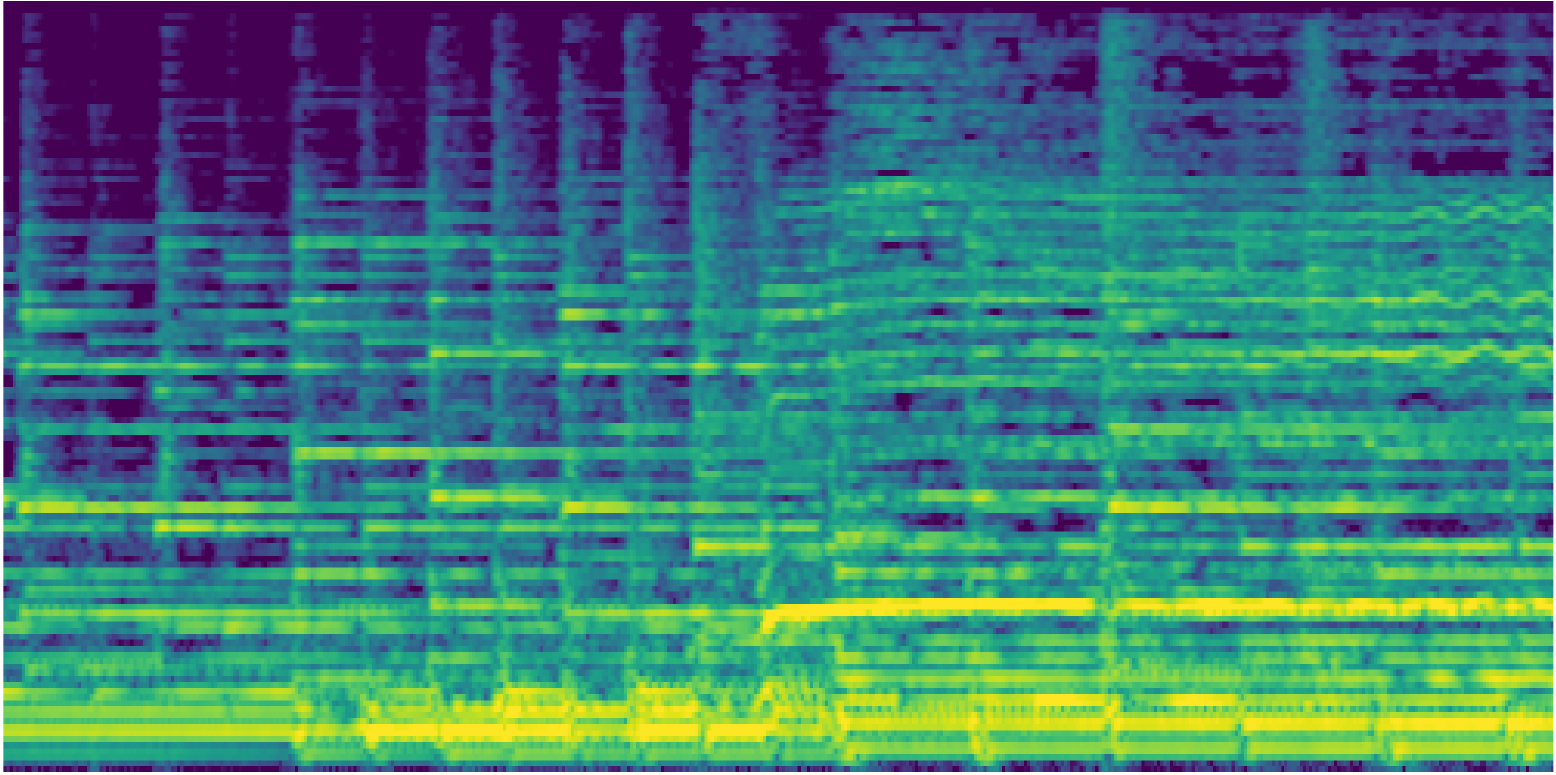}
    \end{subfigure}
    \\
    \begin{subfigure}[b]{0.245\linewidth}
        \includegraphics[width=0.98\linewidth,height=0.4\linewidth]{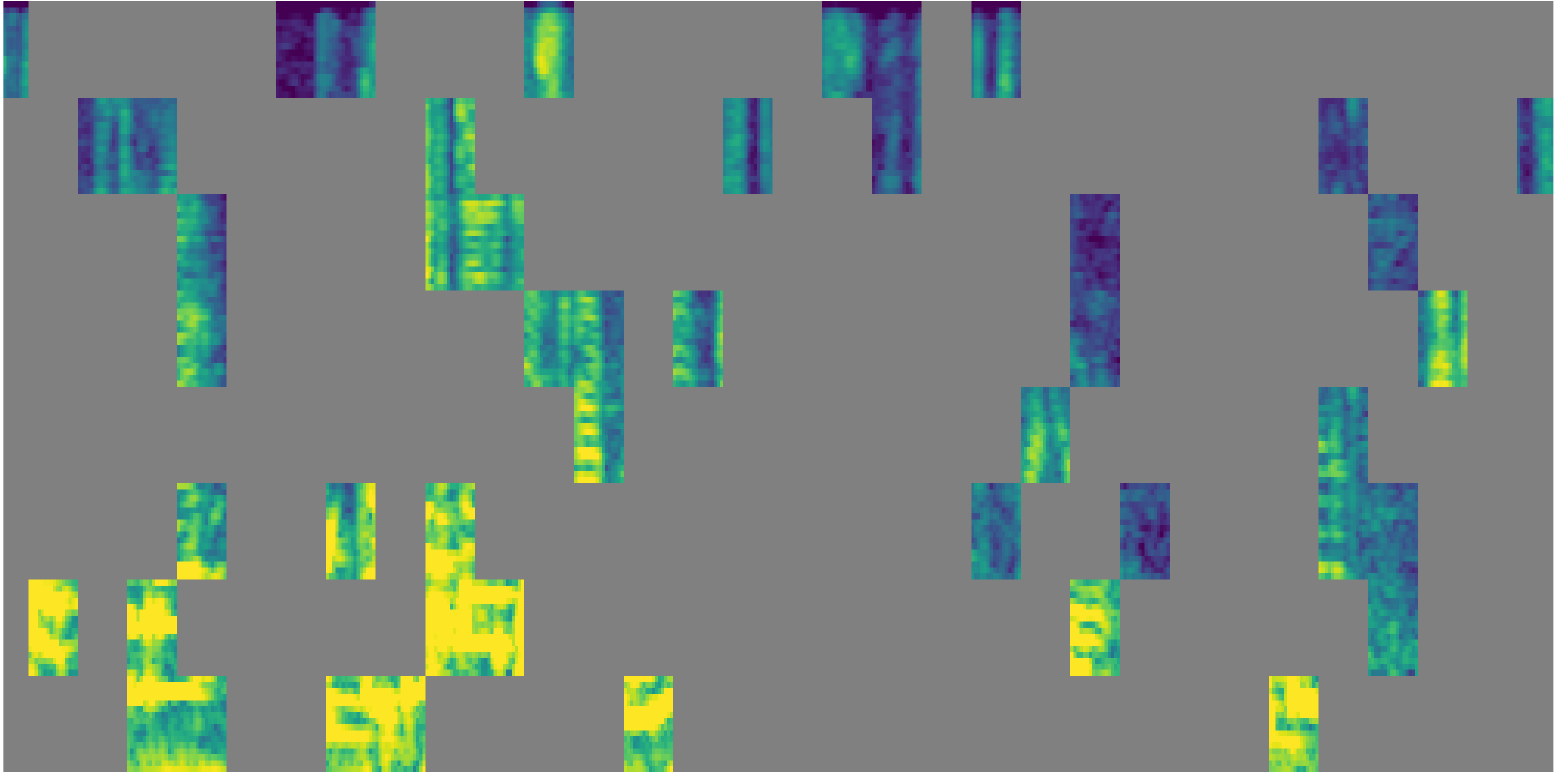}
    \end{subfigure}     
    \begin{subfigure}[b]{0.245\linewidth}
        \includegraphics[width=0.98\linewidth,height=0.4\linewidth]{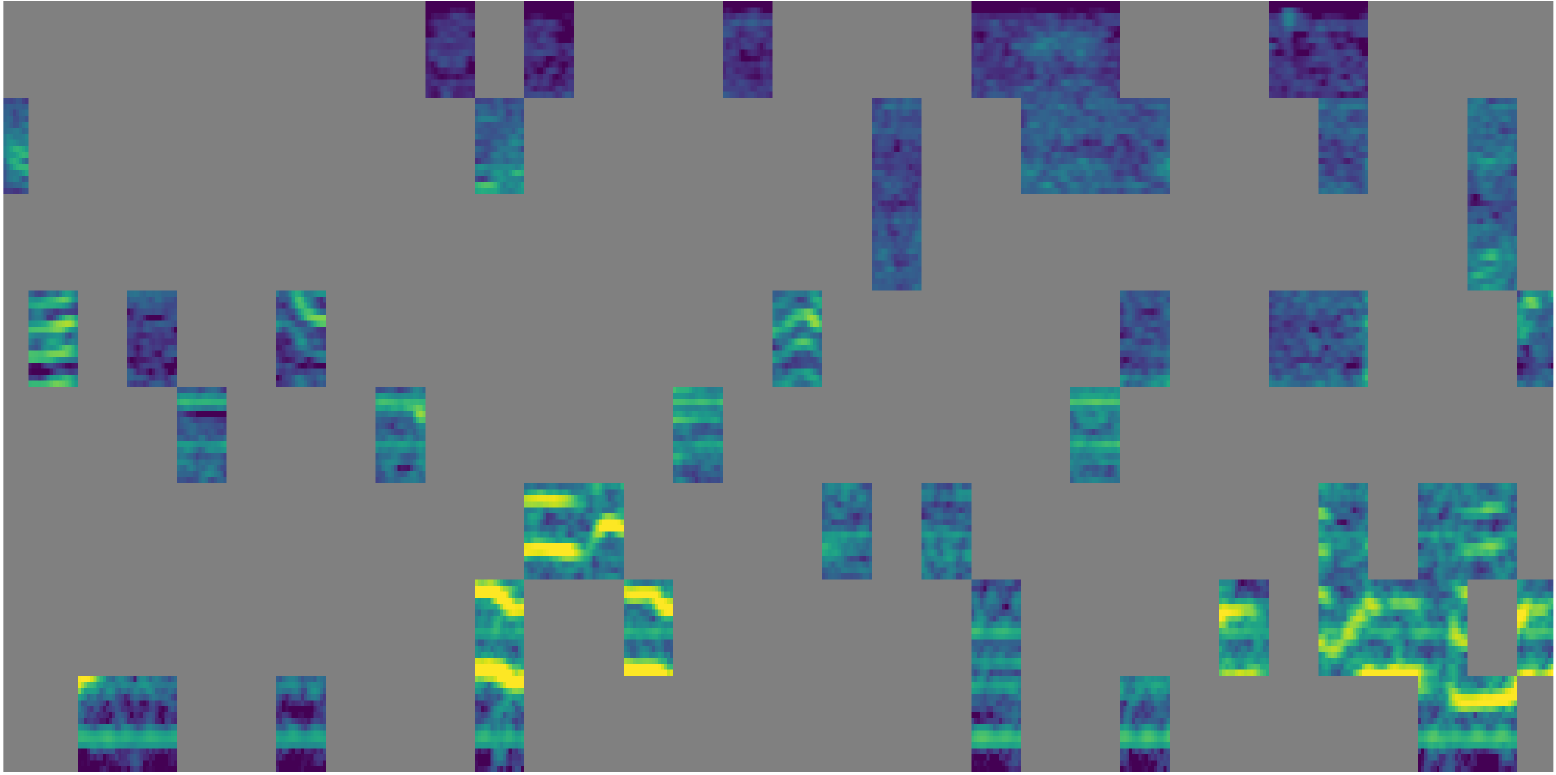}
    \end{subfigure}
    \begin{subfigure}[b]{0.245\linewidth}
        \includegraphics[width=0.98\linewidth,height=0.4\linewidth]{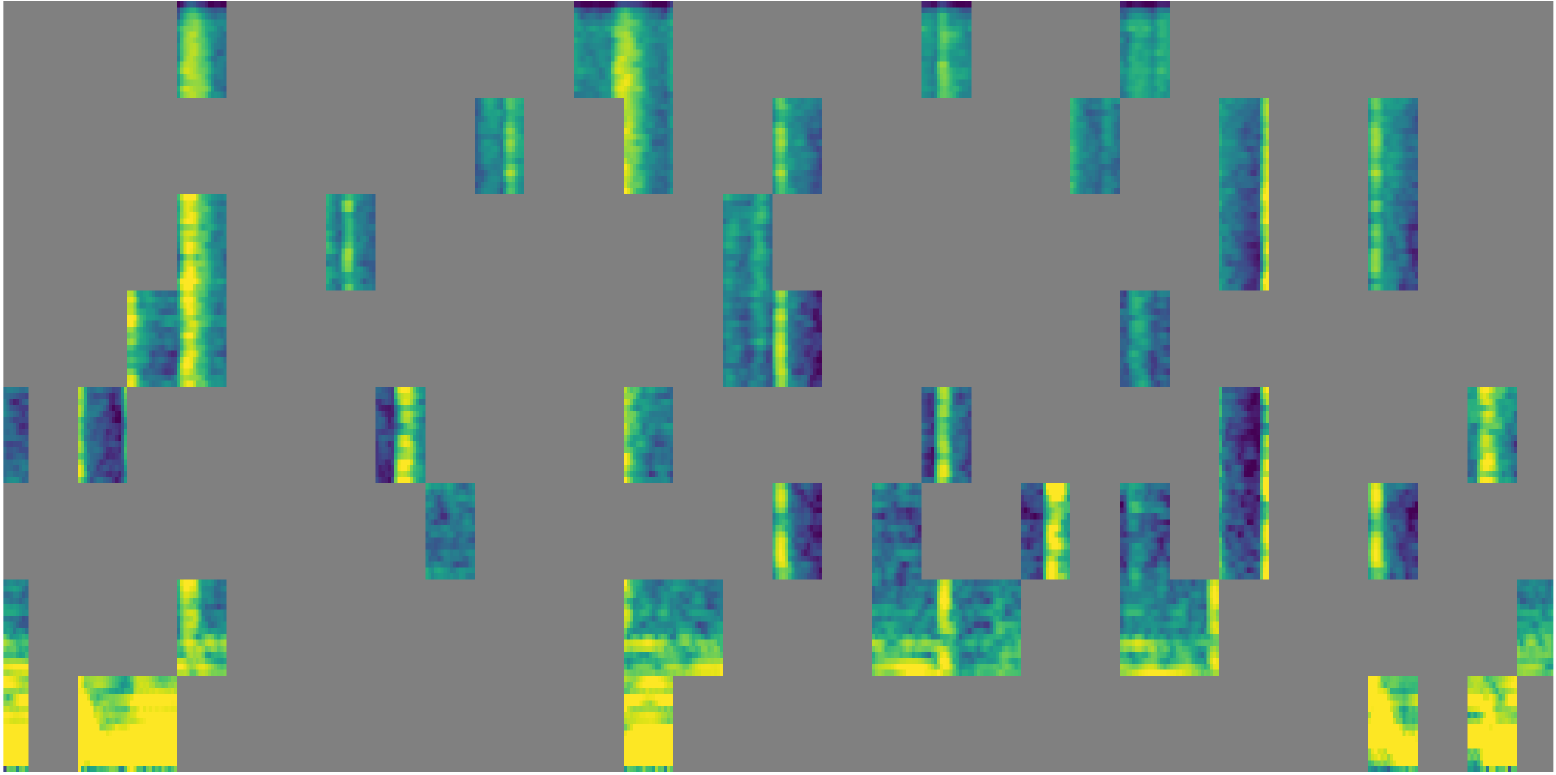}
    \end{subfigure}
    \begin{subfigure}[b]{0.245\linewidth}
        \includegraphics[width=0.98\linewidth,height=0.4\linewidth]{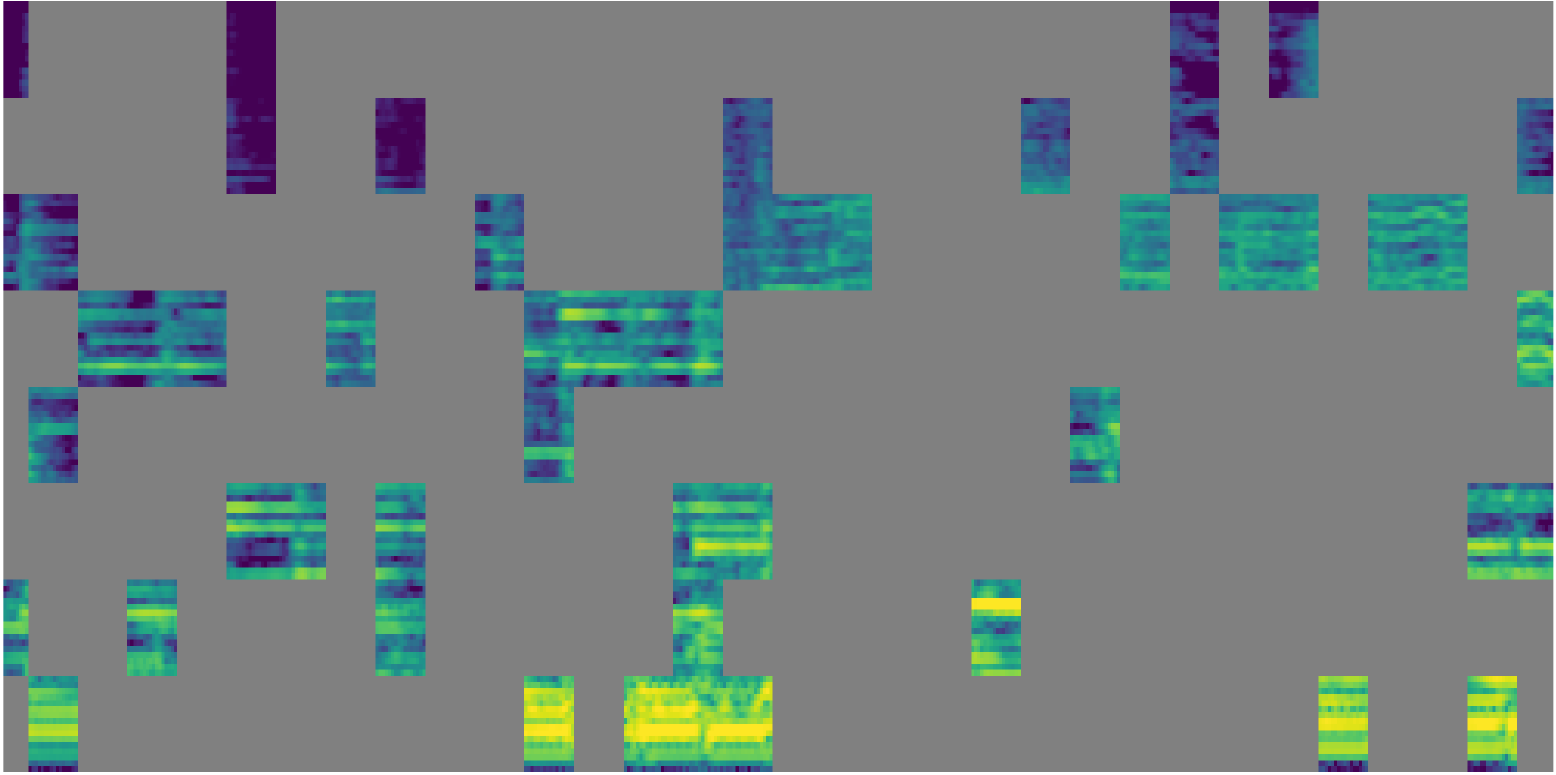}
    \end{subfigure}
    \\ 
    \begin{subfigure}[b]{0.245\linewidth}
        \includegraphics[width=0.98\linewidth,height=0.4\linewidth]{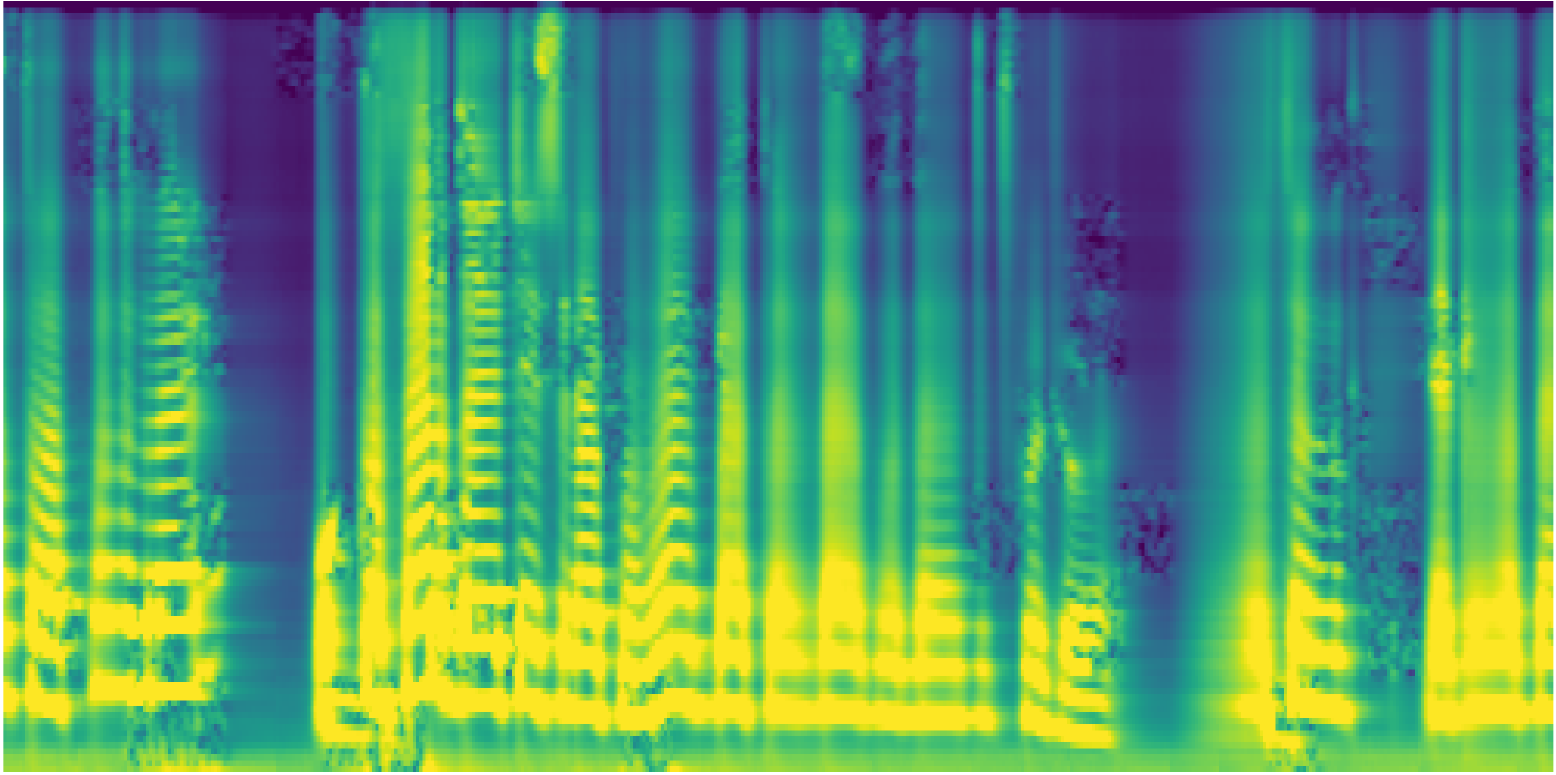}
        %\vspace{-0.2em}
        \subcaption{
            80\% Unstructured
            \href{https://www.dropbox.com/s/7l140pvl8mdy8p2/A7wdgVaqhPI_0.8_org.mp4?dl=0}{1}
            \href{https://www.dropbox.com/s/fmp1c6pcz7kctgs/A7wdgVaqhPI_0.8_masked.mp4?dl=0}{2}
            \href{https://www.dropbox.com/s/dx5syovedsoukrg/A7wdgVaqhPI_0.8_restored.mp4?dl=0}{3}
        }
        \label{fig:app:vis:e}
    \end{subfigure}
    \begin{subfigure}[b]{0.245\linewidth}
        \includegraphics[width=0.98\linewidth,height=0.4\linewidth]{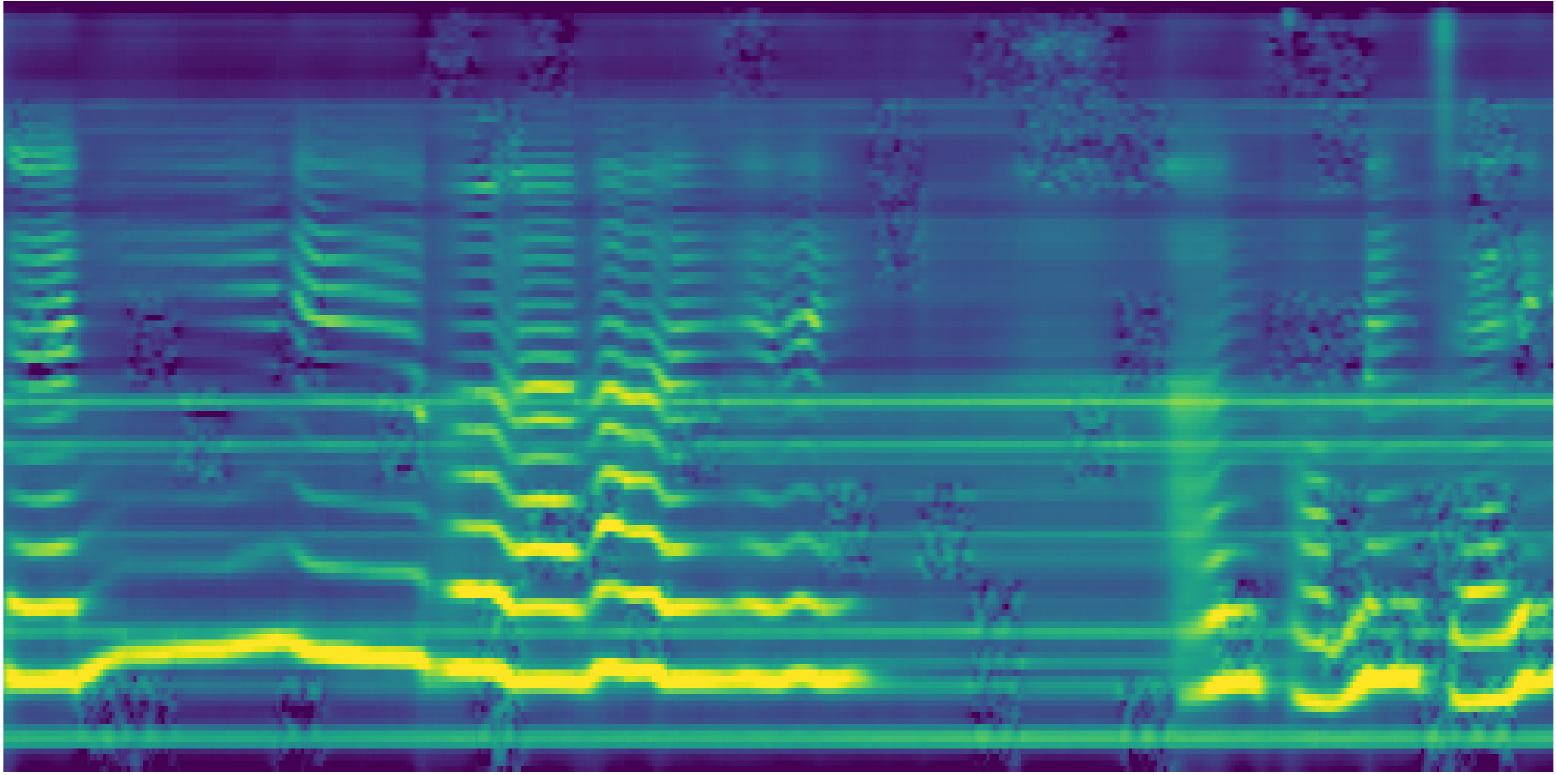}
        %\vspace{-0.2em}
        \subcaption{
            80\% Unstructured
            \href{https://www.dropbox.com/s/yh8vc2x9jwc38k5/-K1BRF6qng8_0.8_org.mp4?dl=0}{1}
            \href{https://www.dropbox.com/s/czjnnx43hidduhg/-K1BRF6qng8_0.8_masked.mp4?dl=0}{2}
            \href{https://www.dropbox.com/s/v2w2qdox6dnlyz5/-K1BRF6qng8_0.8_restored.mp4?dl=0}{3}
        }
        \label{fig:app:vis:f}
    \end{subfigure}    
    \begin{subfigure}[b]{0.245\linewidth}
        \includegraphics[width=0.98\linewidth,height=0.4\linewidth]{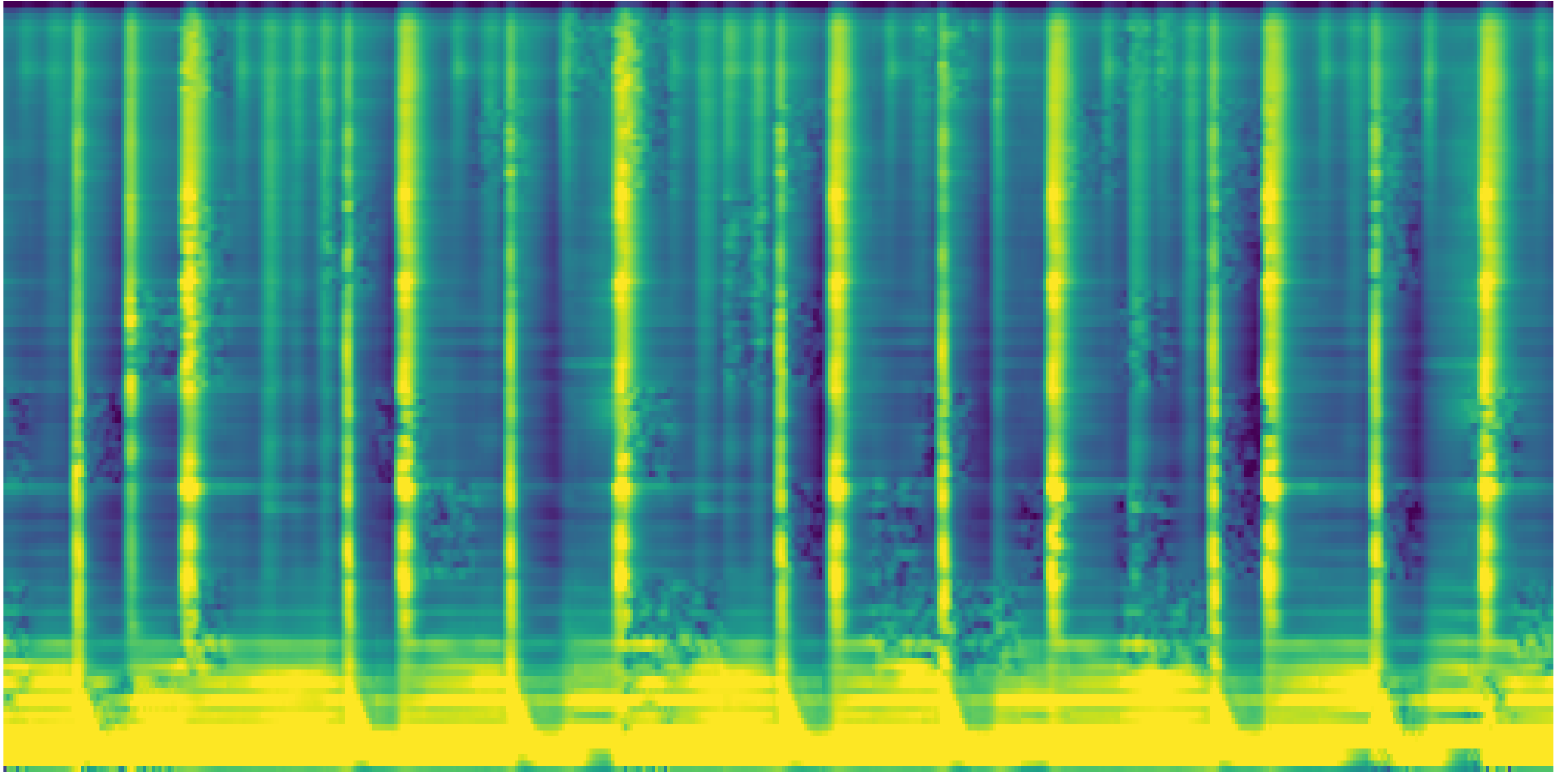}
        %\vspace{-0.2em}
        \subcaption{
            80\% Unstructured 
            \href{https://www.dropbox.com/s/7yk3ljorvth22qc/IFimpFwvbz8_0.8_org.mp4?dl=0}{1}
            \href{https://www.dropbox.com/s/2g1avcbi4fi99qm/IFimpFwvbz8_0.8_masked.mp4?dl=0}{2}
            \href{https://www.dropbox.com/s/e14wh7pcbtkapyh/IFimpFwvbz8_0.8_restored.mp4?dl=0}{3}
        }
        \label{fig:app:vis:g}
    \end{subfigure}
    \begin{subfigure}[b]{0.245\linewidth}
        \includegraphics[width=0.98\linewidth,height=0.4\linewidth]{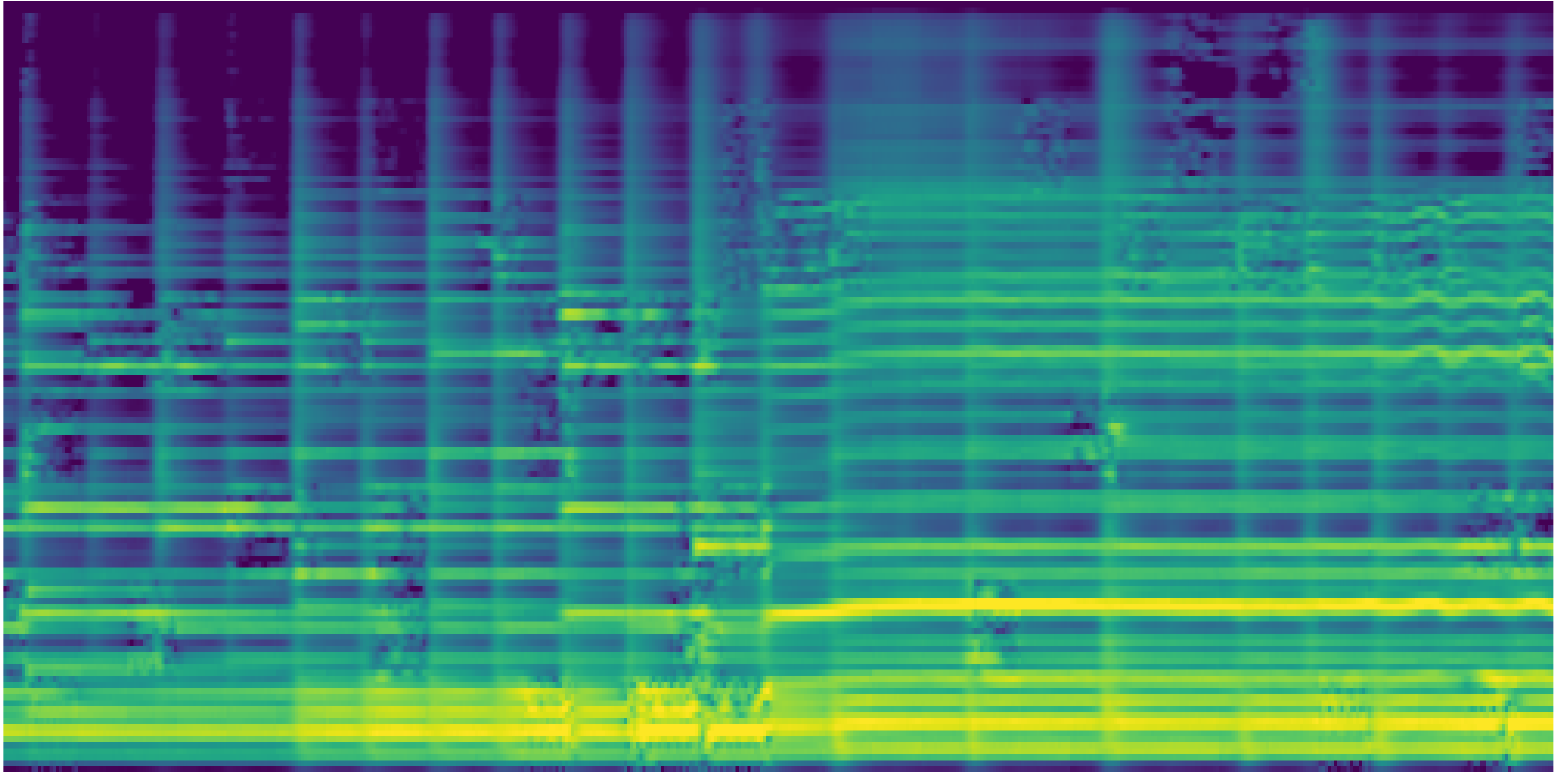}
        %\vspace{-0.2em}
        \subcaption{
            80\% Unstructured 
            \href{https://www.dropbox.com/s/ldyynkzb3iizju8/LCJblaUkkfc_0.8_org.mp4?dl=0}{1}
            \href{https://www.dropbox.com/s/sfi126vd7cb249l/LCJblaUkkfc_0.8_masked.mp4?dl=0}{2}
            \href{https://www.dropbox.com/s/yt5lgxeqnz1brrq/LCJblaUkkfc_0.8_restored.mp4?dl=0}{3}
        }
        \label{fig:app:vis:h}
    \end{subfigure}
    \\ \vspace{5pt}    
    %%%%%%%%%%%%%%%%%%% 3rd row %%%%%%%%%%%%%%%%%%%%%%%%
    \begin{subfigure}[b]{0.245\linewidth}
        \includegraphics[width=0.98\linewidth,height=0.4\linewidth]{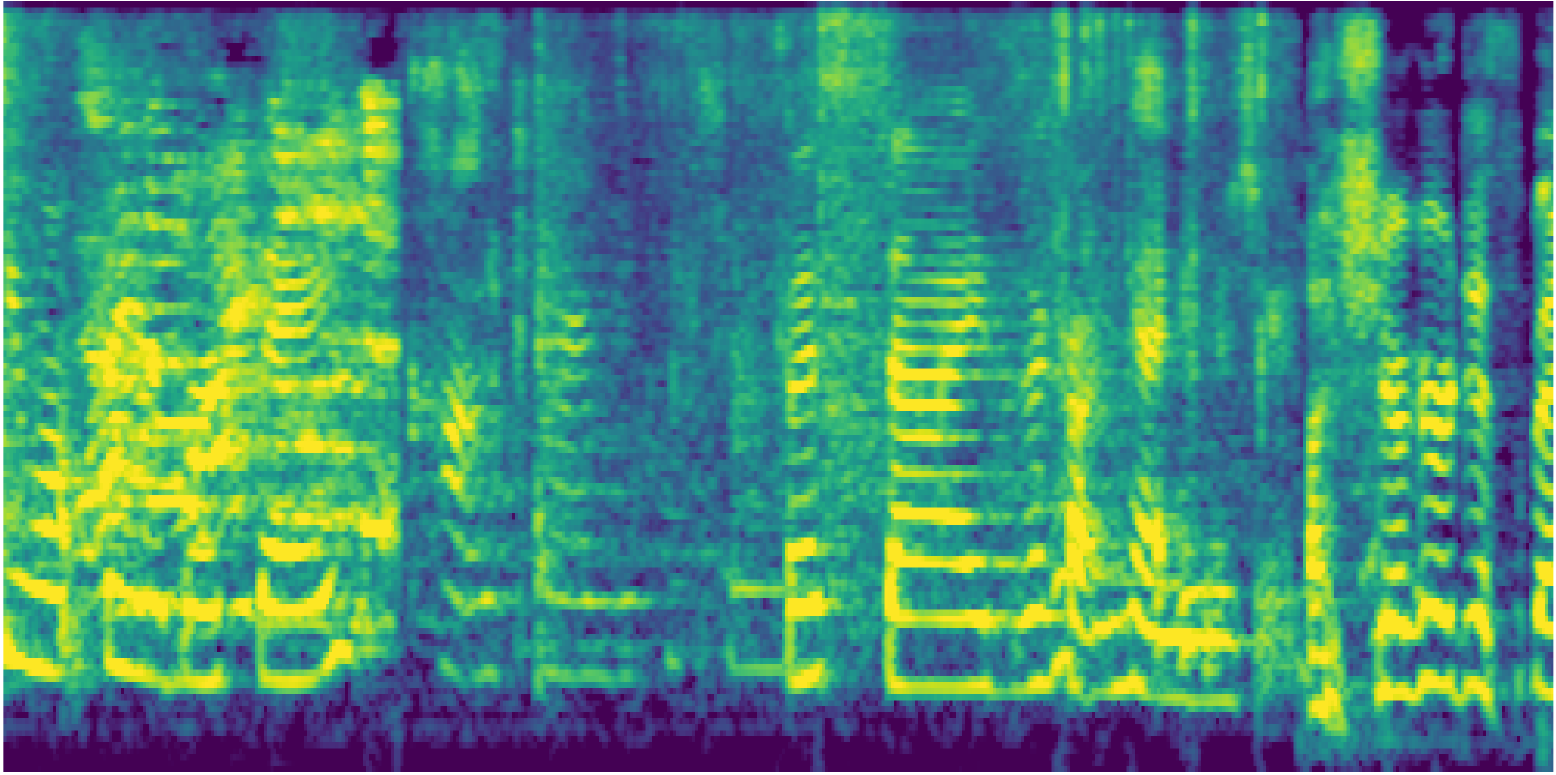}
    \end{subfigure}    
    \begin{subfigure}[b]{0.245\linewidth}
        \includegraphics[width=0.98\linewidth,height=0.4\linewidth]{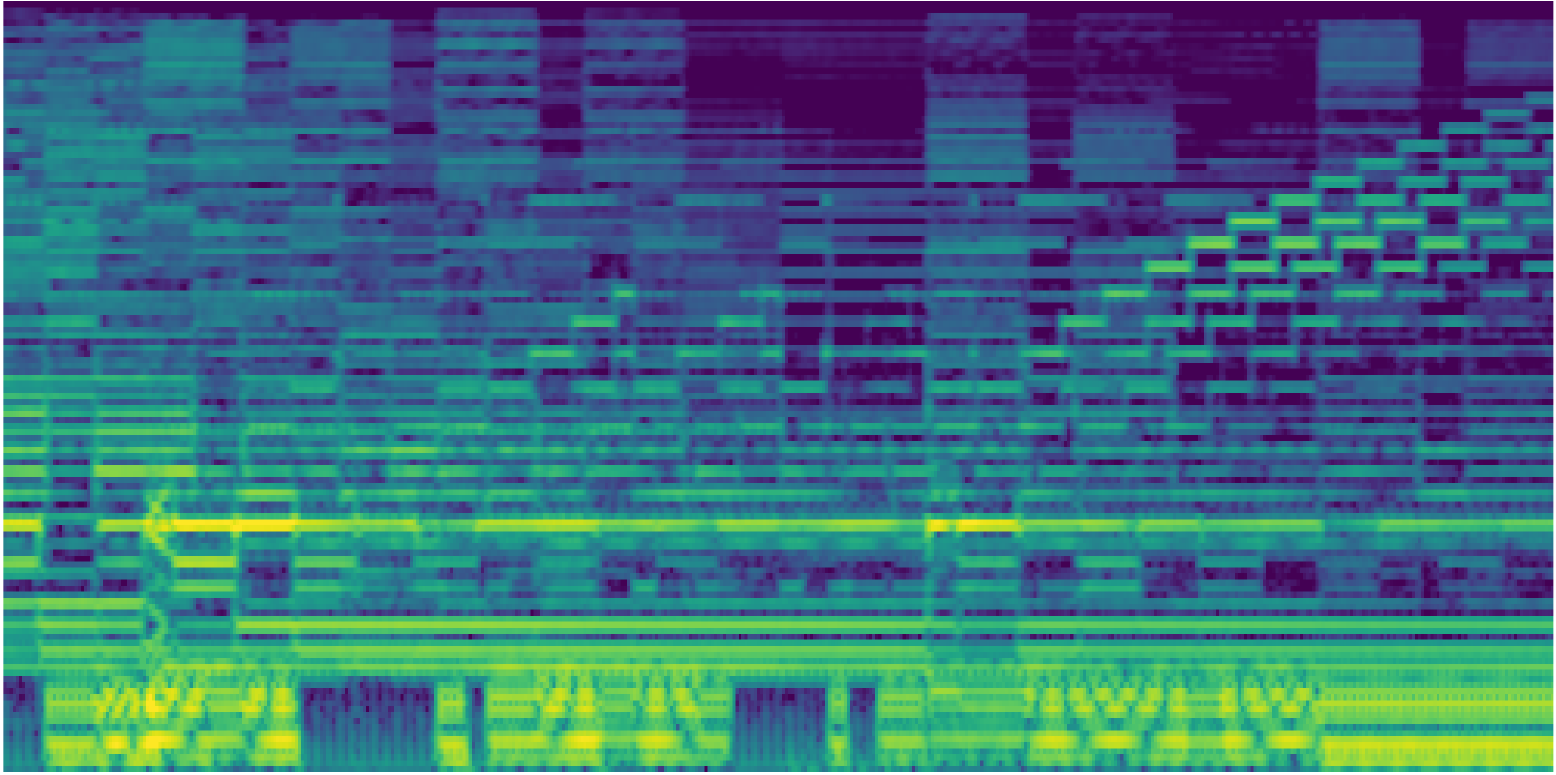}
    \end{subfigure}     
    \begin{subfigure}[b]{0.245\linewidth}
        \includegraphics[width=0.98\linewidth,height=0.4\linewidth]{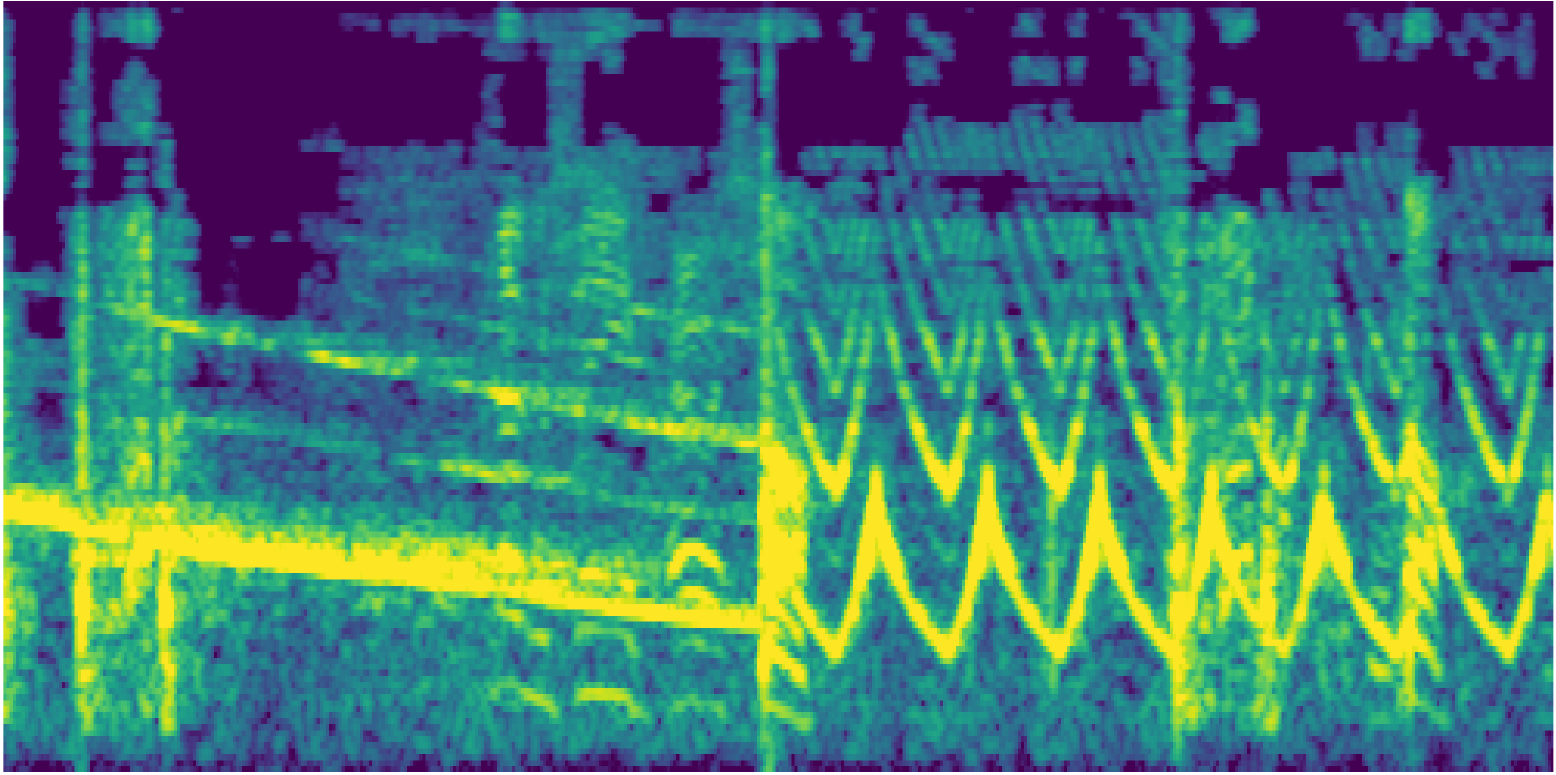}
    \end{subfigure}
    \begin{subfigure}[b]{0.245\linewidth}
        \includegraphics[width=0.98\linewidth,height=0.4\linewidth]{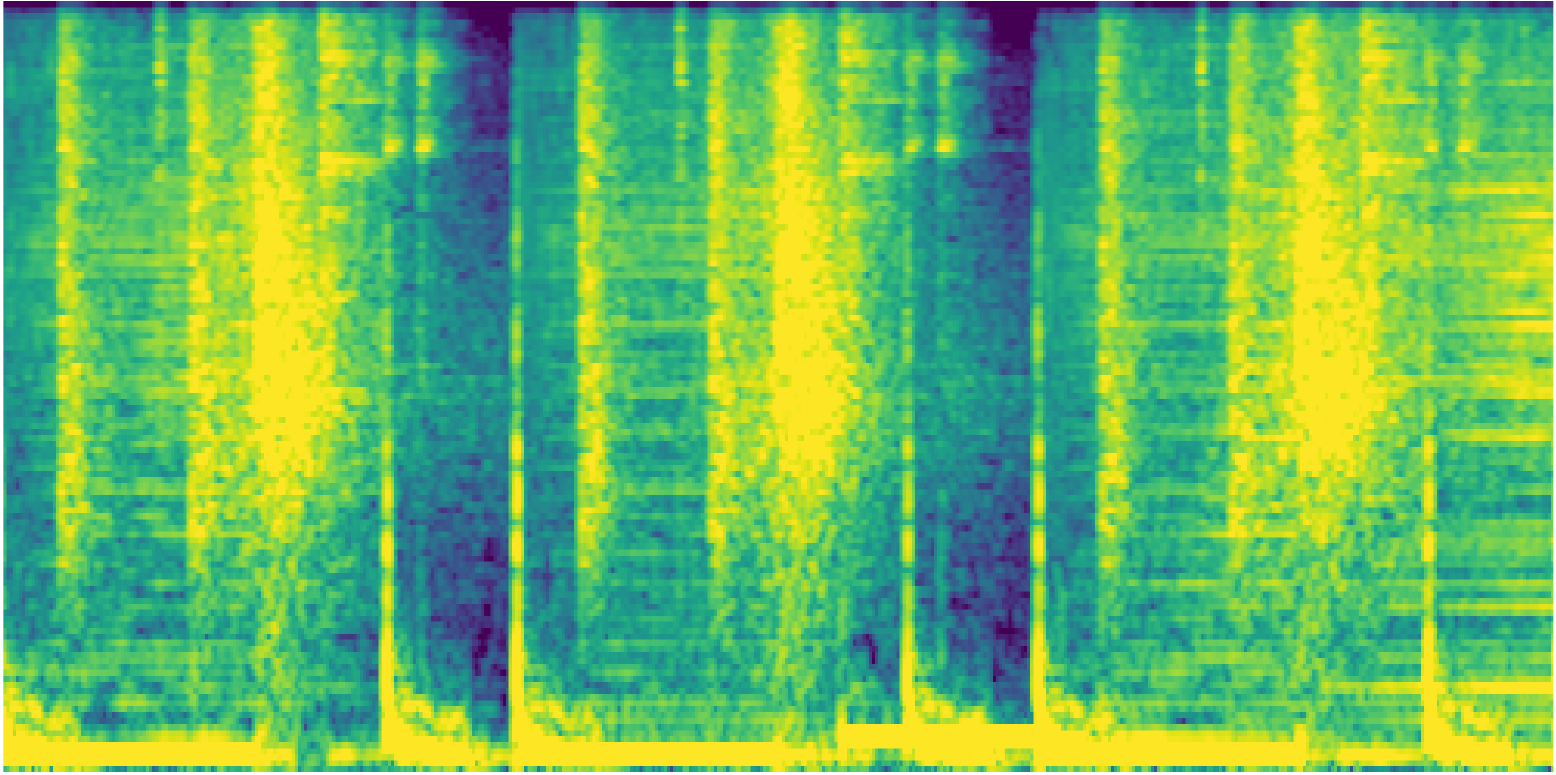}
    \end{subfigure}
    \\
    \begin{subfigure}[b]{0.245\linewidth}
        \includegraphics[width=0.98\linewidth,height=0.4\linewidth]{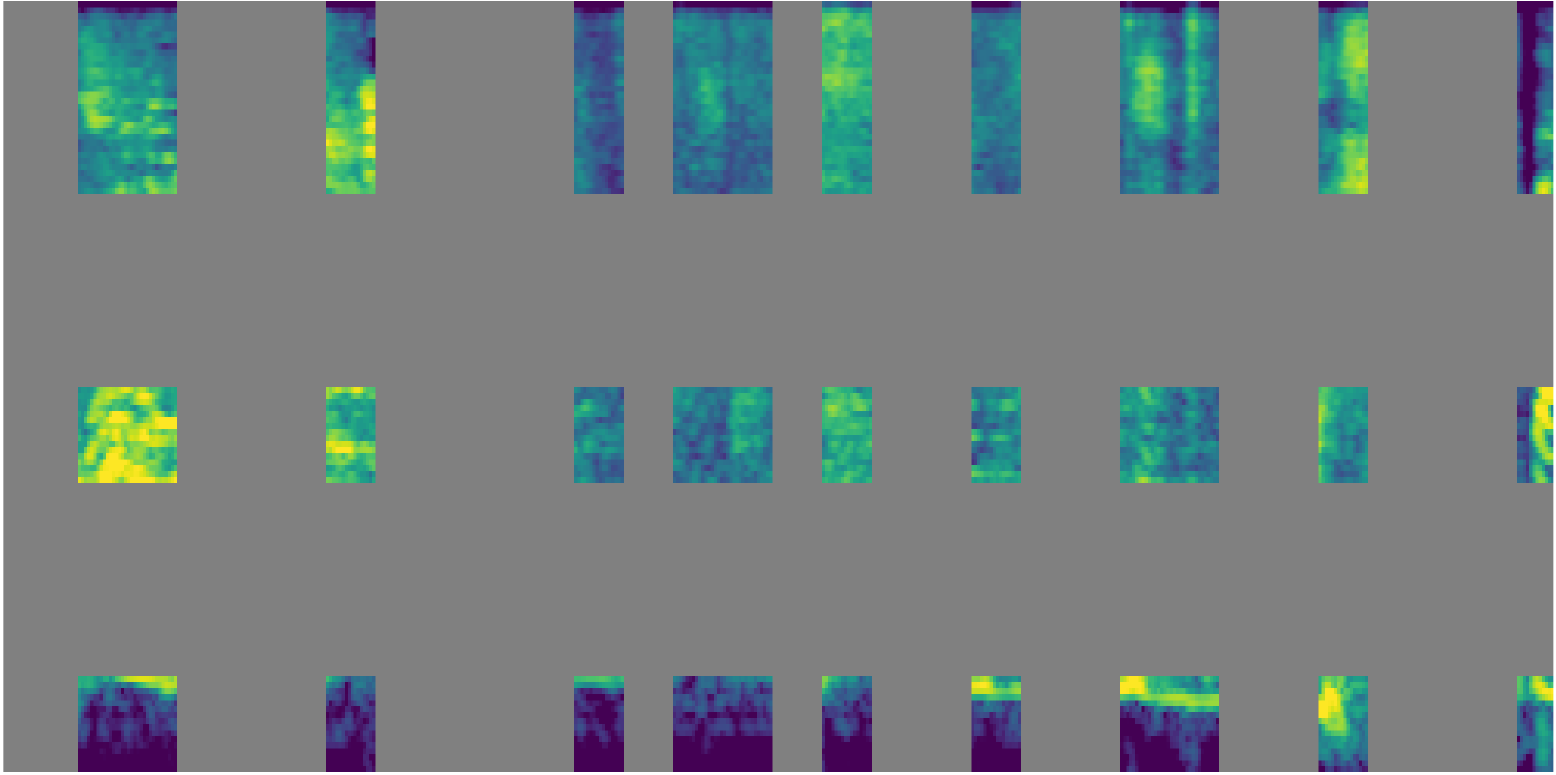}
    \end{subfigure}    
    \begin{subfigure}[b]{0.245\linewidth}
        \includegraphics[width=0.98\linewidth,height=0.4\linewidth]{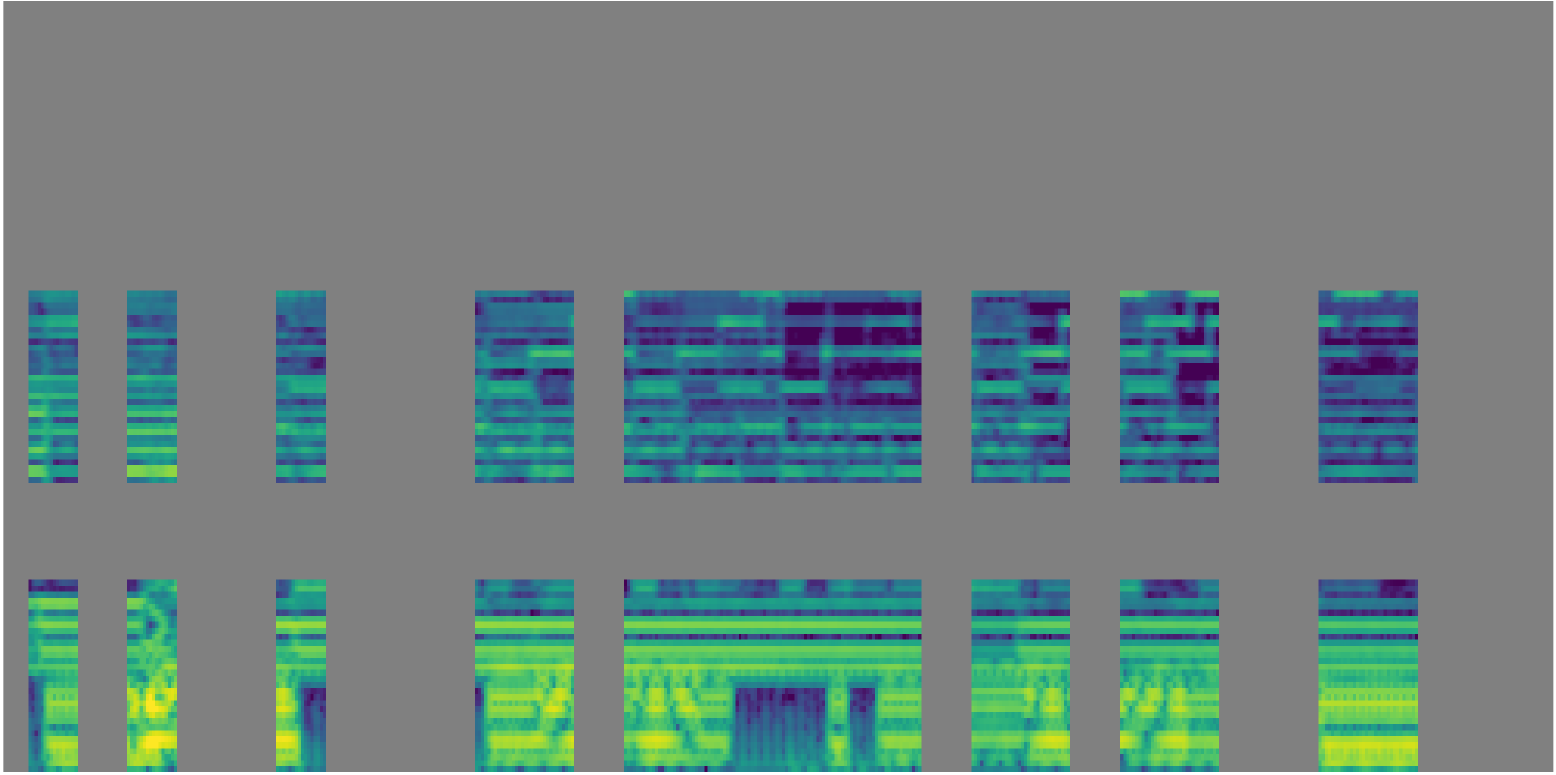}
    \end{subfigure}      
    \begin{subfigure}[b]{0.245\linewidth}
        \includegraphics[width=0.98\linewidth,height=0.4\linewidth]{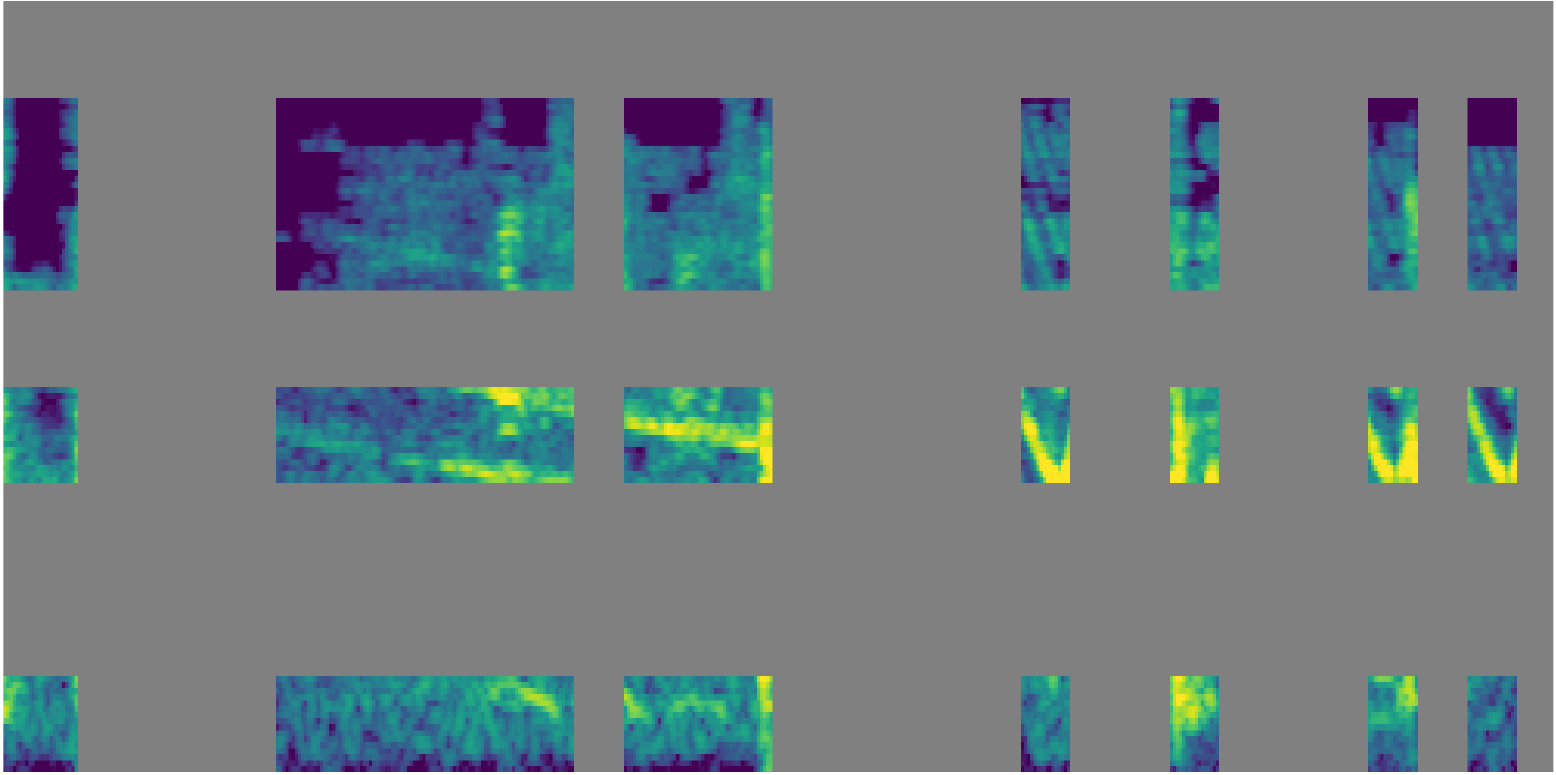}
    \end{subfigure}
    \begin{subfigure}[b]{0.245\linewidth}
        \includegraphics[width=0.98\linewidth,height=0.4\linewidth]{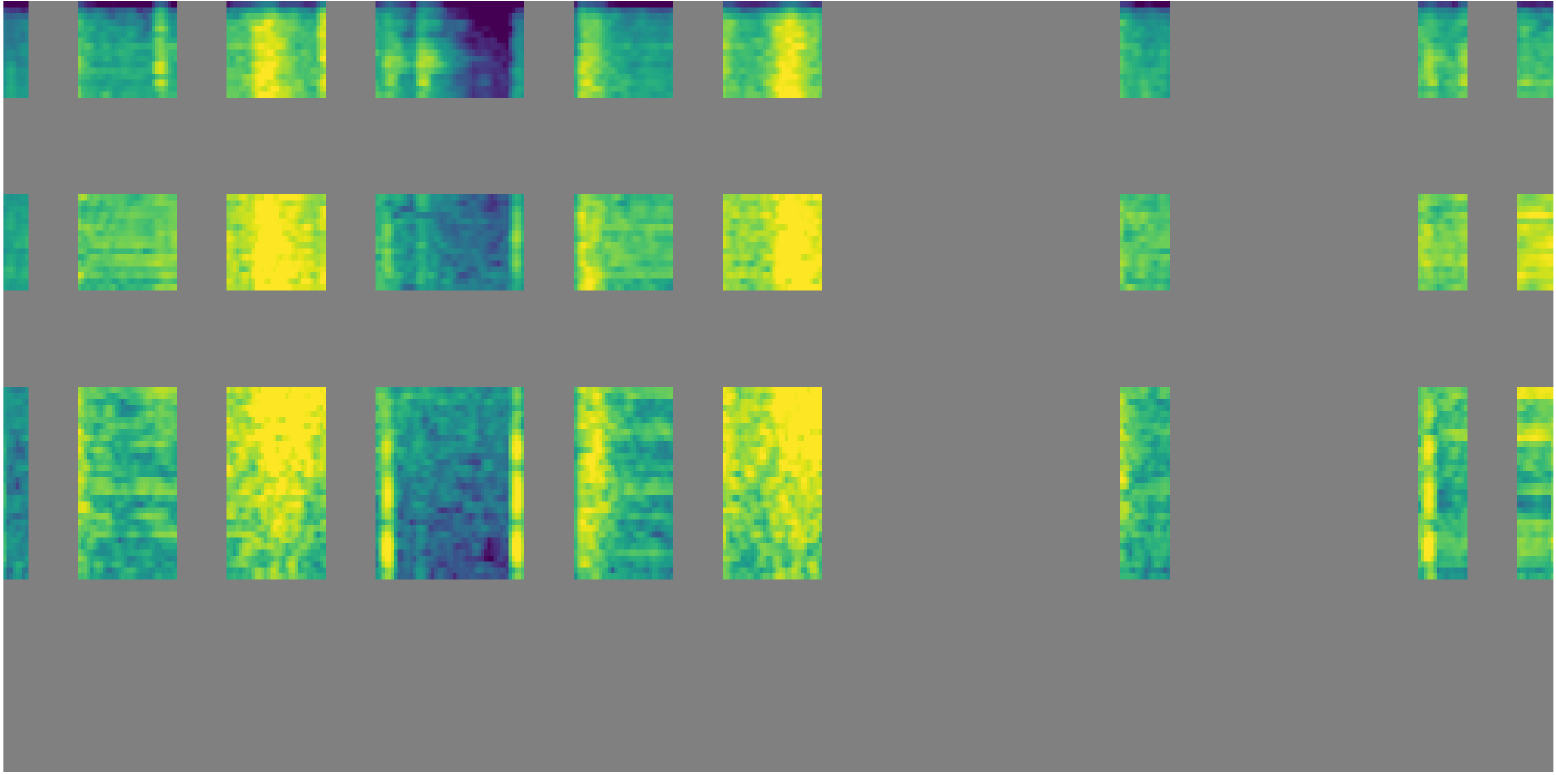}
    \end{subfigure}
    \\
    \begin{subfigure}[b]{0.245\linewidth}
        \includegraphics[width=0.98\linewidth,height=0.4\linewidth]{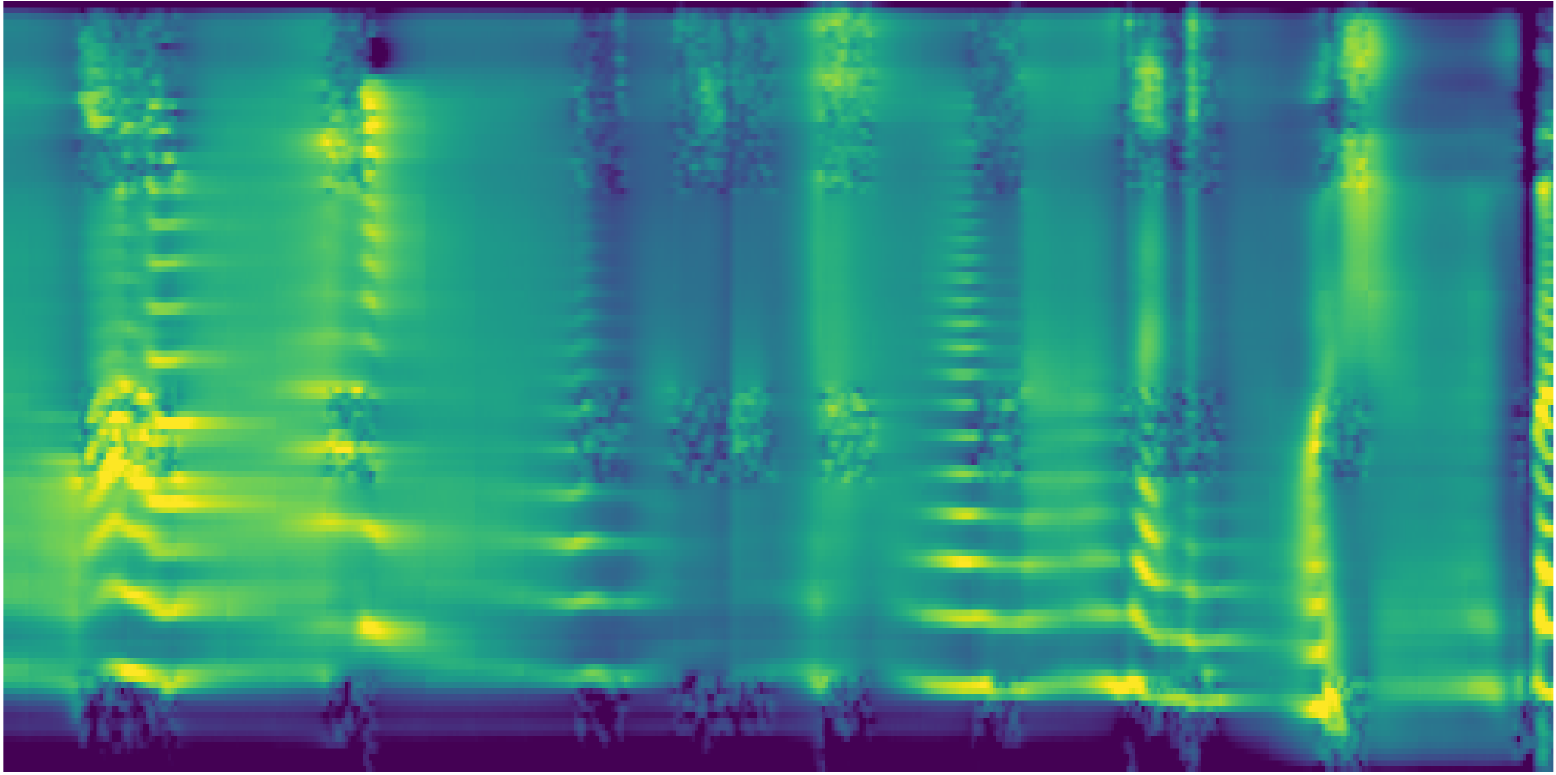}
        %\vspace{-0.2em}
        \subcaption{
            70\% Structured
            %\href{https://www.dropbox.com/s/am0kqnb0tcp6yn6/u4X11MnUWX8_0.8_2d_org.mp4?dl=0}{1}
            %\href{https://www.dropbox.com/s/xt8e63ehof32u7e/u4X11MnUWX8_0.8_2d_masked.mp4?dl=0}{2}
            %\href{https://www.dropbox.com/s/fm6cum2otncvnm3/u4X11MnUWX8_0.8_2d_restored.mp4?dl=0}{3}
            \href{https://www.dropbox.com/s/3oeoniwv8jpxm9x/u4X11MnUWX8_0.7_2d_org.mp4?dl=0}{1}
            \href{https://www.dropbox.com/s/evwh3jgqgz7vz6r/u4X11MnUWX8_0.7_2d_masked.mp4?dl=0}{2}
            \href{https://www.dropbox.com/s/onz83xx2ntycujy/u4X11MnUWX8_0.7_2d_restored.mp4?dl=0}{3}            
        }
        \label{fig:app:vis:i}
    \end{subfigure} 
    \begin{subfigure}[b]{0.245\linewidth}
        \includegraphics[width=0.98\linewidth,height=0.4\linewidth]{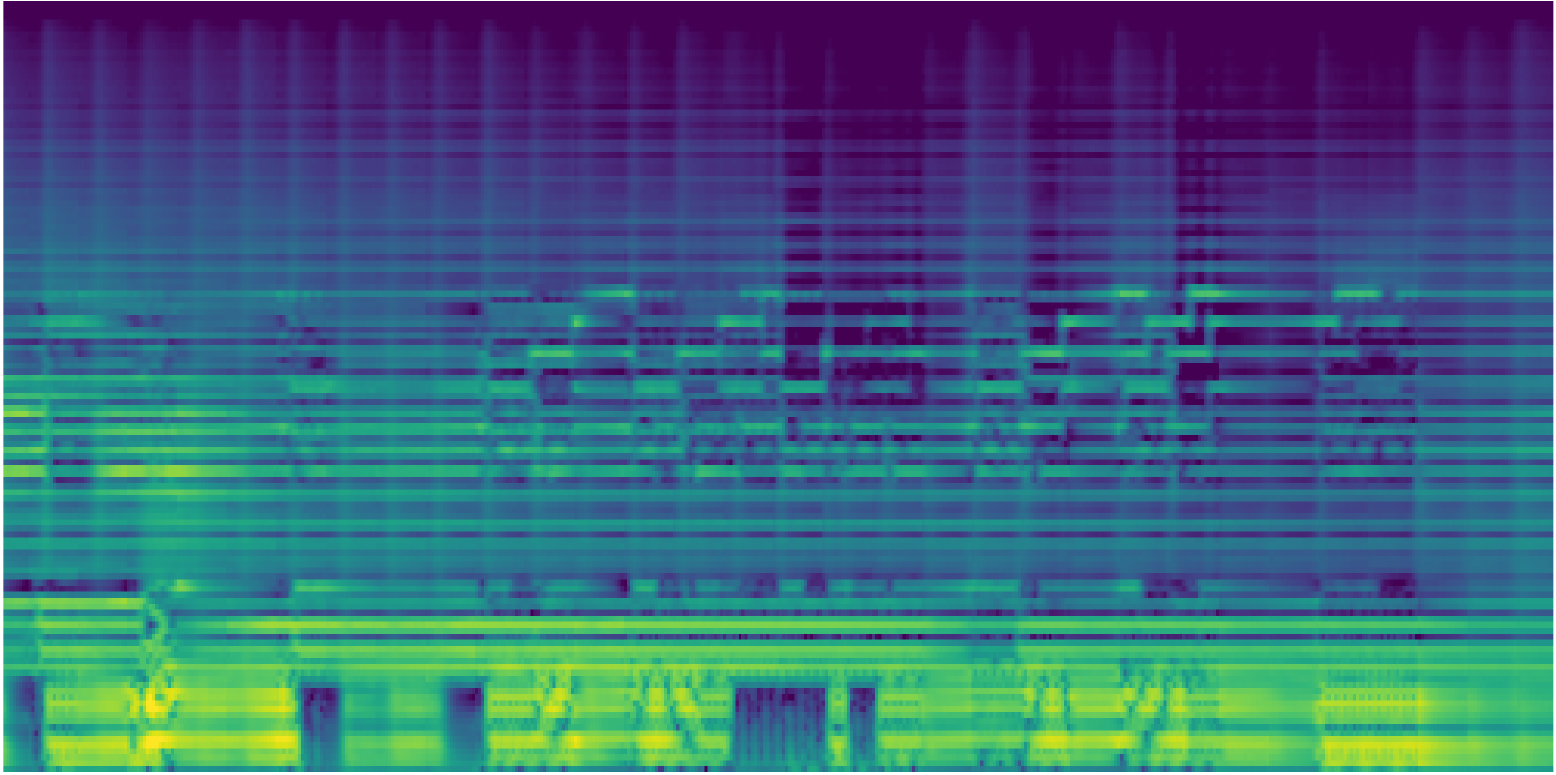}
        %\vspace{-0.2em}
        \subcaption{
            70\% Structured
            \href{https://www.dropbox.com/s/ghjrnpws0qxtczw/NwfEO8cjSK0_0.7_2d_org.mp4?dl=0}{1}
            \href{https://www.dropbox.com/s/i13zxe0vp8fg8r5/NwfEO8cjSK0_0.7_2d_masked.mp4?dl=0}{2}
            \href{https://www.dropbox.com/s/yr13d9k4giha1wa/NwfEO8cjSK0_0.7_2d_restored.mp4?dl=0}{3}            
        }
        \label{fig:app:vis:j}
    \end{subfigure}    
    \begin{subfigure}[b]{0.245\linewidth}
        \includegraphics[width=0.98\linewidth,height=0.4\linewidth]{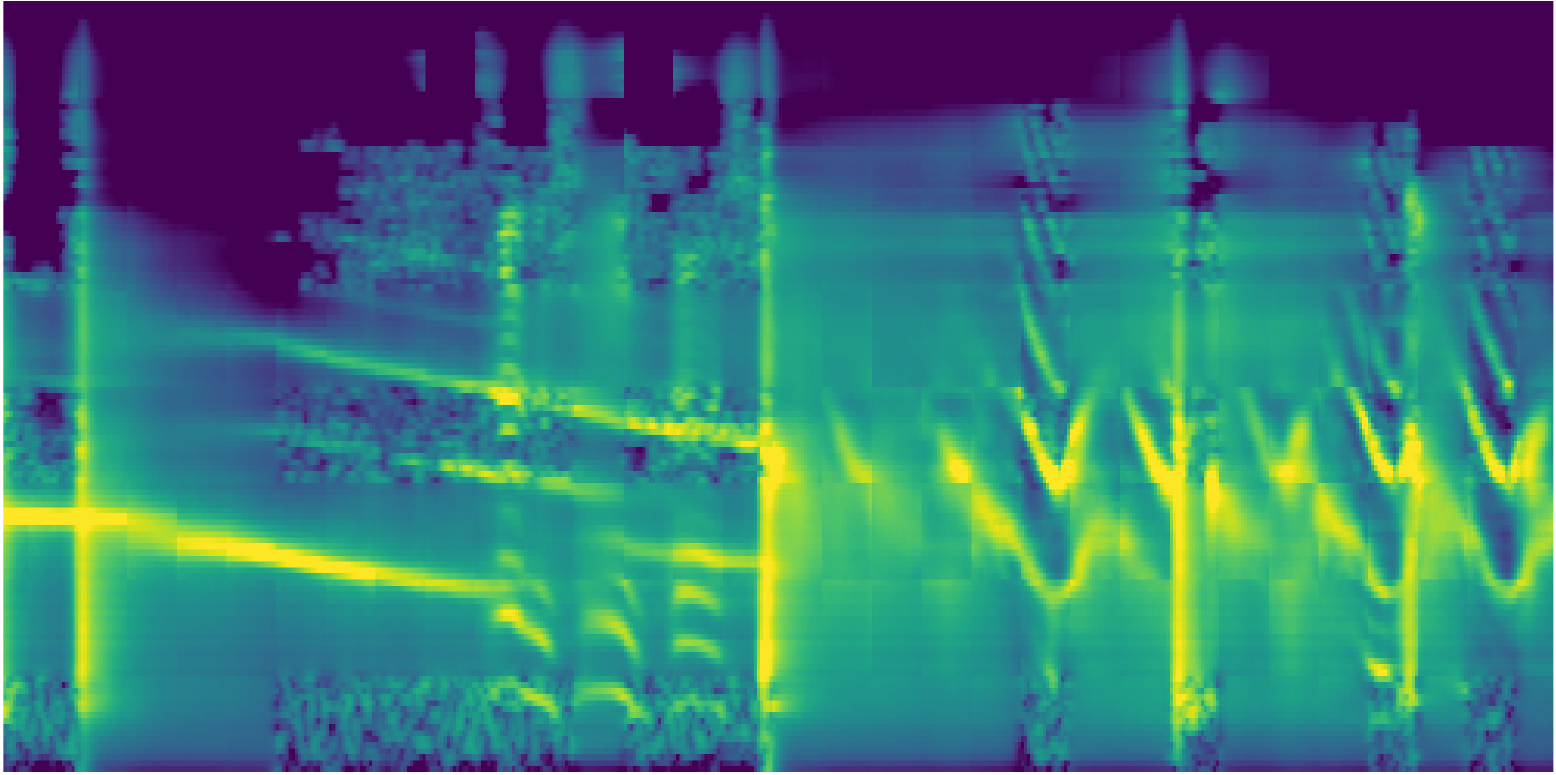}
        %\vspace{-0.2em}
        \subcaption{
            70\% Structured
            \href{https://www.dropbox.com/s/0rkyku68m2p6hmu/o3F3tUpmVaw_0.7_2d_org.mp4?dl=0}{1}
            \href{https://www.dropbox.com/s/hyycbytl6qz59h7/o3F3tUpmVaw_0.7_2d_masked.mp4?dl=0}{2}
            \href{https://www.dropbox.com/s/19n7ajh4nrbqukm/o3F3tUpmVaw_0.7_2d_restored.mp4?dl=0}{3}            
        }
        \label{fig:app:vis:k}
    \end{subfigure}
    \begin{subfigure}[b]{0.245\linewidth}
        \includegraphics[width=0.98\linewidth,height=0.4\linewidth]{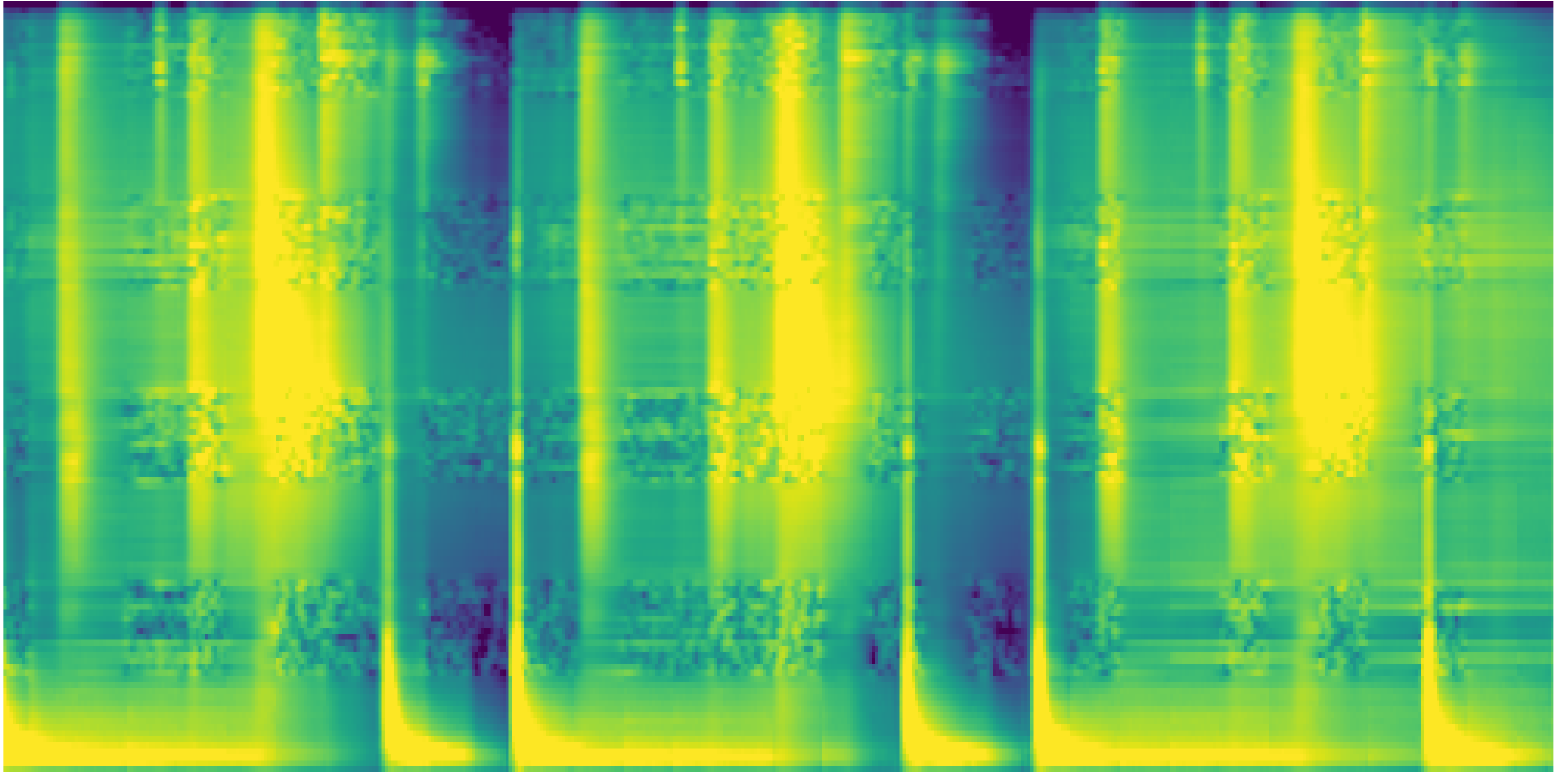}
        %\vspace{-0.2em}
        \subcaption{
            70\% Structured
            \href{https://www.dropbox.com/s/1jelm97ff26p9pg/Q0KwG3ynscI_0.7_2d_org.mp4?dl=0}{1}
            \href{https://www.dropbox.com/s/qp6ly6hal1fouq3/Q0KwG3ynscI_0.7_2d_masked.mp4?dl=0}{2}
            \href{https://www.dropbox.com/s/mzpjpptc7zq3lgx/Q0KwG3ynscI_0.7_2d_restored.mp4?dl=0}{3}
        }
        \label{fig:app:vis:l}
    \end{subfigure}
    \\
    \vspace{5pt}

%%%%%%%%%%%%%%%%%%%%%%%%% 4th Row %%%%%%%%%%%%%%%%%%%%%%%%%%%%%%%%% 
    \begin{subfigure}[b]{0.245\linewidth}
        \includegraphics[width=0.98\linewidth,height=0.4\linewidth]{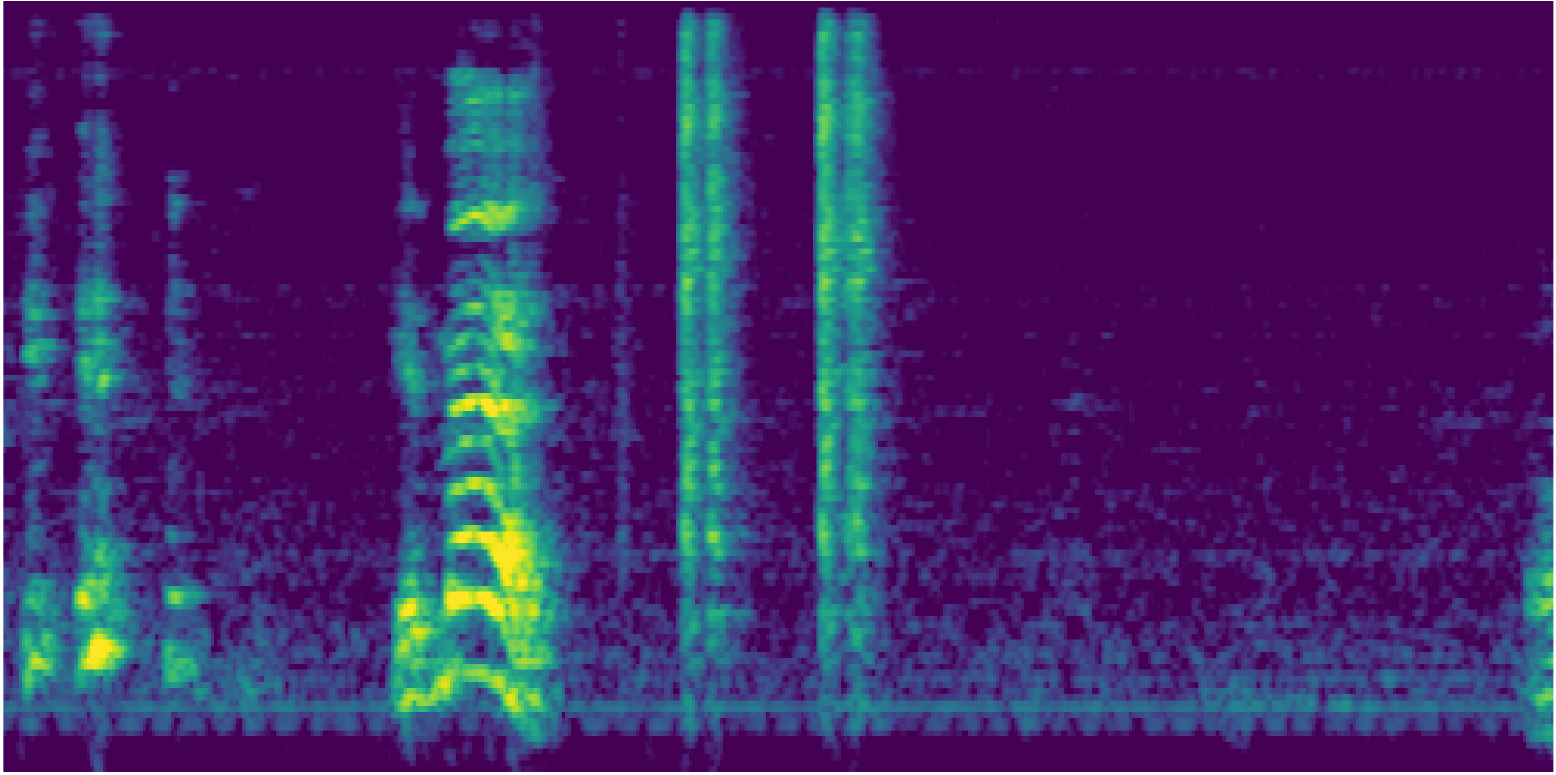}
    \end{subfigure}    
    \begin{subfigure}[b]{0.245\linewidth}
        \includegraphics[width=0.98\linewidth,height=0.4\linewidth]{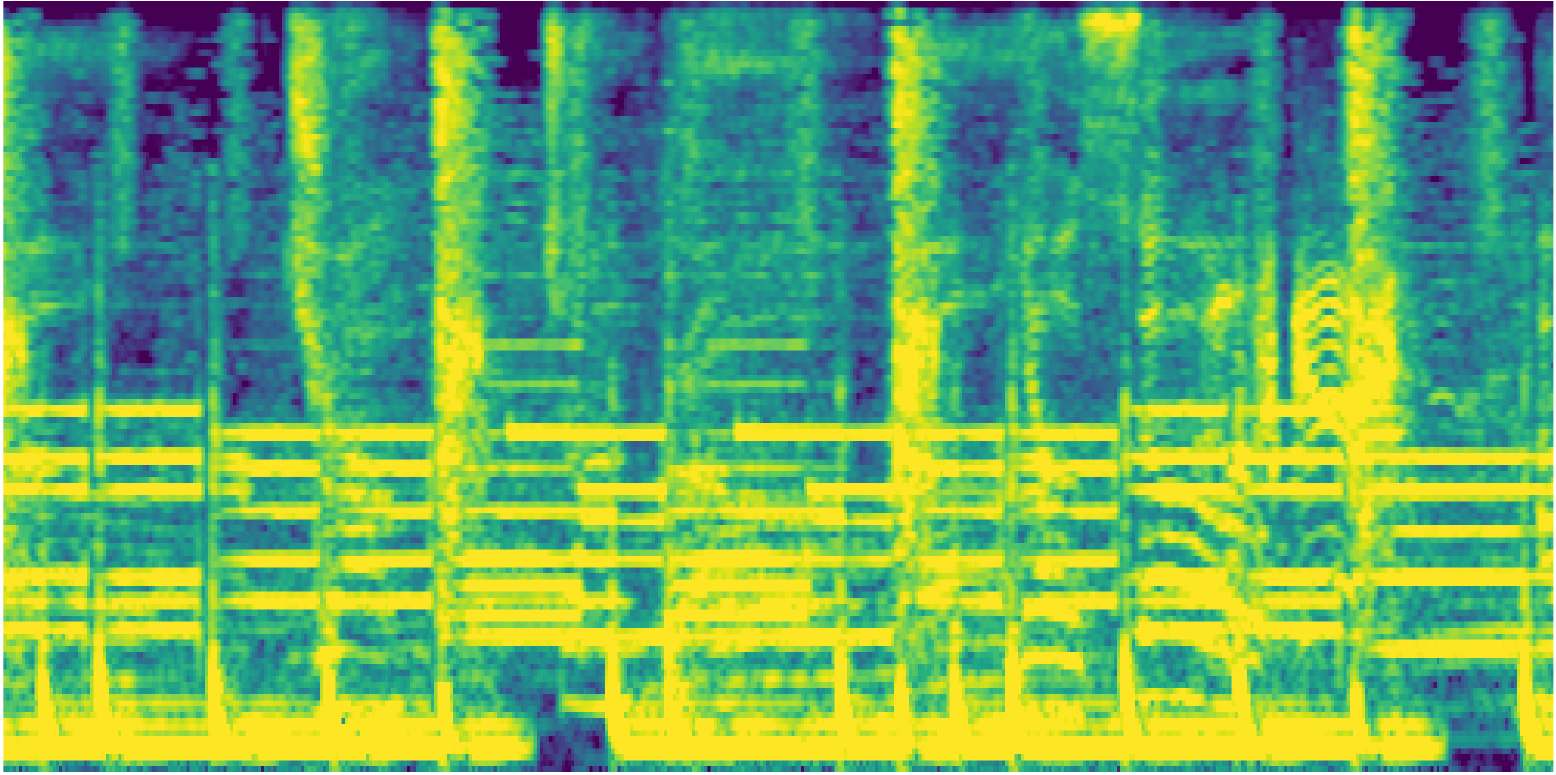}
    \end{subfigure}     
    \begin{subfigure}[b]{0.245\linewidth}
        \includegraphics[width=0.98\linewidth,height=0.4\linewidth]{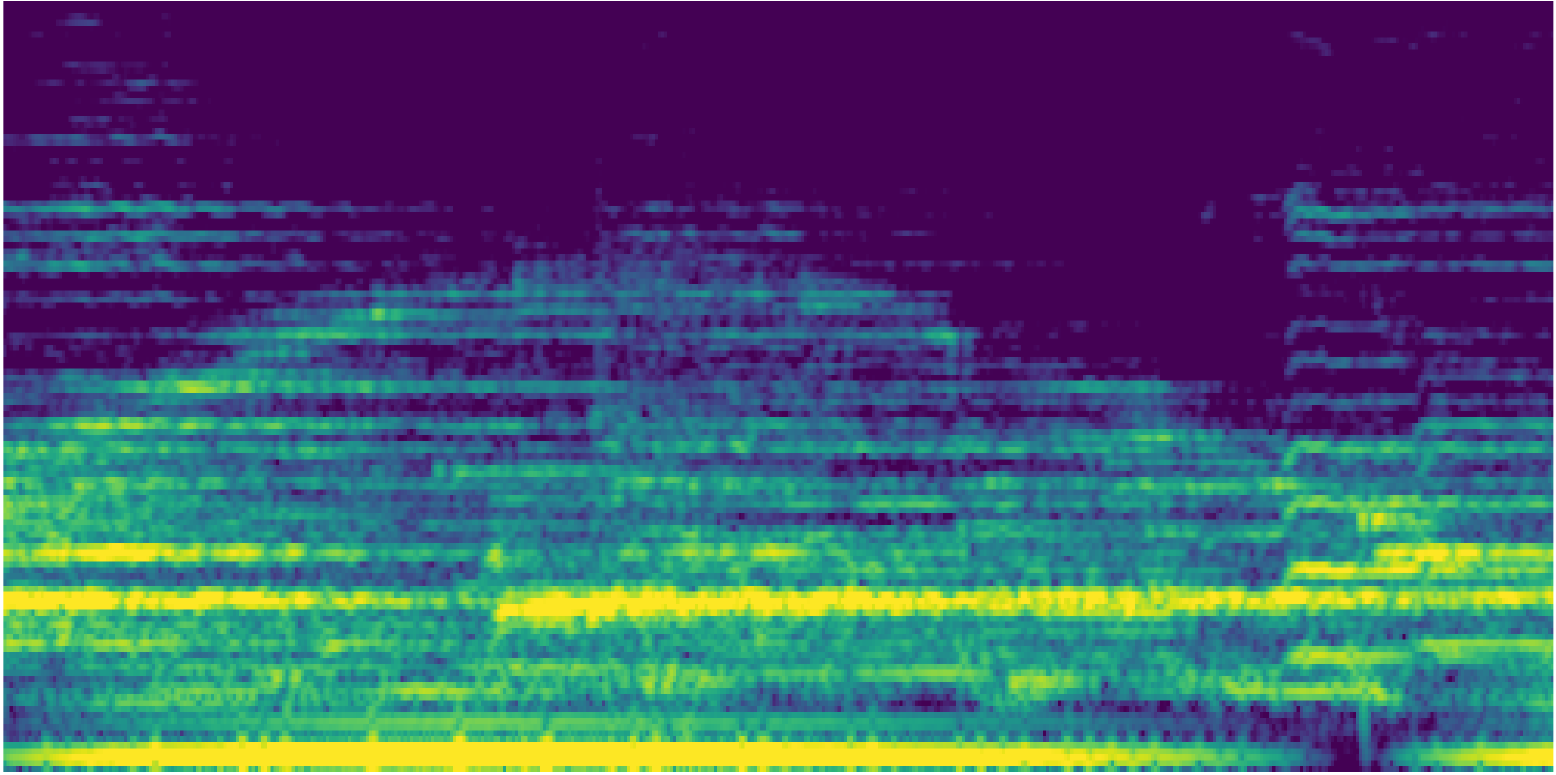}
    \end{subfigure}
    \begin{subfigure}[b]{0.245\linewidth}
        \includegraphics[width=0.98\linewidth,height=0.4\linewidth]{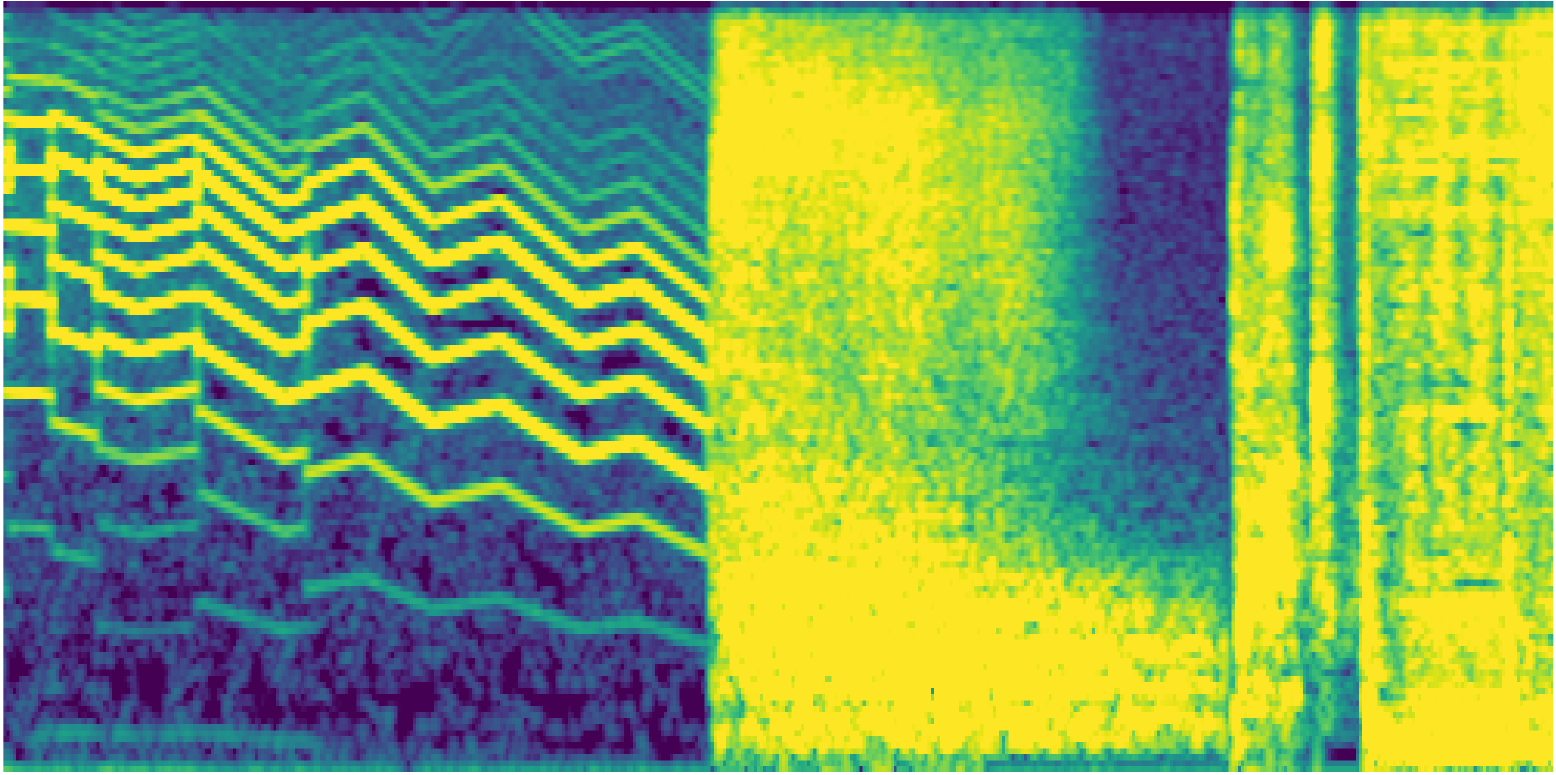}
    \end{subfigure}   
    \\
    \begin{subfigure}[b]{0.245\linewidth}
        \includegraphics[width=0.98\linewidth,height=0.4\linewidth]{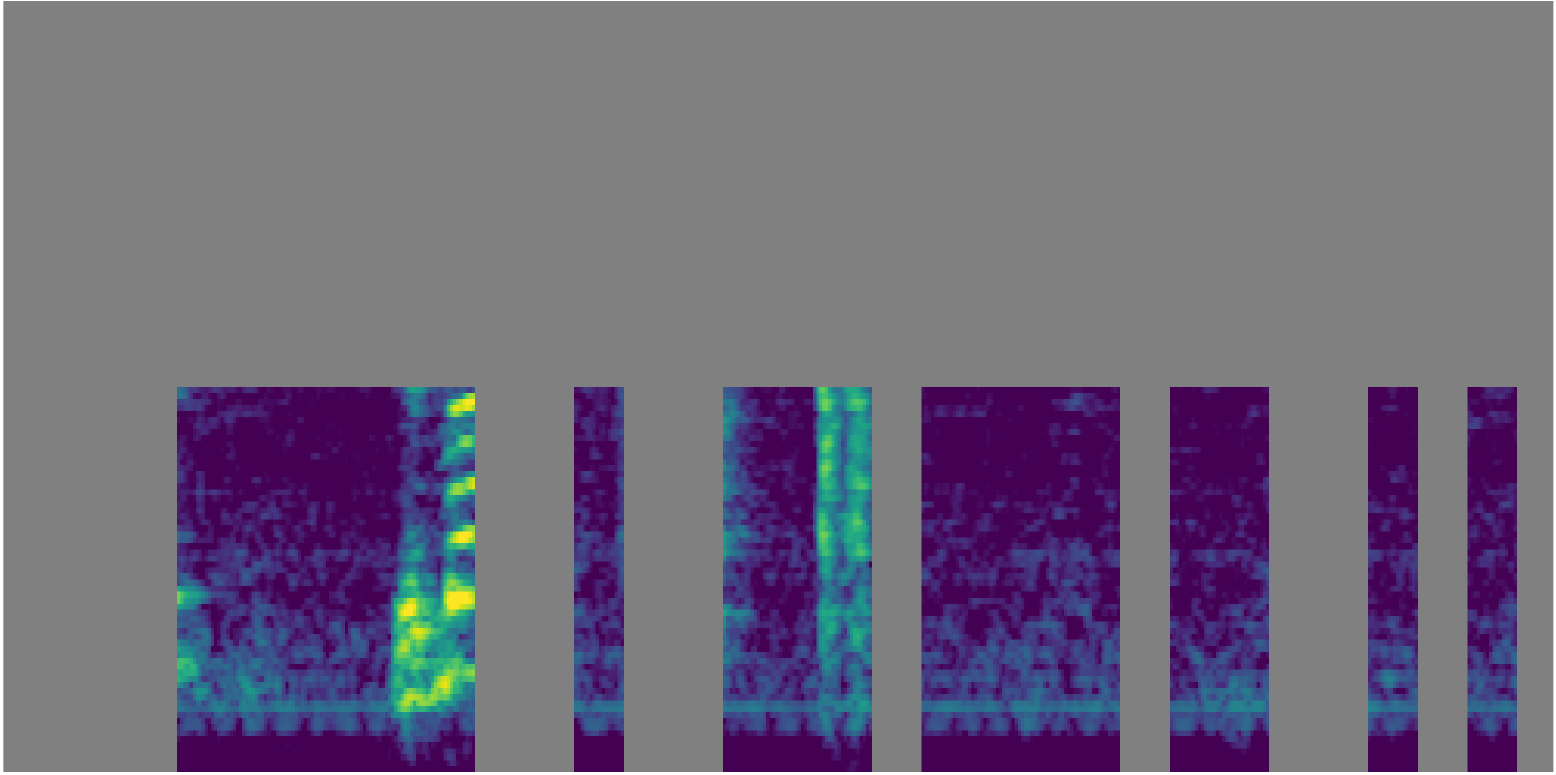}
    \end{subfigure}    
    \begin{subfigure}[b]{0.245\linewidth}
        \includegraphics[width=0.98\linewidth,height=0.4\linewidth]{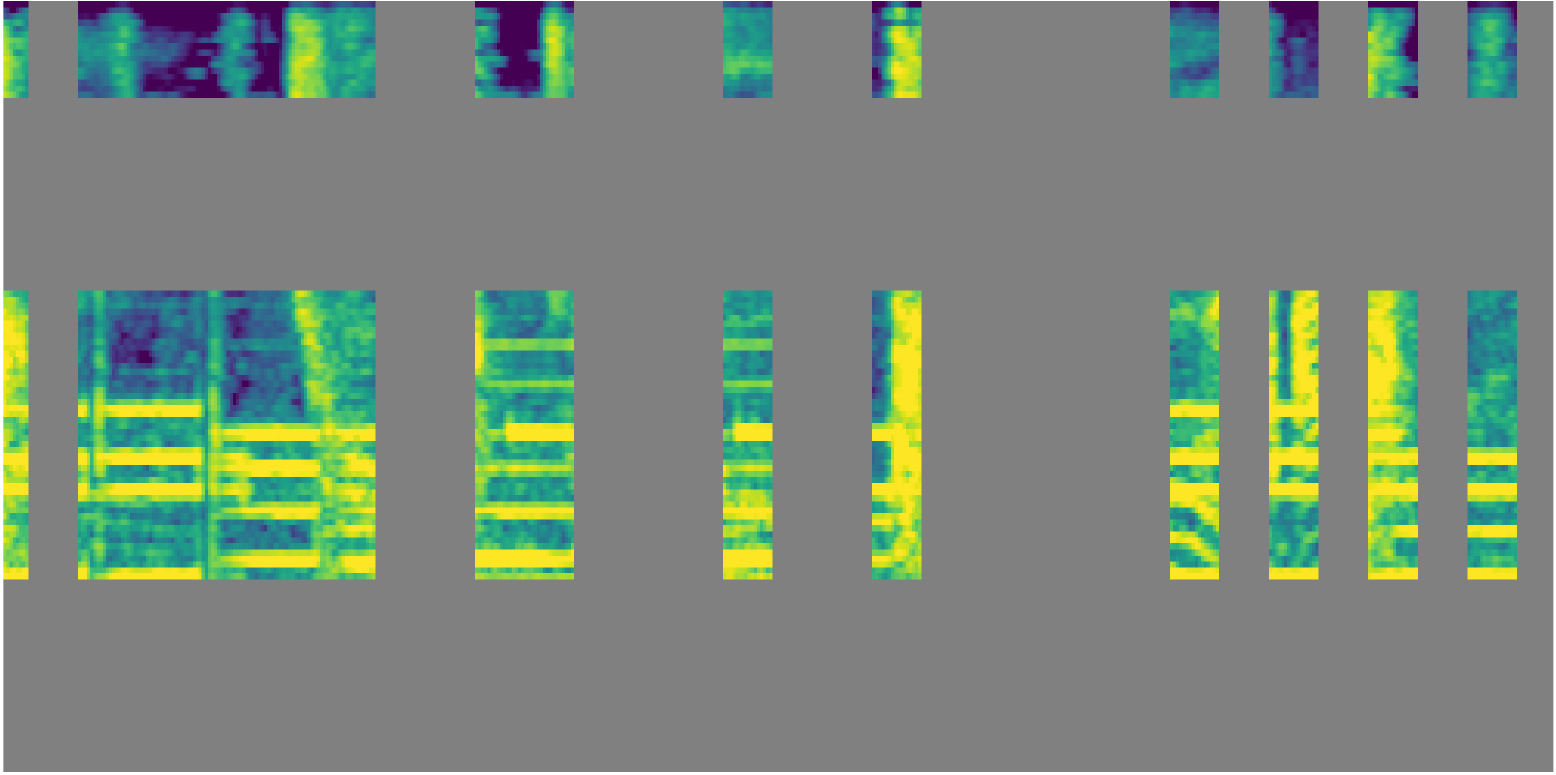}
    \end{subfigure}      
    \begin{subfigure}[b]{0.245\linewidth}
        \includegraphics[width=0.98\linewidth,height=0.4\linewidth]{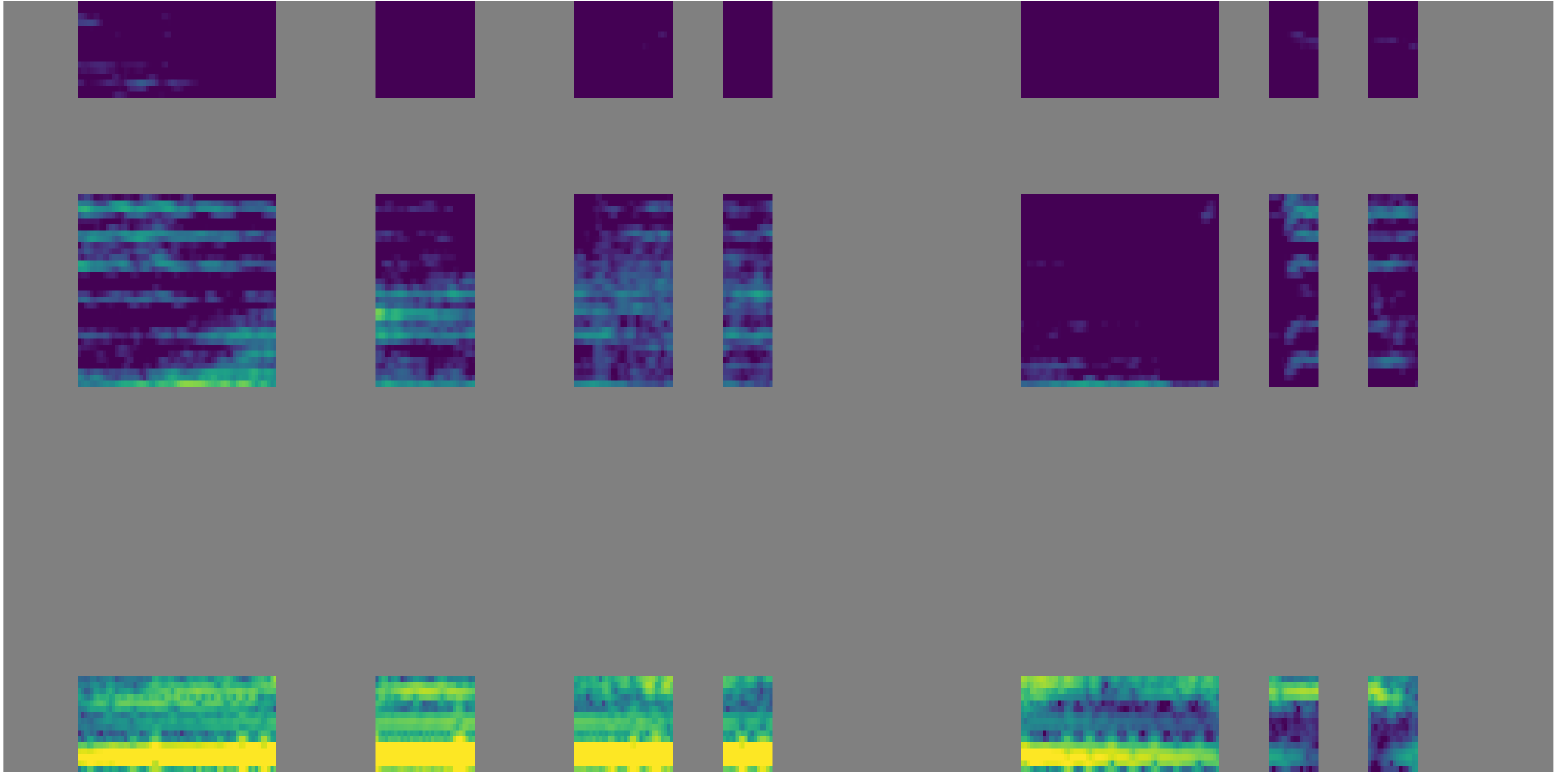}
    \end{subfigure}
    \begin{subfigure}[b]{0.245\linewidth}
        \includegraphics[width=0.98\linewidth,height=0.4\linewidth]{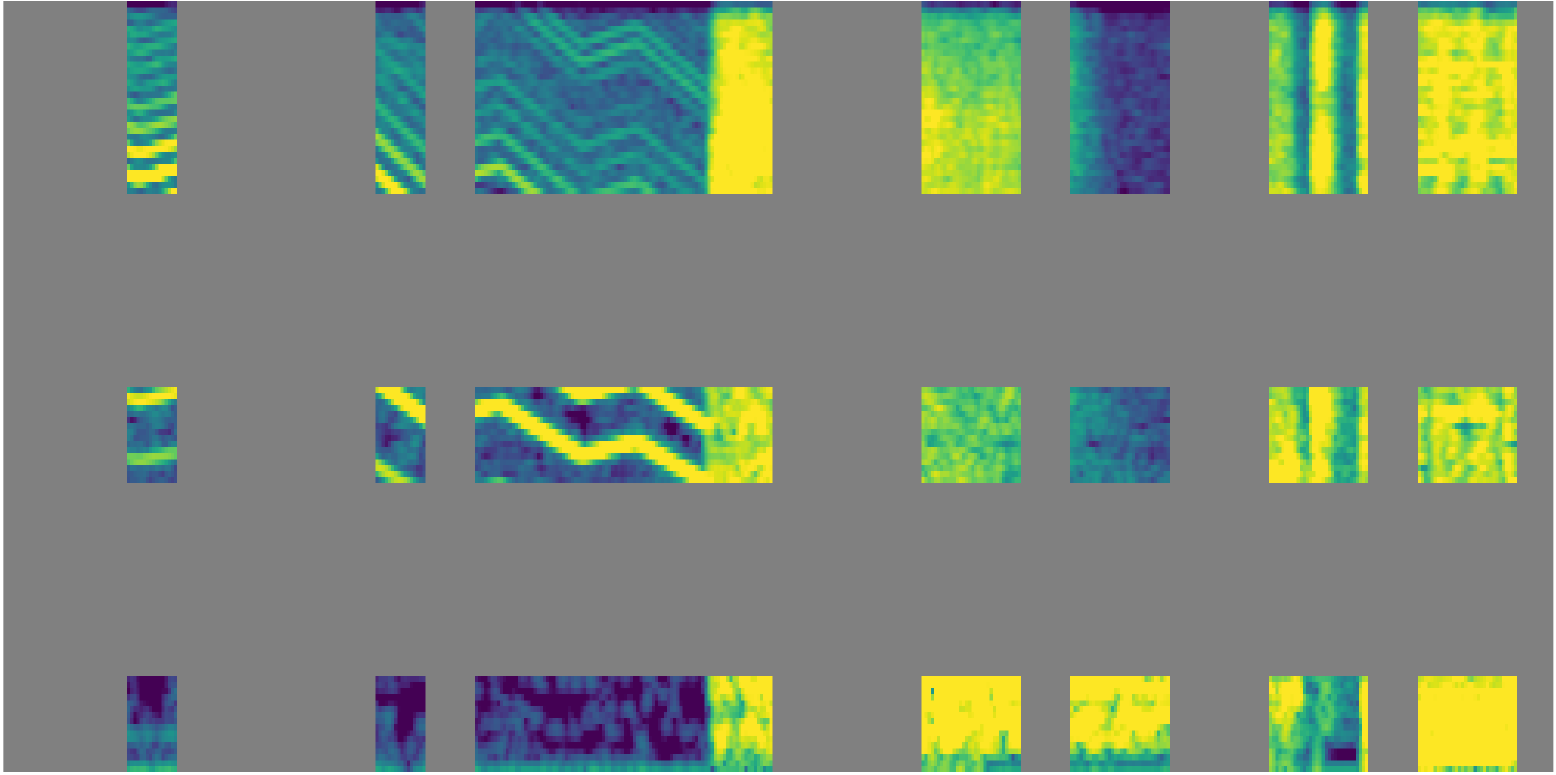}
    \end{subfigure}
    \\
    \begin{subfigure}[b]{0.245\linewidth}
        \includegraphics[width=0.98\linewidth,height=0.4\linewidth]{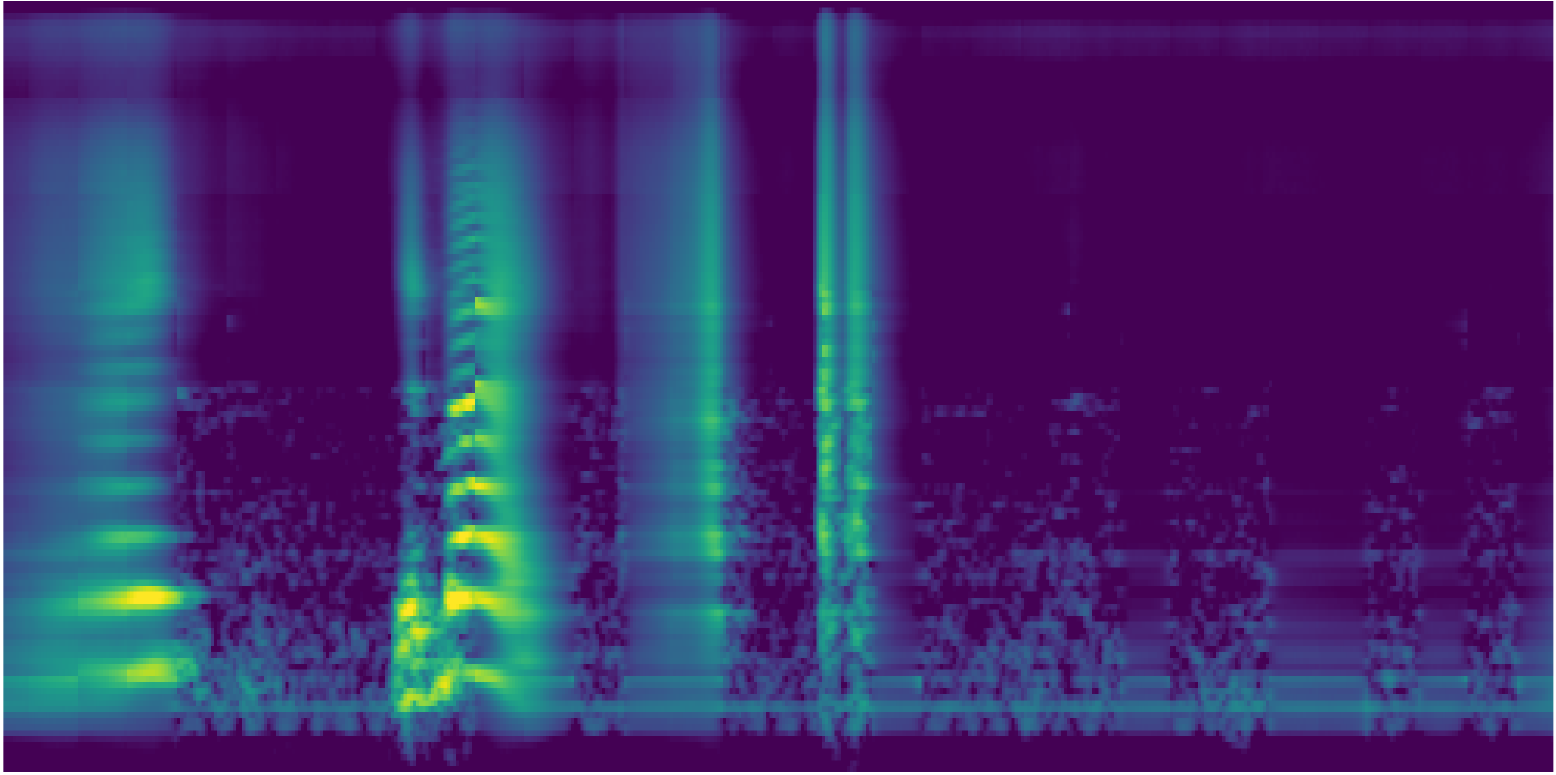}
        %\vspace{-0.2em}
        \subcaption{
            70\% Structured
            \href{https://www.dropbox.com/s/gph4eolsa1y6t1j/uh6GGA3qGOo_0.7_2d_org.mp4?dl=0}{1}
            \href{https://www.dropbox.com/s/0yvif3wpvnr8my3/uh6GGA3qGOo_0.7_2d_masked.mp4?dl=0}{2}
            \href{https://www.dropbox.com/s/m5y3fkuvqdk5bly/uh6GGA3qGOo_0.7_2d_restored.mp4?dl=0}{3}
        }
        
        \label{fig:app:vis:m}
    \end{subfigure} 
    \begin{subfigure}[b]{0.245\linewidth}
        \includegraphics[width=0.98\linewidth,height=0.4\linewidth]{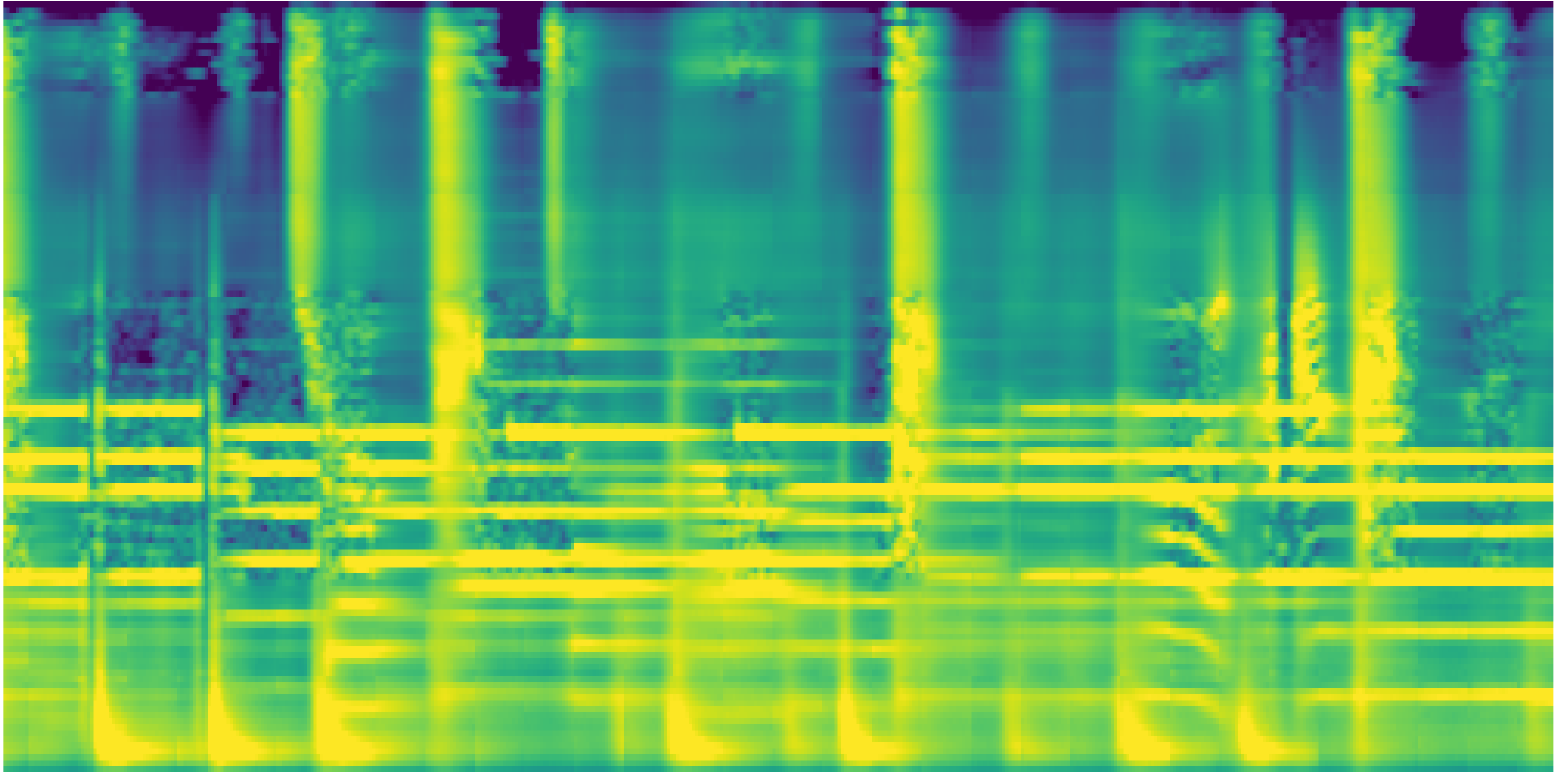}
        %\vspace{-0.2em}
        \subcaption{
            70\% Structured
            \href{https://www.dropbox.com/s/cgpwzcp30k4qffy/KiFQFxJphjI_0.7_2d_org.mp4?dl=0}{1}
            \href{https://www.dropbox.com/s/q6k9wtd1k53iq8i/KiFQFxJphjI_0.7_2d_masked.mp4?dl=0}{2}
            \href{https://www.dropbox.com/s/jw3nwj2es2265dr/KiFQFxJphjI_0.7_2d_restored.mp4?dl=0}{3}
        }
        
        \label{fig:app:vis:n}
    \end{subfigure}    
    \begin{subfigure}[b]{0.245\linewidth}
        \includegraphics[width=0.98\linewidth,height=0.4\linewidth]{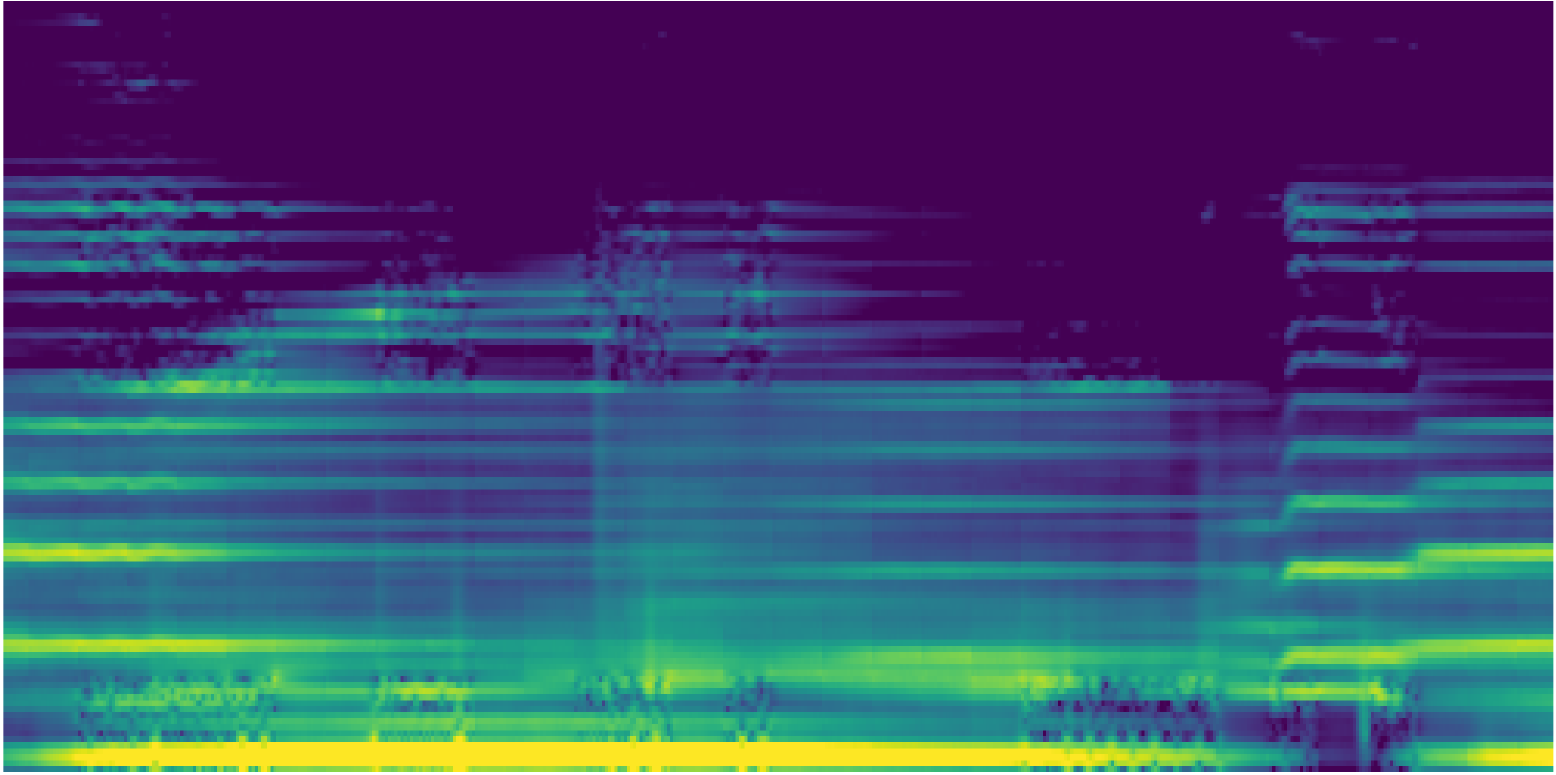}
        %\vspace{-0.2em}
        \subcaption{
            70\% Structured
            \href{https://www.dropbox.com/s/05mzy0zces9fp7l/MipnqUXgpOA_0.7_2d_org.mp4?dl=0}{1}
            \href{https://www.dropbox.com/s/he7jjc46vlux0jt/MipnqUXgpOA_0.7_2d_masked.mp4?dl=0}{2}
            \href{https://www.dropbox.com/s/cvy3dlrl0yts98x/MipnqUXgpOA_0.7_2d_restored.mp4?dl=0}{3}
        }
        
        \label{fig:app:vis:o}
    \end{subfigure}
    \begin{subfigure}[b]{0.245\linewidth}
        \includegraphics[width=0.98\linewidth,height=0.4\linewidth]{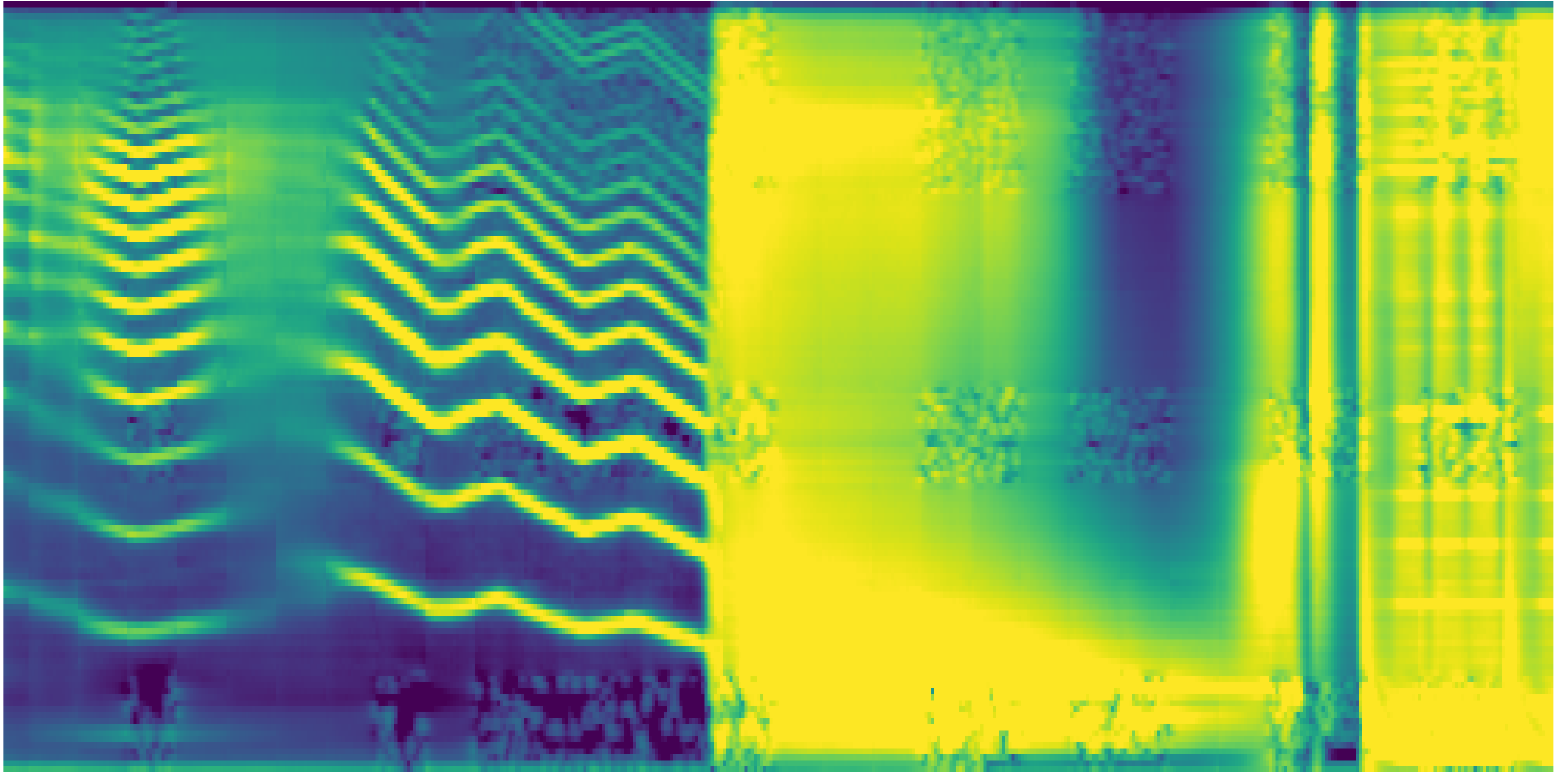}
        \subcaption{
            70\% Structured
            \href{https://www.dropbox.com/s/c7wayy0l3665ef9/bvapjUmC7bY_0.7_2d_org.mp4?dl=0}{1}
            \href{https://www.dropbox.com/s/vnbdltc5y9ltwd5/bvapjUmC7bY_0.7_2d_masked.mp4?dl=0}{2}
            \href{https://www.dropbox.com/s/t8815j9s5xoq2xr/bvapjUmC7bY_0.7_2d_restored.mp4?dl=0}{3}
        }
        \label{fig:app:vis:p}
        
    \end{subfigure}   
    \\
    \vspace{1pt}
    \caption{
    \textbf{Additional spectrogram reconstruction visualizations on the AudioSet \textit{eval} set}. 
    Column-wise type: speech, music, event, others.
    Masking type: (a-h) unstructured (random); (i-p) structured (time$+$frequency).
    Masking ratio: 80\% for (e-h) and the rest are 70\% .
    In each group, we show the original spectrogram ({\color{hrefcolor} 1}, top), masked input ({\color{hrefcolor} 2}, middle), and Audio-MAE output ({\color{hrefcolor} 3}, bottom). 
    The spectrogram size is 1024$\times$128; patch size is 16$\times$16. Each sample has 64$\times$8=512 patches with either 154 (for 70\% masked) or 102 (for 80\% masked) patches being visible to Audio-MAE. 
    Please click on corresponding ({\color{hrefcolor} 1 2 3}) for audible \emph{.wav}s.
    }
    \label{fig:app:vis}
    %\vspace{-1.5em}
\end{figure}
%102 (80\% masked)
%154 (70\% masked)
\clearpage
}

\section{Additional Reconstruction Details and Results by Audio-MAE Decoder}
\label{sec:app:vis}

Fig.~\ref{fig:app:vis} illustrates additional reconstruction results on the AudioSet-2M \textit{eval} set.
Audible examples are under the anonymous links, accessible by clicking on respective {\color{hrefcolor} 1 2 3}. ({\color{hrefcolor} 1} is the ground truth reference, 
{\color{hrefcolor} 2} is the masked input for Audio-MAE, and {\color{hrefcolor} 3} is the reconstruction output by Audio-MAE.) 

We use an Audio-MAE model with a ViT-L encoder and a 16-layer decoder with local attention for visualization.
The model is trained under 80\%  unstructured (random) masking on AudioSet.
We inverse Mel-spectrograms and exploit the Griffin-Lim~\cite{griffin_lim} algorithm to reconstruct waveform.
There could be perceivable artifacts due to imperfect phase estimation in~\cite{griffin_lim}. 
Note that the default masking ratio in Fig.~\ref{fig:app:vis} is 70\% for better visualization.
We also show reconstruction results under 80\% masking ratio in Fig.~\ref{fig:app:vis:e}-\ref{fig:app:vis:h} for comparison.

Comparing {\color{hrefcolor} 2} and {\color{hrefcolor} 3} under the each caption in Fig.~\ref{fig:app:vis}, even with 70\%-80\% masking ratio, Audio-MAE can still create reasonable reconstructions. 
Music and event sound are easier for Audio-MAE due to their relatively predictable spectrogram patterns. 
For example, the repeating tempos across time domain (\eg, the music in Fig.~\ref{fig:app:vis:b} and Fig.~\ref{fig:app:vis:l}) 
and the harmonics across frequency domain (\eg, the siren in Fig.~\ref{fig:app:vis:c} and the trumpeting elephant in Fig.~\ref{fig:app:vis:d}) are very well reconstructed.
Speech recordings are more challenging as shown in Fig.~\ref{fig:app:vis:a} and Fig.~\ref{fig:app:vis:e}.

In most cases, Audio-MAE successfully restores audio from masked/corrupted inputs. With these encouraging results, we envision that Audio-MAE can also be applied to other speech generation tasks and qualitatively case-study an application in \S\ref{sec:app:exp:plc}.

\section{Experimental Details and Hyperparameter Settings}
\label{sec:app:hyper}

In this section we provide additional experimental details.
For audio recordings in each dataset, we pre-process all of them into mono channel under 16K sampling rate for simplicity and consistency between pre-training and fine-tuning tasks. 
Note that their native sampling rate may not be 16K (there are many 8K or higher sampling rate recordings in AudioSet. Also, video compression by YouTube may up-samples or down-samples the audio tracks of user-uploaded videos).
During data loading, we pad or trim the audio length (in seconds) on each dataset as follows: AudioSet: 10, ESC: 5, SPC-1 and SPC-2: 1, SID: 10 seconds. 
%For the input we use mono-channel audio under 16K sampling rate. 
We use a window of 25 ms with a hop length of 10 ms to transform waveform into 128 mel-bank features. 
The resulting input shapes are: AudioSet: $1\times1024\times128$, ESC: $1\times512\times128$, SPC: $1\times128\times128$, SID: $1\times1024\times128$. 
With different input shapes and audio types, we adjust the hyperparameters and data augmentation for each task respectively.
We summarize the pre-training (AS-2M PT) and fine-tuning details on each dataset in Table~\ref{tab:app:train}.

\begin{table*}[t]\centering%
    %\small
    \tablestyle{2pt}{1.1}
    \setlength\tabcolsep{3.0pt}
    \begin{tabular}{l|c|cccccc}
        & pre-training & \multicolumn{6}{c}{fine-tuning} \\
        Configuration & AS-2M PT  & AS-2M  & AS-20K  & ESC~\cite{piczak2015dataset}  & SPC-2~\cite{speechcommandsv2}  & SPC-1 & SID~\cite{Nagrani2020VoxcelebLS} \\
        \toprule
        Optimizer & \multicolumn{7}{c}{AdamW~\cite{adamw}}\\
        Optimizer momentum & \multicolumn{7}{c}{$\beta_1=0.9$, $\beta_2=0.95$}\\
        Weight decay & \multicolumn{7}{c}{0.0001} \\
        Base learning rate & 0.0002 & 0.0002\textsuperscript{$\dagger$} & 0.001& 0.001 & 0.001 & 0.001 & 0.001\\
        Learning rate schedule & \multicolumn{7}{c}{half-cycle cosine decay~\cite{sgdr}}\\
        Minimum learning rate & \multicolumn{7}{c}{0.000001}\\
        Gradient clipping & \multicolumn{7}{c}{None}\\
        Warm-up epochs & 3 & 20 & 4  & 4 &  4 & 1 & 4\\
        Epochs & 32 & 100 & 60 & 60 & 60 & 10 & 60 \\
        Batch size & 512 & 512 & 32 & 64 & 256 & 256 & 64\\
        GPUs & 64 & 64 & 4 & 4 & 4 & 4 & 4\\
        Weighted sampling  & False & True & False & False & False & False\textsuperscript{*} & False\\
        Weighted sampling  size & - & 200,000 & - & -& - & - & -\\
        %Augmentation & Rolling & Rolling & Rolling & Rolling & Rolling & Rolling & Rolling \\
        Augmentation & R & R & R & R & R+N & R+N & R+N \\
        SpecAug~\cite{Park2019SpecAugmentAS} (time/frequency) & - & 192/48 & 192/48 &  96/24 & 48/48 & 48/48 & 192/48 \\
        Drop path~\cite{droppath} & 0.0 & 0.1 & 0.1 & 0.1 & 0.1 & 0.1 & 0.1 \\
        Dropout~\cite{dropout} & 0.0 & 0.0 & 0.0 & 0.0 & 0.0 & 0.0 & 0.0 \\
        Mixup~\cite{mixup} & 0.0 & 0.5 & 0.5 & 0.0 & 0.5 & 0.5 & 0.0 \\
        Multilabel &n/a & True& True& False & False & False & False\\
        Loss Function & MSE & BCE & BCE & CE & BCE & BCE & CE \\
        Dataset Mean for Normalization & -4.268 & -4.268 & -4.268 & -6.627 & -6.846 & -6.702 & -6.370 \\
        Dataset Std for Normalization & 4.569 & 4.569 & 4.569 & 5.359 & 5.565 & 5.448 & 3.074 \\
    \end{tabular}
    \caption{\textbf{Pre-training (PT) and Fine-tuning (FT) hyperparameters}. For augmentation, R: sampling random starting points with cyclic rolling in time; N: adding random noise (signal-to-noise ratio (SNR): 20dB) to spectrograms. For loss functions, BCE: binary cross entropy loss (for multi-label datasets or when using mixup~\cite{mixup}); CE: cross-entropy loss, MSE: mean square error loss.
    \textsuperscript{*}: We repeat and balance each class to 50\% of the size of the unknown class.    
    \textsuperscript{$\dagger$}: For ViT-S, We use a learning rate of 0.0005 on AS-2M FT and 0.002 on AS-20K FT as we find larger learning rates work better for ViT-S encoder.
    \label{tab:app:train}}
    \vspace{-1em}
\end{table*}

We adopt most of the default hyper-parameters used in MAE~\cite{mae}.
Note that the effective learning rate ($lr_{\text{eff}}$) depends on the base learning rate ($lr_{\text{base}}$) and the batch size. Precisely, $lr_{\text{eff}}=lr_{\text{base}}*\frac{\text{batch size}}{256}$.
When the dataset is multi-label or the mixup~\cite{mixup} augmentation is enabled, we use binary cross-entropy loss (BCE) as the fine-tuning objective without label smoothing~\cite{smoothing}.% otherwise cross-entropy (CE).
We also experimented using strong data augmentations (\eg, mixup~\cite{mixup}, SpecAug~\cite{mixup}, and CutMix~\cite{cutmix}) for pre-training but found the resulting performance similar or worse (especially for CutMix which resulted in $\app$0.5 mAP degrade in AudioSet-2M). Therefore we discard these strong data augmentations in the pre-training phase by default.

To perform importance sampling when fine-tuning on the unbalanced AudioSet-2M, 
following prior works, we apply a weighted sampler. 
We set the probability of sampling a sample proportional to the inverse frequency of its labels, where the label frequency is estimated over the training set. Specifically, for a instance $i$ in a dataset $\mathbf{D}$ with a label pool $\mathbf{C}$, its sampling weight is proportional to $\sum_{c_i \in \mathbf{C}}{w_c}$, where $w_c=\frac{1000}{\sum_{i\in\mathbf{D}}{c_i}+\epsilon}$ and $\epsilon=0.01$ is set to avoid underflow in majority classes as in~\cite{gong2021ast}.
In each fine-tuning epoch on AS-2M, we sample 200K instances ($\app$10\% of AudioSet-2M) without replacement in avoidance of duplicated samples in a batch and repeating samples within an epoch.
We fine-tune for 100 epochs, which aggregate to $\app$10 full epochs of AudioSet-2M.
Proper normalization for audio is important to avoid pre-training fine-tuning discrepancy. We use the training split of each end task to estimate dataset-wise mean and standard deviation 
The code, scripts, and pre-trained models for reproducibility are at \url{https://github.com/facebookresearch/AudioMAE}.

\section{Additional Experiments}
\label{sec:app:exp}

In this section, we extend our experimental investigation of Audio-MAE to include additional results that are not covered in the main paper.
First (\S\ref{sec:app:exp:esc}), on ESC-50, we report and compare model performance under an additional round of supervised pre-training on labeled AudioSet-2M (models marked with $\dagger$ in Table 2 of the main paper).
Second (\S\ref{sec:app:exp:plc}), we include additional qualitative results on packet loss concealment (PLC) 
%and acoustic bandwidth expansion (BWE) 
as a preliminary case study on practically useful downstream tasks for the \textit{decoder} in Audio-MAE, and demonstrate its potential impact for generative applications.
Third (\S\ref{sec:app:exp:negative}), we share some negative results when we tried incorporating contrastive objectives for Audio-MAE. Our findings suggest that using reconstruction objective alone is sufficient.

\subsection{ESC-50 with AudioSet-2M Supervised Pre-training}
\label{sec:app:exp:esc}

\paragraph{ESC-50} is designed for environmental sound classification. Besides the pre-training setup introduced in the original paper, we further study a widely compared setup where the models are additionally supervisedly pre-trained with AudioSet data and labels before fine-tuning on ESC-50. Table~\ref{tab:supp:ft_esc} summarizes the results under this setup where our Audio-MAE achieves state of the art accuracy with the additional AudioSet-2M supervised pre-training. Note that our model is still audio-only and uses \emph{no} ImageNet data (IN-SL).

\begin{table}[h!]
\setlength\tabcolsep{3.0pt}
\small
\begin{center}
\begin{tabular}{llll} 
\multicolumn{1}{l}{Model} & \multicolumn{1}{l}{Backbone} & Pre-training & ESC-50 FT\\
\midrule
ERANN~\cite{verbitskiy2021eranns} & CNN & AS-SL & 96.1\\
PANN~\cite{kong2019panns} & CNN & AS-SL & 94.7  \\
\color{gray}AST~\cite{gong2021ast} & \color{gray}DeiT-B & \color{gray}IN-SL, AS-SL & \color{gray}95.6 \\ 
\color{gray}HTS-AT~\cite{chen2022hts} & \color{gray}Swin-B & \color{gray}IN-SL, AS-SL & \color{gray}97.0 \\
\color{gray}PASST~\cite{paast} & \color{gray}DeiT-B & \color{gray}IN-SL, AS-SL &  \color{gray}96.8\\
\rowcolor{gray!20} \textbf{Audio-MAE} (global)  & ViT-B & AS-SSL, AS-SL & 96.9 \\
\rowcolor{gray!30} \textbf{Audio-MAE} (local) & ViT-B & AS-SSL, AS-SL & \textbf{97.4}\\
\end{tabular}
\vspace{5pt}
\caption{\textbf{Comparison with other state-of-the-art models on ESC-50} with an additional round of supervised pre-training on AudioSet (AS-SL). SSL: self-supervised learning. We {\color{gray}gray-out} the models with out-of-domain pre-training on ImageNet (IN).}
\label{tab:supp:ft_esc}
\end{center}
\end{table}

%\subsection{AS per-class analysis}

\subsection{Qualitative Results for a practical generation task}
\label{sec:app:exp:plc}

\paragraph{Packet Loss Concealment} (PLC) is a widely deployed technique to alleviate side effects from missing or corrupted packets in Voice over IP (VoIP) applications (\eg, video conferencing, Bluetooth earbuds, wireless virtual reality headset, \etc) When an encoded speech is sent as a sequence of VoIP packets over a network, these packets may get lost or be corrupted during the transmission, resulting in undesirable low quality speech. To this end, various PLC techniques has been developed. The recent approaches substitute the corrupted waveform segments by either replacing the corrupted waveform segments with other intact segments base on the acoustic pitch detected, or via inpainting with RNN-based~\cite{LeeC16}, CNN-based~\cite{plc_wang}, or autoencoding-based~\cite{audio_inpainting,MarafiotiPHM19} reconstruction.

\begin{figure}[h!]
    \centering
    \begin{subfigure}[b]{0.49\linewidth}
        \includegraphics[width=0.49\linewidth,height=0.25\linewidth]{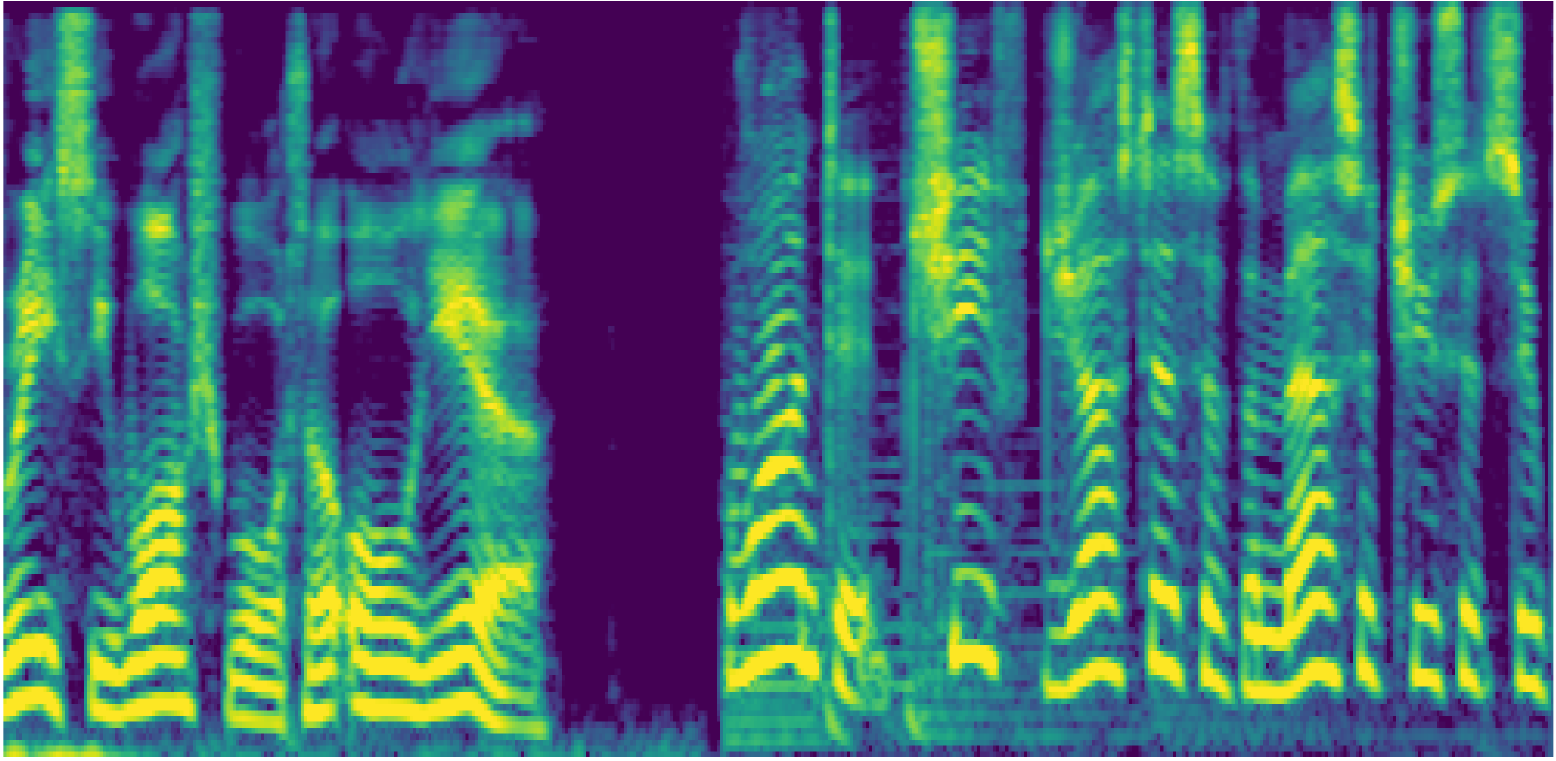}
        \hfill
        \includegraphics[width=0.49\linewidth,height=0.25\linewidth]{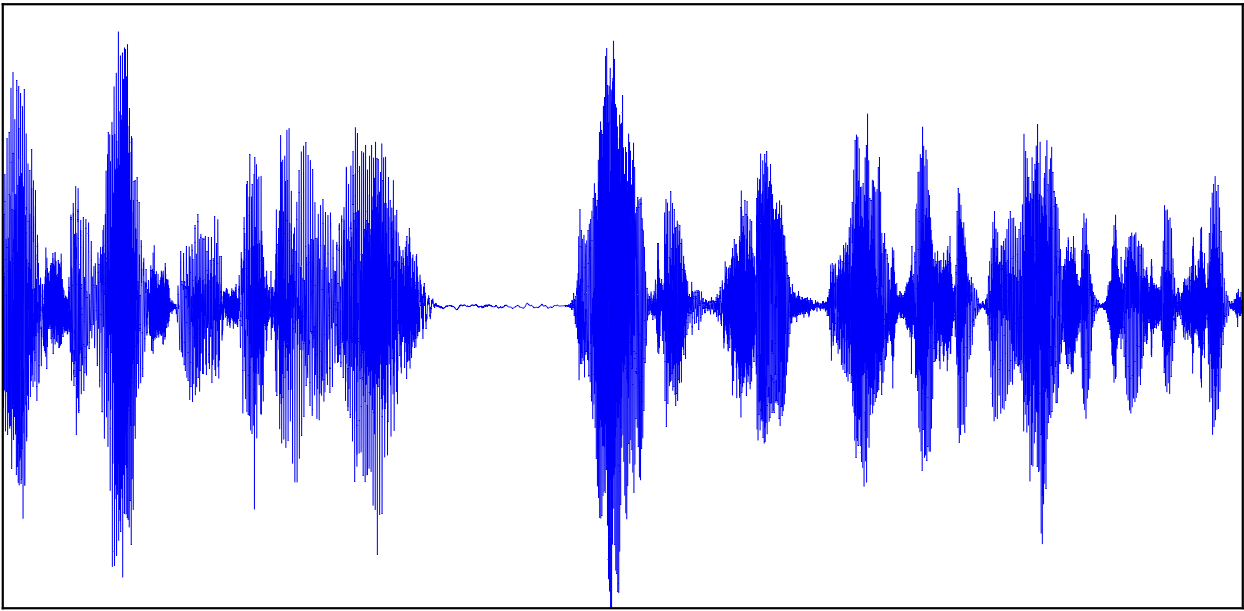}
    \end{subfigure}    
    \begin{subfigure}[b]{0.49\linewidth}
        \includegraphics[width=0.49\linewidth,height=0.25\linewidth]{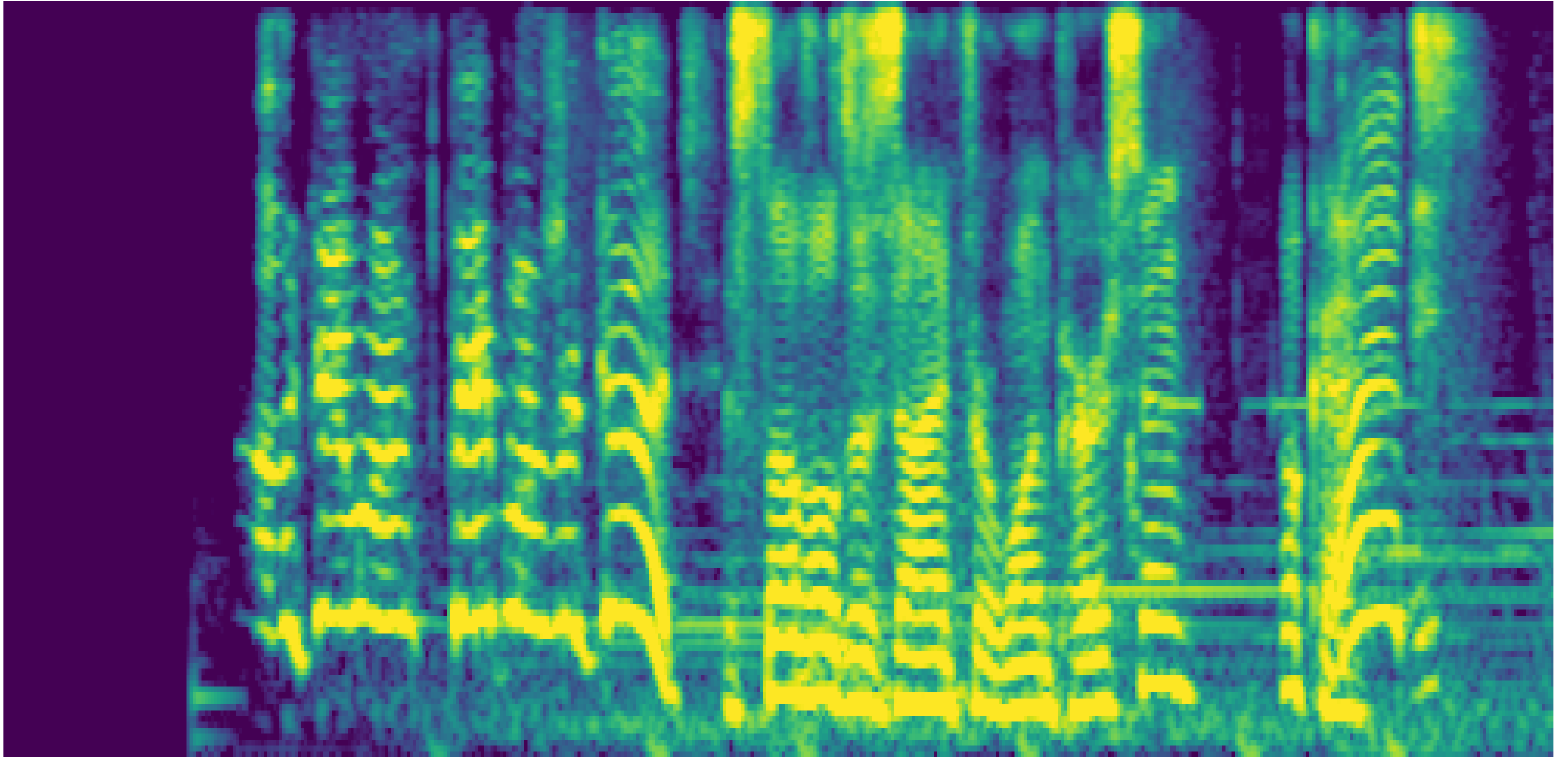}
        \hfill
        \includegraphics[width=0.49\linewidth,height=0.25\linewidth]{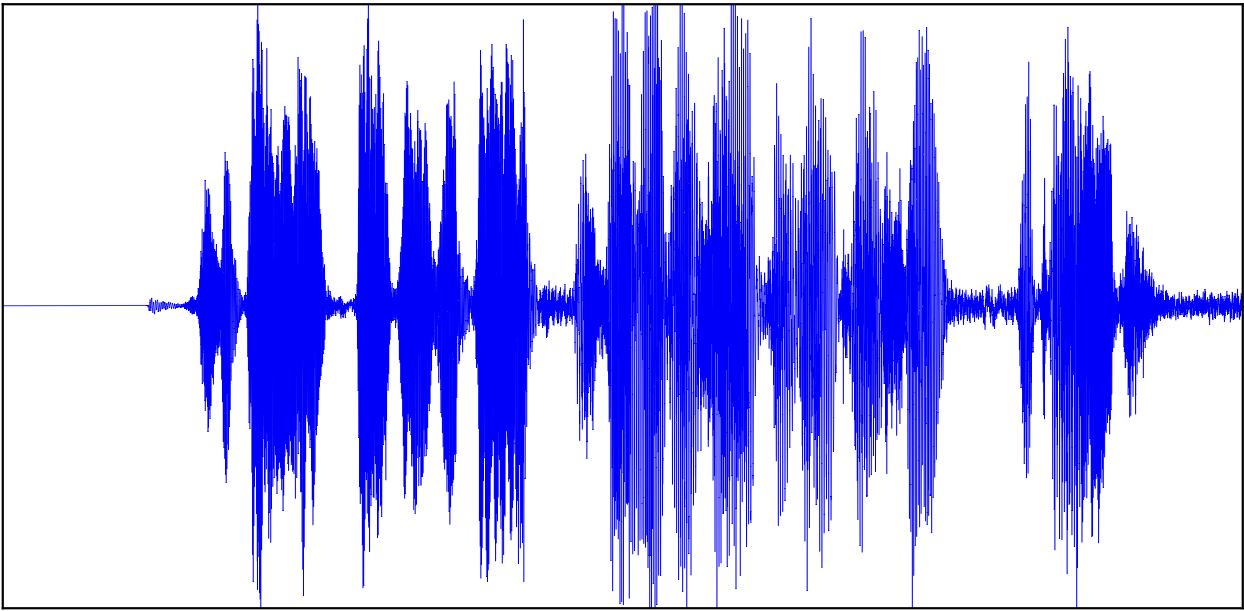}
    \end{subfigure}
    \\
    \begin{subfigure}[b]{0.49\linewidth}
        \includegraphics[width=0.49\linewidth,height=0.25\linewidth]{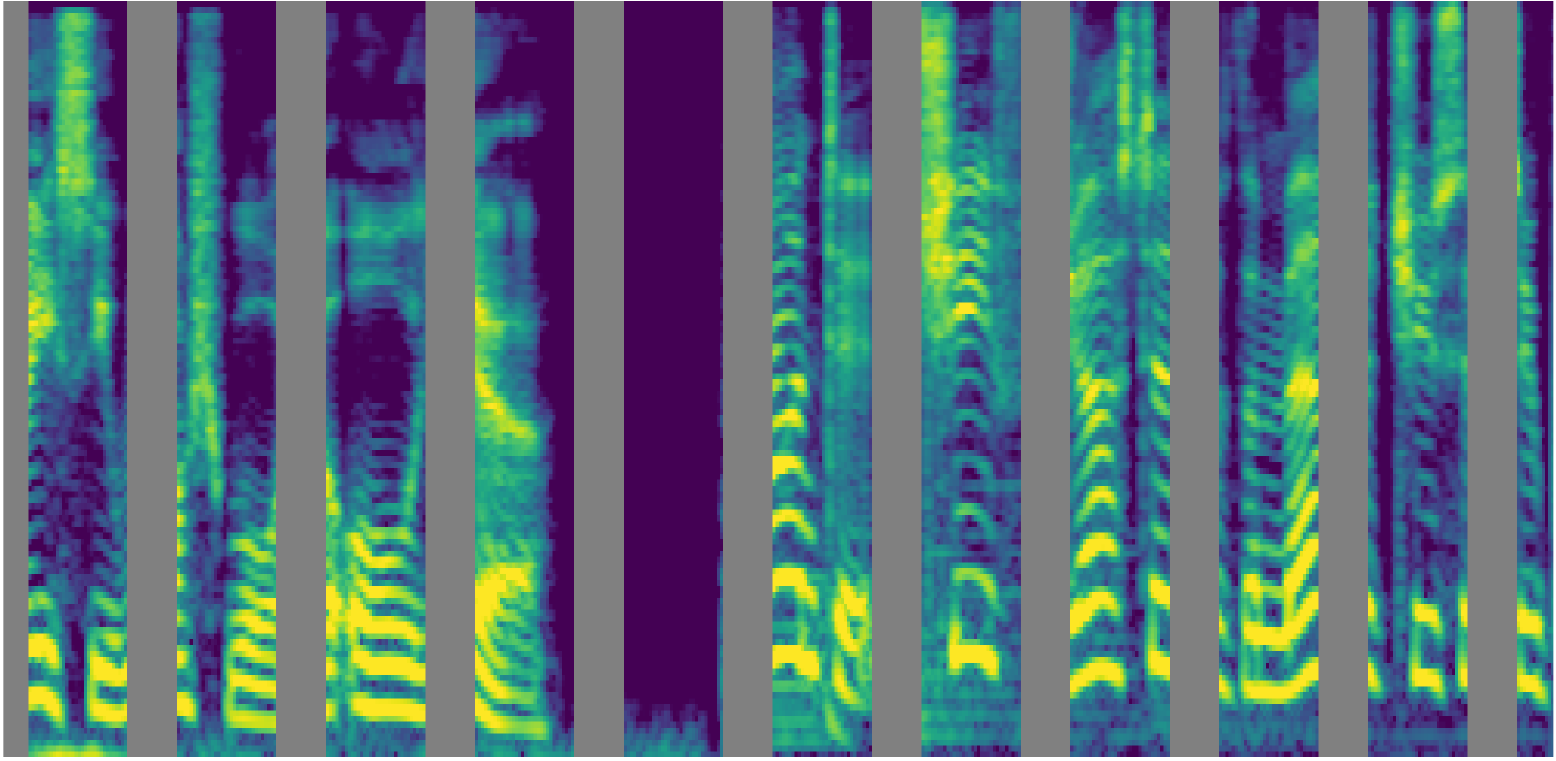}
        \hfill
        \includegraphics[width=0.49\linewidth,height=0.25\linewidth]{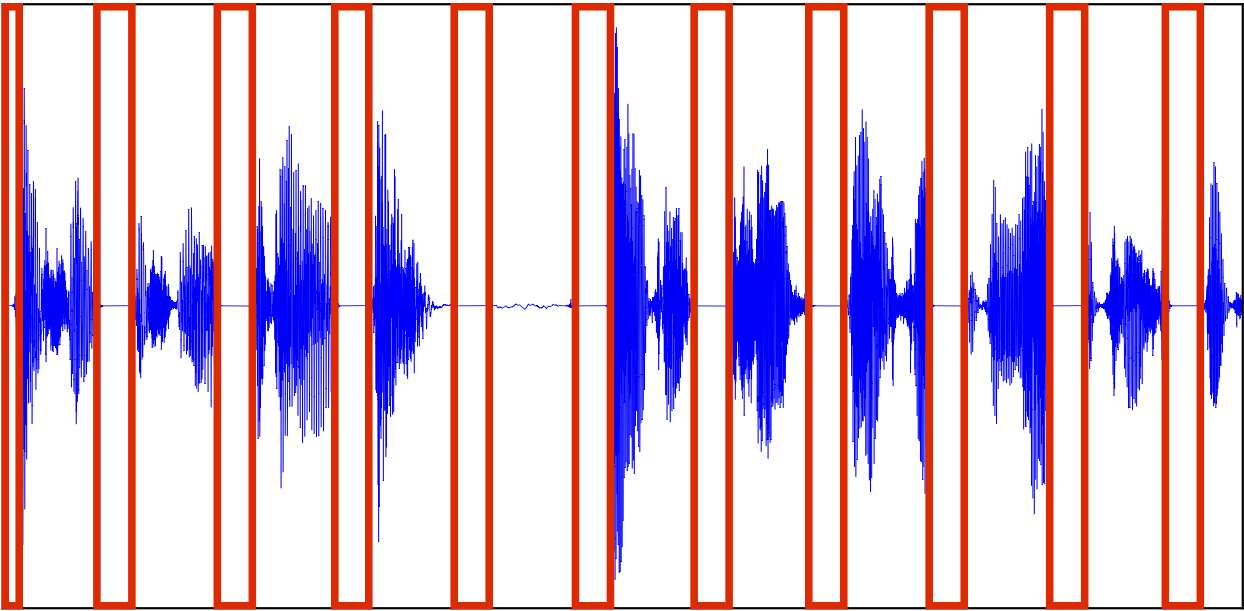}
    \end{subfigure}    
    \begin{subfigure}[b]{0.49\linewidth}
        \includegraphics[width=0.49\linewidth,height=0.25\linewidth]{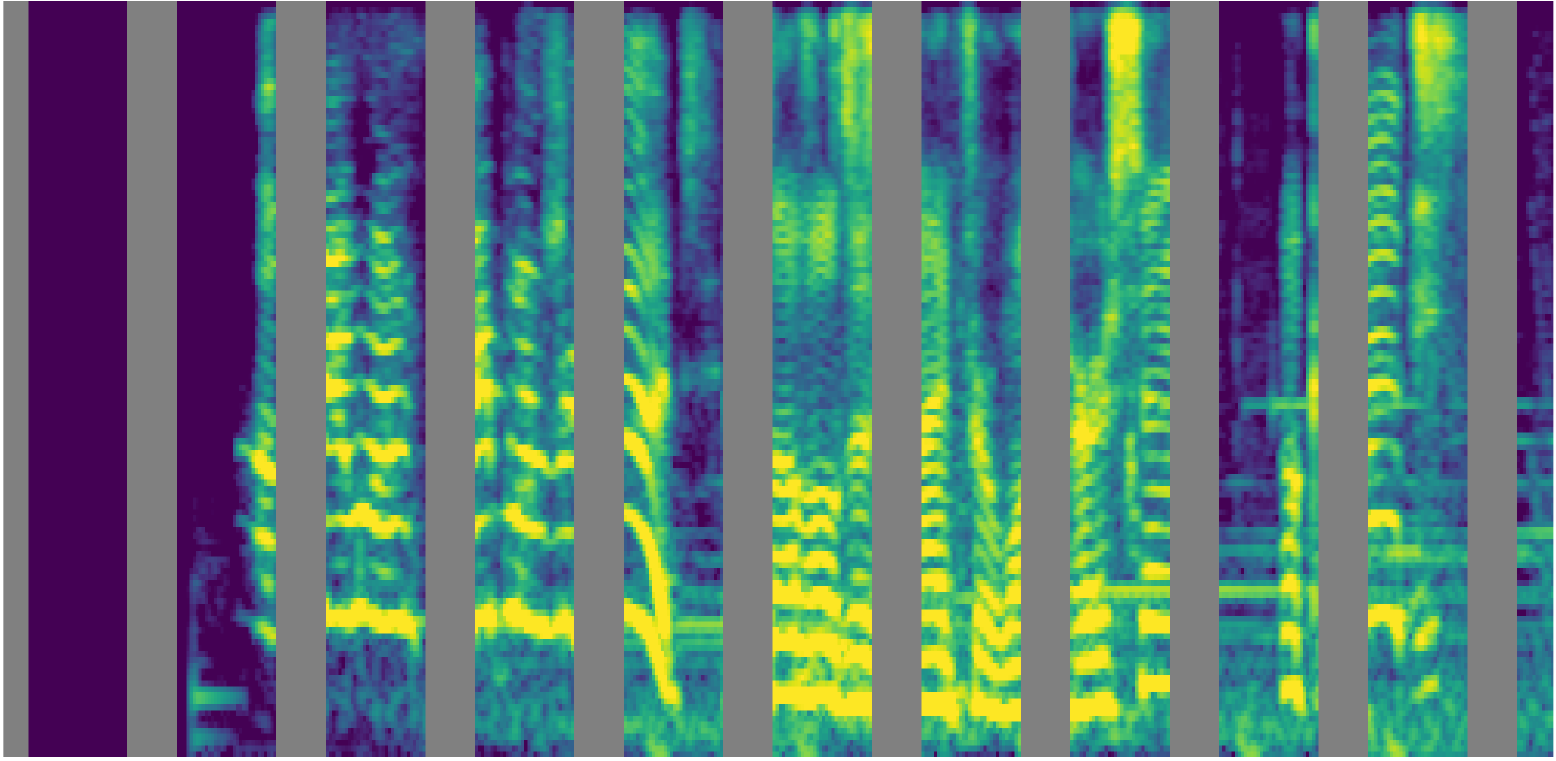}
        \hfill
        \includegraphics[width=0.49\linewidth,height=0.25\linewidth]{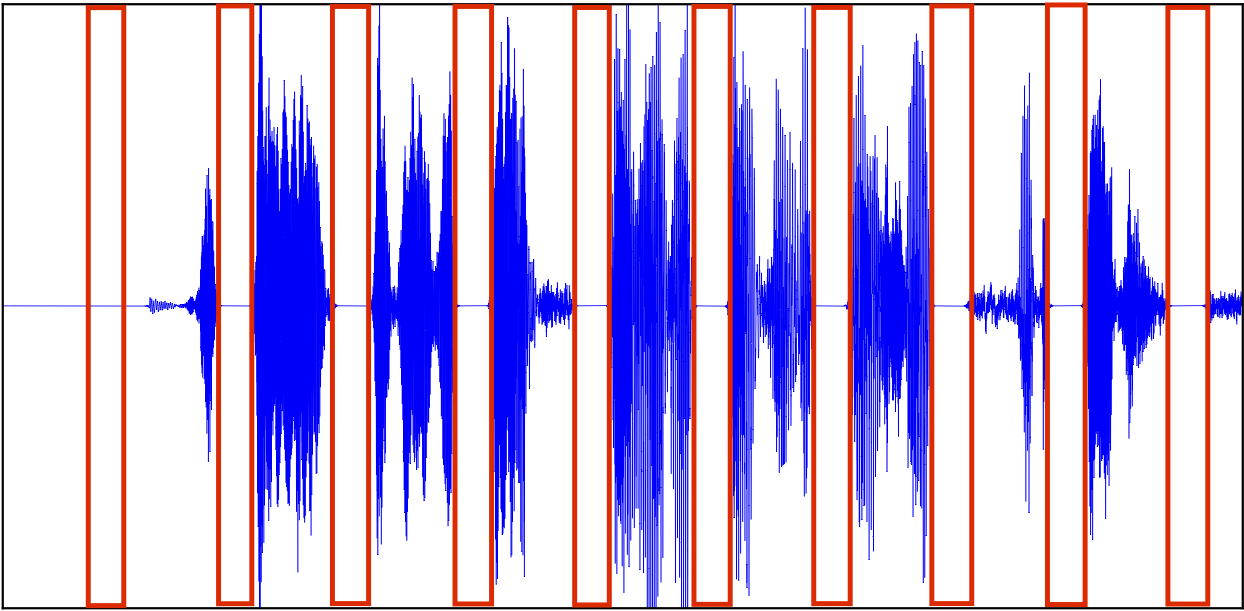}
    \end{subfigure}
    \\
    \begin{subfigure}[b]{0.49\linewidth}
        \includegraphics[width=0.49\linewidth,height=0.25\linewidth]{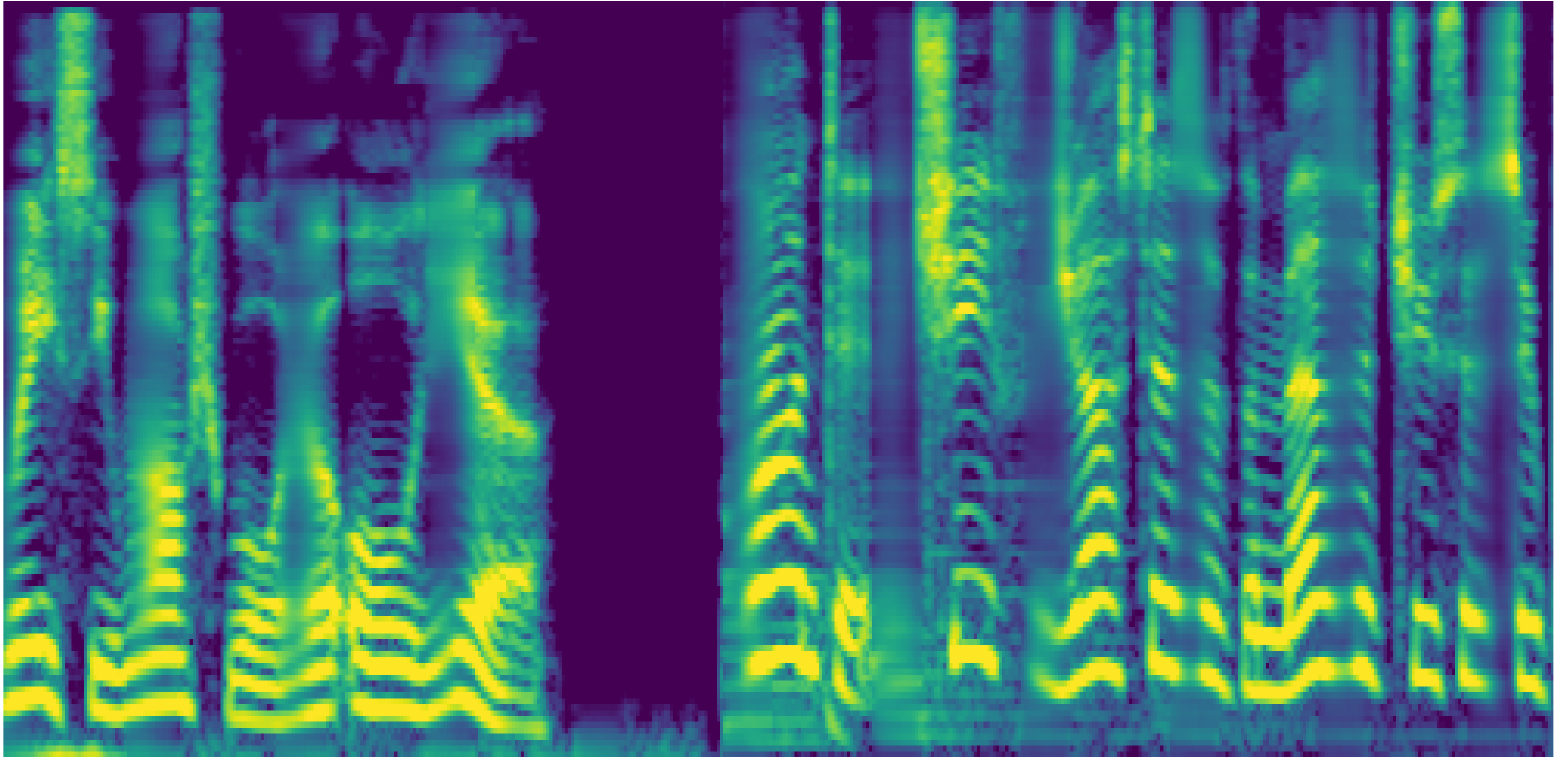}
        \hfill
        \includegraphics[width=0.49\linewidth,height=0.25\linewidth]{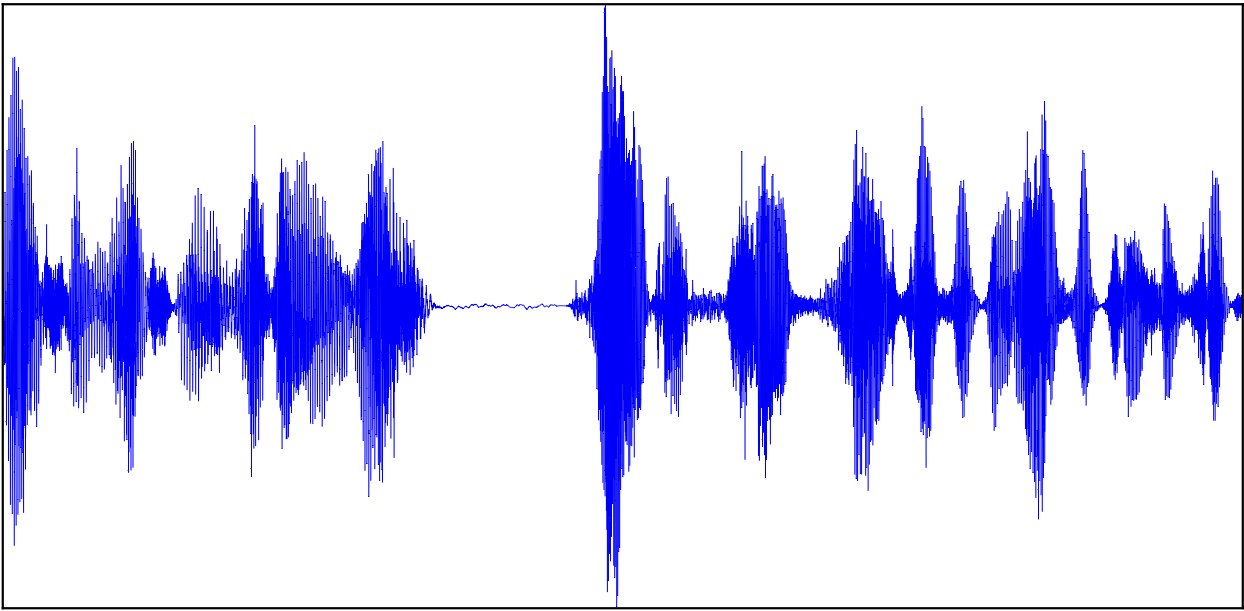}
        %\vspace{-0.2em}
        \subcaption{
            Speech one (Freq./Time)
            \href{https://www.dropbox.com/s/rl96iu6sy7aydox/demo_org.mp4?dl=0}{1}
            \href{https://www.dropbox.com/s/00ouizfitwnlbyi/demo_masked.mp4?dl=0}{2}
            \href{https://www.dropbox.com/s/b05uxgt6sg4bw0h/demo_restored.mp4?dl=0}{3}
        }
        \label{fig:app:plc:a}
    \end{subfigure} 
    \begin{subfigure}[b]{0.49\linewidth}
        \includegraphics[width=0.49\linewidth,height=0.25\linewidth]{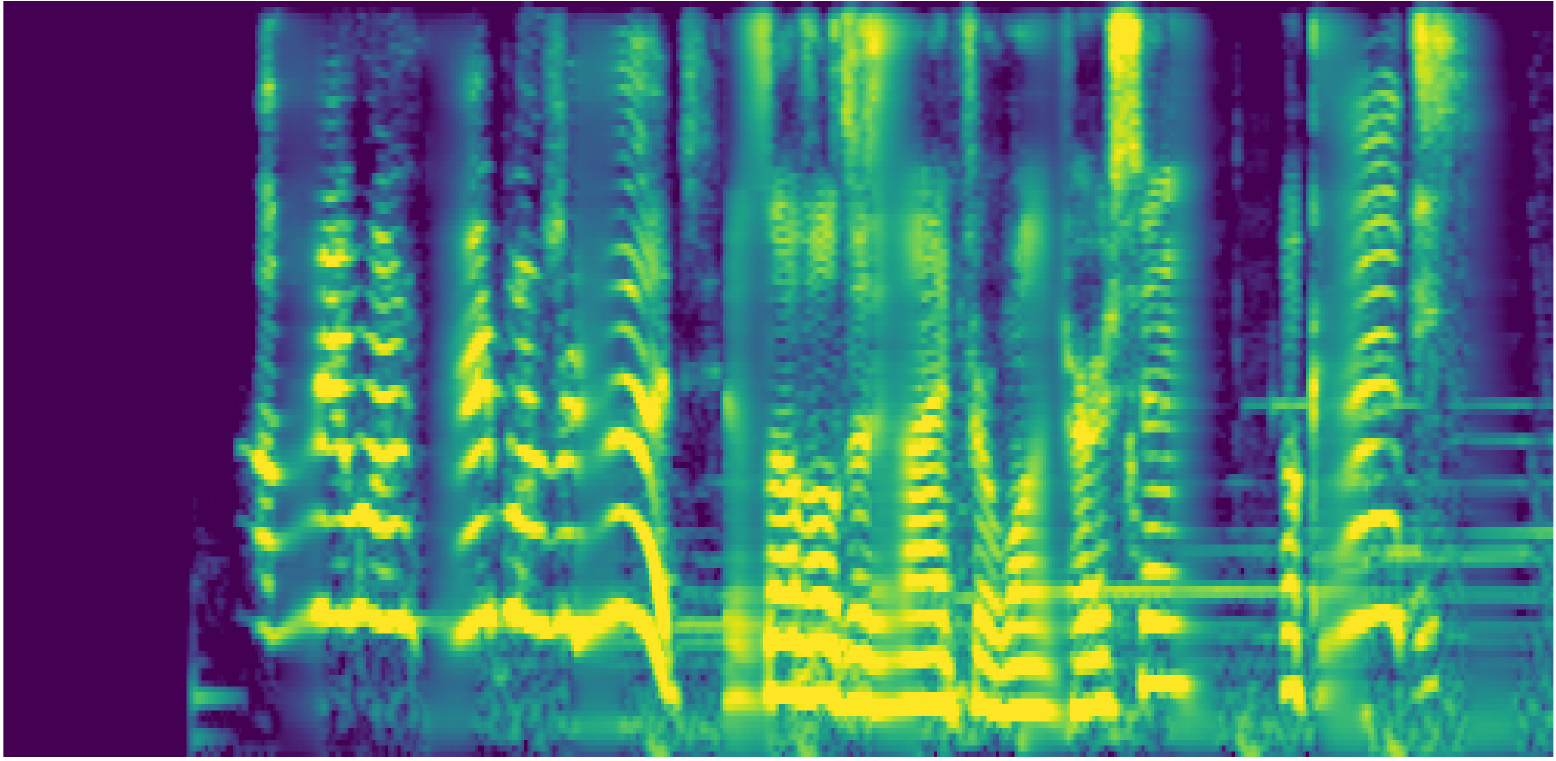}
        \hfill
        \includegraphics[width=0.49\linewidth,height=0.25\linewidth]{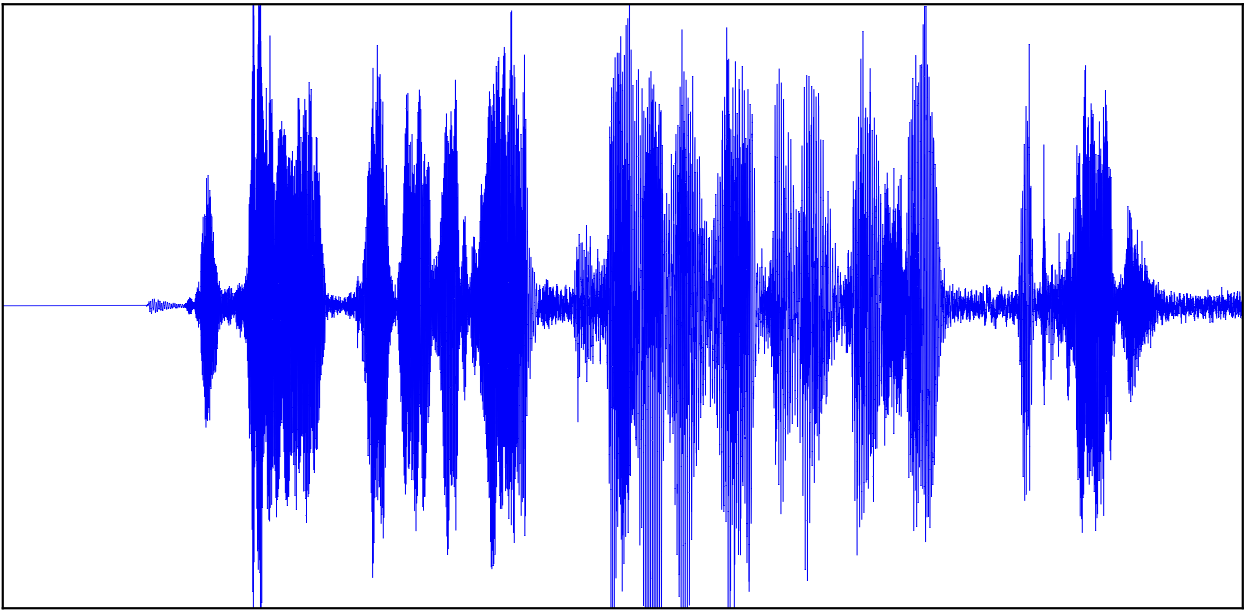}
        %\vspace{-0.2em}
        \subcaption{
            Speech two (Freq./Time)
            \href{https://www.dropbox.com/s/3b7fgzxv71tafg2/1IrYZhVhN1s_org.mp4?dl=0}{1}
            \href{https://www.dropbox.com/s/sv7lt5fzgpxrwcn/1IrYZhVhN1s_masked.mp4?dl=0}{2}
            \href{https://www.dropbox.com/s/jwz8kwgjrdrq308/1IrYZhVhN1s_restored.mp4?dl=0}{3}
        }
        \label{fig:app:plc:c}
    \end{subfigure}
    \caption{
    \textbf{Qualitative Results for Packet Loss Concealment with Audio-MAE Decoder}. 
    Simulations of 25\% packet loss rate in time for two speech recordings.
    In each group, we show the original spectrogram(left) and time(right) sequence ({\color{hrefcolor}1}, top), corrupted input with packet loss ({\color{hrefcolor}2}, middle), and Audio-MAE restoration ({\color{hrefcolor}3}, bottom). 
    The spectrogram size is 1024$\times$128; patch size is 16$\times$16. 
    Please click ({\color{hrefcolor} 1 2 3})
    for audible \emph{.wav}s.
    }
    \label{fig:app:plc}
    
\end{figure}

In this section, we qualitatively demonstrate how Audio-MAE could potentially be applied for PLC to recover corrupted waveform segments with its encoder-decoder architecture. In Fig.~\ref{fig:app:plc}, we simulate two time-corrupted speech recordings by masking speech in time and perform reconstruction with Audio-MAE.
In practice, a PLC system may exploit packet checksums to identify corrupted or missing packets and mask them. The PLC problem then can be viewed as a special case (time-only, structured masking) of Audio-MAE. As shown in both cases, the Audio-MAE decoder produces reasonable speech reconstruction.
We leave the in-depth study and analysis of generative tasks (\eg. PLC and speech bandwidth expansion (BWE)~\cite{LiuTWLB15,Nagrani2020VoxcelebLS}) as the future work.

%\paragraph{Speech Bandwidth Expansion}
%Speech bandwidth extension (BWE)~\cite{LiuTWLB15,DBLP:conf/icassp/LiL15} aims to expand the bandwidth of low-sampling rate speech and improve the speech quality. It is used in many applications, including speech enhancement~\cite{speech_enhancement}, speech synthesis~\cite{synthesis} and speaker identification~\cite{Nagrani2020VoxcelebLS}. In this study, we exploit Audio-MAE to expand narrow-band (8K) speech into wide-band (16K) speech by masking the 8K-16K frequency components. As can be seen and heard in Fig.~\ref{fig:supp:bew}, Audio-MAE decoder can be applied for BWE and generate reasonable speech wide-band speech.

\subsection{Negative Results: Directions that did not work well }
\label{sec:app:exp:negative}
\paragraph{Additional Contrastive Objective}
We examined using additional contrastive objectives in the pre-training phase but do not find them helpful empirically.
Similar to SS-AST~\cite{ssast} and Wave2vec 2.0~\cite{wav2vec2}, we apply InfoNCE~\cite{info_nce} loss over masked tokens of an instance. Specifically, let $\mathbf{x}_i, i=1\dots N$ denotes the values of $i$-th masked spectrogram patch where $N$ is the number of masked patches in an instance. (\eg, rounded $N=102$ under 80\% masking over $64\times8$ spectrogram patches of a 10-second audio recording.)
And let $\mathbf{c}_i$ denotes its corresponding contextualized embedding projected by a separated decoder head. We investigate the following contrastive objective:
%billy: Should we mention MoCo or other approach and offer some intuition why reconstruction is the best pretext task empirically for Audio?

\begin{equation}
    L_c = -\frac{1}{N}\sum_{i=1}^{N}\text{log}\frac{e^{\mathbf{c}^T_i\mathbf{x}_i}}{\sum_{j=1}^N e^{\mathbf{c}^T_i\mathbf{x}_j}}.
\end{equation}

Intuitively, $L_c$ draws closer patches with their contextualized embeddings (positive pairs) at each masked position while contrasting and pushing away mismatched ones (negative pairs) from all masked patches.
For the reconstructive objective, let $\hat{\mathbf{x}_i}, i=1\dots N$ be the reconstruction of $i$-th masked spectrogram patch generated by the reconstruction head of our Audio-MAE decoder. The original reconstruction objective $L_r$ in Audio-MAE is formally defined as:

\begin{equation}
    L_r = \frac{1}{N}\sum_{i=1}^N(\hat{\mathbf{x}}_i-\mathbf{x}_i)^2.
\end{equation}

We consider three setups: (1) Using the reconstructive objective ($L_r$) alone (the default setup); (2) using the contrastive objective ($L_c$) alone; (3) multi-tasking with both the reconstructive and contrastive objectives ($L_r + \alpha L_c$), where $\alpha$ is the hyper-parameter that balances two objectives.

Table~\ref{tab:supp:contrastive} shows the results: We see that the reconstruction objective $L_r$ alone is sufficient and yields the best performance. 
Empirically, we do not observe improvement with contrastive objectives alone or under the multi-task setup (the best $\alpha$ is 0.2 in our experiments). $L_c$ and $L_r$ do not work complementarily in Audio-MAE.

\begin{table}[h!]
\small
\tablestyle{2pt}{1.05}
\setlength\tabcolsep{1.0pt}
\begin{tabular}{ccc}
    Objective & AS-20K & AS-2M  \\
    \toprule
    Reconstruction ($L_r$)&  \textbf{37.1} & \textbf{47.3}  \\
    Contrastive ($L_c$)& 36.4 & 46.6  \\
    Contrastive + Reconstruction ($L_r+\alpha L_c$)& 36.8 & 46.8 \\
\end{tabular}
\vspace{3pt}
\caption{\textbf{Impact of contrastive objective}.\label{tab:supp:contrastive}}
\end{table}

\section{Limitations}
\label{sec:app:limitations}

% data size, length, distribution, model size
We think there are few direct limitations of this work.
The data scale is one of them. 
AudioSet used by Audio-MAE is around two orders of magnitude smaller than the text corpus used in the language~\cite{bert,roberta,gpt}
%or image~\cite{mae,jft}
counterparts.
Another limitation is duration of each sample: the 10-second recordings in AudioSet are short and thus distant temporal dependencies in audio may not be properly learned yet.
Further, as AudioSet is unbalanced and there are many audio types beyond the 527 classes annotated in AudioSet, Audio-MAE could be sub-optimal when transferring to tasks concerning rare or unseen audio events.
Lastly, while Audio-MAE has greatly improved the efficiency of large-scale self-supervised learning, modeling lengthy audio and high-dimensional data with Transformers is computationally demanding.

\end{document}